\documentclass[11pt]{article}
\pdfoutput=1

\usepackage{cite}

\usepackage[nointlimits,reqno]{amsmath}
\usepackage{amssymb}
\usepackage{setspace}
\usepackage{mathtools}
\usepackage{simplewick}
\usepackage{slashed}
\numberwithin{equation}{section} 
\usepackage{mathtools}
\usepackage{graphicx}
\usepackage{dsfont}
\usepackage{mathrsfs}
\usepackage[mathscr]{euscript}
\usepackage{jheppub}
\usepackage{hyperref}
\usepackage[usenames,dvipsnames]{xcolor}
\usepackage{calligra}
\DeclareMathAlphabet{\mathcalligra}{T1}{calligra}{m}{n}
\DeclareFontShape{T1}{calligra}{m}{n}{<->s*[2.2]callig15}{}
\usepackage[vcentermath]{youngtab}
\newcommand{\myng}[1]{\,{\tiny\yng #1}\,}
\newcommand*{\uniq}{\raisebox{-0.7ex}{\scalebox{1.8}{$\cdot$}}}
\usepackage{tikz}
\usetikzlibrary{trees}
\usetikzlibrary{decorations.pathmorphing}
\usetikzlibrary{decorations.markings}
\usetikzlibrary{shapes.misc}

\tikzset{
  threept/.style={
    circle,
    draw,
    inner sep=2pt,
  },
  twopt/.style={
    circle,
    draw,
    fill=black,
    inner sep=1pt,
    minimum size=1pt
  },
  cross/.style={
    cross out,
    draw=black, 
    minimum size=7pt, 
    inner sep=0pt,
    outer sep=0pt
  },
  scalar/.style={
    thick,
    dashed,
    postaction={
      decorate,
      decoration={
        markings,
        mark=at position 0.5 with {\arrow{>}}
      }
    }
  },
  spinning/.style={
    thick,
    postaction={
      decorate,
      decoration={
        markings,
        mark=at position 0.5 with {\arrow{>}}
      }
    }
  },
  spinning no arrow/.style={
    thick,
  },
  finite with arrow/.style={
    decoration={
      snake,
      amplitude=1pt,
      segment length=6pt,
      post length=2pt
    },
    decorate,
    thick,->
  },
  finite/.style={
    decoration={
      snake,
      amplitude=1pt,
      segment length=6pt,
    },
    decorate,
    thick
  }
}

\newcommand{\diagramEnvelope}[1]{#1}

\def\[#1\]{\begin{align}#1\end{align}}

\def\(#1\){{\begin{align}\begin{split}{#1}\end{split}\end{align}}}

\newcommand{\beq}{\begin{equation}}
\newcommand{\eeq}{\end{equation}}
\newcommand{\bal}{\begin{align}}
\newcommand{\eal}{\end{align}}

\newcommand{\eql}[1]{\label{eq:#1}}

\newcommand{\Eq}[1]{Eq.~\eq{#1}}

\newcommand{\eq}[1]{(\ref{eq:#1})}  

\newcommand{\scr}[1]{\ensuremath{\mathcal{#1}}}

\newcommand{\ggap}{\hspace{0.1em}}

\newcommand{\bra}[1]{\langle #1 |}
\newcommand{\ket}[1]{| #1 \rangle}

\newcommand{\vev}[1]{\langle #1 \rangle}

\newcommand{\nop}[1]{:\kern-.3em#1\kern-.3em:}

\newcommand{\myint}{\int\mkern-5mu}

\renewcommand{\d}{\partial}

\newcommand{\sfrac}[2]{{\textstyle\frac{#1}{#2}}}

\newcommand{\al}{\ensuremath{\alpha}}
\newcommand{\be}{\ensuremath{\beta}}
\newcommand{\ga}{\ensuremath{\gamma}}
\newcommand{\Ga}{\ensuremath{\Gamma}}
\newcommand{\de}{\ensuremath{\delta}}
\newcommand{\De}{\ensuremath{\Delta}}

\newcommand{\ka}{\ensuremath{\kappa}}
\newcommand{\la}{\ensuremath{\lambda}}

\newcommand{\si}{\ensuremath{\sigma}}

\makeatletter
\def\@fpheader{\relax}
\makeatother

\title{\boldmath Recursion relations for 5-point conformal blocks}

\author[a]{David Poland,}
\author[b, c]{Valentina Prilepina}

\affiliation[a]{Department of Physics, Yale University, New Haven, CT 06520, USA}
\affiliation[b]{D\'epartement de Physique, de G\'enie Physique et d'Optique,\\Universit\'e Laval, Qu\'ebec, QC G1V 0A6, Canada}

\affiliation[c]{
Institute of Theoretical and Mathematical Physics (ITMP)\\Lomonosov Moscow State University\\Leninskie Gory, GSP-1, 119991 Moscow, Russian Federation
}
\emailAdd{david.poland@yale.edu}
\emailAdd{valentina.prilepina.1@ulaval.ca}

\abstract{We consider 5-point functions in conformal field theories in $d>2$ dimensions. Using weight-shifting operators, we derive recursion relations which allow for the computation of arbitrary conformal blocks appearing in 5-point functions of scalar operators, reducing them to a linear combination of blocks with scalars exchanged. We additionally derive recursion relations for the conformal blocks which appear when one of the external operators in the 5-point function has spin 1 or 2. Our results allow us to formulate positivity constraints using 5-point functions which describe the expectation value of the energy operator in bilocal states created by two scalars.}

\begin{document}
\maketitle

\section{Introduction}

Conformal field theories (CFTs) are remarkable quantum field theories (QFTs) endowed with an enhanced symmetry under the conformal group. Such special theories represent fixed points in renormalization group flows and at the same time describe second-order phase transitions in statistical and condensed matter systems. Strikingly, they also shed light on the structure of the landscape of quantum field theories and elucidate aspects of quantum gravity and black hole physics via the AdS/CFT correspondence. In recent years, a modern revival of the conformal bootstrap program~\cite{Rattazzi:2008pe} has spurred many impressive advances in this field~\cite{Poland:2018epd}. The bootstrap program furnishes a systematic non-perturbative approach to reconstructing the CFT data associated with a given theory by exploiting conformal symmetry and imposing stringent consistency conditions, such as crossing symmetry and unitarity. This powerful method empowers us to gradually carve out the complete allowed space of CFTs, thus extending our understanding of the structure of the full landscape of theories.

Most results so far have been extracted through the consideration of 4-point correlation functions of operators, as reviewed in~\cite{Poland:2018epd}. For example, well-known explicit expressions or recursion relations exist for the conformal blocks appearing in 4-point functions of scalars in arbitrary spacetime dimensions. Moreover, a rich variety of techniques have been developed for handling 4-point conformal blocks involving external and internal exchanged operators in arbitrary Lorentz representations. In addition to the tremendous profusion of progress with respect to 4-point functions, there exists some recent stimulating work on higher-point functions. Some of this work focuses on 2d CFTs~\cite{Alkalaev:2015fbw,Kusuki:2019gjs,Alkalaev:2020kxz,Anous:2020vtw,Fortin:2020zxw} or on constructing correlators in the context of the AdS/CFT correspondence~\cite{Paulos:2011ie,Goncalves:2019znr,Jepsen:2019svc,Carmi:2019ocp,Meltzer:2019nbs,Eberhardt:2020ewh}. 

There are a number of reasons to desire a more precise understanding of five- and higher-point correlation functions. Some motivation includes the compelling idea that it might be possible to get more leverage from examining these objects in the context of the conformal bootstrap. While in principle crossing symmetry of all 4-point functions is sufficient, it may be easier to extract constraints from simple higher-point point functions than from arbitrary spinning 4-point functions. Relatedly, the analysis of higher-point functions may also enable us to more easily probe different physical regimes of a conformal field theory, e.g.~by taking various lightcone and Regge limits. The behavior of correlators in such regimes gives alternative ways of packaging information on the structure of the CFT. 

In holographic CFTs, the elusive 3-point functions of heavy operators $\vev{\scr{O}_H \scr{O}_H \scr{O}_H}$ can be probed by studying the 5-point object $\vev{\scr{O}_L \scr{O}_L \scr{O}_H \scr{O}_L \scr{O}_L}$ involving mostly light operators. Moreover, additional motivation comes from the desire to understand the CFT implications of the averaged null energy condition (ANEC)~\cite{Hofman:2008ar,Buchel:2009sk,Chowdhury:2012km,Hartman:2016dxc,Hofman:2016awc,Faulkner:2016mzt,Hartman:2016lgu,Chowdhury:2017vel,Cordova:2017zej,Meltzer:2018tnm,Chowdhury:2018uyv}. For example, one can require positivity of the expectation value of the energy flux operator $\scr{E}$ in a bilocal state created by two scalar operators, which is encoded in a 5-point function of the form $\vev{\phi_i \phi_j T^{\mu\nu} \phi_i \phi_j}$. It may be interesting to understand if this positivity condition leads to new constraints on the CFT spectrum.

Literature on higher-point conformal blocks for general CFTs in $d>2$, which may be required for the above analyses, is relatively scarce. A noteworthy contribution is the analysis by Rosenhaus~\cite{Rosenhaus:2018zqn}, where the author computed a series expansion for the 5-point scalar exchange conformal block in a 5-point function of external scalar operators. Holographic representations of higher-point conformal blocks were constructed in the works~\cite{Parikh:2019ygo,Parikh:2019dvm,Hoback:2020pgj}, and dimensional reduction formulae for higher-point scalar exchange blocks were recently derived in~\cite{Hoback:2020syd}. In addition, the null polygon limit of multi-point correlators was later explored in~\cite{Vieira:2020xfx}. Further, a connection to Lauricella systems was made in~\cite{Pal:2020dqf}, while a connection to Gaudin integrable models was proposed in~\cite{Buric:2020dyz}. A different perspective was introduced by~\cite{Fortin:2019fvx}, and general representations of higher-point blocks were developed using the operator product expansion (OPE) in embedding space~\cite{Fortin:2019fvx,Fortin:2019dnq,Fortin:2019zkm,Fortin:2020ncr, Fortin:2020yjz,Fortin:2020bfq}. In particular, let us highlight that both~\cite{Parikh:2019dvm} and~\cite{Fortin:2019zkm} derived alternate (but equivalent) expansions for the scalar 5-point block with scalars exchanged.

As of now, few explicit results for higher-point conformal blocks capturing the exchange of spinning operators exist. A notable exception is~\cite{Goncalves:2019znr}, which developed a series expansion for such blocks with identical external scalars. However, this result involves a summation over 9 variables, with coefficients that must be determined recursively by solving the Casimir differential equations. A simpler approach to computing such blocks would clearly be desirable. In this work, we endeavor to improve our understanding of 5-point blocks, by deriving simple recursion relations which can be used to compute 5-point blocks for the exchange of arbitrary symmetric traceless tensors. Our results may be regarded as a natural generalization of recursion relations for 4-point blocks obtained by Dolan and Osborn~\cite{Dolan:2011dv} to the 5-point case. 

We will perform our analysis in the context of the weight-shifting operator formalism developed in~\cite{Karateev:2017jgd}. We begin by considering a purely scalar 5-point function $\vev{\phi_{\Delta_1} \phi_{\Delta_2} \Phi_{\Delta_3} \phi_{\Delta_4} \phi_{\Delta_5}}$. Here we seek to compute the general conformal block for this object, namely that of arbitrary symmetric traceless tensor exchange in the $(12)$ and $(45)$ OPEs. We may express this block in terms of the following $5$-point conformal integral:
\begin{align}
&\vev{\phi_{\Delta_1}(X_1) \phi_{\Delta_2}(X_2) \Phi_{\Delta_3}(X_3) \phi_{\Delta_4}(X_4) \phi_{\Delta_5}(X_5)} =
\sum_{\scr{O} \in \phi_{\Delta_1}\times \phi_{\Delta_2}} 
\sum_{\scr{O}^\prime{} \in \phi_{\Delta_4}\times \phi_{\Delta_5}} \,\,
\frac{1}{\scr{N}_{\scr{O}} \scr{N}_{\scr{O}^\prime{}}}\myint D^d X \myint D^d Y \nonumber\\
 &\vev{\phi_{\Delta_1}(X_1) \phi_{\Delta_2}(X_2) \scr{O}_{A_1\dots A_\ell}(X)} 
 \vev{\tilde{\scr{O}}^{A_1\dots A_\ell}(X)  \Phi_{\Delta_3}(X_3) \scr{O}^\prime_{B_1\dots B_{\ell'}}(Y)} \vev{\tilde{\scr{O}}^{\prime B_1\dots B_{\ell'}}(Y)\phi_{\Delta_4}(X_4) \phi_{\Delta_5} (X_5)}
 \bigg|_{M}\,, 
\end{align}
where $\tilde{\scr{O}}$, $\tilde{\scr{O}}^\prime$ denote the shadow transforms of the operators $\scr{O}$, $\scr{O}^\prime$, and $\scr{N}_{\scr{O}^\ell}$, $\scr{N}_{\scr{O}^{\ell'}}$ are the appropriate shadow normalization factors. The $\big|_{M}$ indicates that we must project onto the appropriate monodromy-invariant subspace to extract the physical block. Here we choose to work in a channel where the middle scalar operator occupies a special place in the correlator, which is often referred to as the comb channel. The 3-point functions on either side are then treated on the same footing, both being of the type (scalar)-(scalar)-(spin), while the central 3-point correlator is distinct and takes the form (spin)-(scalar)-(spin) or more generally (spin)-(spin)-(spin) if we promote $\Phi$ to a spinning operator. 

This paper is organized as follows. In section~\ref{sec:5pt}, we review the result for the scalar exchange 5-point block, summarizing the general properties of 5-point functions and their cross-ratios as well as the calculation of scalar-exchange blocks performed in \cite{Rosenhaus:2018zqn}. We next give an overview of the weight-shifting operator formalism in section~\ref{sec:weight}. Here we describe the weight-shifting differential operators, characterizing their properties and their action on correlation functions. Along the way, we recall the convenient diagrammatic notation introduced by \cite{Karateev:2017jgd}. We then summarize the 2- and 3-point crossing relations, which constitute the main computational tools of the weight-shifting formalism, along with the construction of conformal blocks.  We additionally discuss how to derive recursion relations which lower the spin of the exchanged operators, reviewing the derivation of analogous recursion relations for 4-point functions.  

In section~\ref{sec:recursion} we give our derivation of new recursion relations which can compute arbitrary conformal blocks appearing in 5-point functions of scalar operators, capturing general symmetric traceless tensor exchange. We present two distinct kinds of recursion relations, representing two different ways of encoding the same information.  We also perform several explicit checks that our  5-point recursion relations are consistent with known properties of 4-point functions by taking appropriate limits. In section~\ref{sec:conservation}, we discuss the special case of conserved tensor exchange, focusing on the case of conserved vector and spin-2 tensor operators. 

In section~\ref{sec:phivec}, we consider promoting the middle scalar operator $\Phi$ to a spin-$1$ vector. We detail how to encode the corresponding blocks for symmetric traceless tensor exchange in terms of appropriate combinations of weight-shifting operators acting on lower-spin blocks for a 5-point function of purely scalar external operators. The latter objects act as the seed blocks and may in turn be computed with the aid of the recursion relations described in the previous section. Next, in section~\ref{sec:phitensor}, we discuss how to further promote $\Phi$ from a vector to a spin-$2$ tensor operator. Here the procedure is completely analogous to the previous one, with the difference that the seed blocks are now the ones for symmetric traceless tensor exchange in the corresponding 5-point function featuring a spin-$1$ operator. 

In section~\ref{sec:anec}, we discuss a possible application of these results in the context of the averaged null energy condition (ANEC). In particular, they empower us to apply the OPE to compute the expectation value of the ANEC operator in bilocal states. Invoking the ANEC positivity condition, one may extract constraints on the OPE coefficients. Here we provide an initial discussion of the resulting constraints. We conclude in section~\ref{sec:conclusion}, giving a discussion of directions for future work. A brief description of the box tensor basis and as well as various coefficients derived in this work are included in a number of appendices.

\section{5-point functions}
\label{sec:5pt}

In this section, we give a broad overview of what is known about 5-point correlation functions up to this point. We begin by describing the general form of a 5-point correlation function of symmetric traceless primary operators, then proceed to discuss 5-point conformal blocks, and last review the prototypical case of scalar exchange in the purely scalar 5-point correlator. Throughout, we work in the index-free embedding formalism of~\cite{Costa:2011mg,Costa:2011dw}. 

Let us label spin-$\ell$ primaries by their dimension and spin as $\chi\equiv [\De, \ell]$. Then, conformal invariance fixes a generic $5$-point function of spin-$\ell$ primaries to be of the form
\[
\tilde G_{\chi_1,\dots,\chi_5}  = \vev{\scr{O}_1(X_1; Z_1)\cdots \scr{O}_5(X_5; Z_5)} =  \prod_{i<j}^5 X_{ij}^{-\alpha_{ij}} \,
\sum_k  f_k(u_a) \,Q^{(k)}_{\chi_1, \dots ,\chi_5}  (\{ X_i;Z_i\})\,,
\eql{npf}
\]
where 
\[
\alpha_{ij}=\frac{1}{3}\left(\tau_i+\tau_j - \frac{1}{4} \sum_{k=1}^5 \tau_k\right) 
\eql{alpha_ij}\,,
\]
with $\tau_i=\De_i+ \ell_i$. Here $f_k(u_a)$ represents some function of the conformal cross-ratios $u_a$. In this case, there are five such invariants which we will explicitly describe below.\footnote{See e.g.~\cite{Costa:2011mg,Kravchuk:2016qvl,Fortin:2019dnq,Irges:2020lgp} for general discussions of $n$-point functions and their cross ratios.}

The factors $X_{ij}$ in the overall prefactor in \Eq{npf} carry powers that are fixed by the homogeneity requirement of having proper behavior under scale transformations. To be precise, the transformation of a primary field $\scr{O}_{\De, \ell}$ of scaling dimension $\De$ and spin $\ell$ may be encoded by a polynomial $F(X, Z)$ in its position $X$ and the polarization vector $Z$, subject to the constraint 
\[
F(\lambda X; \alpha Z + \beta X ) = \lambda^{-\De} \alpha^{\ell} F(X;Z)\,.
\eql{EmbField}
\]
In general, the powers $\alpha_{ij}$ in \Eq{alpha_ij} depend on the external operator scaling dimensions $\De_i$ and the spins $\ell_i$; in the scalar case, they depend exclusively on the $\De_i$. By extracting the above prefactor, we have ensured that the $Q^{(k)}$ have weight $\ell_i$ in each point $X_i$. Moreover, they are required to be identically transverse:
\[
Q^{(k)}_{\chi_1,\dots,\chi_5}(\{\lambda_i X_i; \alpha_i Z_i+\beta_i X_i\})=
 Q^{(k)}_{\chi_1,\dots,\chi_5}(\{X_i;Z_i\})\  
\prod_i \left(\lambda_i  \alpha_i \right)^{\ell_i} \,.
\]
We may construct these polynomials from the basic building blocks $V_{i,  jk} $ and $H_{ij}$ of the standard box tensor basis, which we review in Appendix~\ref{app:A}. 

In a nutshell, using the embedding formalism the most general form of the $5$-point correlator $\vev{\scr{O}_1(X_1; Z_1)\cdots \scr{O}_5(X_5; Z_5)}$ compatible with conformal invariance is simply a linear combination of homogeneous polynomials of degree $\ell_i$ in each $Z_i$. These constituent polynomials are in turn constructed from products of appropriate powers of $V_{i, jk}$ and $H_{ij}$. Finally, the $X_i$ dependence is fixed by the scaling requirement of \Eq{EmbField}. 

We may expand the $\sum_k \big[\dots\big]$ in \Eq{npf} in a basis of so-called conformal blocks, which capture the exchange of specific primary operators in the operator product expansion (OPE). Let us briefly recall the fundamental concept of the OPE. This is the statement that, inside an appropriate correlation function, the product of any two local primary operators at two distinct spacetime points may be replaced by an infinite sum over primaries at a single point. In position space, we may express the OPE of two scalar operators $\phi_{\Delta_1}$ and $\phi_{\Delta_2}$ as 
\[
\eql{eq:12OPE}
\phi_{\Delta_1}(x_1)\phi_{\Delta_2}(x_2)=\sum_{\scr{O}} \lambda_{\phi_{\Delta_1} \phi_{\Delta_2} \scr{O}}  
C(x_{12},\d_{x_2})_{e_1\ldots e_{\ell}} \scr{O}^{e_1\ldots e_{\ell}}(x_2)\,,
\]
where the sum ranges over the infinite set of primary operators which appear in the $\phi_{\Delta_1}\times\phi_{\Delta_2}$ OPE. In the case of two scalars, these operators are just the traceless symmetric tensors of arbitrary spin $\ell$. Here the function $C(x_{12},\d_{x_2})^{e_1\ldots e_{\ell}}$ is a power series in $\d_{x_2}$ that encodes the contribution of the infinite tower of descendant operators corresponding to each primary. It is fixed entirely by conformal invariance in terms of the operator scaling dimensions. Finally, the OPE coefficients $ \lambda_{\phi_{\Delta_1} \phi_{\Delta_2} \scr{O}}$ are undetermined numerical coefficients (effectively the structure constants of the operator algebra).

While in a general QFT the OPE converges only in the asymptotic short distance limit, in a CFT it can give a convergent series expansion at finite separation, owing to the enhanced symmetry of the theory. This absolute convergence renders the OPE a well-defined quantity in a conformally invariant theory and lends it additional power. In particular, in a CFT, the OPE may be applied to recursively reduce $n$-point correlation functions to $(n-1)$-point functions, all the way down to 2-point and 3-point functions.

For simplicity, let us restrict our attention to the 5-point function with external scalar operators $\vev{\phi_{\Delta_1}(x_1) \phi_{\Delta_2}(x_2) \Phi_{\Delta_3}(x_3) \phi_{\Delta_4}(x_4) \phi_{\Delta_5}(x_5)}$. Here we label each scalar by its scaling dimension $\Delta_i$, and the notation is meant to highlight the distinct role of the middle operator $\Phi_{\Delta_3}$ in the OPE channels we will consider. A single application of the OPE in the channel $(12)$ enables us to cast this object as a sum of 4-point functions. In particular,
\[
\vev{\phi_{\Delta_1}(x_1)& \phi_{\Delta_2}(x_2) \Phi_{\Delta_3}(x_3) \phi_{\Delta_4}(x_4) \phi_{\Delta_5}(x_5)} \nonumber\\ =& \sum_{\mathcal{O}_{\Delta,\ell}}\la_{\phi_{\Delta_1} \phi_{\Delta_2} \mathcal{O}_{\Delta,\ell}} C(x_{12}, \partial_{x_2})_{e_1\ldots e_{\ell}} \langle \mathcal{O}_{\Delta,\ell}^{e_1\ldots e_{\ell}}(x_2) \Phi_{\Delta_3}(x_3)\phi_{\Delta_4}(x_4) \phi_{\Delta_5}(x_5)\rangle\,,
\]
where the sum runs over primary operators $\mathcal{O}_{\Delta,\ell}$.

Alternatively, exploiting the OPE twice, for example, in the double OPE channel $(12)(45)$, permits us to express this same object as a double sum over derivatives of 3-point functions. That is, we have
\[
\vev{\phi_{\Delta_1}(x_1) &\phi_{\Delta_2}(x_2) \Phi_{\Delta_3}(x_3) \phi_{\Delta_4}(x_4) \phi_{\Delta_5}(x_5)} \nonumber\\ =&\sum_{\mathcal{O}_{\Delta,\ell},\mathcal{O}'_{\Delta',\ell'}}\la_{\phi_{\Delta_1} \phi_{\Delta_2} \mathcal{O}_{\Delta,\ell}} \la_{\phi_{\Delta_4} \phi_{\Delta_5} \mathcal{O}'_{\Delta',\ell'}} \nonumber\\ 
&\times C(x_{12}, \partial_{x_2})_{e_1\ldots e_{\ell}} C(x_{45}, \partial_{x_5})_{f_1\ldots f_{\ell'}} \langle \mathcal{O}_{\Delta,\ell}^{e_1\ldots e_{\ell}}(x_2)\Phi(x_3)\mathcal{O}_{\Delta',\ell'}^{' f_1\ldots f_{\ell'}}(x_5) \rangle \,,
\label{eq:doubleOPE}\]
where now we sum over two sets of primary operators $\mathcal{O}_{\Delta,\ell}$ and $\mathcal{O}'_{\Delta',\ell'}$.

In an analogous manner as for the familiar 4-point function, this double OPE expansion recasts the 5-point object in terms of an expansion in conformal blocks. The conformal blocks are the building blocks of CFT correlation functions that effectively encode the kinematical contribution of the descendant operators in terms of the primary operators, which is fixed by the conformal algebra. In this paper, we primarily choose to compute the conformal blocks in the double OPE channel $(12)(45)$ applied to a 5-point function of scalar operators, as shown above.

Individual terms in this sum could be picked out by inserting a projector $|\mathcal{O}_{\Delta,\ell}|$ onto the conformal multiplet of $\mathcal{O}_{\Delta,\ell}$ (and similarly for $\mathcal{O}'_{\Delta',\ell'}$) into the 5-point function. Note that each 3-point function appearing in (\ref{eq:doubleOPE}) can in turn be expanded in a basis of tensor structures, which are readily described in the embedding space formalism. Labeling the tensor structures by an index $a$, each comes with an independent coefficient $\lambda^a_{\mathcal{O}_{\Delta,\ell} \Phi_{\Delta_3} \mathcal{O}'_{\Delta',\ell'}}$, and a 5-point conformal block will similarly be labeled by the index $a$. Thus, moving to embedding space notation, we have:
\begin{align}
\eql{form 5-point}
\vev{\phi_{\Delta_1}(X_1) \phi_{\Delta_2}(X_2) |\scr{O}_{\De, \ell}|&\Phi_{\Delta_3}(X_3) |
\scr{O}^\prime_{\De', \ell'}|\phi_{\Delta_4}(X_4) \phi_{\Delta_5}(X_5)} = \nonumber\\
&\sum_a\la_{\phi_{\Delta_1} \phi_{\Delta_2} \scr{O}_{\De, \ell}}
\la^a_{\scr{O}_{\De, \ell} \Phi_{\Delta_3} \scr{O}^\prime_{\De', \ell'}}
\la_{\phi_{\Delta_4} \phi_{\Delta_5} \scr{O}^\prime_{\De', \ell'}}
W_{\De, \ell, \De', \ell'; \De_i}^{(a)}(X_i)\,,
\end{align}
where
\[
\eql{eq:5ptblock}
W_{\De, \ell, \De', \ell'; \De_i}^{(a)}(X_i)= P_{\De_i}(X_i) G^{(a)}_{\De, \ell, \De', \ell'}(u_i)\,.
\] 
The object $W_{\De, \ell, \De', \ell'; \De_i}^{(a)}(X_i)$ is comprised of an external-dimension-dependent prefactor $P_{\De_i}(X_i)$ and the 5-point conformal block $G^{(a)}_{\De, \ell, \De', \ell'}(u_i)$, which is a function exclusively of the conformally-invariant cross-ratios $u_i$. In the case of the 5-point function, there are generically five independent cross-ratios $u_i$,\footnote{More precisely, there are $5$ independent cross-ratios for sufficiently large values of the spacetime dimension $d$, in particular for $d\geq 3$. For lower $d$, some of the cross-ratios become dependent, leaving us with only $5d - (d+2)(d+1)/2$ independent cross-ratios.} and different choices of basis can be made for them. In addition, there are multiple forms of the prefactor $P_{\De_i}(X_i)$ that are consistent with homogeneity, leading to another convention choice. Sometimes referred to as the ``leg factor", this prefactor encodes the scaling properties in an explicit coordinate dependence of the 5-point function, but it is ambiguous because it can be multiplied by various combinations of the cross-ratios. We will intentionally choose to write these coordinates and functions in a convention-independent way as much as possible.

Various sets of conventions for these quantities exist in the literature. For instance, in~\cite{Parikh:2019dvm} the external prefactor is given by
\[
\eql{convention 1 prefactor}
P_{\De_i}(X_i)=
\left( { X_{25} \over X_{15} X_{12}} \right)^{\Delta_1 \over 2} \left( { X_{14} \over X_{15} X_{45}} \right)^{\Delta_5 \over 2} 
\left( { X_{15} \over X_{12} X_{25}} \right)^{\Delta_2 \over 2}
\left( { X_{15} \over X_{13} X_{35}} \right)^{\Delta_3 \over 2}
 \left( { X_{15} \over X_{14} X_{45}} \right)^{\Delta_4 \over 2},
\]
where $X_{ij} = -2 X_i\cdot X_j$. This is coupled to the basis of cross-ratios 
\[
\eql{convention 1 cross-ratios}
u_1 = { X_{12} X_{35} \over X_{25} X_{13}} ,\qquad
u_2= { X_{13} X_{45} \over X_{35} X_{14}}, 
\qquad 
w_{ 2;3 } = {X_{15} X_{23} \over X_{25} X_{13}}, 
\qquad 
w_{ 2;4 } = { X_{15} X_{24} \over X_{25} X_{14}}, 
\qquad
w_{ 3;4 } = { X_{15} X_{34} \over X_{35} X_{14}}.
\]  

Another convention is that of~\cite{Rosenhaus:2018zqn}, where the prefactor and cross-ratios are given by
\[
\eql{convention 2 prefactor}
P_{\De_i}(X_i)=\dfrac{1}{(X_{12})^{\frac{\De_1+\De_2}{2}}(X_{34})^{\frac{\De_3}{2}}
(X_{45})^{\frac{\De_4+\De_5}{2}}}\left(\dfrac{X_{23}}{X_{13}}\right)^{\frac{\De_{12}}{2}}
\left(\dfrac{X_{24}}{X_{23}}\right)^{\frac{\De_3}{2}}
\left(\dfrac{X_{35}}{X_{34}}\right)^{\frac{\De_{45}}{2}}
\]
and
\[
\eql{convention 2 cross-ratios}
u_1'= \dfrac{X_{12} X_{34}}{X_{13} X_{24}}, \qquad
v_1' =\dfrac{X_{14} X_{23}}{X_{13} X_{24}}, \qquad u_2' = \dfrac{X_{23} X_{45}}{X_{24} X_{35}}, \qquad v_2' = \dfrac{X_{25} X_{34}}{X_{24} X_{35}}, \qquad 
w' = \dfrac{X_{15} X_{23} X_{34}}{X_{24} X_{13} X_{35}},
\]
respectively.  

Lastly, according to the conventions of~\cite{Fortin:2019zkm}, the external prefactor is given by 
\[
\eql{convention 3 prefactor}
P_{\De_i}(X_i)=\left(\dfrac{X_{4 5}}{X_{14}X_{15}}\right)^{\De_1/2}\left(\dfrac{X_{34}}{X_{23} X_{24}}\right)^{\De_2/2}
\left(\dfrac{X_{2 4}}{X_{2 3} X_{3 4}}\right)^{\De_3/2}
\left(\dfrac{X_{3 5}}{X_{3 4} X_{4 5}}\right)^{\De_4/2}
\left(\dfrac{X_{1 4 }}{X_{4 5} X_{1 5}}\right)^{\De_5/2},
\]
while the conformal cross-ratios are
\[
\eql{convention 3 cross-ratios}
u_1^5 = \dfrac{X_{2 3}X_{4  5}}{X_{2  4}X_{3  5}},\qquad
u_2^5=\dfrac{X_{3  4}X_{15 }}{X_{3  5}X_{14}},\qquad
v_{11}^5=\dfrac{X_{2 5} X_{34}}{X_{3 5} X_{24}}, \qquad
v_{12}^5=\dfrac{X_{13} X_{4 5}}{X_{14}X_{3 5}}, \qquad
v_{22}^5=\dfrac{X_{12} X_{4 5} X_{3 4} }{X_{1 4} X_{3 5} X_{2 4}}.
\]

Once the conventions are chosen, the next issue is how to explicitly compute the conformal blocks. Of these, two prominent approaches are to solve the conformal Casimir equation and express the block as a conformal integral.\footnote{Another approach, which we do not pursue in this paper, is to develop Zamolodchikov-like recursion relations for the block, making use of its expansion in poles in the exchanged operator dimension~\cite{Kos:2013tga,Kos:2014bka,Penedones:2015aga,Yamazaki:2016vqi,Kravchuk:2017dzd,Erramilli:2019njx,Erramilli:2020rlr}.}  We summarize each of these in turn. 

The conformal Casimir approach relies on the property that 3-point functions of primary operators furnish natural eigenvectors of the quadratic conformal Casimir, defined as $\dfrac{1}{2} L^{AB} L_{AB}$, with $L_{AB}$ denoting the generators of the conformal algebra. Let us represent the action of $L_{AB}$ on an operator $\mathcal{O}(X_i)$ by the differential operator $\scr{L}^i_{AB}$.
If we consider the scalar 3-point function $\vev{\phi_{\Delta_1} \phi_{\Delta_2} \scr{O}}$, conformal covariance of this object leads to
\[
(\scr{L}^1_{AB}+\scr{L}^2_{AB})\vev{\phi_{\Delta_1}(X_1)\phi_{\Delta_2}(X_2)\scr{O}(X_3)} 
&=
-\scr{L}^3_{AB}\vev{\phi_{\Delta_1}(X_1)\phi_{\Delta_2}(X_2)\scr{O}(X_3)}\,.
\]
The content of this statement is that the action of the conformal Casimir on $X_1$, $X_2$ is equivalent to its action on $X_3$, which simply yields the eigenvalue $C_{\mathcal{O}}$. That is,
\[
\dfrac{1}{2} (\scr{L}^{1AB}+\scr{L}^{2AB})(\scr{L}^{1}_{AB}+\scr{L}^{2}_{AB})\vev{\phi_{\Delta_1}(X_1)\phi_{\Delta_2}(X_2)\scr{O}(X_3)}
&= \dfrac{1}{2} \scr{L}^{3AB}\scr{L}^3_{AB}\vev{\phi_{\Delta_1}(X_1)\phi_{\Delta_2}(X_2)\scr{O}(X_3)} \nonumber
\\
&= C_{\scr{O}}\vev{\phi_{\Delta_1}(X_1)\phi_{\Delta_2}(X_2)\scr{O}(X_3)}\,.
\]
By solving the Casimir eigenvalue equation, Dolan and Osborn succeeded in deriving compact expressions for even-dimensional symmetric traceless exchange conformal blocks for scalar 4-point functions~\cite{Dolan:2003hv,Dolan:2011dv}.

Another leading method for determining conformal blocks is the conformal integral approach, as described e.g.~in~\cite{SimmonsDuffin:2012uy}. In this approach, one introduces a projector onto the conformal multiplet of $\scr{O}$, defined by
\[
|\scr{O}|\equiv \dfrac{1}{\scr{N}_{\scr{O}}}\myint D^d X \ket{\scr{O}(X)}\bra{\tilde{\scr{O}}(X)}\,,
\]
where $\tilde{\scr{O}}$ denotes the corresponding shadow operator with scaling dimension $d-\De$:
\[
\eql{shadowdefinition}
\tilde{\scr{O}}(X) &= \int D^d Y\frac{1}{(-2 X\cdot Y)^{d-\De}}\scr{O}(Y)\,.
\]
The approach then entails inserting the projector $|\scr{O}|$ inside a correlator $\vev{\phi_{\Delta_1}\dots \phi_{\Delta_m} \phi_{\Delta_{m+1}}\dots \phi_{\Delta_n}}$, thus forming a conformal integral of a product of correlators, supplemented by the appropriate monodromy projections
\[
\vev{\phi_{\Delta_1}\dots \phi_{\Delta_m}|\scr{O}|\phi_{\Delta_{m+1}}\dots \phi_{\Delta_n}}\equiv
\dfrac{1}{\scr{N}_{\scr{O}}}\myint D^d X \vev{\phi_{\Delta_1}\dots \phi_{\Delta_m} \scr{O}(X)}\vev{\tilde{\scr{O}}(X)\phi_{\Delta_{m+1}}\dots \phi_{\Delta_n}}\bigg|_{M = e^{2\pi i \varphi}}\,.
\]
Here $M$ denotes the monodromy factor that arises under the mapping $X_{ij}\to e^{4\pi i} X_{ij}$ for $i, j\leq m$, while keeping the other $X_{ij}$ invariant. The value of $\varphi$ is fixed by demanding consistency with the OPE, which gives $\varphi = \De- \sum_{i\leq m} \De_i$. For example, the 4-point conformal block for scalar exchange may be expressed as the following conformal integral: 
\[
W_{\De, 0; \De_i}(X_i)=\dfrac{1}{\scr{N}_{\scr{O}}} \myint 
D^d X \vev{\phi_{\Delta_1}(X_1)\phi_{\Delta_2}(X_2)\scr{O}(X)}\vev{\tilde{\scr{O}}(X) \phi_{\Delta_3}(X_3)\phi_{\Delta_4}(X_4)}\bigg|_{M = e^{2\pi i \varphi}}\,.
\]
The purpose of the $M$ projection is to restrict the conformal integral onto the proper monodromy invariant subspace, thus removing the contribution of the unphysical shadow block.

Each of these two methods may be exploited in the context of the scalar 5-point function in order to compute 5-point conformal blocks. From the perspective of the conformal Casimir, we would need to simultaneously solve two eigenvalue equations obeyed by the block, subject to appropriate boundary conditions. 

In particular, the 5-point function satisfies
\[
0 &=\bigg[ \dfrac{1}{2}(\scr{L}^1_{AB}+\scr{L}^2_{AB})(\scr{L}^{1,AB}+\scr{L}^{2,AB})-C_{\De, \ell}\bigg]W_{\De, \ell, \De', \ell'; \De_i}^{(a)}(X_i)\,, \\
0 &= \bigg[\dfrac{1}{2}(\scr{L}^4_{AB}+\scr{L}^5_{AB})(\scr{L}^{4,AB}+\scr{L}^{5,AB})-C_{\De', \ell'}\bigg]W_{\De, \ell, \De', \ell'; \De_i}^{(a)}(X_i)\,,
\] 
where
\[
C_{\De,\ell}=\De(\De-d)+\ell(\ell+d-2)\,.
\]
This directly leads to a system of differential equations for the conformal blocks which take the form
\[
\scr{D}_{12} G^{(a)}_{\De, \ell, \De', \ell'}(u_i)&=C_{\De,\ell}G^{(a)}_{\De, \ell, \De', \ell'}(u_i)\,,\\
\scr{D}_{45} G^{(a)}_{\De, \ell, \De', \ell'}(u_i)&=C_{\De',\ell'}G^{(a)}_{\De, \ell, \De', \ell'}(u_i)\,,
\]
where $\scr{D}_{12}$ and $\scr{D}_{45}$ are appropriate second-order differential operators in the conformal cross-ratios. This approach was used in~\cite{Parikh:2019ygo} when developing holographic representations of 5-point blocks, and in~\cite{Goncalves:2019znr}, in the context of developing a series expansion for 5-point blocks. An interesting related approach was recently proposed in~\cite{Buric:2020dyz}, which advocated for combining these equations with those arising from fourth-order differential operators and mapping the resulting system to a Gaudin model.

Alternatively, we may express our object of interest as a conformal integral of a product of 3-point functions in the following fashion:
\[
&W_{\De, \ell, \De', \ell'; \De_i}^{(a)}(X_i) 
= \,\,
\frac{1}{\scr{N}_{\scr{O}_{\De, \ell}} \scr{N}_{\scr{O}'_{\De', \ell'}}}\myint D^d X_I \myint D^d X_J\\
& \vev{\phi_{\Delta_1}(X_1) \phi_{\Delta_2}(X_2) \scr{O}_{\De,\ell}(X_I)}
 \vev{\tilde{\scr{O}}_{\De,\ell}(X_I)   \Phi_{\Delta_3}(X_3)
 \scr{O}^\prime_{\De', \ell'}(X_J)}^{(a)} \vev{\tilde{\scr{O}}^{\prime}_{\De',\ell'}(X_J)\phi_{\Delta_4} (X_4) \phi_{\Delta_5}(X_5)}
 \bigg|_{M}\,, \nonumber
\] 
where we have expanded the 5-point correlator in terms of the exchanged operators $\scr{O}_{\De, \ell}$ and $\scr{O}'_{\De', \ell'}$, both symmetric traceless tensors. The indices have been suppressed for brevity. Here $\big|_M$ denotes an appropriate monodromy projection which removes the shadow contributions. Note that the integrand contains two kinds of 3-point functions, namely the familiar (scalar)-(scalar)-(spin-$\ell$) correlator, which features just a single 3-point structure, and the (spin-$\ell$)-(scalar)-(spin-$\ell'$) correlator, which consists of multiple independent structures. 

For the case of scalar exchange, it is convenient to exploit the conformal integral approach to obtain the 5-point block. A straightforward application of the method renders the extraction of the block quite effortless owing to the appearance of Mellin-Barnes type integrals, which are simple to evaluate. This was first carried out in~\cite{Rosenhaus:2018zqn}, where the author employed the conformal integral technique to obtain
the 5-point conformal block for scalar exchange. The result is given by
\[
\eql{Rosenhaus form}
&G_{\De, 0, \De', 0}(u_1', v_1', u_2', v_2', w') =\,
\sum_{n_i\,=\,0}^\infty 
\dfrac{u_1'{}^{\frac{\De}{2} + n_1}}{n_1!}\dfrac{(1-v_1')^{n_2}}{n_2!}
\dfrac{(1-w')^{n_3}}{n_3!}\dfrac{(1-v_2')^{n_4}}{n_4!}
\dfrac{u_2'{}^{\frac{\De'}{2} + n_5}}{n_5!}\nonumber\\
& \times \bigg(\dfrac{\De+\De'-\De_3}{2}\bigg)_{\sum_{i=1}^5 n_i}
\dfrac{\bigg(\dfrac{\De+\De_{12}}{2}\bigg)_{n_1+n_2+n_3}
\bigg(\dfrac{\De-\De_{12}}{2}\bigg)_{n_1+n_4}}{(\De)_{2n_1+n_2+n_3+n_4}}
\dfrac{\bigg(\dfrac{\De'-\De_{45}}{2}\bigg)_{n_3+n_4+n_5}
\bigg(\dfrac{\De'+\De_{45}}{2}\bigg)_{n_2+n_5}}{(\De')_{2n_5+n_2+n_3+n_4}}\nonumber\\
& \times \dfrac{
\bigg(\dfrac{\De-\De'+\De_3}{2}\bigg)_{n_1}
\bigg(\dfrac{\De'-\De+\De_3}{2}\bigg)_{n_5}}
{\bigg(\De-d/2+1\bigg)_{n_1}\bigg(\De'-d/2+1\bigg)_{n_5}}
{}_3F_2\bigg[\begin{matrix}
-n_1,   -n_5,  \frac{2-d+\De+\De'-\De_3}{2}
\\
\frac{\De-\De'-\De_3}{2}+1-n_5,  
\frac{\De'-\De-\De_3}{2}+1-n_1 
\end{matrix}; 1\bigg]\,.
\]
Subsequent studies employed the AdS/CFT correspondence~\cite{Parikh:2019dvm} 
or the embedding space OPE formalism~\cite{Fortin:2019zkm}, confirming and extending this result, thus placing the form of the five- and higher-point blocks corresponding to scalar exchange on a strong footing.

The explicit result obtained by~\cite{Parikh:2019dvm} (see Eq.~(4.2)) is given by
\[
&G_{\De, 0, \De', 0}(u_1,u_2,w_{2;3},w_{2;4},w_{3;4})=
\sum_{\substack{j_{\langle 2|3\rangle},\:j_{\langle 2|4\rangle},\:j_{\langle 3|4\rangle},\\k_1,\:k_2 \,=\, 0}}^\infty
\dfrac{u_1^{\frac{\Delta }{2}+k_1}}{k_1! }
\dfrac{u_2^{\frac{\Delta '}{2}+k_2}}{k_2!}
\dfrac{(1-w_{2;3})^{j_{\langle 2|3\rangle}}}{j_{\langle 2|3\rangle}! }
\dfrac{(1-w_{2;4})^{j_{\langle 2|4\rangle}} }{j_{\langle 2|4\rangle}!}
\dfrac{ (1-w_{3;4})^{j_{\langle 3|4\rangle}}}{j_{\langle 3|4\rangle}!}\nonumber\\
&\times\left(\frac{1}{2} \left(\Delta -\Delta _1-\Delta _2\right)+1\right)_{k_1} \left(\frac{1}{2} \left(\Delta +\Delta _1-\Delta _2\right)\right)_{k_1} \left(\frac{1}{2} \left(-\Delta +\Delta _1+\Delta_2\right)\right)_{-k_1} \left(\frac{1}{2} \left(\Delta -\Delta _3-\Delta '\right)+1\right)_{k_1}\nonumber\\ 
   &\times \left(\frac{1}{2} \left(\Delta'-\Delta_4-\Delta _5\right)+1\right)_{k_2} \left(\frac{1}{2} \left(\Delta'-\Delta_4+\Delta _5\right)\right)_{k_2}  \left(\frac{1}{2} \left(-\Delta'+\Delta _4+\Delta_5\right)\right)_{-k_2} \left(\frac{1}{2} \left(-\Delta -\Delta _3+\Delta'\right)+1\right)_{k_2} 
 \nonumber\\
  &\times \left(\frac{1}{2} \left(\Delta +\Delta _3-\Delta '\right)\right)_{j_{\langle 2|3\rangle}+k_1-k_2}
    \left(\frac{1}{2} \left(\Delta -\Delta _3+\Delta '\right)\right)_{j_{\langle 2|4\rangle}+k_1+k_2}
   \left(\frac{1}{2} \left(-\Delta +\Delta _3+\Delta '\right)\right)_{j_{\langle 3|4\rangle}-k_1+k_2} 
\nonumber\\
  &\times \frac{\left(\frac{1}{2} \left(\Delta -\Delta _1+\Delta_2\right)\right)_{j_{\langle 2|3\rangle}+j_{\langle 2|4\rangle}+k_1} \left(\frac{1}{2} \left(\Delta'+\Delta _4-\Delta _5\right)\right)_{j_{\langle 2|4\rangle}+j_{\langle 3|4\rangle}+k_2}}{\left(\Delta \right)_{2 k_1+j_{\langle 2|3\rangle}+j_{\langle 2|4\rangle}} \left(\Delta '\right)_{2 k_2+j_{\langle 2|4\rangle}+j_{\langle 3|4\rangle}} \left(\Delta -\frac{d}{2}+1\right)_{k_1} \left( \Delta '-\frac{d}{2}+
 1\right)_{k_2}} {}_3F_2\bigg[\begin{matrix}
-k_1,   -k_2,  \frac{\De + \De'+\De_3 -d}{2} 
\\ \frac{\De-\De'+\De_3}{2} - k_2
,  \frac{\De'-\De+\De_3}{2} - k_1
\end{matrix}; 1\bigg]\,.
\]
In~\cite{Parikh:2019dvm} it was checked that this form is equivalent to \Eq{Rosenhaus form}. In particular, the two forms match to arbitrary numerical precision in the mutual regime of convergence of the two series expansions. 

An alternate form of the series expansion was also obtained in~\cite{Fortin:2019zkm}. This is given by
\[
&G_5^{(d,h_2,h_3,h_4;p_2,p_3,p_4)}(u_1^5,u_2^5,v_{11}^5,v_{12}^5,v_{22}^5)=\sum_{\{m_a,m_{ab}\}\,\geq\,0}(p_3)_{m_1+m_{11}+m_{22}}(p_4-m_1)_{m_2}\nonumber\\
&\phantom{=}\qquad\times\frac{(p_2+h_2)_{m_1+m_{12}}(\bar{p}_3+\bar{h}_3)_{m_1+m_2+m_{11}+m_{12}+m_{22}}(\bar{p}_4+\bar{h}_4)_{m_2+m_{12}+m_{22}}}{(\bar{p}_3+h_2)_{2m_1+m_{11}+m_{12}+m_{22}}(\bar{p}_4+\bar{h}_3)_{2m_2+m_{11}+m_{12}+m_{22}}}\nonumber\\
&\phantom{=}\qquad\times\frac{(-h_3)_{m_1}(-h_4)_{m_2}(-h_4+m_2)_{m_{11}}}{(\bar{p}_3+h_2+1-d/2)_{m_1}(\bar{p}_4+\bar{h}_3+1-d/2)_{m_2}}{}_3F_{2}\left[\begin{array}{c}-m_1,-m_2,-\bar{p}_3-h_2+d/2-m_1\\p_4-m_1,h_3+1-m_1\end{array};1\right]\nonumber\\
&\phantom{=}\qquad\times\frac{(u_1^5)^{m_1}}{m_1!}\frac{(u_2^5)^{m_2}}{m_2!}\frac{(1-v_{11}^5)^{m_{11}}}{m_{11}!}\frac{(1-v_{12}^5)^{m_{12}}}{m_{12}!}\frac{(1-v_{22}^5)^{m_{22}}}{m_{22}!},
\]
where 
\begin{align}
p_2=\Delta_{i_3}\,,&\qquad2p_3=\Delta_{i_2}+\Delta_{k_1}-\Delta_{i_3}\,,&2p_4=\Delta_{i_4}+\Delta_{k_{2}}-\Delta_{k_{1}}\,,&
\nonumber\\
2h_2=\Delta_{k_1}-\Delta_{i_2}-\Delta_{i_3}\,,&\qquad2h_3=\Delta_{k_{2}}-\Delta_{k_{1}}-\Delta_{i_{4}}\,,&
2h_4=\Delta_{k_{3}}-\Delta_{k_{2}}-\Delta_{i_{5}}\,,&
\end{align}
and 
$\bar{p}_a=\sum_{b=2}^ap_b$ and $\bar{h}_a=\sum_{b=2}^ah_b$. Here the $\Delta_{i_j}$ for $j=1, \dots, 5$ denote the scaling dimensions of the five external scalar operators, while $\Delta_{k_{1}}$ and $\Delta_{k_{2}}$ represent the dimensions of the two exchanged scalar primaries. The authors analytically showed this form to be equivalent to the one of \Eq{Rosenhaus form} using a sequence of hypergeometric identities.

If we attempt to extend the conformal integral approach to the case of the exchange of arbitrary operators of spin $\ell$, $\ell'$, we find almost immediately that the natural generalization by introducing shifts in the powers of the $X_{ij}$ is of little advantage to us, as the integrals quickly become unmanageable. Instead, we turn to the convenient and elegant framework afforded by the weight-shifting operator formalism~\cite{Karateev:2017jgd}. This approach empowers us to derive a set of efficient recursion relations for generating the 5-point conformal blocks for the case of arbitrary symmetric traceless exchange. In the next section, we give a brief overview of this formalism, highlighting some of its essential features and laying out the basic method for obtaining recursion relations for conformal blocks.

\section{Review of the weight-shifting operator formalism}
\label{sec:weight}

We begin by giving a brief description of the weight-shifting operator formalism, a framework originally presented in~\cite{Karateev:2017jgd} and further developed in \cite{Costa:2018mcg,Kravchuk:2018htv,Karateev:2018oml,Albayrak:2020rxh}. In this formalism, a large class of conformally-covariant differential operators is introduced along with a crossing equation that they respect. Referred to as weight-shifting operators, these operators may be used to relate correlation functions of operators in different representations of the conformal group. A crucial advantage of this approach is the substantial simplification it lends to the computation of conformal blocks involving external operators with spin. Notably, this method allows one to find expressions for seed conformal blocks as well as for more general blocks. This is in contrast to the differential basis approach~\cite{Costa:2011dw}, which gives a prescription for computing more general blocks from a set of simpler seed blocks but does not allow one to determine the seed blocks themselves, which are left to be extracted using other methods~\cite{SimmonsDuffin:2012uy,Iliesiu:2015akf,Echeverri:2016dun}. The weight-shifting formalism may therefore be regarded as an extension of the differential basis approach, since it enables one to alter the exchanged representation when desired. 

Another convenient aspect of the framework is that it allows one to derive recursion relations for conformal blocks quite efficiently, a feature which is essentially built into the formalism. An inherent property of this formalism is that it naturally packages the conformal blocks into the form of a linear combination of some differential operators acting on some set of seed blocks, e.g.~the conformal blocks for symmetric traceless exchange in a purely scalar 4-point function. Thus, the conformal blocks are naturally expressed in a differential operator basis in the context of this framework. 

We next briefly introduce the weight-shifting differential operators themselves.

\subsection{Weight-shifting operators}
 
The general construction of the weight-shifting operator formalism reveals a large class of conformally covariant differential operators, which correspond to tensor products with different finite-dimensional representations $\scr{W}$. That is, each set of operators $\{\scr{D}_x^{(v)A}\}$ is associated with a particular representation $\scr{W}$, where $A = 1,\dots,\text{dim}\, \scr{W}$ is an index for $\scr{W}$, while $v$ refers to a weight vector of $\scr{W}$ (i.e., a common eigenvector of the Cartan subalgebra). We may regard the finite-dimensional representation $\scr{W}$ as a vector space with basis $e^A$. For example, we may consider $\scr{W}$ to be the fundamental vector representation, $\scr{W} =\scr{V}=\square$. Following~\cite{Karateev:2017jgd}, we denote the representation of a given operator by the pairing $[\De, \rho]$, where $\De$ labels the scaling dimension, while $\rho$ stands for the Lorentz quantum numbers of the representation, often simply the spin.

In particular, the operators $\scr{D}_x^{(v)A}: [\De, \rho] \to [\De- \delta\De_v, \la]$ associated with $\scr{W}$ for generic $\De$ are in one-to-one correspondence with the irreducible components in the tensor product decomposition of $\scr{W}^\ast \otimes V_{\De, \rho}$, where $V_{\De, \rho}$ is the representation in which a given operator $\scr{O}(x)$ transforms. Aptly named weight-shifting operators, the operators $\scr{D}_x^{(v)A}$ act on $\scr{O}(x)$ by shifting the weights of $\scr{O}$ by the weights of $v$, while introducing a free $A$ index. For example, they may increase or decrease the spin or dimension of $\scr{O}$. Roughly speaking, when $\scr{D}_x^{(v)A}$ lowers the spin of $\scr{O}$, its missing degrees of freedom are transferred to the index $A$ for $\scr{W}$. 

Such operators can be constructed explicitly using the embedding space formalism. The weight-shifting formalism, however, is completely general and can be used without reference to the embedding space framework. Here we will be primarily interested in the case of symmetric traceless tensors of $SO(d)$ in general spacetime dimensions. For the vector representation, we can build differential operators $\{\scr{D}_X^{(\de\De, \de\ell)A}\}$ with a vector index in the embedding space, which map
\[
\scr{D}_X^{(-0)A}: [\De, \ell] &\to [\De-1, \ell]\,, \nonumber\\
\scr{D}_X^{(0+)A}: [\De, \ell] &\to [\De, \ell +1]\,, \nonumber\\
\scr{D}_X^{(0-)A}: [\De, \ell] &\to [\De, \ell -1]\,, \nonumber\\
\scr{D}_X^{(+0)A}: [\De, \ell] &\to [\De+1, \ell ]\,. 
\]

As explained in~\cite{Karateev:2017jgd}, the differential operators $\scr{D}_X^{(\de\De, \de\ell)A}$ may be determined by assuming a suitable ansatz and then fixing the coefficients by demanding that they preserve the ring of functions $R$ of $X, Z\in \mathbb{R}^{d+1, 1}$ that are invariant under $Z\to Z+\lambda X$ (ie. they must map $R\to R$) and also preserve the ideal $R\cap I$, where $I$ is the ideal generated by $\{X^2, X\cdot Z, Z^2\}$. That is, the operators must be well-defined on $R/(R\cap I)$.

In particular, there are four weight-shifting operators for the vector representation $\scr{W} =\scr{V}$. These are explicitly given by\footnote{We follow the normalization conventions of~\cite{Karateev:2017jgd}.}
\[
\eql{vector WS}
\scr{D}_X^{(-0)A} ={}& X^A \,,\nonumber\\
\scr{D}_X^{(0+)A} ={}& \Big((\ell+\De) \delta_B^A + X^A \d_{X_B}\Big) Z^B \,,\nonumber\\
\scr{D}_X^{(0-)A} ={}&\Big( (\De -d +2 - \ell) \delta_B^A + X^A \d_{X_B} \Big)
\Big( (d-4+2 \ell) \delta_C^B -Z^B \d_{Z_C} \Big) \d_{Z}^C \,,\nonumber\\
\scr{D}_X^{(+0)A} ={}& \frac{1}{2}\Big( c_1 \delta^A_B + X^A \d_{X_B} \Big)
\Big( c_2 \delta^B_C + Z^B \d_{Z_C} \Big)
\Big( c_3 \delta^C_D - \partial^C_{Z} Z_{D}  \Big) \partial^D_{X} \,,
\]
where
\[
c_1 = 2 - d + 2 \De\,, \qquad
c_2 = 2 - d + \De - \ell\,, \qquad
c_3 = \De + \ell\,,
\eql{c_i}
\]
and $\De$ and $\ell$ label the dimension and spin, respectively, of the operator in question.

A crucial aspect of this construction is that differential operators of this type obey a type of crossing relation, which we may schematically denote by
\[
\eql{schematic crossing}
\scr{D}_{X_1}^{(v)A} \vev{\scr{O}_1^\prime(X_1) \scr{O}_2(X_2)\scr{O}_3(X_3)}^{(a)}=
\sum_{\scr{O}_2^\prime, v^\prime, b} \{\dots\} \scr{D}_{X_2}^{(v')A} 
 \vev{\scr{O}_1(X_1) \scr{O}_2^\prime(X_2)\scr{O}_3(X_3) }^{(b)}\,,
\]
where $a$ and $b$ label conformally-invariant 3-point structures that appear in a correlation function of the given operators. The $\{\dots\}$ represent group-theoretic expansion coefficients, which are examples of $6j$ symbols (or Racah-Wigner coefficients) for the conformal group. The function of \Eq{schematic crossing} is to relate weight-shifting operators acting on 3-point structures at different points to each other, thus enabling us to re-express a covariant differential operator acting on $X_1$ as a linear combination of operators acting on $X_2$. There is also a variety of this crossing relation for 2-point structures, which leads to a convenient integration-by-parts rule in the context of the conformal integral. 

By judiciously selecting suitable combinations of weight-shifting operators and applying the two- and three-point crossing relations where appropriate, we are able to derive various recursion relations for conformal blocks quite efficiently. The bulk of the computation involves extracting the relevant $6j$ symbols, which may be time consuming for more complicated cases. However, this is the only potentially challenging part of the calculation. 

Having given a broad overview of the formalism, we next take a closer look at the crossing relations for weight-shifting operators.

\subsection{Crossing relations for weight-shifting operators}

We now proceed to describe the action of the weight-shifting operators on conformally-invariant correlation functions of local operators. Acting with such a differential operator on an arbitrary $n$-point correlator produces a conformally-covariant $n$-point function. The resulting object has an additional interpretation as a conformally-invariant $(n+1)$-point function that includes a basis element of a given finite-dimensional representation $\scr{W}$ of the conformal group $SO(d+1, 1)$. In particular, we may represent this invariant $(n+1)$-point function as a $n$-point correlation function with the formal insertion of a basis element $e^A$ of $\scr{W}$: 
\[
\scr{D}_X^{(a)A} \vev{\scr{O}(X)\dots} \sim  \vev{e^A\scr{O}^\prime(X) \dots}\,.
\]

Here we find it convenient to employ the diagrammatic language introduced in \cite{Karateev:2017jgd}. Following \cite{Karateev:2017jgd}, we symbolize such a conformally-covariant differential operator by 
\begin{equation}
\scr{D}_X^{(a)A} \quad=\quad
\diagramEnvelope{\begin{tikzpicture}[anchor=base,baseline]
	\node (vert) at (0,0) [threept] {$a$};
	\node (opO) at (-0.5,-1) [below] {$\scr{O}$};
	\node (opOprime) at (-0.5,1) [above] {$\scr{O}'$};
	\node (opFinite) at (1,0) [right] {$\scr{W}$};	
	\draw [spinning] (vert)-- (opOprime);
	\draw [spinning] (opO) -- (vert);
	\draw [finite with arrow] (vert) -- (opFinite);
\end{tikzpicture}},
\label{eq:diffoppicture}
\end{equation}
where a wavy line is used to denote the finite-dimensional representation $\scr{W}$. 

We next consider the action of such an operator on two- and three-point correlation functions in turn. We represent a conformally-invariant 2-point structure by 
\[
\vev{\scr{O}_1 (X_1)\scr{O}_2 (X_2)} \quad=\quad
\diagramEnvelope{\begin{tikzpicture}[anchor=base,baseline]
	\node (opO) at (-1,0) [left] {$\scr{O}_1$};
	\node (opOprime) at (1,0) [right] {$\scr{O}_2$};
	\node (vert) at (0,0) [twopt] {};
	\draw [spinning] (vert) -- (opO);
	\draw [spinning] (vert) -- (opOprime);
\end{tikzpicture}},
\]
where the dot may be seen as separating the representation at the first point from that at the second point. That is, we regard the part of the diagram to the left (right) of the dot in the direction indicated by the arrow as being at point $X_1$ ($X_2$). It is well known that there is at most one such structure consistent with conformal invariance. In particular, this object is nonvanishing provided that $\scr{O}_1$ and $\scr{O}_2$ are in what are sometimes referred to as the contragredient or dual-reflected representations with respect to each other, which we denote by $\rho_ 1 =(\rho_2^P)^\ast$. In Lorentzian signature, this is the same as the complex conjugate representation, $(\rho_2^P)^\ast= \rho_2^\dag$. Moreover, the dimensions of the operators should match, $\De_1=\De_2$; otherwise, the 2-point function vanishes.   

We now consider acting with a weight-shifting operator on the 2-point function. It is straightforward to infer and verify a suggestive crossing relation that encodes the relation between 2-point correlators of the operators $\scr{O}$ and $\scr{O}'$. This relation assumes the form \cite{Karateev:2017jgd} 
\[
\diagramEnvelope{\begin{tikzpicture}[anchor=base,baseline]
	\node (vertU) at (0,0) [twopt] {};
	\node (vertD) at (1,-0.08) [threept] {$m$};
	\node (opO1) at (-1,0) [left] {$\scr{O}^\dag$};
	\node (opO3) at (2,0) [right] {$\scr{O}'$};
	\node (opW) at (1,-1) [below] {$\scr{W}$};	
	\node at (0.6,0.1) [above] {$\scr{O}$};
	\draw [spinning] (vertU)-- (opO1);
	\draw [spinning] (vertU)-- (vertD);
	\draw [spinning] (vertD)-- (opO3);
	\draw [finite with arrow] (vertD)-- (opW);
\end{tikzpicture}}
	\quad=\quad
\bigg\{\begin{matrix}
\scr{O}^\dag\\ \scr{O}'
\end{matrix} \bigg\}^{(m)}_{(\bar{m})} 
\diagramEnvelope{\begin{tikzpicture}[anchor=base,baseline]
	\node (vertL) at (1,0) [twopt] {};
	\node (vertR) at (0,-0.12) [threept] {$\bar{m}$};
	\node (opO1) at (-1,0) [left] {$\scr{O}^\dag$};
	\node (opO3) at (2,0) [right] {$\scr{O}'$};
	\node (opW) at (0,-1) [below] {$\scr{W}$};
	\node at (0.6,0.1) [above] {$\scr{O}'^\dag$};
	\draw [spinning] (vertR)-- (opO1);
	\draw [spinning] (vertL)-- (vertR);
	\draw [spinning] (vertL)-- (opO3);
	\draw [finite with arrow] (vertR)-- (opW);
\end{tikzpicture}}.
\label{eq:twoptcrossing}
\]
For symmetric traceless operators, this diagrammatic statement corresponds to the equation
\beq
\scr{D}_{X_2}^{(m)A}
\vev{ \scr{O} (X_1) \scr{O}(X_2) }
=
\left\{\begin{matrix}
\scr{O}^\dag\\ \scr{O}'
\end{matrix} \right\}^{(m)}_{(\bar{m})} 
\scr{D}_{X_1}^{(\bar{m})A}
\vev{\scr{O}' (X_1) \scr{O}'(X_2)}\,,  
\label{eq:twoptcrossing_formula}
\eeq
where $\bar{m}$ labels the relevant weight-shifting operator carrying a shift opposite to $m$. This equation is referred to as the ``2-point crossing relation". Here $\scr{O}^\dag$ represents the operator with which $\scr{O}$ has a nonvanishing 2-point function, which is the complex conjugate in Lorentzian signature. 
In the special case of CFT weight-shifting operators in the vector representation given in \Eq{vector WS}, one finds that \Eq{twoptcrossing} reads \cite{Karateev:2017jgd} 
\begin{align}
{}& \scr{D}_{X_2}^{(\de\De, \de\ell)A}
\vev{ \scr{O}_{\De,\ell} (X_1) \scr{O}_{\De,\ell} (X_2) }=
\left\{ \begin{matrix}
\scr{O}_{\De,\ell}\\ \scr{O}_{\De+\delta\De,\ell+\delta\ell}
\end{matrix} \right\}^{(\delta\De, \delta\ell)}_{(-\delta\De, -\delta\ell)}
\scr{D}_{X_1}^{(-\de\De, -\de\ell)A}
\vev{ \scr{O}_{\De+\delta\De,\ell+\delta\ell} (X_1) \scr{O}_{\De+\delta\De,\ell+\delta\ell} (X_2) } \,.
\nonumber
\end{align}
In this case, we have that $\scr{D}_{X_i}^{(m)A}=\scr{D}_{X_i}^{(-0)A}$ implies $\scr{D}_{X_i}^{(\bar{m})A}=\scr{D}_{X_i}^{(+0)A}$ and vice versa; similarly, $\scr{D}_{X_i}^{(m)A}=\scr{D}_{X_i}^{(0+)A}$ corresponds to $\scr{D}_{X_i}^{(\bar{m})A}=\scr{D}_{X_i}^{(0-)A}$ and vice versa.

The $6j$ symbols appearing in the 2-point crossing relation for the vector case are explicitly given by
\begin{align}
\left\{ \begin{matrix}
\scr{O}_{\De,\ell}\\ \scr{O}_{\De,\ell+1}
\end{matrix} \right\}^{(0+)}_{(0-)}
&=
\left(
\left\{ \begin{matrix}
\scr{O}_{\De,\ell+1}\\ \scr{O}_{\De,\ell}
\end{matrix} \right\}^{(0-)}_{(0+)}
\right)^{-1}
= \frac{\De+\ell}{(\ell+1) (d + 2 \ell - 2) (\De + 1 - d - \ell)}\,,
\nonumber\\
\left\{ \begin{matrix}
\scr{O}_{\De,\ell}\\ \scr{O}_{\De+1,\ell}
\end{matrix} \right\}^{(+0)}_{(-0)}
&=
\left(
\left\{ \begin{matrix}
\scr{O}_{\De+1,\ell}\\ \scr{O}_{\De,\ell} 
\end{matrix} \right\}^{(-0)}_{(+0)}
\right)^{-1}
= 2 (d - \De+  \ell  -2) (\De-1) (\De+\ell) \big(d - 2 (\De+1)\big) \,.
\label{eq:6j_CFT}
\end{align}
These will be used repeatedly in the analysis that follows.

We now turn to the 3-point structures. A conformally invariant 3-point function is a linear combination of 3-point tensor structures, with each individual structure weighted by a different OPE coefficient $\lambda_a$: 
\beq\label{eq:physicalthreept}
\vev{\scr{O}_1(X_1)\scr{O}_2(X_2)\scr{O}_3(X_3)}=\sum_{a=1}^{N_{123}}\lambda_a\,\vev{\scr{O}_1(X_1)\scr{O}_2(X_2)\scr{O}_3(X_3)}^{(a)}\,,
\eeq
where the sum runs from 1 to the total number of structures $N_{123}$, which corresponds to the dimension of the tensor product of the representations of the three operators $(\rho_1\otimes \rho_2\otimes \rho_3)^{SO(d-1)}$, $N_{123}=\text{dim}(\rho_1\otimes \rho_2\otimes \rho_3)^{SO(d-1)}$. The index $a$ is in place to label the particular 3-point structure of interest. In the case that there is only a single such structure, the label $a$ will be omitted.

We represent such a conformally invariant 3-point structure by the vertex
\begin{equation}
 \vev{\scr{O}_1 (X_1)\scr{O}_2 (X_2) \scr{O}_3 (X_3)}^{(a)} \quad=\quad
\diagramEnvelope{\begin{tikzpicture}[anchor=base,baseline]
	\node (vert) at (0,-0.08) [threept] {$a$};
	\node (opO1) at (-0.5,-1) [below] {$\scr{O}_1$};
	\node (opO2) at (-0.5,1) [above] {$\scr{O}_2$};
	\node (opO3) at (1,0) [right] {$\scr{O}_3$};	
	
	\draw [spinning] (vert)-- (opO1);
	\draw [spinning] (vert)-- (opO2);
	\draw [spinning] (vert)-- (opO3);
\end{tikzpicture}}
\label{eq:3pt_function},
\end{equation}
with the label $a$ enumerating all the singlets in the tensor product decomposition of $(\rho_1\otimes \rho_2 \otimes \rho_3)^{SO(d-1)}$. Just as for the 2-point functions, here the arrows diverge from the center of the diagram, indicating that each operator is inserted at a different point.

Let us next consider acting with some weight-shifting operator $\scr{D}^{(b)A}_{X_3}$ on an invariant 3-point structure $\vev{\scr{O}_1 \scr{O}_2 \scr{O}_3^\prime}^{(a)}$, with $\scr{O}_3^\prime$ transforming in the representation $[\De_3+i, \lambda]$ so that the label $a$ runs over the singlets in $(\rho_1\otimes \rho_2 \otimes \lambda)^{SO(d-1)}$. This generates an associated covariant 3-point structure for $\vev{\scr{O}_1\scr{O}_2 \scr{O}_3 e^A}$. We may represent the action of $\scr{D}_{X_3}^{(b)A}$ symbolically by
\beq  
\label{eq:covariantthreeptdiagrams}
\scr{D}_{X_3}^{(b)A}\vev{\scr{O}_1(X_1)\scr{O}_2(X_2)\scr{O}_3'(X_3)}^{(a)} \quad =\quad
\diagramEnvelope{\begin{tikzpicture}[anchor=base,baseline]
	\node (vertL) at (0,0) [threept] {$a$};
	\node (vertR) at (2,-0.05) [threept] {$b$};
	\node (opO1) at (-0.5,-1) [below] {$\scr{O}_1$};
	\node (opO2) at (-0.5,1) [above] {$\scr{O}_2$};
	\node (opO3) at (2.5,1) [above] {$\scr{O}_3$};
	\node (opW) at (2.5,-1) [below] {$\scr{W}$};	
	\node at (1,0.1) [above] {$\scr{O}_3'$};	
	\draw [spinning] (vertL)-- (opO1);
	\draw [spinning] (vertL)-- (opO2);
	\draw [spinning] (vertL)-- (vertR);
	\draw [spinning] (vertR)-- (opO3);
	\draw [finite with arrow] (vertR)-- (opW);
\end{tikzpicture}}.
\eeq
 
Every such conformally-covariant 3-point structure can be constructed through the action of some differential operators on some conformally-invariant 3-point structures, as established in \cite{Karateev:2017jgd}. A powerful consequence is that the diagrams in \Eq{covariantthreeptdiagrams} comprise a particular basis for the finite-dimensional space of covariant 3-point structures $\vev{\scr{O}_1 \scr{O}_2 \scr{O}_3 e^A}$. This basis is special in that it singles out a specific operator $\scr{O}_3^\prime$ contributing to the $\scr{O}_1\times \scr{O}_2$ OPE. Alternatively, we may instead choose to select an operator in another channel, for instance, by considering the structure $\vev{\scr{O}_1^\prime \scr{O}_2 \scr{O}_3}^{(m)}$ and feeding it to the operator $\scr{D}_{X_1}^{(n)A}: \scr{O}_1^\prime \to \scr{O}_1$. In this way, we would generate another basis of covariant 3-point structures, this time, a set composed of the objects $\scr{D}_{X_1}^{(n)A}\vev{\scr{O}_1'(X_1)\scr{O}_2(X_2)\scr{O}_3(X_3)}^{(m)}$.

These two bases are linked by a linear transformation, which leads to a crossing relation for differential operators. Diagrammatically, this is encoded by
\beq
\diagramEnvelope{\begin{tikzpicture}[anchor=base,baseline]
	\node (vertL) at (0,0) [threept] {$a$};
	\node (vertR) at (2,-0.05) [threept] {$m$};
	\node (opO1) at (-0.5,-1) [below] {$\scr{O}_1$};
	\node (opO2) at (-0.5,1) [above] {$\scr{O}_2$};
	\node (opO3) at (2.5,1) [above] {$\scr{O}_3$};
	\node (opW) at (2.5,-1) [below] {$\scr{W}$};	
	\node at (1,0.1) [above] {$\scr{O}_3'$};	
	\draw [spinning] (vertL)-- (opO1);
	\draw [spinning] (vertL)-- (opO2);
	\draw [spinning] (vertL)-- (vertR);
	\draw [spinning] (vertR)-- (opO3);
	\draw [finite with arrow] (vertR)-- (opW);
\end{tikzpicture}}
	\quad=\quad
	\sum_{\scr{O}_1',b,n}
	\left\{
		\begin{matrix}
		\scr{O}_1 & \scr{O}_2 & \scr{O}_1' \\
		\scr{O}_3 & \scr{W} & \scr{O}_3'
		\end{matrix}
	\right\}^{(a)(m)}_{(b)(n)}
\diagramEnvelope{\begin{tikzpicture}[anchor=base,baseline]
	\node (vertU) at (0,0.7) [threept] {$b$};
	\node (vertD) at (0,-0.7) [threept] {$n$};
	\node (opO1) at (-1,-1.5) [below] {$\scr{O}_1$};
	\node (opO2) at (-1,1.5) [above] {$\scr{O}_2$};
	\node (opO3) at (1,1.5) [above] {$\scr{O}_3$};
	\node (opW) at (1,-1.5) [below] {$\scr{W}$};	
	\node at (0.1,0) [right] {$\scr{O}_1'$};	
	\draw [spinning] (vertD)-- (opO1);
	\draw [spinning] (vertU)-- (opO2);
	\draw [spinning] (vertU)-- (vertD);
	\draw [spinning] (vertU)-- (opO3);
	\draw [finite with arrow] (vertD)-- (opW);
\end{tikzpicture}}.
\label{eq:6jdefinition}
\eeq
This diagrammatic statement corresponds to the equation
\[ 
{}&
\scr{D}_{X_3}^{(m)A}\vev{\scr{O}_1 (X_1)\scr{O}_2 (X_2) \scr{O}'_3 (X_3)}^{(a)}=
\sum_{\scr{O}_1',b, n}
	\left\{
		\begin{matrix}
		\scr{O}_1 & \scr{O}_2 & \scr{O}_1' \\
		\scr{O}_3 & \scr{W} & \scr{O}_3'
		\end{matrix}
	\right\}^{(a)(m)}_{(b)(n)} 
\scr{D}_{X_1}^{(n)A}\vev{\scr{O}'_1 (X_1)\scr{O}_2 (X_2) \scr{O}_3 (X_3)}^{(b)}\,.
\eql{3ptCFTcrossing}
\]
Effectively, this relation constitutes a change-of-basis equation between different bases of covariant 3-point structures, each generated by the action of some particular weight-shifting operator at a given point $X_1$ or $X_3$. Here the sum over the operator $\scr{O}_1^\prime$ is finite, with its value ranging over the operators in the tensor product $\scr{O}_1\otimes \scr{W}$. 

The transformation coefficients in this equation are referred to as the Racah coefficients or $6j$ symbols. Since one of the representations is finite-dimensional here, these coefficients have a degenerate form. It can be shown explicitly that they are intimately connected to the algebra of conformally covariant differential operators. 

The above 3-point crossing equation empowers us to move weight-shifting operators from one leg (or, equivalently, operator) to another. As such, this relation constitutes the primary computational tool in the weight-shifting operator formalism. In practice, we may apply it in a variety of settings, such as in the derivation of conformal block recursion relations. 

Note that this equation reduces to the 2-point crossing relation if we take the operator $\scr{O}_2$ to be the identity. In particular, the 3-point $6j$ symbols coincide with the corresponding 2-point $6j$ symbols if $\scr{O}_2 = \mathds{1}$: 
\[
	\bigg\{
		\begin{matrix}
		\scr{O}_1 & \mathds{1} & \scr{O}_3 \\
		\scr{O}_3 & \scr{W} & \scr{O}_1
		\end{matrix}
	\bigg\}^{\uniq (m)}_{\uniq (\bar{m})}
=\bigg\{ \begin{matrix}
\scr{O}_1 \\ \scr{O}_3
\end{matrix} \bigg\}^{(m)}_{(\bar{m})}\,.
\label{eq:shorthand_6j_2pt}
\]

Further, if we consider contracting both sides of the 3-point relation \Eq{3ptCFTcrossing} with the weight-shifting operator $\scr{D}_{X_1\, A}^{(n)}$, we find 
\[
\scr{D}_{X_1\, A}^{(n)}
\scr{D}_{X_3}^{(m)A}\vev{\scr{O}_1 &(X_1)\scr{O}_2 (X_2) \scr{O}'_3 (X_3)}^{(a)} \nonumber\\
=&\sum_{\scr{O}_1', b, p}
	\left\{
		\begin{matrix}
		\scr{O}_1 & \scr{O}_2 & \scr{O}_1' \\
		\scr{O}_3 & \scr{W} & \scr{O}_3'
		\end{matrix}
	\right\}^{(a)(m)}_{(b)(p)} 
\scr{D}_{X_1\, A}^{(n)}\scr{D}_{X_1}^{(p)A}
\vev{\scr{O}'_1 (X_1)\scr{O}_2 (X_2) \scr{O}_3 (X_3)}^{(b)}\,.
\eql{bubble defn}
\]
We remark here that the right-hand side of this equation features two contracted weight-shifting operators acting on a single leg (equivalently, at the same point). This composition $\scr{D}_{X_1\, A}^{(n)}\scr{D}_{X_1}^{(p)A}$ corresponds to a bubble diagram:
\[
\scr{D}_{X_1\, A}^{(n)}\scr{D}_{X_1}^{(p)A}\ =\ 
\diagramEnvelope{\begin{tikzpicture}[anchor=base,baseline]
\node (top) at (0,1.8) [above] {$\scr{O}_1'$};
\node (topmid) at (0,0.9) [threept] {$p$};
\node (botmid) at (0,-0.9) [threept] {$n$};
\node (bot) at (0,-1.8) [below] {$\scr{O}_1''$};
\node () at (0.7,0) [right] {$\scr{W}$};
\node () at (-0.7,0) [left] {$\scr{O}_1$};
\draw [spinning] (top) -- (topmid);
\draw [spinning] (botmid) -- (bot);
\draw [finite with arrow] (topmid) to[out=-20,in=20] (botmid);
\draw [spinning] (topmid) to[out=-160,in=160] (botmid);
\end{tikzpicture}}
&=
\begin{pmatrix}
\scr{O}_1'\\
\scr{O}_1\ \scr{W}
\end{pmatrix}^{(n) (p)} \de_{\scr{O}_1'\scr{O}_1''}\,,
\label{eq:bubble}
\]
which gives a contribution proportional to the identity in the case that the representations $\scr{O}_1^\prime$ and $\scr{O}_1^{\prime\prime}$ coincide and vanishes otherwise. 

Such bubble coefficients may be determined for a specific set of weight-shifting operators of our choice by acting with the composition $\scr{D}_{X_1\, A}^{(n)}\scr{D}_{X_1}^{(p)A}$ on some 2-point function under consideration. For the case of interest here, we consider the operators in the fundamental vector representation and act with \Eq{bubble} on the 2-point function of symmetric traceless operators, which gives
\[
{}& 
\scr{D}_{X_1\, A}^{(-\delta\De, -\delta\ell)}
\scr{D}_{X_1}^{(\delta\De, \delta\ell)A}
 \vev{\scr{O}_{\De,\ell} (X_1) \scr{O}_{\De,\ell} (X_2)}=
\begin{pmatrix}
\scr{O}_{\De,\ell}\\
\scr{O}_{\De+\delta\De,\ell+\delta\ell} \ \scr{V}
\end{pmatrix}^{(-\delta\De, -\delta\ell)(\delta\De, \delta\ell)}
\vev{\scr{O}_{\De,\ell} (X_1) \scr{O}_{\De,\ell} (X_2)}\,.
\]
We find that only the coefficients corresponding to equal and opposite shifts are nonzero; the rest vanish. 
The nonzero coefficients are given by
\begin{align}
\begin{pmatrix}
\scr{O}_{\De,\ell}\\
\scr{O}_{\De+1,\ell}\ \scr{V}
\end{pmatrix}^{\!\!(-0)(+0)}
\hspace{-10pt}
&= -(\De-1)(\De+\ell)(d+\ell-\De-2)(d-2\De-2)\,,
\nonumber
\\
\begin{pmatrix}
\scr{O}_{\De,\ell}\\
\scr{O}_{\De-1,\ell}\ \scr{V}
\end{pmatrix}^{\!\!(+0)(-0)}
\hspace{-10pt}
&= (2-\De-\ell)(\De-\ell-d)(d-\De-1)(d+2-2\De)\,,
\nonumber
\\
\begin{pmatrix}
\scr{O}_{\De,\ell}\\
\scr{O}_{\De,\ell+1}\ \scr{V}
\end{pmatrix}^{\!\!(0-)(0+)}
\hspace{-10pt}
&= (\De+\ell)(d+2\ell)(d+\ell-\De)(2-d-\ell)\,,
\nonumber
\\
\begin{pmatrix}
\scr{O}_{\De,\ell}\\
\scr{O}_{\De,\ell-1}\ \scr{V}
\end{pmatrix}^{\!\!(0+)(0-)}
\hspace{-10pt}
&=\ell(d-4+2\ell)(2-\ell-\De)(d-2+\ell-\De)\,.
\label{eq:bubble_symbols_bulk}
\end{align}
These formulas provide the bubble coefficients for the vector representation. They are ubiquitous in the computations performed here and will be used extensively throughout this work. 

In addition, there is a variant of the 3-point crossing relation for special cases where the 3-point correlation function of interest features a unique 3-point structure, such as the (scalar)-(scalar)-(spin-$\ell$) 3-point structure. Such a relation involves changing only the dimension $\De$ but not the spin $\ell$. In particular, it is given by 
\[
\eql{alternative crossing}
&\scr{D}_X^{(\de\De,\de\ell)A}
\vev{ \scr{O}_{\De_1}(X_1) \scr{O}_{\De_2}(X_2) \scr{O}_{\De-\de\De,\ell-\de\ell}(X)  }\nonumber\\
&=
\sum_{\delta{\De_1} = \pm 1} 
\left\{
\begin{matrix}
\scr{O}_{\De_1} & \scr{O}_{\De_2} & \scr{O}_{\De_1-\delta{\De_1}} \\
\scr{O}_{\De,\ell} & \scr{V} & \scr{O}_{\De-\de\De,\ell-\de\ell}
\end{matrix}
\right\}^{(\de\De,\de\ell)}_{(\delta{\De_1}, 0)}
 \scr{D}_{X_1}^{(\delta{\De_1}, 0)A }\vev{ \scr{O}_{\De_1-\delta{\De_1}}(X_1) \scr{O}_{\De_2}(X_2) \scr{O}_{\De,\ell}(X)}\nonumber\\
&+ (-1)^{\de\ell}
\sum_{\delta{\De_2} = \pm 1}
\left\{
\begin{matrix}
\scr{O}_{\De_2} & \scr{O}_{\De_1} & \scr{O}_{\De_2-\delta{\De_2}}  \\
\scr{O}_{\De,\ell} & \scr{V} & \scr{O}_{\De-\de\De,\ell-\de\ell}
\end{matrix}
\right\}^{(\de\De,\de\ell)}_{(\delta{\De_2}, 0)}
 \scr{D}_{X_2}^{(\delta{\De_2}, 0)A}\vev{ \scr{O}_{\De_1}(X_1) \scr{O}_{\De_2-\delta{\De_2}}(X_2) \scr{O}_{\De,\ell}(X)} \,.
\]
Diagrammatically this is the statement 
\beq
\diagramEnvelope{\begin{tikzpicture}[anchor=base, baseline]
	\node (vertL) at (0,0) [twopt] {};
	\node (vertR) at (1.5,-0.12) [threept] {$m$};
	\node (opO1) at (-0.5,-1) [below] {$\scr{O}_1$};
	\node (opO2) at (-0.5,1) [above] {$\scr{O}_2$};
	\node (opO3) at (2,1) [above] {$\scr{O}_3$};
	\node (opW) at (2,-1) [below] {$\scr{W}$};	
	\node at (0.75,0.1) [above] {$\scr{O}_3'$};	
	\draw [scalar] (vertL)-- (opO1);
	\draw [scalar] (vertL)-- (opO2);
	\draw [spinning] (vertL)-- (vertR);
	\draw [spinning] (vertR)-- (opO3);
	\draw [finite with arrow] (vertR)-- (opW);
\end{tikzpicture}}
	\hspace{-0.3cm}=
	\sum_{\scr{O}_1',n}
	\left\{
		\begin{matrix}
		\scr{O}_1 & \scr{O}_2 & \scr{O}_1' \\
		\scr{O}_3 & \scr{W} & \scr{O}_3'
		\end{matrix}
	\right\}^{\uniq (m)}_{\uniq (n)}
\hspace{-0.8cm}
\diagramEnvelope{\begin{tikzpicture}[anchor=base,baseline]
	\node (vertU) at (0,0.7) [twopt] {};
	\node (vertD) at (0,-0.7) [threept] {$n$};
	\node (opO1) at (-1,-1.5) [below] {$\scr{O}_1$};
	\node (opO2) at (-1,1.5) [above] {$\scr{O}_2$};
	\node (opO3) at (1,1.5) [above] {$\scr{O}_3$};
	\node (opW) at (1,-1.5) [below] {$\scr{W}$};	
	\node at (0.1,0) [right] {$\scr{O}_1'$};	
	\draw [scalar] (vertD)-- (opO1);
	\draw [scalar] (vertU)-- (opO2);
	\draw [scalar] (vertU)-- (vertD);
	\draw [spinning] (vertU)-- (opO3);
	\draw [finite with arrow] (vertD)-- (opW);
\end{tikzpicture}}
\hspace{-0.3cm}+  (-1)^{\ell-\ell'}
	\sum_{\scr{O}_2',n}
	\left\{
		\begin{matrix}
		\scr{O}_2 & \scr{O}_1 & \scr{O}_2' \\
		\scr{O}_3 & \scr{W} & \scr{O}_3'
		\end{matrix}
	\right\}^{\uniq (m)}_{\uniq (n)}
\hspace{-0.8cm}
\diagramEnvelope{\begin{tikzpicture}[anchor=base,baseline]
	\node (vertU) at (0,-0.7) [twopt] {};
	\node (vertD) at (0,0.7) [threept] {$n$};
	\node (opO1) at (-1,1.5) [above] {$\scr{O}_2$};
	\node (opO2) at (-1,-1.5) [below] {$\scr{O}_1$};
	\node (opO3) at (1,-1.5) [below] {$\scr{O}_3$};
	\node (opW) at (1,1.5) [above] {$\scr{W}$};	
	\node at (0.1,0) [right] {$\scr{O}_2'$};	
	\draw [scalar] (vertD)-- (opO1);
	\draw [scalar] (vertU)-- (opO2);
	\draw [scalar] (vertU)-- (vertD);
	\draw [spinning] (vertU)-- (opO3);
	\draw [finite with arrow] (vertD)-- (opW);
\end{tikzpicture}}
\hspace{-0.3cm},
\label{eq:6jdefinition_single_ts_appendix} 
\eeq
where dashed lines have been used to indicate the scalar operators $\scr{O}_1$ and $\scr{O}_2$, while $\ell$ and $\ell'$ denote the spin of $\scr{O}_3$ and $\scr{O}_3'$, respectively.

It is straightforward to obtain the respective $6j$ coefficients for the vector representation $\scr{W}= \scr{V}$. These are given by
\begin{align}
\eql{special 6j symbols}
\left\{
\begin{matrix}
\scr{O}_{\De_1} & \scr{O}_{\De_2} & \scr{O}_{\De_1-1} \nonumber\\
\scr{O}_{\De,\ell} & \scr{V} & \scr{O}_{\De-1,\ell}
\end{matrix}
\right\}^{\uniq (+0)}_{\uniq(+0)} 
={}&
\frac{(\Delta -2) \left(\Delta -\Delta _1+\Delta _2+\ell-1\right) \left(d-\Delta +\Delta _1-\Delta _2+\ell-1\right)}{2 \left(\Delta
   _1-2\right) \left(d-2 \Delta _1\right) \left(d-\Delta _1-1\right)}\,,\nonumber\\
\left\{
\begin{matrix}
\scr{O}_{\De_1} & \scr{O}_{\De_2} & \scr{O}_{\De_1+1} \\
\scr{O}_{\De,\ell} & \scr{V} & \scr{O}_{\De-1,\ell}
\end{matrix}
\right\}^{\uniq (+0)}_{ \uniq (-0)} 
={}& 
-\frac{(\Delta -2) \left(\Delta +\Delta _1-\Delta _2+\ell-1\right) \left(-d+\Delta +\Delta _1-\Delta _2-\ell+1\right) }{2 \left(d-2 \Delta _1\right)}\nonumber\\
&\times
\left(-2
   d+\Delta +\Delta _1+\Delta _2-\ell+1\right) \left(-d+\Delta +\Delta _1+\Delta _2+\ell-1\right)\,, \nonumber\\
\left\{
\begin{matrix}
\scr{O}_{\De_1} & \scr{O}_{\De_2} & \scr{O}_{\De_1-1} \\
\scr{O}_{\De,\ell} & \scr{V} & \scr{O}_{\De+1,\ell}
\end{matrix}
\right\}^{\uniq (-0)}_{\uniq (+0)} 
={}&
\frac{1}{2 (\Delta -1) \left(\Delta _1-2\right) \left(d-2 \Delta _1\right) \left(d-\Delta _1-1\right)}, \nonumber\\
\left\{
\begin{matrix}
\scr{O}_{\De_1} & \scr{O}_{\De_2} & \scr{O}_{\De_1+1} \\
\scr{O}_{\De,\ell} & \scr{V} & \scr{O}_{\De+1,\ell}
\end{matrix}
\right\}^{\uniq (-0)}_{\uniq (-0)} 
={}& 
-\frac{\left(-\Delta +\Delta _1+\Delta _2+\ell-1\right) \left(d+\Delta -\Delta _1-\Delta _2+\ell-1\right)}{2 (\Delta -1) \left(d-2
   \Delta _1\right)}\,, 
 \nonumber\\
\left\{
\begin{matrix}
\scr{O}_{\De_1} & \scr{O}_{\De_2} & \scr{O}_{\De_1-1} \\
\scr{O}_{\De,\ell} & \scr{V} & \scr{O}_{\De,\ell-1}
\end{matrix}
\right\}^{\uniq (0+)}_{\uniq (+0)} 
={}&
\frac{\Delta -\Delta _1+\Delta _2+\ell-1}{2 \left(\Delta _1-2\right) \ell \left(d-2 \Delta _1\right) \left(d-\Delta _1-1\right)}\,, \nonumber\\
\left\{
\begin{matrix}
\scr{O}_{\De_1} & \scr{O}_{\De_2} & \scr{O}_{\De_1+1} \\
\scr{O}_{\De,\ell} & \scr{V} & \scr{O}_{\De,\ell-1}
\end{matrix}
\right\}^{\uniq (0+)}_{ \uniq (-0)} 
={}& 
-\frac{\left(\Delta +\Delta _1-\Delta _2+\ell-1\right) \left(-\Delta +\Delta _1+\Delta _2+\ell-1\right) }{2 \ell \left(d-2 \Delta _1\right)}\nonumber\\
&\times
\left(-d+\Delta +\Delta
   _1+\Delta _2+\ell-1\right)\,,\nonumber\\
\left\{
\begin{matrix}
\scr{O}_{\De_1} & \scr{O}_{\De_2} & \scr{O}_{\De_1-1} \\
\scr{O}_{\De,\ell} & \scr{V} & \scr{O}_{\De,\ell+1}
\end{matrix}
\right\}^{\uniq (0-)}_{\uniq (+0)} 
={}&
\frac{(\ell+1) \left(d-\Delta +\Delta _1-\Delta _2+\ell-1\right)}{2 \left(\Delta _1-2\right) \left(d-2 \Delta _1\right)
   \left(d-\Delta _1-1\right)}\,,
\nonumber
\end{align}
\begin{align}
\left\{
\begin{matrix}
\scr{O}_{\De_1} & \scr{O}_{\De_2} & \scr{O}_{\De_1+1} \\
\scr{O}_{\De,\ell} & \scr{V} & \scr{O}_{\De,\ell+1}
\end{matrix}
\right\}^{\uniq (0-)}_{ \uniq (-0)} 
={}& 
-\frac{ \left(2 d-\Delta -\Delta _1-\Delta _2+\ell-1\right) \left(d+\Delta -\Delta _1-\Delta _2+\ell-1\right) }{2 \left(d-2 \Delta _1\right)}\nonumber\\
&\times
(\ell+1)\left(d-\Delta -\Delta _1+\Delta _2+\ell-1\right)\,.
\end{align}
These will also prove useful in the analysis below.

We next review how to construct conformal blocks in the context of the weight-shifting formalism.

\subsection{Gluing 3-point functions to form conformal blocks}

A standard way to encode a general conformal block is to express it as the conformal integral of a product of 3-point functions. For concreteness, let us consider the conformal block for scalar exchange in a purely scalar 4-point function. As discussed above, this object has the form
\[
W_{\De, 0; \De_i}(X_i)=\dfrac{1}{\scr{N}_{\scr{O}}} \myint 
D^d X D^d Y\vev{\phi_{\Delta_1}(X_1)\phi_{\Delta_2}(X_2)\scr{O}(X)} \frac{1}{(-2 X\cdot Y)^{d-\De}}\vev{\scr{O}(Y) \phi_{\Delta_3}(X_3)\phi_{\Delta_4}(X_4)}\bigg|_{M=e^{2\pi i\varphi}}\,,
\]
where $\De= \De_{\scr{O}}$ and $M=e^{2\pi i\varphi}$ denotes the projection onto the appropriate monodromy invariant subspace. By expressing the conformal integral in a manifestly conformally invariant way, it becomes transparent that this object is indeed a proper solution to the conformal Casimir equation with the desired transformation properties expected for a conformal block. 

In~\cite{Karateev:2017jgd}, the operation which fuses or ``glues" the 3-point correlators $\vev{\phi_{\Delta_1}(X_1)\phi_{\Delta_2}(X_2)\scr{O}(X)}$ and $\vev{\scr{O}(Y) \phi_{\Delta_3}(X_3)\phi_{\Delta_4}(X_4)}$ together (i.e.~two correlators containing $\scr{O}$) is symbolized by
\[
\eql{gluing}
\ket{\scr{O}}\bowtie\bra{\scr{O}}\equiv \dfrac{1}{\scr{N}_{\scr{O}}}
 \myint D^d X D^d Y \ket{\scr{O}(X)} \frac{1}{(-2 X\cdot Y)^{d-\De}}\bra{\scr{O}(Y)}
  \quad=\quad
\diagramEnvelope{\begin{tikzpicture}[anchor=base,baseline]
	\node (opO) at (-1,0) [left] {$\scr{O}$};
	\node (opOprime) at (1,0) [right] {$\scr{O}$};
	\node (vert) at (0,0) [cross] {};
	\draw [scalar] (opO) -- (vert);
	\draw [scalar] (opOprime) -- (vert);
\end{tikzpicture}}.
\]
The normalization factor $\scr{N}_{\scr{O}}$ here is fixed by requiring that the action of the shadow integral on a 2-point function $\vev{\scr{O}(X_1)\scr{O}(X_2)}$ yield the identity transformation:
\[
\eql{eq:shadownormalization}
\diagramEnvelope{\begin{tikzpicture}[anchor=base,baseline]
	\node (opO) at (-1,0) [left] {$\scr{O}$};
	\node (vert) at (0,0) [twopt] {};
	\node (shad) at (1,0) [cross] {};
	\node (opOprime) at (2,0) [right] {$\scr{O}$};
	\draw [scalar] (vert) -- (opO);
	\draw [scalar] (vert) -- (shad);
	\draw [scalar] (opOprime) -- (shad);
\end{tikzpicture}}
&\quad=\quad
\diagramEnvelope{\begin{tikzpicture}[anchor=base,baseline]
	\node (opO) at (-1,0) [left] {$\scr{O}$};
	\node (opOprime) at (0,0) [right] {$\scr{O}$};
	\draw [scalar] (opOprime) -- (opO);
\end{tikzpicture}}.
\]

For the case of scalar exchange, this condition determines the normalization factor to be 
\[
\scr{N}_{\scr{O}} =\dfrac{\pi^d \Ga(\De-\frac{d}{2})\Ga(\frac{d}{2}-\De)}{\Ga(\De)\Ga(d-\De)}\,.
\]
More generally, for spinning operators, the operator $\scr{O}_{\De, \rho}$ is to be glued to the representation with which it has a nonvanishing 2-point function, i.e.~the dual-reflected one, $\ket{\scr{O}_{\De, \rho}}\bowtie\bra{\scr{O}^\dagger_{\De, \rho^\dagger} }$. Again, all we need to determine is the normalization condition for the spinning operators in question: 
\[
\diagramEnvelope{\begin{tikzpicture}[anchor=base,baseline]
	\node (opO) at (-1,0) [left] {$\scr{O}$};
	\node (vert) at (0,0) [twopt] {};
	\node (shad) at (1,0) [cross] {};
	\node (opOprime) at (2,0) [right] {$\scr{O}$};
	\draw [spinning] (vert) -- (opO);
	\draw [spinning] (vert) -- (shad);
	\draw [spinning] (opOprime) -- (shad);
\end{tikzpicture}}
&\quad=\quad
\diagramEnvelope{\begin{tikzpicture}[anchor=base,baseline]
	\node (opO) at (-1,0) [left] {$\scr{O}$};
	\node (opOprime) at (0,0) [right] {$\scr{O}$};
	\draw [spinning] (opOprime) -- (opO);
\end{tikzpicture}}.
\label{eq:shadownormalizationtwo}
\]
In terms of this notation, a general conformal block is given by the following form in this framework: 
\begin{equation}
W^{ab} \equiv \vev{\scr{O}_1 \scr{O}_2 \scr{O}}^{(a)} \bowtie {}^{(b)}\vev{\scr{O}^\dagger
\scr{O}_3 \scr{O}_4}=
\diagramEnvelope{\begin{tikzpicture}[anchor=base,baseline]
	\node (vertL) at (-2,0) [threept] {$a$};
	\node (vertR) at (2,-0.05) [threept] {$b$};
	\node (opO1) at (-2.5,-1) [below] {$\scr{O}_1$};
	\node (opO2) at (-2.5,1) [above] {$\scr{O}_2$};
	\node (opO3) at (2.5,1) [above] {$\scr{O}_3$};
	\node (opW) at (2.5,-1) [below] {$\scr{O}_4$};
	\node (shad) at (0,.085) [cross] {};
	\node at (1,0.1) [above] {$\scr{O}^\dag$};	
	\node at (-1,0.1) [above] {$\scr{O}$};	
	\draw [spinning] (vertL)-- (opO1);
	\draw [spinning] (vertL)-- (opO2);
	\draw [spinning] (vertL)-- (shad);
	\draw [spinning] (vertR) -- (shad);
	\draw [spinning] (vertR)-- (opO3);
	\draw [spinning] (vertR)-- (opW);
\end{tikzpicture}}.
\end{equation}

In a nutshell, the weight-shifting technique involves acting with specific combinations of weight-shifting operators on a given conformal block and then applying the two- and three-point crossing relations as needed in order to re-express the original block in terms of either (1) linear combinations of compositions of differential operators acting on blocks of fixed spin with shifted external scaling dimensions, or (2) linear combinations of lower-spin blocks with shifted external and, potentially, internal dimensions. 

The precise form of the compositions of the weight-shifting operators is determined by the 3-point tensor structures in question. In particular, the results for conformal blocks of the weight-shifting method may be grouped into two broad classes:
\[
&(1) \quad W^{a b}_{\De, \ell; \De_i}(X_i)= \sum_j  A^{a b}_j(\De, \De_i,\ell) 
\scr{D}_{j} W^{\text{seed}}_{\De, \ell; \De_i+ \de\De_{ij} }(X_i)\,, \nonumber\\
&(2) \quad W^{a b}_{\De, \ell; \De_i}(X_i)= \sum_j B_j(\De, \De_i,\ell) W^{a b}_{\De+  \de\De_j, \ell+ \de\ell_j; \De_i+ \de\De_{ij}  }(X_i)\,,
\]
where we have focused on 4-point conformal blocks for concreteness. Here 
$\de\De_{ij}$, $\de\De_{j}$, and $\de\ell_j$ denote finite shifts in $\De_i$, $\De$, and $\ell$, respectively, while $A_j$ and $B_j$ are coefficients built from products of $6j$ symbols. The first relation re-expresses a generic block in some arbitrary representation in terms of a set of differential operators $\scr{D}_{j}$ acting on a seed block with shifted external dimensions but fixed spin, e.g.~a block with spin-$\ell$ exchange. Here $\scr{D}_{j}$ would be built out of the operators $\scr{D}^{(m)A}_{X_i}$. Meanwhile, the second relation expands a higher-spin block in terms of lower-spin ones (assuming $\delta\ell_j < 0$) with shifted external and possibly exchanged dimensions. This kind of expression is a recursion relation for conformal blocks.

To implement these types of forms, we require a mechanism for moving differential operators onto the other side of the shadow integral, that is, effectively for integrating by parts. For this, there exists a convenient rule that naturally arises in the context of this formalism. Diagrammatically, this is the statement that  
\[
\diagramEnvelope{\begin{tikzpicture}[anchor=base,baseline]
	\node (vertU) at (1,0) [cross] {};
	\node (vertD) at (0,-0.08) [threept] {$c$};
	\node (opO1) at (-1,0) [left] {$\scr{O}$};
	\node (opO3) at (2,0) [right] {$\scr{O}'^\dag$};
	\node (opW) at (0,-1) [below] {$\scr{W}$};	
	\node at (0.6,0.1) [above] {$\scr{O}'$};
	\draw [spinning] (opO1)-- (vertD);
	\draw [spinning] (vertD)-- (vertU);
	\draw [spinning] (opO3)-- (vertU);
	\draw [finite with arrow] (vertD)-- (opW);
\end{tikzpicture}}
	\quad=\quad
	\sum_{m}
	\left\{
		\begin{matrix}
		\scr{O}^\dag & \mathds{1} & \scr{O}'^\dag \\
		\scr{O}' & \scr{W} & \scr{O}
		\end{matrix}
	\right\}^{\uniq (c)}_{\uniq (m)}
\diagramEnvelope{\begin{tikzpicture}[anchor=base,baseline]
	\node (vertL) at (0,0) [cross] {};
	\node (vertR) at (1,-0.08) [threept, inner sep=1pt] {$m$};
	\node (opO1) at (-1,0) [left] {$\scr{O}$};
	\node (opO3) at (2,0) [right] {$\scr{O}'^\dag$};
	\node (opW) at (1,-1) [below] {$\scr{W}$};
	\node at (0.4,0.1) [above] {$\scr{O}^\dag$};
	\draw [spinning] (opO1)-- (vertL);
	\draw [spinning] (vertR)-- (vertL);
	\draw [spinning] (opO3)-- (vertR);
	\draw [finite with arrow] (vertR)-- (opW);
\end{tikzpicture}},
\label{eq:integrationbyparts}
\]
which essentially encodes two integrations by parts in \Eq{gluing}, thus empowering us to move the operators from one side of the $\bowtie$ to the other.

Symbolically, this has the form
\[
\eql{by parts rule}
|\scr{D}^{(c)A}\scr{O}\rangle \bowtie  \langle \scr{O}'^\dag|=\sum_{m}
\left\{
\begin{matrix}
\scr{O}^\dag & \mathds{1} &  \scr{O}'^\dag \\
\scr{O}' & \scr{W} & \scr{O}
\end{matrix}
\right\}^{\uniq (c)}_{\uniq (m)}
|\scr{O}\rangle \bowtie \, \langle \scr{D}^{(m)A}\scr{O}'^\dag| \,.
\]
This integration-by-parts rule is none other than the aforementioned 2-point crossing relation \Eq{twoptcrossing} featuring the operators $\scr{O}$ and $\scr{O}^\prime$. It will prove to be a powerful tool in the analysis that follows.

\section{Recursion relations in the weight-shifting operator formalism}
\label{sec:recursion}

In this section, we derive a set of recursion relations that enable us to build up the 5-point conformal block for the exchange of a pair of arbitrary symmetric traceless operators $\scr{O}_{\De, \ell}, \scr{O}'_{\De', \ell'}$ in terms of the blocks for the exchange of scalar operators $\scr{O}_{\De+m}, \scr{O}_{\De'+n}$ with shifted dimensions. We begin by describing the procedure for extracting recursion relations within the context of the weight-shifting operator formalism.

\subsection{Recursion relations for 4-point conformal blocks}
 
To begin, let us review how to derive the familiar recursion relation for scalar conformal blocks originally obtained by Dolan and Osborn in \cite{Dolan:2011dv}. The derivation using weight-shifting operators was given in~\cite{Karateev:2017jgd}. The 4-point scalar conformal block is defined as
\[
\eql{4-point}
\vev{\phi_{\Delta_1}(X_1)
\phi_{\Delta_2}(X_2)|\scr{O}_{\De, \ell}
|\phi_{\Delta_3}(X_3)\phi_{\Delta_4}(X_4)}
=\frac{1}{(X_{12})^{\sfrac{1}{2}(\De_1+\De_2)}
(X_{34})^{\sfrac{1}{2}(\De_3+\De_4)}}\nonumber\\
\times \bigg(\frac{X_{24}}{X_{14}}\bigg)^{\De_{12}/2}
\bigg(\frac{X_{14}}{X_{13}}\bigg)^{\De_{34}/2}
G_{\De,\ell}(u, v)\,, 
\] 
where $X_{ij}= -2 X_i\cdot X_j$ and $\De_{ij}= \De_i -\De_j$. 
 
We now consider acting on \Eq{4-point} with the combination of operators
\[
-2 (\scr{D}_{X_1}^{(-0)}  \cdot \scr{D}_{X_4}^{(-0)}) =  -2 X_1\cdot X_4= X_{14}\,,
\] 
which is formed by contracting $\scr{D}_{X_1}^{(-0)}{}^{A}$ with $ \scr{D}_{X_4}^{(-0)}{}_A$.

This yields a 4-point function with the operator scaling dimensions at positions $1$ and $4$ shifted by $-1$, that is, $\De_1\to \De_1-1$ and $\De_4 \to \De_4-1$.
These shifts in $\De_1$ and $\De_4$ in turn require a shifted external prefactor. In order to work out the action of the operator $-2 (\scr{D}_{X_1}^{(-0)}  \cdot \scr{D}_{X_4}^{(-0)})$ on the block $G_{\De,\ell}(u, v)$, we may accordingly first multiply by $X_{14}$ and then remove the external prefactor for the new set of scaling dimensions. This leads to a new function of cross-ratios given by $u^{-1/2}G_{\De,\ell}(u, v)$, which can in turn be expanded in conformal blocks with shifted weights. This operation can be represented diagramatically by
\beq\label{eq:DOsimple}
\diagramEnvelope{\begin{tikzpicture}[anchor=base,baseline]
	\node (opO1) at (-1,-2) [below, inner sep=0pt] {$[\De_1-1,0]$};
	\node (opO1p) at (-0.5,-0.5) [left, inner sep=0pt] {$[\De_1,0]$};
	\node (opO2) at (-0.5,1) [above, inner sep=0pt] {$[\De_2,0]$};
	\node (vertWL) at (-0.5,-1) [threept] {};
	\node (vertW) at (2.5,-1) [threept] {};
	\node (vertL) at (0,0) [twopt] {};
	\node (shad) at (1,0) [cross] {};
	\node (vertR) at (2,0) [twopt] {};
	\node (opW) at (1,-1.1) [below] {$\myng{(1)}$};
	\node at (1,0.1) [above] {$[\De,\ell]$};
	\node (opO3) at (2.5,1) [above, inner sep=0pt] {$[\De_3,0]$};
	\node (opO4) at (3,-2) [below] {$[\De_4-1,0]$};
	\node (opO4p) at (2.5,-0.5) [right, inner sep=0pt] {$[\De_4,0]$};
	\draw [spinning] (vertL)-- (shad);
	\draw [spinning] (vertR)-- (shad);
	\draw [scalar] (vertR)-- (vertW);
	\draw [finite with arrow] (vertWL)-- (vertW);
	\draw [scalar] (vertW)-- (opO4);
	\draw [scalar] (vertR)--(opO3);
	\draw [scalar] (vertWL)-- (opO1);
	\draw [scalar] (vertL)--(vertWL);
	\draw [scalar] (vertL) -- (opO2);
	\end{tikzpicture}}.
\eeq
It is evident that the weight-shifting operators ``dress" the conformal block such that a component of the tensor product $\scr{W} \otimes \scr{O}_{\De, \ell}=\square \otimes \scr{O}_{\De, \ell}$ propagates from left to right. With this, one finds that the spin-$\ell$ block may be expanded in terms of scalar conformal blocks with shifted dimensions and spins, where the internal representations that appear are the symmetric traceless tensors appearing in the tensor product decomposition
\[
\square \otimes [\De, \ell]  = [\De-1, \ell] \oplus  [\De, \ell+1] \oplus [\De, \ell-1]
\oplus  [\De+1, \ell] \oplus\dots\,.
\]

An application of the three- and two-point crossing relations then gives rise to the following recursion relation: 
\[
\eql{D and O}
G_{\De,\ell}(u, v)= \frac{1}{s^{(14)} }\bigg(u^{-1/2} G_{\De, \ell-1}(u, v)\bigg|_{\De_1\to\De_1+1,\De_4\to\De_4+1}
- G_{\De-1,\ell-1}(u, v)\nonumber\\ -t^{(14)}  G_{\De,\ell-2}(u, v) - u^{(14)}   G_{\De+1,\ell-1}(u, v)\bigg)\,,
\]
where the coefficients $s^{(14)}  $, $t^{(14)}  $, $u^{(14)} $ are formed from appropriate combinations of $6j$ symbols and are explicitly given by
\[ 
\eql{coeffs standard}
s^{(14)}  &=\frac{\left(\De -\De _{12}+\ell \right) \left(\De
   +\De _{34}+\ell \right)}{2 (\De +\ell -1) (\De +\ell
   )}\,,\nonumber\\
t^{(14)}  &=\frac{\ell  (d+\ell -3) \left(d-\De +\De _{12}+\ell
   -2\right) \left(d-\De -\De _{34}+\ell -2\right)}{2 (d+2
   \ell -4) (d+2 \ell -2) (d-\De +\ell -2) (d-\De +\ell -1)}\,,\nonumber\\
u^{(14)} &=-\frac{(\De -1) (d-\De -2) \left(\De -\De _{12}+\ell
   \right) \left(\De +\De _{34}+\ell \right) \left(d-\De
   +\De _{12}+\ell -2\right) \left(d-\De -\De _{34}+\ell -2\right)}{4 (\De +\ell )(\De +\ell-1)
   (d-2 \De ) (d-2 (\De +1))(d-\De +\ell -2) (d-\De +\ell -1)}\,.
\]
After converting to the same conventions,\footnote{To match Eq.~(4.18) in~\cite{Dolan:2011dv}, we need to change to their conformal block normalization conventions, which introduces a factor $\mathcal{N}_{\delta\ell}= \dfrac{(-2)^{-\delta {\ell}} (d+\ell -2)_{\delta{\ell
   }}}{\left(\frac{d}{2}+\ell -1\right)_{\delta{\ell }}}$. This gives   
$ s = \mathcal{N}_1 s^{(14)}$, 
$ t = \mathcal{N}_{-1} t^{(14)}$,
$ u = \mathcal{N}_0 u^{(14)}$.
} we find agreement with the classical result Eq.~(4.18) in \cite{Dolan:2011dv}. In their analysis, Dolan and Osborn determine the values of the parameters $s$, $t$, and $u$ (which are proportional to the $s^{(14)}  $, $t^{(14)}  $, $u^{(14)} $ shown here) with $r=1$ in Eq.~(4.18) by examining the conformal block's behavior in the OPE limit $u\to 0$ and $v\to 1$. Meanwhile, here they arise directly as products of the appropriate expansion coefficients in the two- and 3-point crossing relations.

The above discussion serves as a concise demonstration of the procedure for deriving recursion relations for conformal blocks in the context of the weight-shifting operator formalism. In the analysis that follows, we will apply an analogous method for establishing recursion relations for conformal blocks for symmetric traceless tensor exchange in purely scalar 5-point functions. Indeed, our final results represent natural generalizations of the above relation \Eq{D and O}. 

\subsection{Recursion relations for 5-point conformal blocks}

In this section, we derive a variety of recursion relations applicable to the case of symmetric traceless tensor exchange in a 5-point function of scalar operators. The basic idea is that such a set of relations would allow one to express the block for $([\De,\ell], [\De', \ell'])$ exchange in terms of lower-spin blocks. 

\subsubsection{Recursion relations from weight-shifting operators}
To begin, we consider acting on the scalar 5-point function
\begin{align}
\vev{\phi_{\De_1}(X_1) &\phi_{\De_2}(X_2) |\scr{O}_{\De, \ell}|\Phi_{\De_3}(X_3) |
\scr{O}^\prime_{\De', \ell'}|\phi_{\De_4}(X_4) \phi_{\De_5}(X_5)} 
\nonumber\\
&=\vev{\phi_{\De_1}(X_1) \phi_{\De_2}(X_2) \scr{O}_{\De, \ell}}\bowtie \vev{\scr{O}_{\De, \ell}\Phi_{\De_3}(X_3) \scr{O}^\prime_{\De', \ell'}}\bowtie
\vev{ \scr{O}^\prime_{\De', \ell'}\phi_{\De_4}(X_4) \phi_{\De_5}(X_5)} 
 \nonumber\\
&=\sum_a\sum_{\scr{O}_{\De, \ell}} \sum_{\scr{O}^\prime_{\De', \ell'}} \la_{\phi_{\De_1} \phi_{\De_2}\scr{O}_{\De, \ell}}
\la^a_{\scr{O}_{\De, \ell}\Phi_{\De_3}\scr{O}^\prime_{\De', \ell'}}
\la_{\phi_{\De_4} \phi_{\De_5}\scr{O}^\prime_{\De', \ell'}}
W_{\De, \ell, \De', \ell'; \De_i}^{(a)}(X_i)\,,
\end{align}
where $W_{\De, \ell, \De', \ell'; \De_i}^{(a)}(X_i)$ has the convention-free form
\[
W_{\De, \ell, \De', \ell'; \De_i}^{(a)}(X_i)= P_{\De_i}(X_i) G^{(a)}_{\De, \ell, \De', \ell'}(u_i)\,,
\] 
with the following combination of weight-shifting operators:
\[
\eql{combination 13}
-2(\scr{D}_{X_1}^{(-0)}\cdot\scr{D}_{X_3}^{(-0)})\vev{\phi_{\De_1}(X_1) \phi_{\De_2}(X_2) |\scr{O}_{\De, \ell}|\Phi_{\De_3}(X_3) |
\scr{O}^\prime_{\De', \ell'}|\phi_{\De_4}(X_4) \phi_{\De_5}(X_5)} 
\nonumber\\=X_{13}\vev{\phi_{\De_1}(X_1) \phi_{\De_2}(X_2) |\scr{O}_{\De, \ell}|\Phi_{\De_3}(X_3) |
\scr{O}^\prime_{\De', \ell'}|\phi_{\De_4}(X_4) \phi_{\De_5}(X_5)}\,.
\] 
We set off the procedure by applying the 3-point crossing relation to the leftmost 3-point structure:
\[
 \scr{D}_{X_1}^{(-0)A} \vev{\phi_{\De_1}(X_1)\phi_{\De_2}(X_2)\scr{O}_{\De,\ell}(X_I)} =\sum_{n} \scr{A}_{(n)}^{(-0)}
\scr{D}^{(n)A}_{X_I} \vev{\phi_{\De_1-1}(X_1)\phi_{\De_2}(X_2)\scr{O}_{\De-\delta\Delta_n,\ell-\delta\ell_n}(X_I)}\,,
\]
where the relevant $6j$ symbol is given by
\[
\scr{A}_{(n)}^{(-0)}\equiv 
\bigg\{
\begin{matrix} 
\scr{O}_{\De,\ell} & \phi_{\De_2} & 
\scr{O}_{\De-\delta\De_n,\ell-\delta\ell_n} \\
\phi_{\De_1-1} & \scr{V} & \phi_{\De_1}
\end{matrix}\bigg\}^{\cdot (-0)}_{\cdot (n)}\,,
\]
with the shifted dimensions $[\De-\delta\De_n, \ell-\delta\ell_n]$ taking on values in the tensor product 
\[
\square \otimes [\De,\ell] =
[\De-1,\ell]\oplus
[\De,\ell+1] \oplus
[\De,\ell-1]\oplus
[\De+1,\ell]\oplus \dots\,.
\]
The dots refer to all non-symmetric traceless contributions, which are irrelevant here, as we are considering exclusively scalar external operators. As before, $\scr{V}$ denotes the vector representation $\scr{V} = \square$.

Upon applying the 3-point crossing relation \Eq{3ptCFTcrossing}, we find that the action of $\scr{D}_{X_1}^{(-0)A}$ unwraps into the contributions
\[
\eql{6j symbols A}
 \scr{D}_{X_1}^{(-0)A} \vev{\phi_{\De_1}(X_1)\phi_{\De_2}(X_2)\scr{O}_{\De,\ell}(X_I)} =
\scr{A}_{(+0)}^{(-0)}
\scr{D}_{X_I}^{(+0)A} \vev{\phi_{\De_1-1}(X_1)\phi_{\De_2}(X_2)\scr{O}_{\De-1,\ell}(X_I)}  \nonumber\\
+ \scr{A}_{(0-)}^{(-0)}
\scr{D}_{X_I}^{(0-)A} \vev{\phi_{\De_1-1}(X_1)\phi_{\De_2}(X_2)\scr{O}_{\De,\ell+1}(X_I)} \nonumber\\
+ \scr{A}_{(0+)}^{(-0)}
\scr{D}_{X_I} ^{ (0+)A}\vev{\phi_{\De_1-1}(X_1)\phi_{\De_2}(X_2)\scr{O}_{\De,\ell-1}(X_I)} \nonumber\\
+  \scr{A}_{(-0)}^{(-0)}
\scr{D}_{X_I}^{(-0)A}  \vev{\phi_{\De_1-1}(X_1)\phi_{\De_2}(X_2)\scr{O}_{\De+1,\ell}(X_I)}\,, 
\]
where $X_I$ denotes the position of the exchanged operator $\scr{O}_{\De,\ell}(X_I)$, which is an internal coordinate that is integrated over. 

We may extract a given $6j$ symbol $\scr{A}_{(n)}^{(-0)}$ by acting on both sides of \Eq{6j symbols A} with the weight-shifting operator $\scr{D}_{X_I\,A}^{(\bar{n})}$, which carries a shift opposite to $n$, and then isolating the coefficient. To give an example, let us consider acting with the simplest operator that can isolate a term on the right-hand side, namely $\scr{D}_{X_I\,A}^{(-0)}  = X_{I\,A}$. With this, we find
\[
\scr{D}_{X_I\,A}^{(-0)}  \scr{D}_{X_1}^{(-0)A} \vev{\phi_{\De_1}(X_1)\phi_{\De_2}(X_2)\scr{O}_{\De,\ell}(X_I)} =
\scr{A}_{(+0)}^{(-0)}
\scr{D}_{X_I\,A}^{(-0)}  \scr{D}_{X_I}^{(+0) A}\vev{\phi_{\De_1-1}(X_1)\phi_{\De_2}(X_2)\scr{O}_{\De-1,\ell}(X_I)}  \nonumber\\
+  \scr{A}_{(0-)}^{(-0)}
\scr{D}_{X_I\,A}^{(-0)}  \scr{D}_{X_I}^{(0-)A} \vev{\phi_{\De_1-1}(X_1)\phi_{\De_2}(X_2)\scr{O}_{\De,\ell+1}(X_I)} \nonumber\\
+ \scr{A}_{(0+)}^{(-0)}
\scr{D}_{X_I\,A}^{(-0)}  \scr{D}_{X_I}^{(0+)A} \vev{\phi_{\De_1-1}(X_1)\phi_{\De_2}(X_2)\scr{O}_{\De,\ell-1}(X_I)} \nonumber\\
+  \scr{A}_{(-0)}^{(-0)}
\scr{D}_{X_I\,A}^{(-0)}  \scr{D}_{X_I}^{(-0)A} \vev{\phi_{\De_1-1}(X_1)\phi_{\De_2}(X_2)\scr{O}_{\De+1,\ell}(X_I)}\,, 
\]
where the right-hand side now involves the bubble coefficients for $\scr{V}$, which can be referenced in \Eq{bubble_symbols_bulk}.

The only nonzero bubble coefficient is given by
\[
b_{\De, \ell}^{(-0)(+0)}\equiv \bigg(\begin{matrix} 
\scr{O}_{\De-1,\ell} \\
\scr{O}_{\De,\ell} \,\, \scr{V}
\end{matrix}\bigg)^{(-0)(+0)}=(\De-2) (2 \De-d)
   (\De+\ell-1)(d-\De+\ell-1)\,.
\]
As discussed above, the other three coefficients that do not involve opposite shifts vanish, i.e.
\[
b_{\De, \ell}^{(-0)(0-)} =  0\,,\quad b_{\De, \ell}^{(-0)(0+)} =  0\,, \quad b_{\De, \ell}^{(-0)(-0)} =  0 \,.
\]
With this, we arrive at the statement
\[
 \scr{D}_{X_I\, A}^{(-0)} \scr{D}_{X_1}^{(-0)A} \vev{\phi_{\De_1}(X_1)\phi_{\De_2}(X_2)\scr{O}_{\De,\ell}(X_I)} =
\scr{A}_{(+0)}^{(-0)}
 b_{\De, \ell}^{(-0)(+0)} \vev{\phi_{\De_1-1}(X_1)\phi_{\De_2}(X_2)\scr{O}_{\De-1,\ell}(X_I)}\,. 
\]
We may now use this equation to determine the $6j$ symbol $\scr{A}_{(+0)}^{(-0)}$ by invoking the explicit form of the 3-point structures. This then elucidates the procedure for extracting the $\scr{A}_{(n)}^{(-0)}$ coefficients. 
Proceeding in this fashion and solving for the respective $6j$ symbols, we ultimately obtain
\[ 
\scr{A}_{(+0)}^{(-0)} &= \frac{1}{2 (\De -2) (d-2 \De ) (\De
   +\ell-1) (d-\De +\ell-1)}\,, \nonumber\\
\scr{A}_{(0-)}^{(-0)} &= \frac{\De_{12}-(\De+\ell)}{2 (\ell+1) (d+2 \ell-2)  (\De +\ell-1) (d-\De +\ell-1)}\,,\nonumber\\
\scr{A}_{(0+)}^{(-0)} &= \frac{\ell (d-\De +\De_{12}+\ell-2)}{2 (d+2 \ell-2)
   (\De +\ell-1) (d-\De +\ell-1)}\,, \nonumber\\
\scr{A}_{(-0)}^{(-0)} &= \frac{(\De -1) (\De_{12}-(\De+\ell)) (d-\De
   +\De_{12}+\ell-2)}{2 (d-2 \De ) (\De +\ell-1)(d-\De +\ell-1)}\,.
\]

The next step in our algorithm is to push each of the operators $\scr{D}_{X_I}^{(n) A}$ in \Eq{6j symbols A} through the shadow integral. To do so, we invoke the integration-by-parts rule in \Eq{by parts rule} to move the operator in question across $\bowtie$. For example, for the case of $\scr{D}_{X_I}^{(+0)A}$, the rule in \Eq{by parts rule} reads
\[
\eql{invoke by parts}
\ket{ \scr{D}_{X_I}^{(+0)A}\scr{O}_{\De-1,\ell}}
\bowtie \bra{\scr{O}_{\De,\ell}} = 
 B_{(+0)(-0)}  \ket{ \scr{O}_{\De-1,\ell}}\bowtie 
\bra{\scr{D}_{X_I}^{(-0)A}
\scr{O}_{\De, \ell}}\,,
\]
where $B_{(+0)(-0)}$ is the 2-point $6j$ symbol
\[
 B_{(+0)(-0)}\equiv \bigg\{
\begin{matrix} 
\scr{O}_{\De-1,\ell} \\
\scr{O}_{\De,\ell}
\end{matrix}\bigg\}^{(+0)}_{(-0)}\,.
\] 
As mentioned above, the 2-point $6j$ symbols for the vector representation are given by \Eq{6j_CFT}. We thus have the respective coefficients
\[
 B_{(+0)(-0)}&\equiv \bigg\{
\begin{matrix} 
\scr{O}_{\De-1,\ell} \\
\scr{O}_{\De,\ell}
\end{matrix}\bigg\}^{(+0)}_{(-0)} = 2 (\De -2) (d-2 \De ) (\De +\ell-1)
   (d-\De +\ell-1)\,,\nonumber\\
B_{(0-)(0+)}&\equiv \bigg\{
\begin{matrix} 
\scr{O}_{\De,\ell+1}  \\
\scr{O}_{\De,\ell} 
\end{matrix}\bigg\}^{(0-)}_{(0+)} = \frac{(\ell+1) (d+2 (\ell-1))
   (-d+\De-\ell+1)}{\De+\ell}\,,\nonumber\\
B_{(0+)(0-)}&\equiv \bigg\{
\begin{matrix} 
\scr{O}_{\De,\ell-1}   \\
\scr{O}_{\De,\ell}  
\end{matrix}\bigg\}^{(0+)}_{(0-)} = \frac{\De+\ell-1}{\ell (d+2( \ell-2)) (2-d+\De-\ell)}\,,
\nonumber\\
B_{(-0)(+0)}&\equiv \bigg\{
\begin{matrix} 
\scr{O}_{\De+1,\ell} \\
\scr{O}_{\De,\ell} 
\end{matrix}\bigg\}^{(-0)}_{(+0)} = \frac{1}{2 (\De-1)(\De+\ell) (d-2
   (\De+1))  (d-\De+\ell-2)}\,.
\]
With these coefficients in hand, at this point, we arrive at  
\[
\eql{after 1}
  \scr{D}_{X_1}^{(-0)A}  \vev{\phi_{\De_1}(X_1)\phi_{\De_2}(X_2)\scr{O}_{\De,\ell}(X_I)}\bowtie {} \vev{\scr{O}_{\De,\ell}(X_I)\Phi_{\De_3}(X_3)\scr{O}'_{\De',\ell'}(X_J)}^{(a)}   
 =\nonumber\\
\scr{A}_{(+0)}^{(-0)}B_{(+0)(-0)}
 \vev{\phi_{\De_1-1}(X_1)\phi_{\De_2}(X_2)\scr{O}_{\De-1,\ell}(X_I)}  
\bowtie {} \scr{D}_{X_I}^{(-0)A}\vev{\scr{O}_{\De,\ell}(X_I)\Phi_{\De_3}(X_3)\scr{O}'_{\De',\ell'}(X_J)}^{(a)}   
\nonumber\\
+ \scr{A}_{(0-)}^{(-0)}
 B_{(0-)(0+)} \vev{\phi_{\De_1-1}(X_1)\phi_{\De_2}(X_2)\scr{O}_{\De,\ell+1}(X_I)}
\bowtie {} \scr{D}_{X_I}^{(0+) A}\vev{\scr{O}_{\De,\ell}(X_I)\Phi_{\De_3}(X_3)\scr{O}'_{\De',\ell'}(X_J)}^{(a)}   
 \nonumber\\
+ \scr{A}_{(0+)}^{(-0)} B_{(0+)(0-)}
 \vev{\phi_{\De_1-1}(X_1)\phi_{\De_2}(X_2)\scr{O}_{\De,\ell-1}(X_I)}
\bowtie {} \scr{D}_{X_I}^{(0-) A}\vev{\scr{O}_{\De,\ell}(X_I)\Phi_{\De_3}(X_3)\scr{O}'_{\De',\ell'}(X_J)}^{(a)}   
 \nonumber\\
+  \scr{A}_{(-0)}^{(-0)} B_{(-0)(+0)}
 \vev{\phi_{\De_1-1}(X_1)\phi_{\De_2}(X_2)\scr{O}_{\De+1,\ell}(X_I)}
\bowtie {} \scr{D}_{X_I}^{(+0) A}\vev{\scr{O}_{\De,\ell}(X_I)\Phi_{\De_3}(X_3)\scr{O}'_{\De',\ell'}(X_J)}^{(a)}   \,.
\]
We next exploit the 3-point crossing relation \Eq{3ptCFTcrossing} once again, this time to express each of the $3$-point structures $\scr{D}_{X_I}^{(b)A}\vev{\scr{O}_{\De,\ell}(X_I)\Phi_{\De_3}(X_3)\scr{O}'_{\De',\ell'}(X_J)}^{(a)}$ in terms of a linear combination of objects $\scr{D}_{X_3}^{(n)A}\vev{\scr{O}_{\De+\delta\De_b,\ell+\delta\ell_b}(X_I)\Phi_{\Delta_3-\delta\De_n,-\delta\ell_n}(X_3)\scr{O}'_{\De',\ell'}(X_J)}^{(m)}$. The intention is to transform the action of a differential operator at an internal point ($X_I$) into the appropriate action at an external point $(X_3)$. That is,
\begin{align}
\scr{D}_{X_I}^{(b)A}\vev{&\scr{O}_{\De,\ell}(X_I)\Phi_{\De_3}(X_3)\scr{O}'_{\De',\ell'}(X_J)}^{(a)}\nonumber\\
=& \sum_{m, n}
 \bigg\{
\begin{matrix} 
\Phi_{\De_3} & \scr{O}'_{\De',\ell'} &  \Phi_{\De_3-\delta\De_n,-\delta\ell_n}\\
\scr{O}_{\De+\delta\De_b,\ell+\delta\ell_b} & \scr{V} & \scr{O}_{\De,\ell}
\end{matrix}\bigg\}^{(a) (b)}_{(m) (n)} \nonumber\\
&\times \scr{D}_{X_3}^{(n)A}\vev{\scr{O}_{\De+\delta\De_b,\ell+\delta\ell_b}(X_I)\Phi_{\De_3-\delta\De_n,-\delta\ell_n}(X_3)\scr{O}'_{\De',\ell'}(X_J)}^{(m)}\,,
\end{align}
where the shifts $[\Delta_3 - \delta\De_n, -\delta\ell_n]$ take on values in the tensor product
\[
\square \otimes [\De_3, 0] =
[\De_3-1,0]\oplus
[\De_3, 1] \oplus
[\De_3+1, 0]\,.
\]
Explicitly, we have 
\begin{align}
\eql{3-point reln again}
\scr{D}_{X_I}^{(b)A}\vev{&\scr{O}_{\De,\ell}(X_I)\Phi_{\De_3}(X_3)\scr{O}'_{\De',\ell'}(X_J)}^{(a)}\nonumber\\
&= \sum_m
 \bigg\{
\begin{matrix} 
\Phi_{\De_3} & \scr{O}'_{\De',\ell'} &  \Phi_{\De_3-1}\\
\scr{O}_{\De+\delta\De_b,\ell+\delta\ell_b} & \scr{V} & \scr{O}_{\De,\ell}
\end{matrix}\bigg\}^{(a) (b)}_{(m) (+0)} 
\scr{D}_{X_3}^{(+0)A}\vev{\scr{O}_{\De+\delta\De_b,\ell+\delta\ell_b}(X_I)\Phi_{\De_3-1}(X_3)\scr{O}'_{\De',\ell'}(X_J)}^{(m)}
\nonumber\\
&+ \sum_m
 \bigg\{
\begin{matrix} 
\Phi_{\De_3} & \scr{O}'_{\De',\ell'} &  \Phi_{\De_3, 1}\\
\scr{O}_{\De+\delta\De_b,\ell+\delta\ell_b} & \scr{V} & \scr{O}_{\De,\ell}
\end{matrix}\bigg\}^{(a) (b)}_{(m) (0-)} \scr{D}_{X_3}^{(0-)A}\vev{\scr{O}_{\De+\delta\De_b,\ell+\delta\ell_b}(X_I)\Phi_{\De_3, 1}(X_3)\scr{O}'_{\De',\ell'}(X_J)}^{(m)}\nonumber\\
& +\sum_m
 \bigg\{
\begin{matrix} 
\Phi_{\De_3} & \scr{O}'_{\De',\ell'} &  \Phi_{\De_3+1}\\
\scr{O}_{\De+\delta\De_b,\ell+\delta\ell_b} & \scr{V} & \scr{O}_{\De,\ell}
\end{matrix}\bigg\}^{(a) (b)}_{(m) (-0)} \scr{D}_{X_3}^{(-0)A}\vev{\scr{O}_{\De+\delta\De_b,\ell+\delta\ell_b}(X_I)\Phi_{\De_3+1}(X_3)\scr{O}'_{\De',\ell'}(X_J)}^{(m)}\,.
\end{align}
At this stage, we recall that the operator $\scr{D}_{X_1}^{(-0)A}$ is contracted with $\scr{D}_{X_3 \, A}^{(-0)}$ in our operator combination of choice \Eq{combination 13}. We therefore act on \Eq{3-point reln again} with $\scr{D}_{X_3 \, A}^{(-0)}$. This leads us to
\begin{align}
\scr{D}_{X_3 \, A}^{(-0)} &\scr{D}_{X_I}^{(b)A}\vev{\scr{O}_{\De,\ell}(X_I)\Phi_{\De_3}(X_3)\scr{O}'_{\De',\ell'}(X_J)}^{(a)}\nonumber\\
&= \sum_m
 \bigg\{
\begin{matrix} 
\Phi_{\De_3} & \scr{O}'_{\De',\ell'} &  \Phi_{\De_3-1}\\
\scr{O}_{\De+\delta\De_b,\ell+\delta\ell_b}& \scr{V} & \scr{O}_{\De,\ell}
\end{matrix}\bigg\}^{(a) (b)}_{(m) (+0)} 
\scr{D}_{X_3 \, A}^{(-0)} \scr{D}_{X_3}^{(+0)A}\vev{\scr{O}_{\De+\delta\De_b,\ell+\delta\ell_b}(X_I)\Phi_{\De_3-1}(X_3)\scr{O}'_{\De',\ell'}(X_J)}^{(m)}\nonumber\\
&+ \sum_m
 \bigg\{
\begin{matrix} 
\Phi_{\De_3} & \scr{O}'_{\De',\ell'} &  \Phi_{\De_3, 1}\\
\scr{O}_{\De+\delta\De_b,\ell+\delta\ell_b} & \scr{V} & \scr{O}_{\De,\ell}
\end{matrix}\bigg\}^{(a) (b)}_{(m) (0-)}  \scr{D}_{X_3 \, A}^{(-0)} \scr{D}_{X_3}^{(0-)A}\vev{\scr{O}_{\De+\delta\De_b,\ell+\delta\ell_b}(X_I)\Phi_{\De_3, 1}(X_3)\scr{O}'_{\De',\ell'}(X_J)}^{(m)}\nonumber\\
&+ \sum_m
 \bigg\{
\begin{matrix} 
\Phi_{\De_3} & \scr{O}'_{\De',\ell'} &  \Phi_{\De_3+1}\\
\scr{O}_{\De+\delta\De_b,\ell+\delta\ell_b} & \scr{V} & \scr{O}_{\De,\ell}
\end{matrix}\bigg\}^{(a) (b)}_{(m) (-0)}  \scr{D}_{X_3 \, A}^{(-0)} \scr{D}_{X_3}^{(-0)A}\vev{\scr{O}_{\De+\delta\De_b,\ell+\delta\ell_b}(X_I)\Phi_{\De_3+1}(X_3)\scr{O}'_{\De',\ell'}(X_J)}^{(m)}\,.
\end{align}
As before, we now observe that all of the bubble coefficients on the right-hand side vanish with the exception of a single one, namely
\[
b^{(-0)(+0)}_\Phi\equiv \bigg(\begin{matrix} 
\Phi_{\De_3-1} \\
\Phi_{\De_3} \,\, \scr{V}
\end{matrix}\bigg)^{(-0)(+0)}=(\De_3-2) (2 \De_3-d)
   (\De_3-1)(d-\De_3-1)\,,
\]
which leaves us with 
 \begin{align}
 \eql{remaining contr}
 \scr{D}_{X_3 A}^{(-0)} &\scr{D}_{X_I}^{(b)A}\vev{\scr{O}_{\De,\ell}(X_I)\Phi_{\De_3}(X_3)\scr{O}'_{\De',\ell'}(X_J)}^{(a)}\nonumber\\
 &= \sum_m
 \bigg\{
\begin{matrix} 
\Phi_{\De_3} & \scr{O}'_{\De',\ell'} &  \Phi_{\De_3-1}\\
\scr{O}_{\De+\delta\De_b,\ell+\delta\ell_b} & \scr{V} & \scr{O}_{\De,\ell}
\end{matrix}\bigg\}^{(a) (b)}_{(m) (+0)} 
b^{(-0)(+0)}_\Phi\vev{\scr{O}_{\De+\delta\De_b,\ell+\delta\ell_b}(X_I)\Phi_{\De_3-1}(X_3)\scr{O}'_{\De',\ell'}(X_J)}^{(m)}\,.
\end{align}
Here the label $a$ enumerates the possible constituent 3-point tensor structures in the conformal 3-point function $\vev{\scr{O}_{\De,\ell}\Phi_{\De_3}\scr{O}'_{\De',\ell'}}$. 

Here we will focus on the case where this 3-point structure is even under parity, in which case we can describe the structures using monomials of $V_{ij, k}$, $H_{ij}$ in the box tensor basis (see Appendix~\ref{app:A}). We may parameterize these structures by the index $n_{IJ}$, which takes on values in the range $0\leq n_{IJ} \leq \text{min}(\ell, \ell')$. The label $m$ can be similarly parameterized by some $m_{IJ}$. We further remark that each of the 3-point structures that appears here is a (spin)-(scalar)-(spin) structure. This means that we can label these structures by their respective $n_{IJ}$ and $m_{IJ}$ parameters. 

To illustrate this, let us restrict attention to just one of the contributions in \Eq{remaining contr}, namely the one with $(b)=(0+)$:
\begin{align}
\eql{B2}
  \scr{D}_{X_3\,A}^{(-0)} &\scr{D}_{X_I}^{(0+)A}\vev{\scr{O}_{\De,\ell}(X_I)\Phi_{\De_3}(X_3)\scr{O}'_{\De',\ell'}(X_J)}^{(n_{IJ})}\nonumber\\
  &= \sum_{m_{IJ} = 0}^{\text{min}(\ell+1, \ell')}
\scr{B}^{n_{IJ} (0+)}_{m_{IJ} (+0)} 
b^{(-0)(+0)}_\Phi\vev{\scr{O}_{\De,\ell+1} (X_I)\Phi_{\De_3-1}(X_3)\scr{O}'_{\De',\ell'}(X_J)}^{(m_{IJ})}\,,
\end{align}
where we have defined
\[
\scr{B}^{n_{IJ} (0+)}_{m_{IJ} (+0)}  \equiv
\bigg\{
\begin{matrix} 
\Phi_{\De_3} & \scr{O}'_{\De',\ell'} &  \Phi_{\De_3-1}\\
\scr{O}_{\De,\ell+1} & \scr{V} & \scr{O}_{\De,\ell}
\end{matrix}\bigg\}^{(n_{IJ}) (0+)}_{(m_{IJ}) (+0)}  \,.
\]

We next select a particular 3-point structure labeled by $n_{IJ} = n$, where $n$ is some nonnegative integer $n \in (0, \text{min}(\ell, \ell'))$: 
\begin{align}
 \scr{D}_{X_3\,A}^{(-0)} &\scr{D}_{X_I}^{(0+)A}\vev{\scr{O}_{\De,\ell}(X_I)\Phi_{\De_3}(X_3)\scr{O}'_{\De',\ell'}(X_J)}^{(n)}\nonumber\\
&= \sum_{m_{IJ} = 0}^{n-1}
\scr{B}^{n (0+)}_{m_{IJ} (+0)} 
b^{(-0)(+0)}_\Phi\vev{\scr{O}_{\De,\ell+1} (X_I)\Phi_{\De_3-1}(X_3)\scr{O}'_{\De',\ell'}(X_J)}^{(m_{IJ})}\nonumber\\
&+\scr{B}^{n (0+)}_{n (+0)} 
b^{(-0)(+0)}_\Phi\vev{\scr{O}_{\De,\ell+1} (X_I)\Phi_{\De_3-1}(X_3)\scr{O}'_{\De',\ell'}(X_J)}^{(n)}\nonumber\\
&+\sum_{m_{IJ} = n+1}^{\text{min}(\ell+1, \ell')}
\scr{B}^{n (0+)}_{m_{IJ} (+0)} 
b^{(-0)(+0)}_\Phi\vev{\scr{O}_{\De,\ell+1} (X_I)\Phi_{\De_3-1}(X_3)\scr{O}'_{\De',\ell'}(X_J)}^{(m_{IJ})}\,.
\end{align}
Evaluating the left-hand side explicitly and matching coefficients, we find that only two of these $6j$ symbols are nonvanishing. They are given by
\[
\eql{sample 6j}
\scr{B}^{n (0+)}_{n (+0)}  &= \frac{\De_3-2 n-(\De+\ell)-(\De'-\ell')}{2 (\De_3 -2)
   (\De_3 -1) (d-2 \De_3 )
   (d-\De_3 -1)}\,, \nonumber\\   
\scr{B}^{n (0+)}_{n+1 (+0)} &=\frac{\ell'-n}{2 (\De_3 -2) (\De_3 -1) (d-2 \De_3) (d-\De_3 -1)}\,.
\]
In particular, all other $6j$ symbols for which $m_{IJ} \neq n, n+1$ are identically zero, since the corresponding structures do not appear on the left-hand side. That is,
\[
\scr{B}^{n (0+)}_{m_{IJ} (+0)}  = 0\,, \qquad m_{IJ} \neq n, n+1\,.
\]
We now proceed to check the endpoint values, $n_{IJ} = 0$ and $n_{IJ}= \text{min}(\ell, \ell')$. For $n_{IJ}=0$, the nonzero coefficients are $\scr{B}^{0 (0+)}_{0 (+0)} $ and $\scr{B}^{0 (0+)}_{1 (+0)}$. Next, we observe that for $n_{IJ}= \text{min}(\ell, \ell')$, there are two cases: \\
(1) If $\ell'\leq \ell$, then $n_{IJ}=\ell'$. Since the sum over $m_{IJ}$ ranges from $0$ to $\text{min}(\ell+1, \ell')$, we expect the symbol $\scr{B}^{n_{IJ} (0+)}_{m_{IJ}(+0)}=\scr{B}^{n_{IJ} (0+)}_{n_{IJ}+1 (+0)}$ to vanish, since in that case $m_{IJ}=\ell'+1$. From the form of $\scr{B}^{n (0+)}_{n+1 (+0)}$ in \Eq{sample 6j}, we find that this is indeed the case for $n=\ell'$. \\
(2) If $\ell'>\ell$, then $n_{IJ}=\ell$. In this case, we expect the value of $\scr{B}^{\ell (0+)}_{\ell+1 (+0)}$ to be nonvanishing, which is in fact true.

We may now continue in the same style for the remaining three equations, thus obtaining all the relevant $6j$ symbols. In particular, it transpires that only the following ones are nonzero a priori:
\[
\eql{6j inside interval}
 &\scr{B}^{n_{IJ} (-0)}_{n_{IJ} (+0)}\,, \nonumber\\
 &\scr{B}^{n_{IJ} (0+)}_{n_{IJ} (+0)}\,, \quad \scr{B}^{n_{IJ} (0+)}_{n_{IJ}+1 (+0)}\,, \nonumber\\
 & \scr{B}^{n_{IJ} (0-)}_{n_{IJ} -1 (+0)}\,, \quad \scr{B}^{n_{IJ} (0-)}_{n_{IJ}  (+0)}\,, \quad \scr{B}^{n_{IJ} (0-)}_{n_{IJ}+1 (+0)}\,, \nonumber\\
 &  \scr{B}^{n_{IJ} (+0)}_{n_{IJ}-1 (+0)}\,, \quad  \scr{B}^{n_{IJ} (+0)}_{n_{IJ} (+0)}\,,
 \quad \scr{B}^{n_{IJ} (+0)}_{n_{IJ}+1 (+0)}\,, \quad \scr{B}^{n_{IJ} (+0)}_{n_{IJ}+2 (+0)}\,.
\]
Here the parameter $n_{IJ}$ takes on values within the interval $[0, \text{min}(\ell, \ell')]$, while $m_{IJ}$ takes on values within $[0, \text{min}(\ell+1, \ell')]$. It follows that the $6j$ symbols $\scr{B}^{n_{IJ} (a)}_{m_{IJ} (b)}$ are nonvanishing provided that for any allowed value of $n_{IJ}$, the parameter $m_{IJ}$ satisfies $m_{IJ} \in [0,\text{min}(\ell+1, \ell')]$, i.e.~it does not fall outside its range of definition. Otherwise, the $6j$ symbol vanishes. For example, if $\ell< \ell'$, then for $n_{IJ} =\text{min}(\ell, \ell')= \ell$, the symbol $\scr{B}^{n_{IJ} (+0)}_{n_{IJ}+2 (+0)}=\scr{B}^{\ell (+0)}_{\ell+2 (+0)}$ must be forced to vanish by definition, since $m_{IJ} = \ell+2 \not\in [0,\text{min}(\ell+1, \ell')]$, even if it is nonzero for $n_{IJ}=\ell$ a priori.

Combining all of the above results, we ultimately arrive at the following relation:
\begin{align}
\eql{W recursion relation 1} 
-2(\scr{D}_{X_1}^{(-0)}\cdot\scr{D}_{X_3}^{(-0)})&W_{\De, \ell, \De', \ell'; \De_1, \De_2, \De_3, \De_4, \De_5}^{(n_{IJ})}
 =\nonumber\\
&-2 b^{(-0)(+0)}_\Phi\bigg(
 \scr{A}_{(+0)}^{(-0)}B_{(+0)(-0)} 
\scr{B}^{n_{IJ} (-0)}_{n_{IJ} (+0)} 
W_{\De-1,\ell, \De', \ell'; \De_1-1, \De_2, \De_3-1, \De_4, \De_5}^{(n_{IJ})}  
\nonumber\\
&+ \sum_{m_{IJ} = n_{IJ}}^{n_{IJ}+1}
\scr{A}_{(0-)}^{(-0)}
 B_{(0-)(0+)}\scr{B}^{n_{IJ} (0+)}_{m_{IJ} (+0)} 
W_{\De,\ell+1, \De', \ell'; \De_1-1, \De_2, \De_3-1, \De_4, \De_5}^{(m_{IJ})}      
 \nonumber\\
&+  \sum_{m_{IJ} = n_{IJ}-1}^{n_{IJ}+1}\scr{A}_{(0+)}^{(-0)} B_{(0+)(0-)}\scr{B}^{n_{IJ} (0-)}_{m_{IJ} (+0)} 
W_{\De,\ell-1, \De', \ell'; \De_1-1, \De_2, \De_3-1, \De_4, \De_5}^{(m_{IJ})} 
 \nonumber\\
&+ \sum_{m_{IJ} = n_{IJ}-1}^{n_{IJ}+2} \scr{A}_{(-0)}^{(-0)} B_{(-0)(+0)}\scr{B}^{n_{IJ} (+0)}_{m_{IJ} (+0)} 
 W_{\De+1,\ell, \De', \ell'; \De_1-1, \De_2, \De_3-1, \De_4, \De_5}^{(m_{IJ})}\bigg)\,,
\end{align}
which is evidently a recursion relation in the spin $\ell$, with the spin $\ell'$ held fixed. We note that the scaling dimension of the exchanged operator $\mathcal{O}_{\De, \ell}$ shifts as well, as is clear from the presence of terms involving $\De+1$ and $\De-1$. 

Just like in the case of the 4-point function, the action of the operator combination $-2(\scr{D}_{X_1}^{(-0)}\cdot\scr{D}_{X_3}^{(-0)}) = X_{13}$ on  the 5-point function results in another 5-point function with scaling dimensions at positions $1$ and $3$ shifted by $-1$, that is, $\De_1 \to \De_1 - 1$ and $\De_3 \to \De_3 - 1$. We may accordingly extract the relevant external-dimension-dependent prefactor for these shifted scaling dimensions. The resulting prefactor depends on the conformal cross-ratios and is determined by our choice of the external prefactor $P_{\De_i}(X_i)$ and the basis of cross-ratios.
 
We next proceed to apply an exactly analogous approach to the other spin, this time keeping $\ell$ (and $\De$)  fixed and varying $\ell'$ (and $\De'$). To this end, we consider acting on the 5-point function with the combination of operators:
\[
\eql{combination 35}
-2(\scr{D}_{X_3}^{(-0)}\cdot\scr{D}_{X_5}^{(-0)})\vev{\phi_{\De_1}(X_1) \phi_{\De_2}(X_2) |\scr{O}_{\De, \ell}|\Phi_{\De_3}(X_3) |
\scr{O}^\prime_{\De', \ell'}|\phi_{\De_4}(X_4) \phi_{\De_5}(X_5)} 
\nonumber\\=X_{35}\vev{\phi_{\De_1}(X_1) \phi_{\De_2}(X_2) |\scr{O}_{\De, \ell}|\Phi_{\De_3}(X_3) |
\scr{O}^\prime_{\De', \ell'}|\phi_{\De_4}(X_4) \phi_{\De_5}(X_5)}\,.
\]  
The procedure is in fact the mirror image of the above algorithm and leads to the direct analog of \Eq{W recursion relation 1}, but this time with the spin $\ell$ held fixed, while the spin $\ell'$ (as well as the exchanged operator dimension $\De'$) is allowed to vary. This analogous relation is identical in form to  \Eq{W recursion relation 1} up to the replacements $\De\leftrightarrow \De',\, \ell\leftrightarrow \ell',\, \De_{12}\to -\De_{45}$. In this case, acting with the operators $\scr{D}_{X_3}^{(-0)A}$ and $\scr{D}_{X_5}^{(-0) A}$ again shifts the external operator dimensions $\De_3$ and $\De_5$ down by $1$, resulting in an overall external prefactor. 

We now examine the relations \Eq{W recursion relation 1} and its spin-$\ell'$ analog as a unit. For convenience, we implement several shifts in the spins and dimensions. In particular, we take $\De_1 \to \De_1+ 1$ in \Eq{W recursion relation 1} and $\De_5 \to \De_5+ 1$ in the spin-$\ell'$ analog, as well as $\De_3 \to \De_3+ 1$ in both relations. Further, in order to place the two relations on an equal footing, we shift $\ell \to \ell - 1$ in \Eq{W recursion relation 1} and $\ell' \to \ell' - 1$ in the spin-$\ell'$ relation. For convenience, we adopt some shorthand notation:
\[
G_{(\ell, \ell'; \de_0, \de_0^{\prime})}^{(n)}
\equiv G^{(n)}_{\De+\de_0, \ell, \De^{\prime}+\de_0^{\prime}, \ell'}(u_i)\,.
\]
With this, we are left with the following two recursion relations for the 5-point conformal blocks for $([\De, \ell], [\De', \ell'])$ exchange:
\[
\eql{13 rel}
 G_{(\ell, \ell'; 0, 0)}^{(n_{IJ})}   
  = \frac{1}{s_{n_{IJ}}}\bigg(f(u_i) G_{(\ell-1, \ell'; 0, 0)}^{(n_{IJ})}\bigg|_{\De_1\to \De_1+1, \De_3\to \De_3+1}
-  G_{(\ell-1,\ell';-1,0)}^{(n_{IJ})}  
-  s_{n_{IJ}+1} G_{(\ell, \ell'; 0, 0)}^{(n_{IJ}+1)}       
 \nonumber\\
- t_{n_{IJ}-1}  G_{(\ell-2, \ell'; 0, 0)}^{(n_{IJ}-1)} 
-t_{n_{IJ}} G_{(\ell-2, \ell'; 0,0)}^{(n_{IJ})} 
-t_{n_{IJ}+1} G_{(\ell-2,  \ell'; 0, 0)}^{(n_{IJ}+1)} \nonumber\\
-u_{n_{IJ}-1}   G_{(\ell-1, \ell'; 1, 0)}^{(n_{IJ}-1)} 
-u_{n_{IJ}} G_{(\ell-1, \ell'; 1, 0)}^{(n_{IJ})}
- u_{n_{IJ}+1}G_{(\ell-1, \ell'; 1, 0)}^{(n_{IJ}+1)} 
- u_{n_{IJ}+2}G_{(\ell-1,  \ell'; 1, 0)}^{(n_{IJ}+2)}\bigg)\,,
\] 
and  
\[ 
\eql{35 rel}
G_{(\ell, \ell'; 0, 0)}^{(n_{IJ})} 
 =  \frac{1}{s^\prime_{n_{IJ}}}
 \bigg(f'(u_i)G_{(\ell,  \ell'-1; 0, 0)}^{(n_{IJ})}\bigg|_{\De_3 \to \De_3+1, \De_5\to \De_5+1}
 - G_{(\ell, \ell'-1; 0, - 1)}^{(n_{IJ})}
- s^\prime_{n_{IJ}+1} G_{(\ell, \ell'; 0, 0)}^{(n_{IJ}+1)}
\nonumber\\
-t^\prime_{n_{IJ}-1} G_{(\ell, \ell'-2; 0, 0)}^{(n_{IJ}-1)}
-t^\prime_{n_{IJ}} G_{(\ell, \ell'-2; 0, 0)}^{(n_{IJ})}
-t^\prime_{n_{IJ}+1} G_{(\ell, \ell'-2; 0, 0)}^{(n_{IJ}+1)}
\nonumber\\
- u^\prime_{n_{IJ}-1} G_{(\ell, \ell'-1; 0,1)}^{(n_{IJ}-1)}
-u^\prime_{n_{IJ}} G_{(\ell, \ell'-1; 0,1)}^{(n_{IJ})}
- u^\prime_{n_{IJ}+1} G_{(\ell, \ell'-1; 0, 1)}^{(n_{IJ}+1)}
- u^\prime_{n_{IJ}+2} G_{(\ell, \ell'-1; 0, 1)}^{(n_{IJ}+2)}\bigg).
\] 
Both recursion relations are defined in a convention-independent way, where $f(u_i)$ and $f'(u_i)$ represent the leftover cross-ratio-dependent prefactors that arise upon removing the external prefactor for the relevant shifted scaling dimensions. We expect each of these prefactors to be built from powers of the $u_i$. 
We may readily determine the form of the $f(u_i)$ and $f'(u_i)$ for the three different sets of conventions mentioned above. For the set 
(\ref{eq:convention 1 prefactor}, \ref{eq:convention 1 cross-ratios}), we find
\[
f(u_i)=(u_1)^{-1/2}, \quad f'(u_i)=(u_2)^{-1/2}
\]
Further, for the set (\ref{eq:convention 2 prefactor}, \ref{eq:convention 2 cross-ratios}), we have
\[
f(u_i)=(u_1')^{-1/2}, \quad f'(u_i)=(u_2')^{-1/2}
\]
Lastly, for the set (\ref{eq:convention 3 prefactor}, \ref{eq:convention 3 cross-ratios}), we find 
\[
f(u_i)=v_{12}^5
(u_1^5)^{-1/2}  (u_2^5)^{-1/2}, \quad f'(u_i)=(u_1^5)^{-1/2}  (u_2^5)^{-1/2}.
\]
The coefficients in \Eq{13 rel} are built from products of the various $6j$ symbols. In particular, we have
\[
\eql{13 coeffs}
s_{m_{IJ}} &\equiv -2\scr{A}_{(0-)}^{(-0)} 
 B_{(0-)(0+)}\scr{B}^{n_{IJ} (0+)}_{m_{IJ} (+0)} 
b^{(-0)(+0)}_\Phi\bigg|_{\De_1 \to \De_1+ 1, \De_3 \to \De_3+1, \ell \to \ell-1}, \nonumber\\
t_{m_{IJ}} &\equiv -2\scr{A}_{(0+)}^{(-0)} B_{(0+)(0-)}\scr{B}^{n_{IJ} (0-)}_{m_{IJ} (+0)} 
b^{(-0)(+0)}_\Phi\bigg|_{\De_1 \to \De_1+ 1, \De_3 \to \De_3+1,
\ell \to \ell-1}, \nonumber\\
u_{m_{IJ}}  &\equiv -2\scr{A}_{(-0)}^{(-0)}  B_{(-0)(+0)}\scr{B}^{n_{IJ} (+0)}_{m_{IJ} (+0)} b^{(-0)(+0)}_\Phi\bigg|_{\De_1 \to \De_1+ 1, \De_3 \to \De_3+1,\ell \to \ell-1}\,, 
\]
while in \Eq{35 rel}, the respective coefficients are given by
\[
\eql{35 coeffs}
s^\prime_{m_{IJ}} &= s_{m_{IJ}}\bigg|_{\De\leftrightarrow \De', \ell\leftrightarrow \ell', \De_{12}\to -\De_{45}}, \nonumber\\
t^\prime_{m_{IJ}} &=t_{m_{IJ}}\bigg|_{\De\leftrightarrow \De', \ell\leftrightarrow \ell', \De_{12}\to -\De_{45}}\,, \nonumber\\
u^\prime_{m_{IJ}} &=u_{m_{IJ}}\bigg|_{\De\leftrightarrow \De', \ell\leftrightarrow \ell', \De_{12}\to -\De_{45}}\,, 
\]
where $\De_{ij} \equiv \De_i-\De_j$.

The explicit form of the coefficients in \Eq{13 coeffs} can be found in Appendix~\ref{app:B}. When used in conjunction, the two recursion relations \Eq{13 rel} and \Eq{35 rel} enable us to construct the 5-point blocks for general symmetric traceless tensor exchange in the comb channel, with the exception of the block $G_{(\ell, \ell; 0, 0)}^{(n_{IJ}+1)}=G_{(\ell, \ell; 0, 0)}^{(\ell)}$, where $\ell=\ell'$ and $n_{IJ}=\ell-1$. This special case will be treated separately below. The relations \Eq{13 rel} and \Eq{35 rel} may be regarded as two independent results. One re-expresses a given block for $[\De, \ell]$, $[\De', \ell']$ exchange in terms of a linear combination of $(\ell, \ell')$, $(\ell-1, \ell')$, and $(\ell-2, \ell')$ blocks with $\ell'$ held fixed, and the other does the same for spin-$\ell'$, with $\ell$ held fixed. 

Note that the contributions on the right-hand side of the relations \Eq{13 rel} and \Eq{35 rel} all feature terms involving lower spins with the exception of the ones labeled by the coefficients $s_{n_{IJ}+1}$ and $s^\prime_{n_{IJ}+1}$. The latter terms carry the same spins $(\ell, \ell')$ as the conformal block we are after. However, at the same time, they exhibit a larger value of the 3-point structure index, namely $n_{IJ} + 1$. Closer inspection reveals that the coefficients of these blocks in fact vanish when $n_{IJ}$ takes on a particular maximum value and are nonzero otherwise. To be precise, this is $\max(n_{IJ})=\ell'$ for \Eq{13 rel} and $\max(n_{IJ})=\ell$ for \Eq{35 rel}. 

We now describe how to use these recursion relations to generate conformal blocks. For concreteness, let us consider the spin-$\ell$ relation \Eq{13 rel}. When deriving this form, we shifted $\ell\to\ell-1$ everywhere, which implies that the range of allowed values of the structure label $n_{IJ}$ is $0 \leq n_{IJ}\leq \min(\ell-1, \ell')$. It follows that there are two possible cases here: (1) $\ell' \leq \ell-1$ so that $\max(n_{IJ})= \ell'$ and (2) $\ell-1 < \ell'$ so that $\max(n_{IJ})= \ell-1$. For case (1), we find that if we set $n_{IJ}$ to its maximum value, namely $\max(n_{IJ})= \ell'$, then $s_{n_{IJ}+1}=s_{\ell'+1}=0$. Hence, the block $G_{(\ell, \ell'; 0, 0)}^{(\ell'+1)}$ is absent, exactly as desired, since the corresponding structure is unphysical. Since in this case the range of $n_{IJ}$ is $0 \leq n_{IJ} \leq \ell'$, we can iteratively apply \Eq{13 rel} to increase the value of $n_{IJ}$ and eventually relate the block to a sum of blocks with smaller values of $\ell$.
 
We next remark that this is the only scenario in which $s_{n_{IJ}+1}$ vanishes. In particular, for all other possible values of $n_{IJ}$, it turns out that $s_{n_{IJ}+1} \neq 0$. This implies that for case (2), we have that $\max(n_{IJ})= \ell-1$ and $s_{n_{IJ}+1}=s_{(\ell-1)+1}\neq 0$. Although the block structure $G_{(\ell, \ell'; 0, 0)}^{(\ell)}$ is physical in this case, the resulting relation now involves a linear combination of two different structures of the highest spin ($G_{(\ell, \ell'; 0, 0)}^{(\ell)}$ and $G_{(\ell, \ell'; 0, 0)}^{(\ell-1)}$), which prevents us from expressing $G_{(\ell, \ell'; 0, 0)}^{(\ell-1)}$ purely in terms of lower-spin blocks. 

The upshot of these observations is that the recursion relation \Eq{13 rel} can always be usefully applied for spins subject to $\ell' < \ell$. Similarly, we can always apply the relation \Eq{35 rel} for blocks with $\ell < \ell'$. More precisely, if $\ell' < \ell$ at the outset, we can iteratively apply \Eq{13 rel} to raise the value of $n_{IJ}$ and eventually lower the spin $\ell$ of all generated blocks. When we reach the case $\ell = \ell'$ the two relations degenerate. We can still use the recursion relation to increase $n_{IJ}$ until we evaluate the block with $n_{IJ} = \ell-1 = \ell'-1$, in which case we will need another relation to express $G_{(\ell, \ell; 0, 0)}^{(\ell)}$ in terms of lower-spin blocks. This will be described in \Eq{special case} below. For blocks with $\ell < \ell'$ we can similarly iteratively apply \Eq{35 rel} to lower $\ell'$ until we evaluate the block with $n_{IJ} = \ell-1 = \ell'-1$, where we would again use \Eq{special case}. This process can continue until both $\ell$ and $\ell'$ have been brought down to $0$, and $n_{IJ}$ has also been reduced to its minimum value of $0$. 

In this fashion, these two relations, coupled with \Eq{special case}, ultimately yield the desired block for spin-$\ell$, spin-$\ell'$ exchange in the form $\sum_{ij} \beta_{ij} G_{(0, 0; \de_{0_i}, \de'_{0_j})}^{(0)}$ for all values of the spins $\ell$ and $\ell'$. Note that for the case of $\ell = 0$, $n_{IJ} = 0$ ($\ell' = 0$,  $n_{IJ} = 0$), the coefficients $t_{n_{IJ}-1}$, $t_{n_{IJ}}$, and $t_{n_{IJ}+1}$ ($t'_{n_{IJ}-1}$, $ t'_{n_{IJ}} = 0$, and $t'_{n_{IJ}+1} = 0$) all vanish, exactly as expected. 
 
We remark that the above method treats the middle operator $\Phi$ as special, while the pairs $\{\phi_{\De_1}(X_1)$, $\phi_{\De_2}(X_2)\}$ and $\{\phi_{\De_4}(X_4)$, $\phi_{\De_5}(X_5)\}$ are considered on an equal footing. From our choice of the weight-shifting operator combinations $(\scr{D}_{X_1}^{(-0)}\cdot \scr{D}_{X_3}^{(-0)})$ and $(\scr{D}_{X_3}^{(-0)}\cdot \scr{D}_{X_5}^{(-0)})$, it is evident that $X_3$ serves as a ``junction point", with the two resulting relations obviously symmetric with respect to each other. This symmetry is reflected in the structure of the coefficients. In particular, if we relabel the external and exchanged spins and dimensions appropriately, we obtain the coefficients $\{s^\prime_{m_{IJ}}, t^\prime_{m_{IJ}}, u^\prime_{m_{IJ}} \}$ from the coefficients $\{s_{m_{IJ}}, t_{m_{IJ}}, u_{m_{IJ}} \}$, as is apparent from \Eq{35 coeffs}.

It is straightforward to implement the recursion relations \Eq{13 rel} and \Eq{35 rel} as well as \Eq{special case} within a symbolic computing environment, e.g.~\texttt{Mathematica}. Doing so enables us to effortlessly generate any symmetric traceless conformal block of interest by specifying a handful of input parameters, namely $\{n_{IJ}, \ell, \ell'\}$. To sum up, at this point, we have given an explicit prescription for deriving the spin-$\ell$, spin-$\ell'$ exchange blocks in a purely scalar 5-point function.

Next we will examine if we can further simplify the recursion relations by eliminating terms involving shifts in the exchanged operator dimensions. In order to achieve such a simplification, it is useful to first establish some identities describing how the blocks transform under various permutations.

\subsubsection{Exchange symmetries}

The $5$-point conformal blocks for $([\De, \ell], [\De', \ell'])$ exchange satisfy some simple symmetry relations under the exchanges $1\leftrightarrow 2$ and $4\leftrightarrow 5$. For concreteness, let us follow the conventions \Eq{convention 1 prefactor} and \Eq{convention 1 cross-ratios} for the external prefactor and basis of cross-ratios, respectively. With these choices, we find that the symmetry relations assume the form
\[
\eql{symmetry 12}
G_{\De, \ell, \De', \ell'}^{(n_{IJ})}&(u_1, u_2, w_{ 2;3 }, w_{ 2;4 }, w_{ 3;4 })  \nonumber\\
&=  (-1)^\ell
w_{ 2;3 }^{-\De_3/2}w_{ 2;4 }^{\frac{1}{2}(-\De_4+\De_5)} 
G_{\De, \ell, \De', \ell'}^{(n_{IJ})}\left(\frac{u_1}{w_{ 2;3 }}, \frac{u_2 w_{ 2;3 }}{w_{ 2;4 }}, \frac{1}{w_{ 2;3 }}, \frac{1}{w_{ 2;4 }}, \frac{w_{ 3;4 }}{w_{ 2;4 }}\right)\bigg|_{\De_1\leftrightarrow \De_2},
\]

\[
\eql{symmetry 45}
G_{\De, \ell, \De', \ell'}^{(n_{IJ})}&(u_1, u_2, w_{ 2;3 }, w_{ 2;4 }, w_{ 3;4 }) \nonumber\\
&=  (-1)^{\ell'}
w_{ 2;4 }^{\frac{1}{2}(\De_1-\De_2)} w_{ 3;4 }^{-\De_3/2}
G_{\De, \ell, \De', \ell'}^{(n_{IJ})}\left(\frac{u_1 w_{ 3;4 }}{w_{ 2;4 }}, \frac{u_2}{w_{ 3;4 }}, \frac{w_{ 2;3 }}{w_{ 2;4 }}, \frac{1}{w_{ 2;4 }}, \frac{1}{w_{ 3;4 }}\right)\bigg|_{\De_4\leftrightarrow \De_5},
\]
and
\[
\eql{symmetry 12 45}
G_{\De, \ell, \De', \ell'}^{(n_{IJ})}& (u_1, u_2, w_{ 2;3 }, w_{ 2;4 }, w_{ 3;4 })
\nonumber\\&= (-1)^{\ell+\ell'}
w_{ 2;3 }^{-\De_3/2}w_{ 2;4 }^{\frac{1}{2}(\De_1-\De_2+\De_3-\De_4+\De_5)} w_{ 3;4 }^{-\De_3/2} \nonumber\\
&\qquad\qquad\times G_{\De, \ell, \De', \ell'}^{(n_{IJ})}\left(\frac{u_1 w_{ 3;4 }}{w_{ 2;3 }}, \frac{u_2 w_{ 2;3 }}{w_{ 3;4 }}, \frac{w_{ 2;4 }}{w_{ 2;3 }}, w_{ 2;4 }, \frac{w_{ 2;4 }}{w_{ 3;4 }}\right)\bigg|_{\De_1\leftrightarrow \De_2, \De_4 \leftrightarrow \De_5}.
\]

Moreover, we have the following symmetry relations under $1\leftrightarrow 5$, $2\leftrightarrow 4$:
\[
\eql{symmetry 15 24}
G_{\De, \ell, \De', \ell'}^{(n_{IJ})}(u_1, u_2, w_{ 2;3 }, w_{ 2;4 }, w_{ 3;4 })=
G_{\De, \ell, \De', \ell'}^{(n_{IJ})}(u_2, u_1, w_{ 3;4 }, w_{ 2;4 }, w_{ 2;3 })
\bigg|_{\De_1\leftrightarrow \De_5, \De_2\leftrightarrow \De_4}\,,
\]

\[
\eql{symmetry 15}
G_{\De, \ell, \De', \ell'}^{(n_{IJ})}(u_1, u_2, w_{ 2;3 }, w_{ 2;4 }, w_{ 3;4 })
=
(u_1 u_2)^{\frac{\De_1+\De_5}{2}}
G_{\De, \ell, \De', \ell'}^{(n_{IJ})}
\left(\frac{1}{u_1},\frac{1}{u_2}, \frac{w_{ 2;3 }}{u_1}, \frac{w_{ 2;4 }}{u_1 u_2}, \frac{w_{ 3;4 }}{u_2}\right)
\bigg|_{\De_1\leftrightarrow \De_5}\,,
\]

\[
\eql{symmetry 24}
G_{\De, \ell, \De', \ell'}^{(n_{IJ})}(u_1, u_2, w_{ 2;3 }, w_{ 2;4 }, w_{ 3;4 })
=
(u_1 u_2)^{\frac{\De_1+\De_5}{2}}
G_{\De, \ell, \De', \ell'}^{(n_{IJ})}\left(\frac{1}{u_2},\frac{1}{u_1}, 
\frac{w_{ 3;4 }}{u_2}, \frac{w_{ 2;4 }}{u_1 u_2}, \frac{w_{ 2;3 }}{u_1}\right)
\bigg|_{\De_2\leftrightarrow \De_4}\,.
\]

In deriving the above relations, we recalled that the $5$-point block of interest is related to the conformal integral
\[
\eql{5-point conformal integral}
G_{\De, \ell, \De', \ell'}^{(n_{IJ})}(u_i)\propto
\vev{\phi_{\De_1}(X_1) \phi_{\De_2}(X_2) \scr{O}_{\De, \ell}}\bowtie \vev{\scr{O}_{\De, \ell}\Phi_{\De_3}(X_3) \scr{O}^\prime_{\De', \ell'}}^{(n_{IJ})}\bowtie
\vev{ \scr{O}^\prime_{\De', \ell'}\phi_{\De_4}(X_4) \phi_{\De_5}(X_5)}\,,
\]
along with our knowledge of the transformation properties of the 3-point functions. In particular, one may readily apply the known properties of 3-point functions under permutations to conclude that the three-point functions of the type (scalar)-(scalar)-(spin-$\ell$) pick up a factor of $(-1)^{\ell}$ upon interchanging the two scalar operators. It is then trivial to assign the appropriate factors of $(-1)^\ell$ and $(-1)^{\ell'}$ as needed for the exchanges $1\leftrightarrow 2$; $4\leftrightarrow 5$; and $1\leftrightarrow 2$, $4\leftrightarrow 5$. 

Further, it is immediately apparent that the blocks are invariant under the simultaneous interchanges $1\leftrightarrow 5$, $2\leftrightarrow 4$, which corresponds to a simple relabeling of the operators. That is, under $1\leftrightarrow 5$, $2\leftrightarrow 4$ we have that
\[
G_{\De, \ell, \De', \ell'}^{(n_{IJ})}&(u_i)\bigg|_{1\leftrightarrow 5, 2\leftrightarrow 4} \\
&\propto
\vev{\phi_{\De_5}(X_5) \phi_{\De_4}(X_4) \scr{O}_{\De, \ell}}\bowtie \vev{\scr{O}_{\De, \ell}\Phi_{\De_3}(X_3) \scr{O}^\prime_{\De', \ell'}}^{(n_{IJ})}\bowtie
\vev{ \scr{O}^\prime_{\De', \ell'}\phi_{\De_2}(X_2) \phi_{\De_1}(X_1)}.\nonumber
\]
Furthermore, we observe that we may reduce this conformal integral to the original one in \Eq{5-point conformal integral} if we also relabel $[\De, \ell]\leftrightarrow [\De', \ell']$.

In addition, the symmetry relation \Eq{symmetry 15 24} makes it manifest why the coefficients of the recursion relation \Eq{35 rel} coincide with those featured in \Eq{13 rel} up to the replacements $\De\leftrightarrow \De', \ell\leftrightarrow \ell', \De_{12}\to -\De_{45}$, as exemplified in \Eq{35 coeffs}. That is, the relation \Eq{35 coeffs} between the two sets of coefficients arises precisely due to the symmetry \Eq{symmetry 15 24}. Moreover, we note that the interchanges $1\leftrightarrow 5$, $2\leftrightarrow 4$ indeed exchange
$f(u_i)\leftrightarrow f'(u_i)$, which allows us to map \Eq{13 rel} to \Eq{35 rel}. It therefore emerges that we may obtain \Eq{35 rel} from  \Eq{13 rel} via \Eq{35 rel} $=$  \Eq{13 rel}$\bigg|_{\De\leftrightarrow \De',\, \ell\leftrightarrow \ell',\, \De_{12}\to -\De_{45}}$.

Now, in light of the symmetry relations \Eq{symmetry 12} - \Eq{symmetry 45}, we proceed to write down additional recursion relations similar to \Eq{13 rel} and \Eq{35 rel} in the same spirit as in \cite{Dolan:2011dv}. To begin with, we rewrite \Eq{13 rel} and \Eq{35 rel} in a form analogous to that of Eq. (4.18) for $4$-point symmetric traceless blocks in \cite{Dolan:2011dv}. To emphasize the dependence of the blocks on the external scaling dimensions, we include the explicit dependence next to each block in parentheses, e.g.~$(\De_{12}, \De_3, \De_{45})$. We obtain
\[
\eql{13 rel rewrite}
u_1^{-1/2} G_{(\ell-1, \ell'; 0, 0)}^{(n_{IJ})}
&(\De_{12}+1, \De_3+1, \De_{45})   \nonumber\\
 =\, &
  G_{(\ell-1,\ell';-1,0)}^{(n_{IJ})}(\De_{12}, \De_3, \De_{45})    
+s_{n_{IJ}} G_{(\ell, \ell'; 0, 0)}^{(n_{IJ})}(\De_{12}, \De_3, \De_{45})    \nonumber\\
&+  s_{n_{IJ}+1} G_{(\ell, \ell'; 0, 0)}^{(n_{IJ}+1)}(\De_{12}, \De_3, \De_{45})         
+ t_{n_{IJ}-1}  G_{(\ell-2, \ell'; 0, 0)}^{(n_{IJ}-1)}(\De_{12}, \De_3, \De_{45})    \nonumber\\
&+t_{n_{IJ}} G_{(\ell-2, \ell'; 0,0)}^{(n_{IJ})}(\De_{12}, \De_3, \De_{45})   
+t_{n_{IJ}+1} G_{(\ell-2,  \ell'; 0, 0)}^{(n_{IJ}+1)}(\De_{12}, \De_3, \De_{45})   \nonumber\\
&+u_{n_{IJ}-1}   G_{(\ell-1, \ell'; 1, 0)}^{(n_{IJ}-1)}(\De_{12}, \De_3, \De_{45})   
+u_{n_{IJ}} G_{(\ell-1, \ell'; 1, 0)}^{(n_{IJ})}(\De_{12}, \De_3, \De_{45})   \nonumber\\
&+ u_{n_{IJ}+1}G_{(\ell-1, \ell'; 1, 0)}^{(n_{IJ}+1)}(\De_{12}, \De_3, \De_{45})   
+ u_{n_{IJ}+2}G_{(\ell-1,  \ell'; 1, 0)}^{(n_{IJ}+2)}(\De_{12}, \De_3, \De_{45}) 
\] 
and  
\[ 
\eql{35 rel rewrite}
u_2^{-1/2}G_{(\ell,  \ell'-1; 0, 0)}^{(n_{IJ})}
&(\De_{12}, \De_3+1, \De_{45}-1) \nonumber\\
=\,&  
  G_{(\ell, \ell'-1; 0, - 1)}^{(n_{IJ})}(\De_{12}, \De_3, \De_{45}) 
 +s^\prime_{n_{IJ}}G_{(\ell, \ell'; 0, 0)}^{(n_{IJ})}(\De_{12}, \De_3, \De_{45})  \nonumber\\
&+ s^\prime_{n_{IJ}+1} G_{(\ell, \ell'; 0, 0)}^{(n_{IJ}+1)}(\De_{12}, \De_3, \De_{45}) 
+t^\prime_{n_{IJ}-1} G_{(\ell, \ell'-2; 0, 0)}^{(n_{IJ}-1)}(\De_{12}, \De_3, \De_{45}) \nonumber\\
&+t^\prime_{n_{IJ}} G_{(\ell, \ell'-2; 0, 0)}^{(n_{IJ})}(\De_{12}, \De_3, \De_{45}) 
+t^\prime_{n_{IJ}+1} G_{(\ell, \ell'-2; 0, 0)}^{(n_{IJ}+1)}(\De_{12}, \De_3, \De_{45}) 
\nonumber\\
&+ u^\prime_{n_{IJ}-1} G_{(\ell, \ell'-1; 0,1)}^{(n_{IJ}-1)}(\De_{12}, \De_3, \De_{45}) 
+u^\prime_{n_{IJ}} G_{(\ell, \ell'-1; 0,1)}^{(n_{IJ})}(\De_{12}, \De_3, \De_{45}) \nonumber\\
&+ u^\prime_{n_{IJ}+1} G_{(\ell, \ell'-1; 0, 1)}^{(n_{IJ}+1)}(\De_{12}, \De_3, \De_{45}) 
+ u^\prime_{n_{IJ}+2} G_{(\ell, \ell'-1; 0, 1)}^{(n_{IJ}+2)}(\De_{12}, \De_3, \De_{45}) \,.
\] 
With this, we now consider interchanging $X_1\leftrightarrow X_2$, which takes
\[
u_1\to \frac{u_1}{w_{ 2;3 }}\,, \quad 
u_2 \to \frac{u_2 w_{ 2;3 }}{w_{ 2;4 }}\,, \quad
w_{ 2;3}\to \frac{1}{w_{ 2;3}}\,, \quad
w_{ 2;4 }\to \frac{1}{w_{ 2;4 }}\,, \quad
w_{ 3;4 }\to \frac{w_{ 3;4 }}{w_{ 2;4 }}\,.
\]
This means that each of the conformal block terms of the type $G_{(\ell,  \ell'; 0, 0)}^{(m)}(u_1,u_2, w_{ 2;3},w_{ 2;4 },w_{ 3;4 })$ in the above two equations gets mapped to 
\[
G_{(\ell,  \ell'; 0, 0)}^{(m)}(u_1,u_2, w_{ 2;3},w_{ 2;4 },w_{ 3;4 })\to G_{(\ell,  \ell'; 0, 0)}^{(m)}\left( \frac{u_1}{w_{ 2;3 }},\frac{u_2 w_{ 2;3 }}{w_{ 2;4 }}, \frac{1}{w_{ 2;3}},  \frac{1}{w_{ 2;4 }},  \frac{w_{ 3;4 }}{w_{ 2;4 }}\right).
\]  
We next observe that according to \Eq{symmetry 12}, we have
\[
 G_{(\ell,  \ell'; 0, 0)}^{(m)}&\left( \frac{u_1}{w_{ 2;3 }},\frac{u_2 w_{ 2;3 }}{w_{ 2;4 }}, \frac{1}{w_{ 2;3}},  \frac{1}{w_{ 2;4 }},  \frac{w_{ 3;4 }}{w_{ 2;4 }}\right) \nonumber\\
 =&  (-1)^\ell
w_{ 2;3 }^{\De_3/2}w_{ 2;4 }^{\De_{45}/2} 
G_{(\ell,  \ell'; 0, 0)}^{(m)}(u_1,u_2, w_{ 2;3},w_{ 2;4 },w_{ 3;4 })\bigg|_{\De_{1} \leftrightarrow \De_{2}}.
\]
Lastly, we relabel $\De_{12}\to -\De_{12}$. Noting that the $\{s_{m_{IJ}}, t_{m_{IJ}}, u_{m_{IJ}} \}$ coefficients depend on $\De_{12}$ but not on $\De_{45}$, while the $\{s^\prime_{m_{IJ}}, t^\prime_{m_{IJ}}, u^\prime_{m_{IJ}} \}$ coefficients only feature $\De_{45}$, we just need to explicitly exhibit the dependence on $\De_{12}$ or $\De_{45}$ in the respective sets. In this way, we arrive at 
\[
\eql{symmetry relation 1.1}
- u_1^{-1/2} w_{2;3} &G_{(\ell-1, \ell'; 0, 0)}^{(n_{IJ})}(\De_{12}- 1, \De_3+1, \De_{45}) \nonumber\\
=&  - G_{(\ell-1,\ell';-1,0)}^{(n_{IJ})}(\De_{12}, \De_3, \De_{45})      \nonumber\\
&+s_{n_{IJ}}(- \De_{12})  G_{(\ell, \ell'; 0, 0)}^{(n_{IJ})}(\De_{12}, \De_3, \De_{45})   
+  s_{n_{IJ}+1}(- \De_{12})  G_{(\ell, \ell'; 0, 0)}^{(n_{IJ}+1)}(\De_{12}, \De_3, \De_{45})         \nonumber\\ 
&+ t_{n_{IJ}-1}(-\De_{12})   G_{(\ell-2, \ell'; 0, 0)}^{(n_{IJ}-1)}(\De_{12}, \De_3, \De_{45})    
+t_{n_{IJ}}(- \De_{12})  G_{(\ell-2, \ell'; 0,0)}^{(n_{IJ})}(\De_{12}, \De_3, \De_{45})   \nonumber\\
&+t_{n_{IJ}+1}(-\De_{12})  G_{(\ell-2,  \ell'; 0, 0)}^{(n_{IJ}+1)}(\De_{12}, \De_3, \De_{45}) 
- u_{n_{IJ}-1}(- \De_{12})    G_{(\ell-1, \ell'; 1, 0)}^{(n_{IJ}-1)}(\De_{12}, \De_3, \De_{45})     \nonumber\\
&-u_{n_{IJ}}(-\De_{12})  G_{(\ell-1, \ell'; 1, 0)}^{(n_{IJ})}(\De_{12}, \De_3, \De_{45})   
- u_{n_{IJ}+1}(- \De_{12}) G_{(\ell-1, \ell'; 1, 0)}^{(n_{IJ}+1)}(\De_{12}, \De_3, \De_{45})   \nonumber\\
&- u_{n_{IJ}+2}(- \De_{12})  G_{(\ell-1,  \ell'; 1, 0)}^{(n_{IJ}+2)}(\De_{12}, \De_3, \De_{45}) \,.
\]

We next repeat the identical procedure for the interchange $X_4\leftrightarrow X_5$ using the associated symmetry relation \Eq{symmetry 45} and the coordinate map
\[
u_1\to \frac{u_1 w_{ 3;4 }}{w_{ 2;4 }}\,, \quad u_2\to \frac{u_2}{w_{ 3;4 }}\,, \quad w_{ 2;3 }\to  \frac{w_{ 2;3 }}{w_{ 2;4 }}\,, \quad w_{ 2;4}\to \frac{1}{w_{ 2;4 }}\,, \quad w_{ 3;4}\to \frac{1}{w_{ 3;4 }}\,.
\]
We find
\[ 
\eql{symmetry relation 2.2}
- u_2^{-1/2} w_{3;4} &G_{(\ell,  \ell'-1; 0, 0)}^{(n_{IJ})}
(\De_{12}, \De_3+1, \De_{45}+ 1) \nonumber\\
=\,&  
-  G_{(\ell, \ell'-1; 0, - 1)}^{(n_{IJ})}(\De_{12}, \De_3, \De_{45}) \nonumber\\
 &+s^\prime_{n_{IJ}}( -\De_{45}) G_{(\ell, \ell'; 0, 0)}^{(n_{IJ})}(\De_{12}, \De_3, \De_{45}) 
+ s^\prime_{n_{IJ}+1}( -\De_{45})  G_{(\ell, \ell'; 0, 0)}^{(n_{IJ}+1)}(\De_{12}, \De_3, \De_{45})  \nonumber\\
&+t^\prime_{n_{IJ}-1}( -\De_{45})  G_{(\ell, \ell'-2; 0, 0)}^{(n_{IJ}-1)}(\De_{12}, \De_3, \De_{45})
+t^\prime_{n_{IJ}}( -\De_{45})  G_{(\ell, \ell'-2; 0, 0)}^{(n_{IJ})}(\De_{12}, \De_3, \De_{45})   \nonumber\\
&+t^\prime_{n_{IJ}+1}( -\De_{45})  G_{(\ell, \ell'-2; 0, 0)}^{(n_{IJ}+1)}(\De_{12}, \De_3, \De_{45})  
- u^\prime_{n_{IJ}-1}( -\De_{45})  G_{(\ell, \ell'-1; 0,1)}^{(n_{IJ}-1)}(\De_{12}, \De_3, \De_{45}) \nonumber\\
&-u^\prime_{n_{IJ}}(-\De_{45})  G_{(\ell, \ell'-1; 0,1)}^{(n_{IJ})}(\De_{12}, \De_3, \De_{45}) 
- u^\prime_{n_{IJ}+1}( -\De_{45})  G_{(\ell, \ell'-1; 0, 1)}^{(n_{IJ}+1)}(\De_{12}, \De_3, \De_{45}) \nonumber\\
&- u^\prime_{n_{IJ}+2}( -\De_{45})  G_{(\ell, \ell'-1; 0, 1)}^{(n_{IJ}+2)}(\De_{12}, \De_3, \De_{45})\,.
\] 
These respective relations are directly analogous to the second line in Eq.~(4.22) of \cite{Dolan:2011dv}.

Armed with the additional recursion relations \Eq{symmetry relation 1.1} - \Eq{symmetry relation 2.2}, we may use them to eliminate the $[\Delta-1, \ell]$, $[\Delta'-1, \ell']$ or the $[\Delta+1, \ell]$, $[\Delta'+1, \ell']$ contributions from \Eq{13 rel rewrite} and \Eq{35 rel rewrite}. In particular, we can consider the sums of \Eq{13 rel rewrite} and \Eq{symmetry relation 1.1} as well as of \Eq{35 rel rewrite} and \Eq{symmetry relation 2.2}, which give
\[
\eql{alternative reln 1.1}
u_1^{-1/2} &(G_{(\ell-1, \ell'; 0, 0)}^{(n_{IJ})}
(\De_{12}+1, \De_3+1, \De_{45})   -  w_{2;3}  G_{(\ell-1, \ell'; 0, 0)}^{(n_{IJ})}(\De_{12}- 1, \De_3+1, \De_{45}) )\nonumber\\
  =\,&
\left(s_{n_{IJ}}+s_{n_{IJ}}\big|_{\De_{12}\to - \De_{12}} \right) G_{(\ell, \ell'; 0, 0)}^{(n_{IJ})}   
+\left(s_{n_{IJ}+1} +  s_{n_{IJ}+1}\big|_{\De_{12}\to - \De_{12}} \right) G_{(\ell, \ell'; 0, 0)}^{(n_{IJ}+1)}      \nonumber\\    
&+ \left(t_{n_{IJ}-1} + t_{n_{IJ}-1}\big|_{\De_{12}\to -\De_{12}}\right) G_{(\ell-2, \ell'; 0, 0)}^{(n_{IJ}-1)}  +\left(t_{n_{IJ}} +t_{n_{IJ}}\big|_{\De_{12}\to - \De_{12}}\right)G_{(\ell-2, \ell'; 0,0)}^{(n_{IJ})}    \nonumber\\
&+\left(t_{n_{IJ}+1}  +t_{n_{IJ}+1}\big|_{\De_{12}\to -\De_{12}}\right) G_{(\ell-2,  \ell'; 0, 0)}^{(n_{IJ}+1)}   
+  \left(u_{n_{IJ}-1} -u_{n_{IJ}-1}\big|_{\De_{12}\to-\De_{12}}\right)   G_{(\ell-1, \ell'; 1, 0)}^{(n_{IJ}-1)}    \nonumber\\
&+ \left(u_{n_{IJ}} -u_{n_{IJ}}\big|_{\De_{12}\to -\De_{12}}\right) G_{(\ell-1, \ell'; 1, 0)}^{(n_{IJ})}
+ \left(u_{n_{IJ}+1} - u_{n_{IJ}+1}\big|_{\De_{12}\to- \De_{12}}\right) G_{(\ell-1, \ell'; 1, 0)}^{(n_{IJ}+1)} \nonumber\\
&+ \left(u_{n_{IJ}+2}- u_{n_{IJ}+2}\big|_{\De_{12}\to - \De_{12}}\right)   G_{(\ell-1,  \ell'; 1, 0)}^{(n_{IJ}+2)}\,,
\] 
where the blocks' dependence on $(\De_{12}, \De_3, \De_{45})$ has been suppressed for the unshifted terms. Further, we have
\[ 
\eql{alternative reln 1.2}
u_2^{-1/2}&(G_{(\ell,  \ell'-1; 0, 0)}^{(n_{IJ})}
(\De_{12}, \De_3+1, \De_{45}-1)- w_{3;4} G_{(\ell,  \ell'-1; 0, 0)}^{(n_{IJ})}
(\De_{12}, \De_3+1, \De_{45}+ 1))\nonumber\\
=\,&  
\left(s^\prime_{n_{IJ}}+s^\prime_{n_{IJ}}\big|_{\De_{45}\to -\De_{45}}\right)G_{(\ell, \ell'; 0, 0)}^{(n_{IJ})} 
+ \left(s^\prime_{n_{IJ}+1}  + s^\prime_{n_{IJ}+1}\big|_{\De_{45}\to -\De_{45}}\right) 
G_{(\ell, \ell'; 0, 0)}^{(n_{IJ}+1)} \nonumber\\
&+\left(t^\prime_{n_{IJ}-1} +t^\prime_{n_{IJ}-1}\big|_{\De_{45}\to -\De_{45}} \right)G_{(\ell, \ell'-2; 0, 0)}^{(n_{IJ}-1)} 
+\left(t^\prime_{n_{IJ}} +t^\prime_{n_{IJ}}\big|_{\De_{45}\to -\De_{45}} \right)G_{(\ell, \ell'-2; 0, 0)}^{(n_{IJ})}\nonumber\\
&+\left(t^\prime_{n_{IJ}+1} +t^\prime_{n_{IJ}+1}\big|_{\De_{45}\to-\De_{45}}\right)G_{(\ell, \ell'-2; 0, 0)}^{(n_{IJ}+1)}
+\left(u^\prime_{n_{IJ}-1} - u^\prime_{n_{IJ}-1}\big|_{\De_{45}\to -\De_{45}}\right)G_{(\ell, \ell'-1; 0,1)}^{(n_{IJ}-1)}\nonumber\\
&+\left(u^\prime_{n_{IJ}} -u^\prime_{n_{IJ}}\big|_{\De_{45}\to-\De_{45}}\right)G_{(\ell, \ell'-1; 0,1)}^{(n_{IJ})} 
+ \left(u^\prime_{n_{IJ}+1} - u^\prime_{n_{IJ}+1}\big|_{\De_{45}\to -\De_{45}} \right) G_{(\ell, \ell'-1; 0, 1)}^{(n_{IJ}+1)}\nonumber\\
&+\left( u^\prime_{n_{IJ}+2} - u^\prime_{n_{IJ}+2}\big|_{\De_{45}\to -\De_{45}}\right)G_{(\ell, \ell'-1; 0, 1)}^{(n_{IJ}+2)}\,.
\] 
The two relations \Eq{alternative reln 1.1} - \Eq{alternative reln 1.2} involve blocks with exchanged dimensions $\De$, $\De+1$ or  $\De'$, $\De'+1$ alone. The block contributions carrying dimensions $\De-1$ and $\De'-1$ have been eliminated.

Alternatively, we may consider taking combinations \Eq{13 rel rewrite} + $c_1$ \Eq{symmetry relation 1.1} and \Eq{35 rel rewrite} + $c_2$ \Eq{symmetry relation 2.2}, with $c_1$ and $c_2$ chosen such that block pieces carrying exchanged dimensions $\De+1$ and $\De'+1$, specifically terms of the type $G_{(\ell-1,  \ell'; 1, 0)}^{(m_{IJ})}(\De_{12}, \De_3, \De_{45})$ and $G_{(\ell, \ell'-1; 0, 1)}^{(m_{IJ})}(\De_{12}, \De_3, \De_{45})$, are eliminated. 
We ultimately find 
\[
\eql{no u terms}
u_1^{-1/2} &\left(G_{(\ell-1, \ell'; 0, 0)}^{(n_{IJ})}
(\De_{12}+1, \De_3+1, \De_{45}) - c_1   w_{ 2;3 } G_{(\ell-1, \ell'; 0, 0)}^{(n_{IJ})}(\De_{12}- 1, \De_3+1, \De_{45}) \right)
\nonumber\\
=\,&
 (1-c_1)G_{(\ell-1,\ell';-1,0)}^{(n_{IJ})}   +\left(s_{n_{IJ}} +c_1 s_{n_{IJ}}\big|_{\De_{12}\to- \De_{12}}\right)G_{(\ell, \ell'; 0, 0)}^{(n_{IJ})}  
\nonumber\\
&+ \left( s_{n_{IJ}+1} +c_1s_{n_{IJ}+1}\big|_{\De_{12}\to - \De_{12}} \right)G_{(\ell, \ell'; 0, 0)}^{(n_{IJ}+1)}     
+ \left(t_{n_{IJ}-1} +c_1  t_{n_{IJ}-1}\big|_{\De_{12}\to-\De_{12}}\right) G_{(\ell-2, \ell'; 0, 0)}^{(n_{IJ}-1)}    \nonumber\\
&+\left(t_{n_{IJ}} +c_1t_{n_{IJ}}\big|_{\De_{12}\to - \De_{12}}\right) 
G_{(\ell-2, \ell'; 0,0)}^{(n_{IJ})}  
+\left(t_{n_{IJ}+1} +c_1 t_{n_{IJ}+1}\big|_{\De_{12}\to-\De_{12}}\right)  G_{(\ell-2,  \ell'; 0, 0)}^{(n_{IJ}+1)}\,,   
\]
where
\[
c_1=\frac{\left(\Delta -\Delta _{12}+\ell -2\right) \left(d-\Delta
   +\Delta _{12}+\ell -2\right)}{\left(\Delta +\Delta
   _{12}+\ell -2\right) \left(d-\Delta -\Delta _{12}+\ell
   -2\right)}\,,
\]
and
\[
\eql{no u' terms}
u_2^{-1/2} &\left(G_{(\ell,  \ell'-1; 0, 0)}^{(n_{IJ})}
(\De_{12}, \De_3+1, \De_{45}-1) - c_2  w_{ 3;4 }    G_{(\ell,  \ell'-1; 0, 0)}^{(n_{IJ})}(\De_{12}, \De_3+1, \De_{45}+ 1)\right)\nonumber\\
=\,& (1-c_2) G_{(\ell, \ell'-1; 0, - 1)}^{(n_{IJ})} 
 +\left(s^\prime_{n_{IJ}}+c_2 s^\prime_{n_{IJ}}\big|_{\De_{45}\to  -\De_{45}} \right) G_{(\ell, \ell'; 0, 0)}^{(n_{IJ})}  \nonumber\\
&+\left( s^\prime_{n_{IJ}+1}+ c_2s^\prime_{n_{IJ}+1}\big|_{\De_{45}\to -\De_{45}}  \right) G_{(\ell, \ell'; 0, 0)}^{(n_{IJ}+1)} 
+\left(t^\prime_{n_{IJ}-1} +c_2 t^\prime_{n_{IJ}-1}\big|_{\De_{45}\to  -\De_{45}} \right)G_{(\ell, \ell'-2; 0, 0)}^{(n_{IJ}-1)}  \nonumber\\
&+\left(t^\prime_{n_{IJ}} +c_2 t^\prime_{n_{IJ}}\big|_{\De_{45}\to  -\De_{45}}  \right)G_{(\ell, \ell'-2; 0, 0)}^{(n_{IJ})}
+\left(t^\prime_{n_{IJ}+1}+c_2t^\prime_{n_{IJ}+1}\big|_{\De_{45}\to  -\De_{45}}\right) G_{(\ell, \ell'-2; 0, 0)}^{(n_{IJ}+1)}\,,
\]
with 
\[
c_2=\frac{\left(\Delta '+\Delta _{45}+\ell '-2\right)
   \left(d-\Delta '-\Delta _{45}+\ell '-2\right)}{\left(\Delta
   '-\Delta _{45}+\ell '-2\right) \left(d-\Delta '+\Delta
   _{45}+\ell '-2\right)}\,.
\]

It turns out that the simplest set of recursion relations we are able to write down only involve the exchanged operators $\{[\De, \ell],[\De-1, \ell], [\De, \ell-1], [\De, \ell-2]\}$ and $\{[\De', \ell'], [\De'-1, \ell'], [\De', \ell'-1], [\De', \ell'-2]\}$, without any blocks carrying the exchanged dimensions $\De + 1$ and $\De' +1$. This result only partially achieves the goal of eliminating shifts in the exchanged operator dimensions. Although we would ideally like to obtain a set of recursion relations purely involving the exchanged operators $\{[\De, \ell], [\De, \ell-1], [\De, \ell-2]\}$ and $\{[\De', \ell'], [\De', \ell'-1], [\De', \ell'-2]\}$, we find that, unlike in the 4-point case, this is not possible with the current approach. 

Ultimately, we find that the simplest form for the first recursion relation \Eq{13 rel rewrite} is as follows:\footnote{We are grateful to Petar Tadi\'c for identifying some mistakes in the previous derivation of this formula and for verifying that the current version agrees numerically with \Eq{13 rel}.}
\[
G_{(\ell, \ell'; 0, 0)}^{(n_{IJ})} 
 &=  
   \frac{(\Delta +\ell -2) (\Delta +\ell -1)
   \left(d-\Delta -\Delta _{12}+\ell -2\right)} {\left(\Delta -\Delta _{12}+\ell -2\right) (d-\Delta
   +\ell -2) \left(\Delta '+\Delta -\Delta _3+2 n_{IJ}-\ell '+\ell
   -2\right)}   \times  \nonumber\\&\bigg[
 -u_1^{-1/2}G_{(\ell-1, \ell'; 0, 0)}^{(n_{IJ})}
(\De_{12}+1, \De_3+1, \De_{45}) \nonumber\\
&+  \frac{\left(\Delta -\Delta _{12}+\ell -2\right) \left(d-\Delta
   +\Delta _{12}+\ell -2\right)}{\left(\Delta +\Delta
   _{12}+\ell -2\right) \left(d-\Delta -\Delta _{12}+\ell
   -2\right)} u_1^{-1/2}   w_{ 2;3 } G_{(\ell-1, \ell'; 0, 0)}^{(n_{IJ})}(\De_{12}- 1, \De_3+1, \De_{45}) 
\nonumber\\
&+\frac{2 \Delta _{12} (d-2 \Delta )}{\left(\Delta +\Delta
   _{12}+\ell -2\right) \left(d-\Delta -\Delta _{12}+\ell
   -2\right)}G_{(\ell-1,\ell';-1,0)}^{(n_{IJ})}  
 \nonumber\\
&+ 
\frac{\left(\Delta -\Delta _{12}+\ell -2\right) (d-\Delta
   +\ell -2) \left(\ell '-n_{IJ}\right)}{(\Delta +\ell -2)
   (\Delta +\ell -1) \left(d-\Delta -\Delta _{12}+\ell
   -2\right)}G_{(\ell, \ell'; 0, 0)}^{(n_{IJ}+1)}     \nonumber\\
&+\frac{n_{IJ} (d+2 n_{IJ}-4) (\Delta +\ell -2) \left(d-\Delta +\Delta
   _{12}+\ell -2\right) }{(d+2 \ell -6) (d+2 \ell -4)
   \left(\Delta +\Delta _{12}+\ell -2\right) (d-\Delta +\ell
   -3) (d-\Delta +\ell -2)}
   \times  \nonumber\\&\left(2 d-\Delta '-\Delta +\Delta _3-2
   n_{IJ}+\ell '+3 \ell -6\right)
   G_{(\ell-2, \ell'; 0, 0)}^{(n_{IJ}-1)}    \nonumber\\
&-\frac{(-n_{IJ}+\ell -1) (\Delta +\ell -2) \left(d-\Delta +\Delta
   _{12}+\ell -2\right) }{(d+2 \ell -6)
   (d+2 \ell -4) \left(\Delta +\Delta _{12}+\ell -2\right)
   (d-\Delta +\ell -3) (d-\Delta +\ell -2)}\times\nonumber\\&
   \big(d^2+d \left(-\Delta '-\Delta
   +\Delta _3+n_{IJ}+2 \ell -6\right)+4 \left(\Delta '+\Delta
   -\Delta _3-n_{IJ}+2\right)\nonumber\\&-n_{IJ} \left(\Delta '+\Delta -\Delta _3+4
   n_{IJ}-3 \ell '\right)-\ell  \left(\Delta '+\Delta -\Delta _3-3
   n_{IJ}+\ell '+6\right)+\ell ^2+2 \ell '\big)
G_{(\ell-2, \ell'; 0,0)}^{(n_{IJ})}  \nonumber\\
&-\frac{(-n_{IJ}+\ell -2) (-n_{IJ}+\ell -1) (\Delta +\ell -2)
   \left(d-\Delta +\Delta _{12}+\ell -2\right) \left(n_{IJ}-\ell
   '\right)}{(d+2 \ell -6) (d+2 \ell -4) \left(\Delta +\Delta
   _{12}+\ell -2\right) (d-\Delta +\ell -3) (d-\Delta +\ell
   -2)} G_{(\ell-2,  \ell'; 0, 0)}^{(n_{IJ}+1)}\bigg].
\]
We may then make the replacements $\De\leftrightarrow \De', \ell\leftrightarrow \ell', \De_{12}\to -\De_{45}$ as well as $u_1\to u_2$ and $w_{ 2;3 }\to w_{ 3;4 }$, inside this result to obtain the corresponding form for the second relation, which holds the spin $\ell$ fixed while varying $\ell'$.

In the following subsection, we derive a different recursion relation for the symmetric traceless five-point blocks $G_{\De, \ell, \De', \ell'}^{(n_{IJ})}(u_i)$. As we will see, this relation involves blocks with shifted external dimensions $\De_1$ and $\De_5$ only, while $\De_3$ is held fixed. This feature makes it possible to analyze the ``natural" 4-point limit $\De_3\to 0$ in the context of such a relation. This relation will also be needed in order for us to lower the spin of blocks in the special case $n_{IJ} = \ell = \ell'$.

\subsubsection{Recursion relations at fixed $\Delta_3$}

We may proceed to write down an alternative recursion relation which holds $\Delta_3$ fixed as follows. We consider acting on the 5-point function with the combination of operators
\[
(\scr{D}_{X_1}^{(-0)}\cdot\scr{D}_{X_5}^{(-0)})W_{\De, \ell, \De', \ell'; \De_1, \De_2, \De_3, \De_4, \De_5}^{(n_{IJ})}.
\]
The calculations are identical to the above analysis through \Eq{after 1}. 
We subsequently use the crossing relation for 3-point structures to move the operators $\scr{D}_{X_I}^{(b)A}$ in each of the middle $3$-point structures $\scr{D}^{(b)A}_{X_I}\vev{\scr{O}_{\De,\ell}(X_I)\Phi_{\De_3}(X_3)\scr{O}_{\De^\prime,\ell^\prime}(X_J)}^{(a)}$ onto the 
$\scr{O}_{\De^\prime,\ell^\prime}(X_J)$ via
\[
\scr{D}_{X_I}^{(b) A} &\vev{\scr{O}_{\De,\ell}(X_I)\Phi_{\De_3}(X_3)
\scr{O}'_{\De^\prime,\ell^\prime}(X_J)}^{(a)} \nonumber\\
=&\sum_{m, n}
\bigg\{
\begin{matrix} 
\scr{O}'_{\De^\prime,\ell^\prime} & \Phi_{\De_3} & 
\scr{O}^\prime_{\De^\prime-\de\De_n,\ell^\prime-\de\ell_n} \\
\scr{O}_{\De+\de\De_b,\ell+\de\ell_b} & \scr{V} & \scr{O}_{\De,\ell}
\end{matrix}\bigg\}^{(a) (b)}_{(m) (n)} \nonumber\\
&\times \scr{D}^{(n)}_{X_J}  \vev{\scr{O}_{\De+\de\De_b,\ell+\de\ell_b} (X_I)\Phi_{\De_3}(X_3)
\scr{O}^\prime_{\De^\prime-\de\De_n,\ell^\prime-\de\ell_n}(X_J)}^{(m)}\,.
\]

Upon expanding the $n$ sum and labeling the structures using the box tensor basis, we obtain
\[
\scr{D}_{X_I}^{(b)A} &\vev{\scr{O}_{\De,\ell}(X_I)\Phi_{\De_3}(X_3)
\scr{O}'_{\De^\prime,\ell^\prime}(X_J)}^{(n_{IJ})}\nonumber\\
=&\sum_{m_{IJ}=0}^{\text{min}(\ell+\de\ell_b,\ell^\prime)} 
\scr{E}_{m_{IJ}(+0)}^{n_{IJ}(b)} \scr{D}_{X_J}^{(+0)A}  \vev{\scr{O}_{\De+\de\De_b,\ell+\de\ell_b}(X_I)\Phi_{\De_3}(X_3)
\scr{O}'_{\De^\prime-1,\ell^\prime}(X_J)}^{(m_{IJ})}
\nonumber\\
&+\sum_{m_{IJ}=0}^{\text{min}(\ell+\de\ell_b, \ell^\prime+1)} 
\scr{E}_{m_{IJ}(0-)}^{n_{IJ}(b)}
\scr{D}_{X_J}^{( 0-)A} \vev{\scr{O}_{\De+\de\De_b,\ell+\de\ell_b}(X_I)\Phi_{\De_3}(X_3)
\scr{O}'_{\De^\prime,\ell^\prime+1} (X_J)}^{(m_{IJ})}
\nonumber\\
&+\sum_{m_{IJ}=0}^{\text{min}(\ell+\de\ell_b, \ell^\prime-1)} 
\scr{E}_{m_{IJ}(0+)}^{n_{IJ}(b)} 
\scr{D}_{X_J}^{(0+)A}  \vev{\scr{O}_{\De+\de\De_b,\ell+\de\ell_b}(X_I)\Phi_{\De_3}(X_3)
\scr{O}'_{\De^\prime,\ell^\prime-1}(X_J)}^{(m_{IJ})}
\nonumber\\
&+\sum_{m_{IJ}=0}^{\text{min}(\ell+\de\ell_b, \ell^\prime)}  
\scr{E}_{m_{IJ}(-0)}^{n_{IJ}(b)}\scr{D}_{X_J}^{(-0)A}  \vev{\scr{O}_{\De+\de\De_b,\ell+\de\ell_b} (X_I)\Phi_{\De_3}(X_3)
\scr{O}'_{\De^\prime+1,\ell^\prime}(X_J)}^{(m_{IJ})}\,,
\]
where the $\scr{E}$ coefficients label the relevant 3-point $6j$ symbols.

The next step is to integrate by parts using the rule \Eq{by parts rule} in order to move the $\scr{D}_{X_J}^{(c)A}$ onto the rightmost 3-point structure exactly as before. Once we collect the entire resulting expression, we reach the following result: 
\[
\eql{recursion 15}
G_{(\ell, \ell'; 0, 0)}^{(n_{IJ}+1)}
&=\frac{1}{s_{2\,n_{IJ}+1}^{(0+)} }\bigg(\tilde{f}(u_i) G_{(\ell-1, \ell'-1; 0, 0)}^{(n_{IJ})}\bigg|_{\De_1\to \De_1+1, \De_5\to \De_5+1}
-\bigg[
G_{(\ell-1, \ell'-1; -1, - 1)}^{(n_{IJ})}  \nonumber\\
&+ \sum_{m_{IJ}=n_{IJ}}^{n_{IJ}+1} 
r_{1\,m_{IJ}}^{(-0)}G_{(\ell-1, \ell'; -1, 0)}^{(m_{IJ})} 
+\sum_{m_{IJ}=n_{IJ}-1}^{n_{IJ}+1} 
r_{2\,m_{IJ}}^{(-0)} G_{(\ell-1, \ell'-2; -1, 0)}^{(m_{IJ})}   
+ \sum_{m_{IJ}=n_{IJ}-1}^{n_{IJ}+2}  
r_{3\,m_{IJ}}^{(-0)}G_{(\ell-1, \ell'-1; -1, 1)}^{(m_{IJ})} \nonumber\\
&+   \sum_{m_{IJ}=n_{IJ}}^{n_{IJ}+1} 
s_{1\,m_{IJ}}^{(0+)}G_{(\ell, \ell'-1; 0,  - 1)}^{(m_{IJ})}
+s_{2\,n_{IJ}}^{(0+)}G_{(\ell, \ell'; 0, 0)}^{(n_{IJ})}   +
s_{2\,n_{IJ}+2}^{(0+)}G_{(\ell, \ell'; 0, 0)}^{(n_{IJ}+2)}   
+\sum_{m_{IJ}=n_{IJ}-1}^{n_{IJ}+2} 
s_{3\,m_{IJ}}^{(0+)}G_{(\ell, \ell'-2; 0, 0)}^{(m_{IJ})}   \nonumber\\
&+ \sum_{m_{IJ}=n_{IJ}-1}^{n_{IJ}+3}  
s_{4\,m_{IJ}}^{(0+)}G_{(\ell, \ell'-1; 0, 1)}^{(m_{IJ})}
+ \sum_{m_{IJ}=n_{IJ}-1}^{n_{IJ}+1} 
t_{1\,m_{IJ}}^{(0-)}G_{(\ell-2, \ell'-1; 0, - 1)}^{(m_{IJ})}  
+ 
\sum_{m_{IJ}=n_{IJ}-1}^{n_{IJ}+2} 
t_{2\,m_{IJ}}^{(0-)}G_{(\ell-2, \ell'; 0, 0)}^{(m_{IJ})}    \nonumber\\
&+\sum_{m_{IJ}=n_{IJ}-2}^{n_{IJ}+2} 
t_{3\,m_{IJ}}^{(0-)} G_{(\ell-2, \ell'-2; 0,0)}^{(m_{IJ})}    
+ \sum_{m_{IJ}=n_{IJ}-2}^{n_{IJ}+2}  
t_{4\,m_{IJ}}^{(0-)}G_{(\ell-2, \ell'-1; 0, 1)}^{(m_{IJ})} 
+  \sum_{m_{IJ}=n_{IJ}-1}^{n_{IJ}+2} 
u_{1\,m_{IJ}}^{(+0)}G_{(\ell-1, \ell'-1; 1, - 1)}^{(m_{IJ})}      \nonumber\\
&+\sum_{m_{IJ}=n_{IJ}-1}^{n_{IJ}+3}
u_{2\,m_{IJ}}^{(+0)}G_{(\ell-1, \ell'; 1, 0)}^{(m_{IJ})}    
+
\sum_{m_{IJ}=n_{IJ}-2}^{n_{IJ}+3} 
u_{3\,m_{IJ}}^{(+0)}G_{(\ell-1, \ell'-2; 1, 0)}^{(m_{IJ})}   
+ \sum_{m_{IJ}=n_{IJ}-2}^{n_{IJ}+4}  
u_{4\,m_{IJ}}^{(+0)}G_{(\ell-1, \ell'-1; 1,1)}^{(m_{IJ})}
\bigg]\bigg)\,,
\]
where $\tilde{f}(u_i)$ is again a convention-dependent prefactor. Here both spins vary simultaneously, while $\De_3$ is held fixed. 

For the set of conventions
\Eq{convention 1 prefactor}-\Eq{convention 1 cross-ratios}, this prefactor is given by
\[
\tilde{f}(u_i)=(u_1 u_2)^{-1/2}\,,
\]
while for the set \Eq{convention 2 prefactor}-\Eq{convention 2 cross-ratios}, we have
\[
\tilde{f}(u_i)=w'(u_1' u_2')^{-1/2}\,.
\]
Lastly, for the set \Eq{convention 3 prefactor}-\Eq{convention 3 cross-ratios}, we find
\[
\tilde{f}(u_i) = 1.
\]

In contrast to the previous set of relations \Eq{13 rel} and \Eq{35 rel}, this recursion relation treats the spins $\ell$ and $\ell'$ on an equal footing, allowing both of them to shift. The coefficients appearing in this recursion relation are somewhat cumbersome, but they are all given explicitly in an associated \texttt{Mathematica} file.

As we did previously, we may combine \Eq{recursion 15} with the symmetry relations \Eq{symmetry 12} - \Eq{symmetry 45} to construct a set of similar equations with different shifts. We write these explicitly in Appendix~\ref{app:C}.

In the next subsection, we perform several checks in order to verify that these relations collapse to the expected forms in various 4-point limits.

\subsubsection{Checks: 4-point limits}

Due to the symmetry of the 5-point function under the interchanges $1\leftrightarrow 2$ and $4\leftrightarrow 5$, we expect that there are only three independent four-point limits to consider. In particular, we first check the simple cases of $\phi_{\De_2} \to \mathds{1}$ ($\De_2\to 0, X_2 \to X_3$) and $\phi_{\De_4}\to \mathds{1}$ ($\De_4\to 0, X_4\to X_3$). Further, we seek to analyze the ``natural" 4-point function limit $\Phi_{\De_3} \to \mathds{1}$ (so that $\De_3\to 0, X_3\to X_4$). However, since the relations \Eq{13 rel} and \Eq{35 rel} both involve 5-point blocks with shifted dimension $\De_3$ in one of the terms, taking this limit is less natural in the context of these relations. On the other hand, it is straightforward to apply it to \Eq{recursion 15}.

We begin by considering the case $\phi_{\Delta_2} \to \mathds{1}$:
\[
\lim_{\phi_{\De_2}\to \mathds{1}, X_2 \to X_3} \vev{\phi_{\De_1}(X_1)
\phi_{\De_2}(X_2) \Phi_{\De_3}(X_3) \phi_{\De_4}(X_4)\phi_{\De_5}(X_5)} 
= \vev{\phi_{\De_1}(X_1)\Phi_{\De_3}(X_3)\phi_{\De_4}(X_4)\phi_{\De_5}(X_5)}\,.
\]
To directly derive the recursion relation for the 4-point function $\vev{\phi_{\De_1}(X_1)\Phi_{\De_3}(X_3)\phi_{\De_4}(X_4)\phi_{\De_5}(X_5)}$, we simply act on the correlator with the combination $\scr{D}_{X_3}^{(-0)}\cdot \scr{D}_{X_5}^{(-0)}$ (at positions 2 and 4). This leads to the result
\[
\eql{4-point 2}
G_{\De,\ell}(u, v)= \frac{1}{s^{(24)} }\bigg(u^{-1/2} G_{\De, \ell-1}(u, v)\bigg|_{\De_3\to\De_3+1,\De_5\to\De_5+1}
- G_{\De-1,\ell-1}(u, v)\nonumber\\ -t^{(24)}  G_{\De,\ell-2}(u, v) - u^{(24)}   G_{\De+1,\ell-1}(u, v)\bigg)\,,
\]
where we make the replacements $X_2\to X_3$, $X_3\to X_4$, $X_4\to X_5$ in $u$ and $v$. Explicitly, in this case, we have that
\[
u\to \dfrac{X_{13} X_{45}}{X_{14} X_{35}},\quad
v\to  \dfrac{X_{15} X_{34}}{X_{14} X_{35}}.
\]
It is straightforward to extract the respective coefficients. It turns out that these are related to the coefficients \Eq{coeffs standard} from the standard relation \Eq{D and O} in the following way:
\[
s^{(24)}&=-\bigg(s^{(14)}\bigg|_{\De_{12}\to -\De_{12}}\bigg)\bigg|_{\De_2\to \De_3, \De_3\to \De_4, \De_4\to \De_5}\,,\\
t^{(24)}&=-\bigg(t^{(14)}\bigg|_{\De_{12}\to -\De_{12}}\bigg)\bigg|_{\De_2\to \De_3, \De_3\to \De_4, \De_4\to \De_5}\,,\\
u^{(24)}&= \bigg(u^{(14)}\bigg|_{\De_{12}\to -\De_{12}}\bigg)\bigg|_{\De_2\to \De_3, \De_3\to \De_4, \De_4\to \De_5}\,. 
\]
Equipped with this result, we next demonstrate explicitly that the 5-point recursion relations \Eq{13 rel} and \Eq{35 rel} indeed reduce to this form in the appropriate limit. We will take $\phi_{\Delta_2} \to \mathds{1}$ and $\scr{O}_{\De, \ell} \to \phi_{\De_1}$ in our 4-point limit of interest so that we should also take $\De_2\to 0$, $\De \to \De_1$, $\ell \to 0$, and $n_{IJ} \rightarrow 0$. 

Upon fixing $n_{IJ}= 0$ and $\ell = 0$, we find that \Eq{13 rel} is irrelevant, as $\ell = 0$ already, while \Eq{35 rel} assumes the form
 \[ 
 \eql{limit 2}
G_{(0, \ell'; 0, 0)}^{(0)} 
 =  \frac{1}{s^\prime_{0}}
 \bigg(u_2^{-1/2} G_{(0,  \ell'-1; 0, 0)}^{(0)}(u_2, w_{ 3;4 })\bigg|_{\De_3 \to \De_3+1, \De_5\to \De_5+1}
 - G_{(0, \ell'-1; 0, - 1)}^{(0)}(u_2, w_{ 3;4 })\nonumber\\
-t^\prime_{0} G_{(0, \ell'-2; 0, 0)}^{(0)}(u_2, w_{ 3;4 })
-u^\prime_{0} G_{(0, \ell'-1; 0,1)}^{(0)}(u_2, w_{ 3;4 })
\bigg)\,,
\] 
where we have fixed the choice of conventions to \Eq{convention 1 prefactor} - \Eq{convention 1 cross-ratios}.

We may now directly compare the two relations \Eq{4-point 2} and \Eq{limit 2} and conclude that they do in fact match. Further, upon checking the respective coefficients, we may readily verify that
\[
s^\prime_{0}=s^{(24)}, \quad t^\prime_{0}=t^{(24)}, \quad u^\prime_{0}=u^{(24)}\,,
\] 
after setting $n_{IJ}=0$, $\De_2=0$, $\De =\Delta_1$, $\ell = 0$.

Here each of the contributions $ G_{(0, \ell'; 0, 0)}^{(0)} $ corresponds to a 4-point block. With this, we have shown directly that the 5-point recursion relations indeed reduce to the 4-point relation \Eq{4-point 2} in the limit $\phi_{\Delta_2}\to \mathds{1}$.

One may examine the case of $\phi_{\De_4} \rightarrow \mathds{1}$ in the exact same manner. This time, we consider the limit
\[
\lim_{\phi_{\De_4}\to \mathds{1}, X_4 \to X_3} \vev{\phi_{\De_1}(X_1)
\phi_{\De_2}(X_2) \Phi_{\De_3}(X_3) \phi_{\De_4}(X_4)\phi_{5}(X_5)} 
= \vev{\phi_{\De_1}(X_1)
\phi_{\De_2}(X_2) \Phi_{\De_3}(X_3) \phi_{5}(X_5)}\,.
\]
Here the associated recursion relation is directly extracted by acting with $\scr{D}_{X_1}^{(-0)}\cdot \scr{D}_{X_3}^{(-0)}$. It is effortless to obtain
\[
\eql{4-point 4}
G_{\De,\ell}(u, v)= \frac{1}{s^{(13)} }\bigg(u^{-1/2} G_{\De, \ell-1}(u, v)\bigg|_{\De_1\to\De_1+1,\De_3\to\De_3+1}
- G_{\De-1,\ell-1}(u, v)\nonumber\\ -t^{(13)}  G_{\De,\ell-2}(u, v) - u^{(13)}   G_{\De+1,\ell-1}(u, v)\bigg)\,,
\]
where we replace $X_4\to X_5$ inside $u$ and $v$. In particular, we have
\[
u\to \dfrac{X_{12} X_{35}}{X_{13} X_{25}}\,,\quad
v\to  \dfrac{X_{15} X_{23}}{X_{13} X_{25}}\,.
\]
Here the coefficients are given by
\[
s^{(13)}&=-\bigg(s^{(14)}\bigg|_{\De_{34}\to -\De_{34}}\bigg)\bigg|_{\De_4\to \De_5}\,, \\
t^{(13)} &=-\bigg(t^{(14)}\bigg|_{\De_{34}\to -\De_{34}}\bigg)\bigg|_{\De_4\to \De_5}\,, \\
u^{(13)} &=\bigg(u^{(14)}\bigg|_{\De_{34}\to -\De_{34}}\bigg)\bigg|_{\De_4\to \De_5}\,.  
\]
Turning to the 5-point relations, we note that we now have $\phi_{\De_4} \to \mathds{1}$ and $\scr{O}'_{\De', \ell'} \to \phi_{\De_5}$ 
so that $\De_4\to 0$, $\De' \to \Delta_5$, $\ell' \to 0$, and $n_{IJ} \to 0$. Taking $X_4 \to X_3$ inside the cross-ratios, we find that this time \Eq{35 rel} is irrelevant, while \Eq{13 rel} takes on the form
\[
\eql{limit 4} 
 G_{(\ell, 0; 0, 0)}^{(0)}  = \frac{1}{s_0}\bigg(u_1^{-1/2} G_{(\ell-1, 0; 0, 0)}^{(0)}(u_1, w_{2;3})\bigg|_{\De_1\to \De_1+1, \De_3\to \De_3+1}
-  G_{(\ell-1,0;-1,0)}^{(0)}(u_1, w_{2;3})  \nonumber\\
-t_0 G_{(\ell-2, 0; 0,0)}^{(0)}(u_1, w_{2;3}) 
-u_0 G_{(\ell-1, 0; 1, 0)}^{(0)}(u_1, w_{2;3})
\bigg)\,,
\] 
where we have again chosen the conventions \Eq{convention 1 prefactor} - \Eq{convention 1 cross-ratios}.
It is straightforward to see that the form of the two relations \Eq{4-point 4} and \Eq{limit 4} is identical. Upon matching coefficients, we find that indeed
\[
s_{0}=s^{(13)}, \quad t_{0}=t^{(13)},\quad u_{0}=u^{(13)},
\]
as expected and desired. Once again, we see that our 5-point result in fact reduces to the appropriate 4-point relation in this limit.

Lastly, we wish to check that the alternate recursion relation \Eq{recursion 15} reproduces the original 4-point relation due to Dolan and Osborn, namely \Eq{D and O} for the 4-point function $\vev{\phi_{\De_1}\phi_{\De_2}\phi_{\De_4}\phi_{\De_5}}$ in the limit $\Phi_{\De_3}\rightarrow \mathds{1}$. That is, we expect that
\[
\lim_{\Phi_{\De_3}\to \mathds{1}, X_3 \to X_2} \vev{\phi_{\De_1}(X_1)
\phi_{\De_2}(X_2) \Phi_{\De_3}(X_3) \phi_{\De_4}(X_4)\phi_{\De_5}(X_5)} 
=\vev{\phi_{\De_1}(X_1)
\phi_{\De_2}(X_2) \phi_{\De_4}(X_4)\phi_{\De_5}(X_5)}\,.
\]
In particular, we expect to verify that the five-point relation reduces to 
\[
\eql{D and O 15}
G_{\De,\ell}(u, v)= \frac{1}{s^{(15)} }\bigg(u^{-1/2} G_{\De, \ell-1}(u, v)\bigg|_{\De_1\to\De_1+1,\De_5\to\De_5+1}
- G_{\De-1,\ell-1}(u, v)\nonumber\\ -t^{(15)}  G_{\De,\ell-2}(u, v) - u^{(15)}   G_{\De+1,\ell-1}(u, v)\bigg)\,,
\]
where the coefficients are given by
\[
s^{(15)}&=s^{(14)}\bigg|_{\De_3\to \De_4, \De_4\to \De_5}\,,\\
t^{(15)}&=t^{(14)}\bigg|_{\De_3\to \De_4, \De_4\to \De_5}\,,\\
u^{(15)}&=u^{(14)}\bigg|_{\De_3\to \De_4, \De_4\to \De_5}\,.
\]
Further, we make the replacements  $X_3\to X_4$, $X_4\to X_5$ inside $u$ and $v$. In particular, we have
\[
u\to \dfrac{X_{12} X_{45}}{X_{14} X_{25}}\,,\quad
v\to  \dfrac{X_{15} X_{24}}{X_{14} X_{25}}\,.
\]
We next note that in the 4-point function limit $\Phi_{\De_3}\rightarrow \mathds{1}$, we have $X_3 \to X_2$ and $\De_3 = 0$, while the exchanged operators coincide, i.e.~$\De' =\De$, $\ell' = \ell$.  Choosing the surviving structure $n_{IJ} + 1= \ell$, the original recursion relation \Eq{recursion 15} reduces to
\[
\eql{limit 3}
G_{(\ell, \ell; 0, 0)}^{(\ell)}
=\frac{1}{s_{2 \,\ell}^{(0+)} }\bigg((u_1 u_2)^{-1/2} G_{(\ell-1, \ell-1; 0, 0)}^{(\ell-1)}\bigg|_{\De_1\to \De_1+1, \De_5\to \De_5+1}(u_1 u_2, w_{ 2;4 } )
-G_{(\ell-1, \ell-1; -1, - 1)}^{(\ell-1)}(u_1 u_2, w_{ 2;4 } )  \nonumber\\
- t_{3\,\ell-2}^{(0-)} G_{(\ell-2, \ell-2; 0,0)}^{(\ell-2)} (u_1 u_2, w_{ 2;4 } ) 
-u_{4 \, \ell-1}^{(+0)}G_{(\ell-1, \ell-1; 1,1)}^{(\ell-1)}(u_1 u_2, w_{ 2;4 } )
\bigg)\,,
\]
where we have yet again fixed the conventions to \Eq{convention 1 prefactor} - \Eq{convention 1 cross-ratios}. Once again, a direct comparison between the forms of \Eq{D and O 15} and \Eq{limit 3} reveals a perfect match. Specifically, we verify that
\[ 
s_{2 \,\ell}^{(0+)}=
s^{(15)}\,, \quad
 t_{3\,\ell-2}^{(0-)}= t^{(15)}\,, \quad
 u_{4 \, \ell-1}^{(+0)}=
u^{(15)}\,.
\]
The above checks help establish the validity of the recursion relations \Eq{13 rel}, \Eq{35 rel}, and \Eq{recursion 15}. We have also checked that the five-point blocks obtained by means of these relations indeed reduce to the appropriate four-point blocks in these various four-point limits. 

In the next section, we will combine \Eq{recursion 15} with \Eq{13 rel} in order to obtain a general recursion relation for the block $G_{(\ell, \ell; 0, 0)}^{(\ell)}$, which was not handled by the single-spin relations \Eq{13 rel} and \Eq{35 rel}. 

\subsection{Treating the special case $G_{(\ell, \ell; 0, 0)}^{(\ell)}$}

We next proceed to treat the special case $G_{(\ell, \ell; 0, 0)}^{(\ell)}$. It is straightforward to see that if we set $\ell=\ell'$ and $n_{IJ} =\ell-1$, then \Eq{13 rel} and \Eq{35 rel} collapse to a single relation. In particular, we find
\[
\eql{13 special}
G_{(\ell, \ell; 0, 0)}^{(\ell)}     
  = \frac{1}{s_{\ell} }\bigg(f(u_i) G_{(\ell-1, \ell; 0, 0)}^{(\ell-1)}(\De_{12}+1, \De_3+1,\De_{45})
-  G_{(\ell-1,\ell;-1,0)}^{(\ell-1)}  
- s_{\ell-1}G_{(\ell, \ell; 0, 0)}^{(\ell-1)}     
 \nonumber\\
- t_{\ell-2}  G_{(\ell-2, \ell; 0, 0)}^{(\ell-2)} 
-u_{\ell-2}   G_{(\ell-1, \ell; 1, 0)}^{(\ell-2)} 
-u_{\ell-1} G_{(\ell-1, \ell; 1, 0)}^{(\ell-1)}
\bigg).
\]
It is evident that we require an additional independent relation which may be invoked to replace the extra block $G_{(\ell, \ell; 0, 0)}^{(\ell-1)}$ by lower-spin contributions, leaving only our block of interest, i.e.~$G_{(\ell, \ell; 0, 0)}^{(\ell)}$, as desired. To this end, we may apply our additional  $(15)$ relation \Eq{recursion 15} for the 5-point blocks, which varies both spins simultaneously. Since the latter relation features the structures $G_{(\ell, \ell; 0, 0)}^{( \ell)}$ and $G_{(\ell, \ell; 0, 0)}^{(\ell-1)}$ inside a different linear combination, we can combine the relations and solve for $G_{(\ell, \ell; 0, 0)}^{( \ell)}$. 

In particular, upon setting $\ell'=\ell$ and $n_{IJ}=\ell-1$ inside \Eq{recursion 15}, we find
\[
\eql{recursion 15 special}
G_{(\ell, \ell; 0, 0)}^{(\ell)}
&=\frac{1}{s_{2\,\ell}^{(0+)} }\bigg(\tilde{f}(u_i) G_{(\ell-1, \ell-1; 0, 0)}^{(\ell-1)}(\De_{12}+1, \De_3,\De_{45}-1)
-\bigg[
G_{(\ell-1, \ell-1; -1, - 1)}^{(\ell-1)}  \nonumber\\
&+  
r_{1\,(\ell-1)}^{(-0)}G_{(\ell-1, \ell; -1, 0)}^{(\ell-1)} 
+ r_{2\,(\ell-2)}^{(-0)} G_{(\ell-1, \ell-2; -1, 0)}^{(\ell-2)}   
+  
r_{3\,(\ell-2)}^{(-0)}G_{(\ell-1, \ell-1; -1, 1)}^{(\ell-2)} \nonumber\\
&+
r_{3\,(\ell-1)}^{(-0)}G_{(\ell-1, \ell-1; -1, 1)}^{(\ell-1)}+   s_{1\,(\ell-1)}^{(0+)}G_{(\ell, \ell-1; 0,  - 1)}^{(\ell-1)} +s_{2\,\ell-1}^{(0+)}G_{(\ell, \ell; 0, 0)}^{(\ell-1)}    
+s_{3\,(\ell-2)}^{(0+)}G_{(\ell, \ell-2; 0, 0)}^{(\ell-2)}   \nonumber\\
&+  
s_{4\,(\ell-2)}^{(0+)}G_{(\ell, \ell-1; 0, 1)}^{(\ell-2)}
+s_{4\,(\ell-1)}^{(0+)}G_{(\ell, \ell-1; 0, 1)}^{(\ell-1)}
+ t_{1\,(\ell-2)}^{(0-)}G_{(\ell-2, \ell-1; 0, - 1)}^{(\ell-2)}  
+ t_{2\,(\ell-2)}^{(0-)}G_{(\ell-2, \ell; 0, 0)}^{(\ell-2)}    \nonumber\\
&+ 
t_{3\,(\ell-3)}^{(0-)} G_{(\ell-2, \ell-2; 0,0)}^{(\ell-3)}   
+t_{3\,(\ell-2)}^{(0-)} G_{(\ell-2, \ell-2; 0,0)}^{(\ell-2)}    
+ 
t_{4\,(\ell-3)}^{(0-)}G_{(\ell-2, \ell-1; 0, 1)}^{(\ell-3)} 
+t_{4\,(\ell-2)}^{(0-)}G_{(\ell-2, \ell-1; 0, 1)}^{(\ell-2)}  \nonumber\\
&+ u_{1\,(\ell-2)}^{(+0)}G_{(\ell-1, \ell-1; 1, - 1)}^{(\ell-2)}  
+u_{1\,(\ell-1)}^{(+0)}G_{(\ell-1, \ell-1; 1, - 1)}^{(\ell-1)}  + u_{2\,(\ell-2)}^{(+0)}G_{(\ell-1, \ell; 1, 0)}^{(\ell-2)}    
+u_{2\,(\ell-1)}^{(+0)}G_{(\ell-1, \ell; 1, 0)}^{(\ell-1)}    \nonumber\\
&  
+u_{3\,(\ell-3)}^{(+0)}G_{(\ell-1, \ell-2; 1, 0)}^{(\ell-3)}   
+u_{3\,(\ell-2)}^{(+0)}G_{(\ell-1, \ell-2; 1, 0)}^{(\ell-2)}   
+u_{4\,(\ell-3)}^{(+0)}G_{(\ell-1, \ell-1; 1,1)}^{(\ell-3)}
 \nonumber\\
&  +
u_{4\,(\ell-2)}^{(+0)}G_{(\ell-1, \ell-1; 1,1)}^{(\ell-2)}
+u_{4\,(\ell-1)}^{(+0)}G_{(\ell-1, \ell-1; 1,1)}^{(\ell-1)}
\bigg]\bigg)\,.
\]
Upon solving this relation for $G_{(\ell, \ell; 0, 0)}^{(\ell-1)}$ and inserting the result inside \Eq{13 special}, we eventually obtain our desired relation:
\[
\eql{special case}
G_{(\ell, \ell; 0, 0)}^{(\ell)}
&  = 
(\bar{s}_{2\,\ell}^{(0+)})^{-1}\bigg(f(u_i) G_{(\ell-1, \ell; 0, 0)}^{(\ell-1)}(\De_{12}+1, \De_3+1,\De_{45}) 
\nonumber\\&
- s_{\ell-1} (s_{2\,\ell-1}^{(0+)})^{-1}\tilde{f}(u_i) G_{(\ell-1, \ell-1; 0, 0)}^{(\ell-1)}(\De_{12}+1, \De_3,\De_{45}-1)
\nonumber\\&
+\dfrac{ s_{\ell-1} }{s_{2\,\ell-1}^{(0+)}}
\bigg[
\bar{r}_{0\,(\ell-1)}^{(-0)}
G_{(\ell-1, \ell-1; -1, - 1)}^{(\ell-1)} 
+\bar{r}_{1\,(\ell-1)}^{(-0)} G_{(\ell-1,\ell;-1,0)}^{(\ell-1)}    
\nonumber\\&+ 
\bar{r}_{2\,(\ell-2)}^{(-0)}  G_{(\ell-1, \ell-2; -1, 0)}^{(\ell-2)}   
+  
\bar{r}_{3\,(\ell-2)}^{(-0)} G_{(\ell-1, \ell-1; -1, 1)}^{(\ell-2)} \nonumber\\
&+
\bar{r}_{3\,(\ell-1)}^{(-0)} G_{(\ell-1, \ell-1; -1, 1)}^{(\ell-1)}+\bar{s}_{1\,(\ell-1)}^{(0+)}  G_{(\ell, \ell-1; 0,  - 1)}^{(\ell-1)}
+\bar{s}_{3\,(\ell-2)}^{(0+)}G_{(\ell, \ell-2; 0, 0)}^{(\ell-2)}   \nonumber\\
&+  
\bar{s}_{4\,(\ell-2)}^{(0+)} G_{(\ell, \ell-1; 0, 1)}^{(\ell-2)}
+\bar{s}_{4\,(\ell-1)}^{(0+)}G_{(\ell, \ell-1; 0, 1)}^{(\ell-1)}
\nonumber\\
&+   \bar{t}_{1\,(\ell-2)}^{(0-)}G_{(\ell-2, \ell-1; 0, - 1)}^{(\ell-2)}  
+\bar{t}_{2\,(\ell-2)}^{(0-)}G_{(\ell-2, \ell; 0, 0)}^{(\ell-2)} \nonumber\\&
+ 
\bar{t}_{3\,(\ell-3)}^{(0-)}  G_{(\ell-2, \ell-2; 0,0)}^{(\ell-3)}   
+\bar{t}_{3\,(\ell-2)}^{(0-)} G_{(\ell-2, \ell-2; 0,0)}^{(\ell-2)}    
   \nonumber\\
&+ 
\bar{t}_{4\,(\ell-3)}^{(0-)}G_{(\ell-2, \ell-1; 0, 1)}^{(\ell-3)} 
+\bar{t}_{4\,(\ell-2)}^{(0-)} G_{(\ell-2, \ell-1; 0, 1)}^{(\ell-2)}  \nonumber\\
&+ \bar{u}_{1\,(\ell-2)}^{(+0)} G_{(\ell-1, \ell-1; 1, - 1)}^{(\ell-2)}  
+\bar{u}_{1\,(\ell-1)}^{(+0)}G_{(\ell-1, \ell-1; 1, - 1)}^{(\ell-1)}    \nonumber\\
&  
+\bar{u}_{2\,(\ell-2)}^{(+0)}G_{(\ell-1, \ell; 1, 0)}^{(\ell-2)}   
+\bar{u}_{2\,(\ell-1)}^{(+0)}G_{(\ell-1, \ell; 1, 0)}^{(\ell-1)}   \nonumber\\&
+\bar{u}_{3\,(\ell-3)}^{(+0)}G_{(\ell-1, \ell-2; 1, 0)}^{(\ell-3)}   
+\bar{u}_{3\,(\ell-2)}^{(+0)} G_{(\ell-1, \ell-2; 1, 0)}^{(\ell-2)}   
+\bar{u}_{4\,(\ell-3)}^{(+0)} G_{(\ell-1, \ell-1; 1,1)}^{(\ell-3)}
 \nonumber\\
&  +
\bar{u}_{4\,(\ell-2)}^{(+0)} G_{(\ell-1, \ell-1; 1,1)}^{(\ell-2)}
+\bar{u}_{4\,(\ell-1)}^{(+0)}G_{(\ell-1, \ell-1; 1,1)}^{(\ell-1)}
\bigg]
\bigg),
\]
where the coefficients are listed in Appendix~\ref{app:B.1}.
We remark that each block on the RHS of this relation features a downshifted value of the spin, exactly as desired. Note that an alternate version of this equation eliminating some coefficients can also be obtained using~\Eq{eq:15 simpler} instead of \Eq{recursion 15}.

In the next section, we will apply the results \Eq{13 rel}, \Eq{35 rel} and \Eq{special case} to determine the 5-point conformal blocks for various cases of interest involving exchanged conserved operators.

\subsection{Exchange of conserved operators}
\label{sec:conservation}

Our next objective is to analyze the situation of exchanged conserved operators in the scalar 5-point function. Before examining the 5-point blocks in this context, we first recall some essential features of the treatment of conserved operators in the embedding space formalism. 

It is well known that in unitary conformal field theories, the scaling dimensions of spin-$\ell$ primary operators respect the unitarity bound 
\[
\De \geq \ell+d-2 \quad (\ell \geq 1)\,.
\]
If $\De$ saturates this $d$-dimensional bound for a given $\ell$, i.e.~$\De = \ell+d-2$, the corresponding operator is a conserved spin-$\ell$ primary. Prominent examples of conserved operators include the energy-momentum tensor (spin-2) and global symmetry currents (spin-1). In these cases, imposing the conservation condition serves to restrict the form of three- and higher-point functions beyond the constraints from conformal symmetry alone. The corresponding unitarity bound for scalars is 
\[
\De \geq (d-2)/2\,.
\]
However, there is little motivation for analyzing the case of conserved scalars, since this bound is saturated exclusively by free fields. 

A convenient consequence of the OPE is that requiring the satisfaction of the full set of 3-point function conservation conditions guarantees the automatic conservation of higher-point functions. It is therefore sufficient to impose such constraints just at the level of the 3-point functions. Demanding operator conservation has the consequence that some 3-point coefficients $\la_a$ in the sum \Eq{physicalthreept} are correlated with each other. Such relations among the coefficients imply that particular elementary structures that are a priori independent ultimately get merged into single overall structures. Subsequently, we convert the resulting constraints into relations among the corresponding structures inside the conformal block decomposition.

It was shown in \cite{Costa:2011mg} how the conservation conditions could be fruitfully analyzed in the context of the index-free embedding space formalism. Let us assume that the conserved operator in question is at position $X_i$ (with $i=1$, $2$, or $3$) inside the 3-point function $\vev{\Phi_1(X_1; Z_1)\Phi_2(X_2; Z_2)\Phi_3(X_3; Z_3)}$. 
We define the divergence operator 
\[
\eql{cons operator}
\partial_{X_i} \cdot D \equiv \dfrac{\partial}{\partial X_{iA}} D_A\,,
\]
where $D_A$ is the Todorov operator~\cite{Dobrev:1975ru,Dobrev:1977qv}
\[
D_A\equiv
\left(\frac{d}{2} - 1 + Z_i \cdot \frac{\partial}{\partial Z_i} \right) \frac{\partial}{\partial Z_i^A} - \frac12 Z_{iA} \frac{\partial^2}{\partial Z_i \cdot \partial Z_i}\,.
\]
We may readily accommodate the conservation condition by demanding that the action of the divergence operator $\partial_{X_i} \cdot D$ on the 3-point function $\vev{\Phi_1(X_1; Z_1)\Phi_2(X_2; Z_2)\Phi_3(X_3; Z_3)}$ given by \Eq{general STT 3-point} yields zero, assuming the presence of a conserved operator at the point $(X_i; Z_i)$. The conservation condition acts to restrict the possible structures appearing in a given 3-point function. These types of constraints were investigated in some detail in the embedding space setting in \cite{Costa:2011mg,Costa:2011dw,Fortin:2020des}. 

We expect these constraints to imply that the constituent blocks in the complete linear combination of $5$-point blocks are not all independent. A priori, the full contribution of the 5-point blocks for some particular values of $(\Delta, \ell, \Delta', \ell')$ may be expressed in the following way:
\[
\sum_{n_{IJ} = 0}^{\text{min}(\ell, \ell')} \al_{n_{IJ}} W_{\De, \ell, \De', \ell'; \De_1, \De_2, \De_3, \De_4, \De_5}^{(n_{IJ})},
\]
where $W^{(n_{IJ})}_{\De, \ell, \De', \ell'; \De_1,\De_2, \De_3, \De_4, \De_5}(X_i)\propto G^{(n_{IJ})}_{\De, \ell, \De', \ell'}(u_i)$. Now, if we suppose that one of the exchanged operators is conserved, then there would be relations among the parameters $\al_{n_{IJ}}$, which would force some of these block structures to merge.

With this in mind, we now turn to the 5-point conformal blocks $G^{(a)}_{\De, \ell, \De', \ell'}(u_i)$. We consider the constituent 3-point correlator $\vev{\scr{O}_{\De, \ell}\Phi_{\De_3} \scr{O}'_{\De', \ell'}}$ and study various cases of interest. In particular, we take the exchanged operators $(\scr{O}_{\De, \ell}, \scr{O}'_{\De', \ell'})$ to belong to the set $\{(\phi, v^a), (\phi, T^{ab}), (v^a, v^b), (v^a, T^{bc}), (T^{ab}, T^{cd})\}$, where $\phi$, $v^a$, and $T^{a b}$ represent scalars, spin-$1$ currents, and spin-$2$ tensors, respectively. In our analysis of these cases, we suppose that either one or both of the exchanged operators are conserved. In the examples below, we'll focus exclusively on parity-even structures and we'll adopt the notation $Q_{(\ell,0,\ell')}(X_i; Z_i)$ for the 3-point functions in contracted form.

As a first example, let us begin by examining the case where both $\scr{O}_{\De, \ell}$ and $\scr{O}'_{\De', \ell'}$ are spin-$1$ currents so that $\ell=1$, $\ell'=1$. For simplicity, we consider the symmetric situation, assuming that the currents are identical. Prior to imposing conservation, we expect to find two independent structures appearing in the 3-point function $\vev{v^a(X_1)\Phi_{\De_3}(X_2) v^b(X_3)}$. In particular, we have the two constituent 3-point structures $\{V_1 V_3, H_{13}\}$. With this, the embedding space 3-point function in contracted form is given by
\[
\eql{spin-1, spin-1}
Q_{(1,0,1)}(X_1,X_2, X_3; Z_1, Z_3)= \dfrac{\al V_1 V_3 +\be H_{13}}{
(X_{12})^{\sfrac{1}{2}(\De+\De_3-\De')}
(X_{13})^{\sfrac{1}{2}(\De+\De'-\De_3+2)}(X_{23})^{\sfrac{1}{2}(\De_3+\De'-\De)}}\,,
\]
where $\al$ and $\be$ are a priori independent coefficients. Hereupon we impose conservation, assuming that both spin-$1$ currents are conserved so that $\De = d-1$, $\De'=d-1$. 

By symmetry, it is sufficient to impose conservation at $X_1$. Acting with the divergence operator \Eq{cons operator} on the 3-point function \Eq{spin-1, spin-1}, one obtains
\[
(\d_{X_1}\cdot D_{Z_1})Q_{(1,0,1)}(X_1,X_2, X_3; Z_1, Z_3)\to\bigg(\frac{d}{2}-1\bigg)(\al(d-1-\De_3)+\be \De_3)\dfrac{V_3}{(X_{12})^{\frac{\De_3}{2}}(X_{13})^{d-\frac{\De_3}{2}}
(X_{23})^{\frac{\De_3}{2}}}\,.
\]
After requiring current conservation, namely, 
\[
(\d_{X_1}\cdot D_{Z_1})Q_{(1,0,1)}(X_1,X_2, X_3; Z_1, Z_3)= 0\,,
\]
it transpires that $\al$ and $\be$ must be related to each other in the following way:
\[
\eql{cons relation}
\al(d-1-\De_3)+\be \De_3 =0 \Rightarrow \be=-\dfrac{(d-1-\De_3)}{ \De_3}\al\,.
\]
At this point, we are left with the single structure
\[
Q_{(1,0,1)}(X_1,X_2, X_3; Z_1, Z_3) &= \dfrac{\al \bigg(V_1 V_3 
-\dfrac{(d-1-\De_3)}{ \De_3}H_{13}\bigg)}{(X_{12})^{\frac{\De_3}{2}}(X_{13})^{d-\frac{\De_3}{2}}(X_{23})^{\frac{\De_3}{2}}}\,,
\]
so that the number of independent structures drops from two to one. The two allowed structures $\{V_1 V_3, H_{13}\}$ still both appear, but they are no longer independent. Rather, they merge into one overall structure.
If only one of the spin-$1$ operators is conserved, say $\scr{O}_{\De, 1}=\scr{O}_{d-1, 1}$, the result is the same, but in this case $\scr{O}_{\De', 1}$ carries an unrestricted value of the scaling dimension $\De'$, up to the unitarity bound. It is apparent from this example how the conservation condition can be directly implemented in the embedding space. 

The relation \Eq{cons relation} thus serves to restrict the form of the 5-point conformal block for $(\ell, \ell') = (1,1)$ exchange to
\begin{align}
\sum_{n_{IJ} = 0}^1 \al_{n_{IJ}} W^{(n_{IJ})}_{\De, 1, \De', 1; \De_1,\De_2, \De_3, \De_4, \De_5}(X_i)=&\nonumber\\
\al_0\left( W^{(0)}_{d-1, 1, d-1, 1; \De_1,\De_1, \De_3, \De_4, \De_4}(X_i) 
\,- \right.&\left. \dfrac{(d-1-\De_3)}{ \De_3} W^{(1)}_{d-1, 1, d-1, 1; \De_1,\De_1, \De_3, \De_4, \De_4}(X_i)\right)\,,
\end{align}
where, for convenience we have written our expression in terms of $W^{(n_{IJ})}_{\De, \ell, \De', \ell'; \De_1,\De_2, \De_3, \De_4, \De_5}(X_i)$. Moreover, note that we must have $\De_1=\De_2$, $\De_4 =\De_5$ due to the Ward identity constraints on $\vev{\phi_{\De_1} \phi_{\De_2} v^a}$ and $\vev{v^b \phi_{\De_4} \phi_{\De_5}}$.

Next, we consider the case $\ell=0$, $\ell'=1$, which gives us a (scalar)-(scalar)-(vector) 3-point function $\vev{\phi(X_1)\Phi_{\De_3}(X_2) v^b(X_3)}$. Clearly, there is only a single allowed structure, namely $V_3$. In particular, we have (in contracted form)
\[
Q_{(0,0,1)}(X_1, X_2, X_3; Z_3)= \dfrac{\al V_3}{(X_{12})^{\frac{1}{2}(\De+\De_3-\De'-1)}
(X_{13})^{\frac{1}{2}(\De+\De'-\De_3+1)} (X_{23})^{\frac{1}{2}(\De_3+\De'-\De+1)}}\,.
\]
Assuming that the vector is a conserved current then constrains $\De' = d-1$, $\Delta_3 = \Delta$, and $\Delta_4 = \Delta_5$. In particular, imposing the conservation of $\phi$ forces the dimension of the middle operator $\Phi$ to match that of $\phi$ due to the Ward identity constraint on $\vev{\phi \Phi_{\De_3} v^b}$. The block is then restricted to the form
\[
\al_0 W^{(0)}_{\De, 0, d-1, 1; \De_1,\De_2, \De, \De_4, \De_4}(X_i)\,.
\]

Next, we turn to the case of a (scalar)-(scalar)-(tensor) 3-point function \\$\vev{\phi(X_1)\Phi_{\De_3}(X_2) T^{ab}(X_3)}$, with $\ell=0$, $\ell'=2$. This time, we have the unique structure
\[
Q_{(0,0,2)}(X_1, X_2, X_3; Z_3)= 
\dfrac{\al V_3^2}{(X_{12})^{\frac{1}{2}(\De+\De_3-\De'-2)}
(X_{13})^{\frac{1}{2}(\De+\De'-\De_3+2)} (X_{23})^{\frac{1}{2}(\De_3+\De'-\De+2)}}\,.
\]
If we demand that the tensor is conserved, we require that $\De' =d$. In addition, Ward identity constraints ensure that $\De_3 = \De$ and $\De_4 = \De_5$. The allowed block then takes the form
\[
\al_0 W^{(0)}_{\De, 0, d, 2; \De_1,\De_2, \De, \De_4, \De_4}(X_i)\,.
\]
 
Proceeding to another case of interest, we take $\ell=1$ and $\ell' = 2$. A priori, we find two independent tensor structures appearing, namely $\{V_1 V_3^2, V_3 H_{13}\}$. In particular, we have the (vector)-(scalar)-(tensor) 3-point function $\vev{v^A(X_1)\Phi_{\De_3}(X_2) T^{ab}(X_3)}$, which takes the contracted form
\[
Q_{(1,0,2)}(X_1, X_2, X_3; Z_1, Z_3)= 
\dfrac{\al V_1 V_3^2+ \be V_3 H_{13}}{(X_{12})^{\frac{1}{2}(\De+\De_3-\De'-1)}
(X_{13})^{\frac{1}{2}(\De+\De'-\De_3+3)} (X_{23})^{\frac{1}{2}(\De_3+\De'-\De+1)}}\,.
\]
This time, we find that imposing conservation of the vector $v^a$ gives $\De=d-1$ and requires the linear combination
\[
Q_{(1,0,2)}(X_1, X_2, X_3; Z_1, Z_3)=\dfrac{\al}
{(X_{12})^{\frac{1}{2}(d+\De_3-\De'-2)}
(X_{13})^{\frac{1}{2}(d+\De'-\De_3+2)} (X_{23})^{\frac{1}{2}(\De_3+\De'-d+2)}}\times\nonumber\\
 \left(V_1 V_3^2+ \dfrac{\De'-\De_3}{\De'-\De_3-d}  V_3 H_{13}\right)\,,
\]
which corresponds to the block
\begin{align}
\sum_{n_{IJ}=0}^1 \al_{n_{IJ}} W^{(n_{IJ})}_{\De, 1, \De', 2; \De_1,\De_2, \De_3, \De_4, \De_5}(X_i)=&\nonumber\\ \al_0 \left(W^{(0)}_{d-1, 1 \De', 2; \De_1,\De_1, \De_3, \De_4, \De_4}(X_i) \,+\right.&\left. \dfrac{\De'-\De_3}{\De'-\De_3-d}  W^{(1)}_{d-1, 1, \De', 2; \De_1,\De_1, \De_3, \De_4, \De_4}(X_i)\right)\,.
\end{align}
On the other hand, demanding that the tensor $T^{ab}$ is conserved fixes $\De' =d $ and yields
\[
Q_{(1,0,2)}(X_1, X_2, X_3; Z_1, Z_3)= \dfrac{1}{(X_{12})^{\frac{1}{2}(\De+\De_3-d-1)}
(X_{13})^{\frac{1}{2}(\De-\De_3+d+3)} (X_{23})^{\frac{1}{2}(\De_3-\De+d+1)}}\times\nonumber\\
\al\left( V_1 V_3^2+\dfrac{2((d-1)(\De-\De_3)+1)}{(d-2)(\De-\De_3-d-1)} V_3 H_{13}\right)\,,
\] 
which corresponds to 
\begin{align}
\sum_{n_{IJ}=0}^1 \al_{n_{IJ}} W^{(n_{IJ})}_{\De, 1, \De', 1; \De_1,\De_2, \De_3, \De_4, \De_5}(X_i)=&\nonumber\\
\al_0\left( W^{(0)}_{\De, 1, d, 1; \De_1,\De_1, \De_3, \De_4, \De_4}(X_i)
\,+\right.&\left. \dfrac{2((d-1)(\De-\De_3)+1)}{(d-2)(\De-\De_3-d-1)} W^{(1)}_{\De, 1, d, 1; \De_1,\De_1, \De_3, \De_4, \De_4}(X_i)\right)\,.
\end{align}
Note that requiring both conservation conditions simultaneously is only possible for the special values $\Delta_3 = 2$ or $\Delta_3 = d-2$. 
 
Finally, we consider the nontrivial special case of $\ell=2$, $\ell'=2$, namely $\vev{T^{ab} \Phi_{\De_3} T^{cd}}$. Here $\scr{O}_{\De, \ell}$ and $\scr{O}'_{\De', \ell'}$ are both spin-$2$ tensors so that there are three independent tensor structures to begin with. In particular, we have $\{V_1^2 V_3^2, H_{13} V_1 V_3, H_{13}^2\}$, as these are the only allowed structures that are symmetric under the exchange of $\{X_1, Z_1\}, \{X_3, Z_3\}$. In this case, imposing conservation at $X_1$ (again, this is sufficient, by symmetry) results in the reduction of the three independent structures to just one, namely
\[
Q_{(2,0,2)}(X_1, X_2, X_3; Z_1, Z_3)= \dfrac{\al}{ (d-2)(\De_3 +2)}
 \dfrac{1}{(X_{12})^{\frac{\De_3}{2}}
(X_{13})^{(d+2-\frac{\De_3}{2})}(X_{23})^{\frac{\De_3}{2}}}
\times\nonumber\\\bigg(V_1^2 V_3^2
- 2    \left(\De_3+2 +d(d-(\De_3 +1) )\right)  H_{13} V_1 V_3+
   \left((d-1)   (\De_3 -2d)+\frac{d(d-2)  (d+1)}{\De_3}\right)H_{13}^2\bigg)\,,
\] 
where we have fixed $\De= d$ and $\De' = d$. Again, if only one of these is conserved, then the scaling dimension of the other is arbitrary (up to unitarity), but the final form of the 3-point function is the same.
With both operators conserved, we obtain the block
\begin{align}
\sum_{n_{IJ} = 0}^{2} \al_{n_{IJ}} W_{d, 2; d, 2; \De_1, \De_1, \De_3, \De_4, \De_4}^{(n_{IJ})} &=\nonumber\\
\al_{0}\bigg[ W_{d, 2; d, 2; \De_1, \De_1, \De_3, \De_4, \De_4}^{(0)} & - \frac{2  \left(\De_3+2 +d(d-(\De_3 +1) )\right)}{(\De_3 +2) (d-2)}W_{d, 2; d, 2; \De_1, \De_1, \De_3, \De_4, \De_4}^{(1)} \nonumber\\ &+
\frac{ \left((d-1) \De_3  (\De_3 -2d)+d(d-2)  (d+1)\right)}{\De_3  (\De_3 +2)
   (d-2)} W_{d, 2; d, 2; \De_1, \De_1, \De_3, \De_4, \De_4}^{(2)} \bigg]\,.
\end{align}
 
In the following two sections, we consider promoting the middle external operator $\Phi$ in our 5-point function to a spin-$1$ or a spin-$2$ operator, respectively. This will allow us to obtain conformal blocks for arbitrary symmetric traceless tensor exchange in the correlators $\vev{\phi_{\De_1}(X_1) \phi_{\De_2}(X_2) v^A(X_3) \phi_{\De_4}(X_4) \phi_{\De_5}(X_5)}$ and $\vev{\phi_{\De_1}(X_1) \phi_{\De_2}(X_2)T^{AB}(X_3)\phi_{\De_4}(X_4) \phi_{\De_5}(X_5)} $, where $v^A$ and $T^{AB}$ each carries some dimension $\De_3$ a priori. Throughout, we restrict our attention to parity-even correlators which exist in generic dimensions.

\subsection{Promoting $\Phi$ to a vector operator} 
\label{sec:phivec}

We first undertake the case of the promotion of $\Phi_{\De_3}$ to a vector operator. Our intention is to cast the result for $(\scr{O}_{\De, \ell}, \scr{O}'_{\De', \ell'})$ exchange in this block exclusively in terms of seed blocks we already know, namely those appearing in the scalar 5-point function. In particular, we will write the resulting blocks in terms of a linear combination of some weight-shifting operators acting on the symmetric traceless exchange blocks appearing in scalar 5-point functions with shifted quantum numbers. That is, we expect to cast the final result in a differential basis.

We recall our definition of the 5-point symmetric traceless exchange conformal block, namely
\[
G^{(n_{IJ})}&_{\De, \ell, \De', \ell'}(u_i) \propto W^{(n_{IJ})}_{\De, \ell, \De', \ell'; \De_1,\De_2, \De_3, \De_4, \De_5}(X_i)\nonumber\\
&=\, \vev{\phi_{\De_1}(X_1) \phi_{\De_2}(X_2) \scr{O}_{\De, \ell}}\bowtie\vev{\scr{O}_{\De, \ell}\Phi_{\De_3}(X_3) \scr{O}^\prime_{\De', \ell'}}^{(n_{IJ})}\bowtie
\vev{ \scr{O}^\prime_{\De', \ell'}\phi_{\De_4}(X_4) \phi_{\De_5}(X_5)}, 
\]
where $G^{(n_{IJ})}_{\De, \ell, \De', \ell'}(u_i)$ and $W^{(n_{IJ})}_{\De, \ell, \De', \ell'; \De_1,\De_2, \De_3, \De_4, \De_5}(X_i)$ are related by a $5$-point external leg factor.

If we promote the middle scalar $\Phi_{\De_3}$ to a vector operator $v^A$, we will take
\[
\vev{\scr{O}_{\De, \ell}(X_1)\Phi_{\De_3}(X_2) \scr{O}'_{\De', \ell'}(X_3)}\to \vev{\scr{O}_{\De, \ell}(X_1)v^A(X_2) \scr{O}^\prime_{\De', \ell'}(X_3)}\,,
\]
where, for convenience, we have taken the positions of $\scr{O}_{\De, \ell}$, $\Phi_{\De_3}$, and $\scr{O}^\prime_{\De', \ell'}$ to be $X_1$, $X_2$, and $X_3$, respectively.
 
This tells us that we need to map a (spin)-(scalar)-(spin) 3-point function to a (spin)-(vector)-(spin) one. There are three distinct classes of constituent 3-point tensor structures in this case. In particular, our basis of allowed structures is   
\[ 
\eql{spin-1 structures 1}
Q_{(\ell,1,\ell')}(X_1, X_2, X_3; Z_1, Z_2, Z_3)=\sum_{i=1}^3 \lambda_{i, n_{IJ}} Q^{(i, n_{IJ})}_{(\ell,1,\ell')}\,,
\]
where $n_{IJ}$ is a parameter that enumerates the various tensor structures, with
\[
Q^{(i, n_{IJ})}_{(\ell,1,\ell')}=\dfrac{
q^{(i, n_{IJ})}_{(\ell,1,\ell')}}{(X_{12})^{\sfrac{1}{2}(\De+\De_3-\De'  + \ell- \ell'+ 1)}
(X_{13})^{\sfrac{1}{2}(\De-\De_3+\De' + \ell+ \ell'- 1)}(X_{23})^{\sfrac{1}{2}( -\De+\De_3 +\De'- \ell+ \ell'+ 1)}}\,,
\]
where the structures $q^{(i, n_{IJ})}_{(\ell,1,\ell')}$ are given by
\[
\eql{spin-1 structures 2}
q^{(1, n_{IJ})}_{(\ell,1,\ell')} &= V_1^{\ell- n_{IJ}} V_2 V_3^{\ell^\prime -  n_{IJ}} H_{13}^{n_{IJ}}\,, \nonumber\\
q^{(2, n_{IJ})}_{(\ell,1,\ell')} &= V_1^{\ell -n_{IJ}} V_3^{(\ell^\prime -1) - n_{IJ}}  H_{13}^{n_{IJ}} H_{23}\,, \nonumber\\
q^{(3, n_{IJ})}_{(\ell,1,\ell')} &= V_1^{(\ell -1)-n_{IJ} }  V_3^{\ell^\prime -n_{IJ} } H_{12} H_{13}^{n_{IJ} }\,.
\]
We remark that the structures $q^{(1, n_{IJ})}_{(\ell,1,\ell')}$, 
$q^{(2, n_{IJ})}_{(\ell,1,\ell')}$, and 
$q^{(3, n_{IJ})}_{(\ell,1,\ell')}$  
exist for $n_{IJ} \in[0, \min(\ell, \ell')]$,  $n_{IJ} \in[0, \min(\ell, \ell'-1)]$, and $n_{IJ} \in[0, \min(\ell-1, \ell')]$, respectively.

With this, we consider the quantity
\[
W&_{\De, \ell; \De^\prime, \ell^\prime; \De_1, \De_2, \De_3, \De_4, \De_5}^{(V)(i,n_{IJ})}  \nonumber\\
&\qquad\equiv 
\vev{\phi_{\De_1}(X_1)\phi_{\De_2}(X_2)\scr{O}_{\De,\ell}}\bowtie   
  \vev{\scr{O}_{\De,\ell}v(X_3)\scr{O}'_{\De',\ell'}}^{(i,n_{IJ})} \bowtie
  \vev{\scr{O}'_{\De',\ell'}
\phi_{\De_4}(X_4)\phi_{\De_5}(X_5)}\,,
\]
where we have suppressed indices for brevity. Here $i$ enumerates three classes of 3-point tensor structures for $\vev{\scr{O}_{\De,\ell}v(X_3)\scr{O}'_{\De',\ell'}}$ and hence runs over $i=1,2,3$, while $n_{IJ}$ parametrizes different possible structures within each class. 

We start by expressing the middle 3-point structure for some fixed $i$ in terms of combinations of weight-shifting operators acting on ${} \vev{\scr{O}_{\De,\ell}(X_I)\Phi_{\De_3}(X_3)\scr{O}_{\De^\prime,\ell^\prime}(X_J)}^{(n_{IJ})}$. Since the spin of the middle operator $v=[\De_3, 1]$ is shifted up by 1 with respect to the original scalar $\Phi=[\De_3, 0]$, we may use either one of the combinations
\[
(\scr{D}_{X_I}^{(-0)}\cdot\scr{D}_{X_3}^{(0+)}) &\quad \text{or} \quad (\scr{D}_{X_J}^{(-0)}\cdot\scr{D}_{X_3}^{(0+)})\,, \nonumber\\
(\scr{D}_{X_I}^{(+0)}\cdot\scr{D}_{X_3}^{(0+)}) &\quad \text{or} \quad  (\scr{D}_{X_J}^{(+0)}\cdot\scr{D}_{X_3}^{(0+)})\,, 
\nonumber\\
(\scr{D}_{X_I}^{(0-)}\cdot\scr{D}_{X_3}^{(0+)}) &\quad \text{or} \quad  (\scr{D}_{X_J}^{(0-)}\cdot\scr{D}_{X_3}^{(0+)})\,, 
\nonumber\\
(\scr{D}_{X_I}^{(0+)}\cdot\scr{D}_{X_3}^{(0+)}) &\quad \text{or} \quad (\scr{D}_{X_J}^{(0+)}\cdot\scr{D}_{X_3}^{(0+)})\,.
\]
This list spans all the possible weight-shifting operators for the vector representation given in \Eq{vector WS}. As these four operators form a differential basis for $\scr{W} =\scr{V}$, we expect to be able to construct the complete solution purely in terms of these objects.
Note that it makes no difference whether we act with one of the operators of a given combination $(\scr{D}_X^{(m)}\cdot\scr{D}_{X_3}^{(n)})$ at $X=X_I$ or $X=X_J$. Both choices give equivalent results. 

For example, for $X=X_I$, acting with each of the four independent combinations on the appropriately shifted 3-point structure, we straightforwardly obtain
\[ 
\eql{one}
(\scr{D}_{X_3}^{(0+)}\cdot \scr{D}_{X_I}^{(-0)}) &\vev{\scr{O}_{\De+1,\ell}(X_I)\Phi_{\De_3}(X_3)\scr{O}_{\De^\prime,\ell^\prime}(X_J)}^{(n_{IJ})} \nonumber\\
=\,&  \al_1 Q_{(\ell,1,\ell')}^{(1, n_{IJ})}+\be_1  Q_{(\ell,1,\ell')}^{(2, n_{IJ})} +\ga_1  Q_{(\ell,1,\ell')}^{(3, n_{IJ})}\,,
\]
\[
\eql{two}
 (\scr{D}_{X_3}^{(0+)}\cdot \scr{D}_{X_I}^{(0+)}) &\vev{\scr{O}_{\De,\ell-1}(X_I)\Phi_{\De_3}(X_3) \scr{O}_{\De^\prime,\ell^\prime}(X_J)}^{(n_{IJ})} \nonumber\\
 =\,&  \al_2 Q_{(\ell,1,\ell')}^{(1, n_{IJ})}+\al_3  Q_{(\ell,1,\ell')}^{(1, n_{IJ}+1)} 
 +\be_2  Q_{(\ell,1,\ell')}^{(2, n_{IJ})} \nonumber\\
 &+\be_3 Q_{(\ell,1,\ell')}^{(2, n_{IJ}+1)} 
 +\ga_2 Q_{(\ell,1,\ell')}^{(3, n_{IJ})} +\ga_3 Q_{(\ell,1,\ell')}^{(3, n_{IJ}+1)}\,,
\]
\[
\eql{three}
(\scr{D}_{X_3}^{(0+)}\cdot \scr{D}_{X_I}^{(0-)}){} &\vev{\scr{O}_{\De,\ell+1}(X_I)\Phi_{\De_3}(X_3)\scr{O}_{\De^\prime,\ell^\prime}(X_J)}^{(n_{IJ})}  \nonumber\\
=\,& \al_4  Q_{(\ell,1,\ell')}^{(1, n_{IJ}-1)}+\al_5  Q_{(\ell,1,\ell')}^{(1, n_{IJ})}
 +\al_6  Q_{(\ell,1,\ell')}^{(1, n_{IJ}+1)}
 \nonumber\\
 & +\be_4  Q_{(\ell,1,\ell')}^{(2, n_{IJ}-1)}+\be_5  Q_{(\ell,1,\ell')}^{(2, n_{IJ})}
 +\be_6  Q_{(\ell,1,\ell')}^{(2, n_{IJ}+1)}
  \nonumber\\
  &+\ga_4 Q_{(\ell,1,\ell')}^{(3, n_{IJ}-1)}+\ga_5 Q_{(\ell,1,\ell')}^{(3, n_{IJ})}
 +\ga_6  Q_{(\ell,1,\ell')}^{(3, n_{IJ}+1)}\,,
  \]
and
 \[
 \eql{four}
(\scr{D}_{X_3}^{(0+)}\cdot \scr{D}_{X_I}^{(+0)}){} &\vev{\scr{O}_{\De-1,\ell}(X_I)\Phi_{\De_3}(X_3)\scr{O}_{\De^\prime,\ell^\prime}(X_J)}^{(n_{IJ})} \nonumber\\ 
=\,& \al_7 Q_{(\ell,1,\ell')}^{(1, n_{IJ}-1)}+\al_8  Q_{(\ell,1,\ell')}^{(1, n_{IJ})}
 +\al_9 Q_{(\ell,1,\ell')}^{(1, n_{IJ}+1)} +\al_{10} Q_{(\ell,1,\ell')}^{(1, n_{IJ}+2)}\nonumber\\
  &+\be_7 Q_{(\ell,1,\ell')}^{(2, n_{IJ}-1)}+\be_8 Q_{(\ell,1,\ell')}^{(2, n_{IJ})}
 +\be_9  Q_{(\ell,1,\ell')}^{(2, n_{IJ}+1)} +\be_{10}  Q_{(\ell,1,\ell')}^{(2, n_{IJ}+2)}\nonumber\\
   &+\ga_7  Q_{(\ell,1,\ell')}^{(3, n_{IJ}-1)}+\ga_8 Q_{(\ell,1,\ell')}^{(3, n_{IJ})}
 +\ga_9  Q_{(\ell,1,\ell')}^{(3, n_{IJ}+1)} +\ga_{10} Q_{(\ell,1,\ell')}^{(3, n_{IJ}+2)}\,,
\]
where $\al_i$, $\be_i$, $\ga_i$ are $6j$ coefficients that are functions of the parameters $\{\De, \ell, \De', \ell',\De_3, n_{IJ}\}$.

Since there are only three independent 3-point structures, we just need three equations. For simplicity, we choose to restrict attention to the set \Eq{one} - \Eq{three}. The relevant coefficients $\al_j$, $\be_j$, $\ga_j$ for $j =1,\dots, 6$ are listed in Appendix~\ref{app:D}. We may now reuse these multiple times to generate a system of equations involving $Q_{(\ell,1,\ell')}^{(i, n_{IJ}-1)}, Q_{(\ell,1,\ell')}^{(i, n_{IJ})}$, and $Q_{(\ell,1,\ell')}^{(i, n_{IJ}+1)}$, where $i= 1, 2, 3$.

In particular, we have three equations from \Eq{one}. These feature 
\[
Q_{(\ell,1,\ell')}^{(i, n_{IJ}-1)}\,; \quad
Q_{(\ell,1,\ell')}^{(i, n_{IJ})}\,; \quad
Q_{(\ell,1,\ell')}^{(i, n_{IJ}+1)}\,.
\]
Further, from \Eq{two} we have two equations containing
\[ 
Q_{(\ell,1,\ell')}^{(i, n_{IJ}-1)}\,, Q_{(\ell,1,\ell')}^{(i, n_{IJ})}\,; \quad
Q_{(\ell,1,\ell')}^{(i, n_{IJ})}\,, Q_{(\ell,1,\ell')}^{(i, n_{IJ}+1)}\,.
\]
Next, from \Eq{three}, we have just one equation involving
\[
Q_{(\ell,1,\ell')}^{(i, n_{IJ}-1)}\,, Q_{(\ell,1,\ell')}^{(i, n_{IJ})}\,, Q_{(\ell,1,\ell')}^{(i, n_{IJ}+1)}\,.
\]
We note that there are nine 3-point structures appearing here, namely 
\[
Q_{(\ell,1,\ell')}^{(i, n_{IJ}-1)}\,, Q_{(\ell,1,\ell')}^{(i, n_{IJ})}\,,  Q_{(\ell,1,\ell')}^{(i, n_{IJ}+1)}\,,   \qquad i = 1, 2, 3\,.
\]

To determine the structures of interest, we first apply \Eq{one} to solve for $Q_{(\ell,1,\ell')}^{(1, n_{IJ})}$. We then generate the corresponding relations for $Q_{(\ell,1,\ell')}^{(1, n_{IJ}-1)}$ and $Q_{(\ell,1,\ell')}^{(1, n_{IJ}+1)}$. Upon inserting these relations inside \Eq{two}, many cancellations occur and we are left with an explicit expression for $Q_{(\ell,1,\ell')}^{(3, n_{IJ})}$ in terms of the structures where $\Phi$ is a scalar.

Further, we may then substitute these results inside the last equation, namely \Eq{three}, which leads to a recursive relation for $Q_{(\ell,1,\ell')}^{(2, n_{IJ})}$ in terms of $Q_{(\ell,1,\ell')}^{(2, n_{IJ}+1)}$, which terminates at $n_{IJ} = \min(\ell, \ell'-1)$. Thereafter, we may apply this equation to recursively determine $Q_{(\ell,1,\ell')}^{(2, n_{IJ})}$. 

In particular, we find
\[
\eql{equation Q1}
Q_{(\ell,1,\ell')}^{(1, n_{IJ})}=\,&
\frac{\left(\ell '-n_{IJ}\right)}{\Delta '-\Delta +\Delta _3-\ell '+\ell -1}Q_{(\ell,1,\ell')}^{(2, n_{IJ})}\nonumber\\
&+
a_1(\scr{D}_{X_3}^{(0+)}\cdot \scr{D}_{X_I}^{(0+)}) \vev{\scr{O}_{\De,\ell-1}(X_I)\Phi_{\De_3}(X_3) \scr{O}_{\De^\prime,\ell^\prime}(X_J)}^{(n_{IJ})} \nonumber\\
   &+a_2
   (\scr{D}_{X_3}^{(0+)}\cdot \scr{D}_{X_I}^{(-0)}) \vev{\scr{O}_{\De+1,\ell}(X_I)\Phi_{\De_3}(X_3)\scr{O}_{\De^\prime,\ell^\prime}(X_J)}^{(n_{IJ}+1)}\nonumber\\
   &+a_3
    (\scr{D}_{X_3}^{(0+)}\cdot \scr{D}_{X_I}^{(-0)}) \vev{\scr{O}_{\De+1,\ell}(X_I)\Phi_{\De_3}(X_3)\scr{O}_{\De^\prime,\ell^\prime}(X_J)}^{(n_{IJ})}\,.
 \]
 
Next, we have 
\[
\eql{equation Q3}
Q_{(\ell,1,\ell')}^{(3, n_{IJ})}=\,&
    c_1
    (\scr{D}_{X_3}^{(0+)}\cdot \scr{D}_{X_I}^{(-0)}) \vev{\scr{O}_{\De+1,\ell}(X_I)\Phi_{\De_3}(X_3)\scr{O}_{\De^\prime,\ell^\prime}(X_J)}^{(n_{IJ})} \nonumber\\
    &+ c_2
    (\scr{D}_{X_3}^{(0+)}\cdot \scr{D}_{X_I}^{(-0)}) \vev{\scr{O}_{\De+1,\ell}(X_I)\Phi_{\De_3}(X_3)\scr{O}_{\De^\prime,\ell^\prime}(X_J)}^{(n_{IJ}+1)} 
  \nonumber\\ 
 &+ c_3(\scr{D}_{X_3}^{(0+)}\cdot \scr{D}_{X_I}^{(0+)}) \vev{\scr{O}_{\De,\ell-1}(X_I)\Phi_{\De_3}(X_3) \scr{O}_{\De^\prime,\ell^\prime}(X_J)}^{(n_{IJ})}\,.
\] 
Finally, we have
\[
\eql{equation Q2}
Q_{(\ell,1,\ell')}^{(2, n_{IJ})}=\,&\frac{(n_{IJ}-\ell ) \left(n_{IJ}-\ell '+1\right)}{(n_{IJ}+1) \left(\Delta '-\Delta
   +\Delta _3-2 n_{IJ}+\ell '+\ell -1\right)}Q_{(\ell,1,\ell')}^{(2, n_{IJ}+1)}
   \nonumber\\
   &+b_1 (\scr{D}_{X_3}^{(0+)}\cdot \scr{D}_{X_I}^{(-0)}) \vev{\scr{O}_{\De+1,\ell}(X_I)\Phi_{\De_3}(X_3)\scr{O}_{\De^\prime,\ell^\prime}(X_J)}^{(n_{IJ}+2)} \nonumber\\
   &+b_2
   (\scr{D}_{X_3}^{(0+)}\cdot \scr{D}_{X_I}^{(0-)}){} \vev{\scr{O}_{\De,\ell+1}(X_I)\Phi_{\De_3}(X_3)\scr{O}_{\De^\prime,\ell^\prime}(X_J)}^{(n_{IJ}+1)}  \nonumber\\
   &+b_3(\scr{D}_{X_3}^{(0+)}\cdot \scr{D}_{X_I}^{(0+)}) \vev{\scr{O}_{\De,\ell-1}(X_I)\Phi_{\De_3}(X_3) \scr{O}_{\De^\prime,\ell^\prime}(X_J)}^{(n_{IJ})} \nonumber\\
   &+b_4(\scr{D}_{X_3}^{(0+)}\cdot \scr{D}_{X_I}^{(0+)}) \vev{\scr{O}_{\De,\ell-1}(X_I)\Phi_{\De_3}(X_3) \scr{O}_{\De^\prime,\ell^\prime}(X_J)}^{(n_{IJ}+1)} \nonumber\\
   &+b_5(\scr{D}_{X_3}^{(0+)}\cdot \scr{D}_{X_I}^{(-0)}) \vev{\scr{O}_{\De+1,\ell}(X_I)\Phi_{\De_3}(X_3)\scr{O}_{\De^\prime,\ell^\prime}(X_J)}^{(n_{IJ})} \nonumber\\
&+b_6
 (\scr{D}_{X_3}^{(0+)}\cdot \scr{D}_{X_I}^{(-0)}) \vev{\scr{O}_{\De+1,\ell}(X_I)\Phi_{\De_3}(X_3)\scr{O}_{\De^\prime,\ell^\prime}(X_J)}^{(n_{IJ}+1)}\,. 
\]
The coefficients in these formulas are all given explicitly in Appendix~\ref{app:E}.

We observe that this is an increasing recursion relation in $n_{IJ}$ for $Q_{(\ell,1,\ell')}^{(2, n_{IJ})}$. It is evident that the coefficient of $Q_{(\ell,1,\ell')}^{(2, n_{IJ}+1)}$ vanishes identically for $n_{IJ} =\min(\ell, \ell'-1)$. Hence, we may first determine the form of $Q_{(\ell,1,\ell')}^{(2, \min(\ell, \ell'-1))}$ and then apply \Eq{equation Q2} to extract the remaining structures $Q_{(\ell,1,\ell')}^{(2, n_{IJ})}$ for $0 \leq n_{IJ}< \min(\ell, \ell'-1)$.
 
At this stage, we have expressed each of $Q_{(\ell,1,\ell')}^{(1, n_{IJ})}$, $Q_{(\ell,1,\ell')}^{(2, n_{IJ})}$, and $Q_{(\ell,1,\ell')}^{(3, n_{IJ})}$ in terms of the structures 
\[
(\scr{D}_{X_3}^{(0+)}\cdot \scr{D}_{X_I}^{(-0)}) \vev{\scr{O}_{\De+1,\ell}(X_I)\Phi_{\De_3}(X_3)\scr{O}_{\De^\prime,\ell^\prime}(X_J)}^{(n_{IJ})}\,, \nonumber\\
(\scr{D}_{X_3}^{(0+)}\cdot \scr{D}_{X_I}^{(0+)}) \vev{\scr{O}_{\De,\ell-1}(X_I)\Phi_{\De_3}(X_3) \scr{O}_{\De^\prime,\ell^\prime}(X_J)}^{(n_{IJ})}\,, \nonumber\\ 
(\scr{D}_{X_3}^{(0+)}\cdot \scr{D}_{X_I}^{(0-)}){} \vev{\scr{O}_{\De,\ell+1}(X_I)\Phi_{\De_3}(X_3)\scr{O}_{\De^\prime,\ell^\prime}(X_J)}^{(n_{IJ})}\,.
\]

We next remark that we may invoke the integration-by-parts rule \Eq{by parts rule} to rewrite each of these as 
\[
\vev{\phi_{\De_1}&(X_1) \phi_{\De_2}(X_2) \scr{O}_{\De, \ell}}\bowtie(\scr{D}_{X_3}^{(0+)}\cdot \scr{D}_{X_I}^{(-0)}) \vev{\scr{O}_{\De+1,\ell}(X_I)\Phi_{\De_3}(X_3)\scr{O}_{\De^\prime,\ell^\prime}(X_J)}^{(n_{IJ})}\nonumber\\ 
=\,&C_{(-0)(+0)}\scr{D}_{X_3\, A}^{(0+)}
\vev{\phi_{\De_1}(X_1) \phi_{\De_2}(X_2)
\scr{D}_{X_I}^{(+0)A}\scr{O}_{\De,\ell}(X_I)}\bowtie \vev{ \scr{O}_{\De+1,\ell}
\Phi_{\De_3}(X_3)\scr{O}_{\De^\prime,\ell^\prime}}^{(n_{IJ})}\,,
\]
\[ 
 \vev{\phi_{\De_1}&(X_1) \phi_{\De_2}(X_2) \scr{O}_{\De, \ell}}\bowtie(\scr{D}_{X_3}^{(0+)}\cdot \scr{D}_{X_I}^{(0+)}){} \vev{\scr{O}_{\De,\ell-1}(X_I)\Phi_{\De_3}(X_3)
 \scr{O}_{\De^\prime,\ell^\prime}(X_J)}^{(n_{IJ})} \nonumber\\
 =\,&C_{(0+)(0-)}
\scr{D}_{X_3\, A}^{(0+)}
\vev{\phi_{\De_1}(X_1) \phi_{\De_2}(X_2)
    \scr{D}_{X_I}^{(0-)A}\scr{O}_{\De,\ell}(X_I)}\bowtie \vev{\scr{O}_{\De,\ell-1}
\Phi_{\De_3}(X_3)
 \scr{O}_{\De^\prime,\ell^\prime}}^{(n_{IJ})}\,,
 \]
 \[
 \vev{\phi_{\De_1}&(X_1) \phi_{\De_2}(X_2) \scr{O}_{\De, \ell}}\bowtie(\scr{D}_{X_3}^{(0+)}\cdot \scr{D}_{X_I}^{(0-)}){} \vev{\scr{O}_{\De,\ell+1}(X_I)\Phi_{\De_3}(X_3)
\scr{O}_{\De^\prime,\ell^\prime}(X_J)}^{(n_{IJ})}\nonumber\\
=\,&
 C_{(0-)(0+)}\scr{D}_{X_3\, A}^{(0+)}
\vev{\phi_{\De_1}(X_1) \phi_{\De_2}(X_2)   \scr{D}_{X_I}^{(0+)A}\scr{O}_{\De,\ell}(X_I)}
\bowtie\vev{\scr{O}_{\De,\ell+1}\Phi_{\De_3}(X_3)
\scr{O}_{\De^\prime,\ell^\prime}}^{(n_{IJ})}\,,
\]
where the relevant 2-point $6j$ symbols are given by
\[
C_{(-0)(+0)} &=\frac{1}{2 (\Delta -1) (d-2 (\Delta +1)) (\Delta +\ell ) (d-\Delta +\ell -2)}\,, \nonumber\\
C_{(0+)(0-)} &=-\frac{\Delta +\ell -1}{\ell  (d+2 \ell -4) (d-\Delta +\ell -2)}\,, \nonumber\\
C_{(0-)(0+)} &= -\frac{(\ell +1) (d+2 \ell -2) (d-\Delta +\ell -1)}{\Delta +\ell }\,.
\]
At this point, one of the weight shifting operators acts on the 3-point structure $\vev{\phi_{\De_1}(X_1) \phi_{\De_2}(X_2) \scr{O}_{\De, \ell}}$, which is of the type (scalar)-(scalar)-(spin) and hence unique. We may therefore use the alternative crossing relation \Eq{alternative crossing} to rewrite each of the pieces $\scr{D}_{X_I}^{(\de\De,\de\ell)A}\vev{ \phi_{\De_1}(X_1) \phi_{\De_2}(X_2) \scr{O}_{\De-\de\De,\ell-\de\ell}(X_I)  }$ in the following way:
\[
\scr{D}_{X_I}^{(\de\De,\de\ell)A} &\vev{ \scr{O}_{\De_1}(X_1) \scr{O}_{\De_2}(X_2) \scr{O}_{\De-\de\De,\ell-\de\ell}(X_I)  }\nonumber\\
=\,&
F^{(\de\De,\de\ell)}_{(+ 0)}\big(
 \scr{D}_{X_1}^{(+ 0)A }\vev{ \phi_{\De_1-1}(X_1) \phi_{\De_2}(X_2) \scr{O}_{\De,\ell}(X_I)}
+ (-1)^{\de\ell}
 \scr{D}_{X_2}^{(+  0)A}\vev{\phi_{\De_1}(X_1) \phi_{\De_2-1}(X_2) \scr{O}_{\De,\ell}(X_I)} 
 \big)
\nonumber\\
&+F^{(\de\De,\de\ell)}_{(- 0)}\big(
 \scr{D}_{X_1}^{(- 0)A }\vev{ \phi_{\De_1+1}(X_1) \phi_{\De_2}(X_2) \scr{O}_{\De,\ell}(X_I)} 
 + (-1)^{\de\ell}
 \scr{D}_{X_2}^{(- 0)A}\vev{ \phi_{\De_1}(X_1) 
 \phi_{\De_2+1}(X_2) \scr{O}_{\De,\ell}(X_I)}  \big)\,.
\]
Equipped with this, we conclude that in order to obtain the corresponding conformal blocks from \Eq{equation Q1} - \Eq{equation Q2}, we must make the replacements
 \[
 \eql{replacements 1}
(\scr{D}_{X_3}^{(0+)}\cdot \scr{D}_{X_I}^{(-0)}) &\vev{\scr{O}_{\De+1,\ell}(X_I)\Phi_{\De_3}(X_3)\scr{O}_{\De^\prime,\ell^\prime}(X_J)}^{(n_{IJ})}\nonumber\\ 
\to\,& C_{(-0)(+0)}\bigg[
 F^{(+ 0)}_{(+ 0)}\bigg(
 (\scr{D}_{X_1}^{(+ 0) }\cdot \scr{D}_{X_3}^{(0+)})
 W^{(n_{IJ})}_{\De, \ell, \De', \ell'; \De_1-1,\De_2, \De_3, \De_4, \De_5} \nonumber\\
&+(\scr{D}_{X_2}^{(+ 0)}\cdot \scr{D}_{X_3}^{(0+)})
W^{(n_{IJ})}_{\De, \ell, \De', \ell'; \De_1,\De_2-1, \De_3, \De_4, \De_5} 
\bigg)
\nonumber\\
&+
F^{(+ 0)}_{(- 0)}\bigg(
 (\scr{D}_{X_1}^{(- 0)}\cdot\scr{D}_{X_3}^{(0+)})
 W^{(n_{IJ})}_{\De, \ell, \De', \ell'; \De_1+1,\De_2, \De_3, \De_4, \De_5} \nonumber\\
&+  (\scr{D}_{X_2}^{(-0)}\cdot\scr{D}_{X_3}^{(0+)})
W^{(n_{IJ})}_{\De, \ell, \De', \ell'; \De_1,\De_2+1, \De_3, \De_4, \De_5} 
\bigg)\bigg]\,,
\]
\[
 \eql{replacements 2}
 (\scr{D}_{X_3}^{(0+)}\cdot \scr{D}_{X_I}^{(0+)}){}& \vev{\scr{O}_{\De,\ell-1}(X_I)\Phi_{\De_3}(X_3) \scr{O}_{\De^\prime,\ell^\prime}(X_J)}^{(n_{IJ})}\nonumber\\
\to\,&C_{(0+)(0-)}\bigg[
F^{(0+)}_{(+ 0)}\bigg(
 (\scr{D}_{X_1}^{(+ 0) }\cdot \scr{D}_{X_3}^{(0+)})
 W^{(n_{IJ})}_{\De, \ell, \De', \ell'; \De_1-1,\De_2, \De_3, \De_4, \De_5}\nonumber\\
&- (\scr{D}_{X_2}^{(+ 0)}\cdot \scr{D}_{X_3}^{(0+)})
W^{(n_{IJ})}_{\De, \ell, \De', \ell'; \De_1,\De_2-1, \De_3, \De_4, \De_5} \bigg)
\nonumber\\
&+F^{(0+)}_{(- 0)}\bigg(
( \scr{D}_{X_1}^{(- 0) }\cdot \scr{D}_{X_3}^{(0+)})
W^{(n_{IJ})}_{\De, \ell, \De', \ell'; \De_1+1,\De_2, \De_3, \De_4, \De_5}\nonumber\\
&-( \scr{D}_{X_2}^{(- 0)}\cdot  \scr{D}_{X_3}^{(0+)})
W^{(n_{IJ})}_{\De, \ell, \De', \ell'; \De_1,\De_2+1, \De_3, \De_4, \De_5}
 \bigg)\bigg]\,,
\]
\[
 \eql{replacements 3}
(\scr{D}_{X_3}^{(0+)}\cdot \scr{D}_{X_I}^{(0-)}){} &\vev{\scr{O}_{\De,\ell+1}(X_I)\Phi_{\De_3}(X_3)\scr{O}_{\De^\prime,\ell^\prime}(X_J)}^{(n_{IJ})}
\nonumber\\
\to\,&C_{(0-)(0+)}\bigg[F^{(0-)}_{(+ 0)}\bigg(
 (\scr{D}_{X_1}^{(+  0) }\cdot \scr{D}_{X_3}^{(0+)})
 W^{(n_{IJ})}_{\De, \ell, \De', \ell';\De_1-1,\De_2, \De_3, \De_4, \De_5}\nonumber\\
&- (\scr{D}_{X_2}^{(+ 0)}\cdot \scr{D}_{X_3}^{(0+)})
W^{(n_{IJ})}_{\De, \ell, \De', \ell'; \De_1,\De_2-1, \De_3, \De_4, \De_5}\bigg)
\nonumber\\
&+F^{(0-)}_{(- 0)}
\bigg( (\scr{D}_{X_1}^{(- 0) }\cdot \scr{D}_{X_3}^{(0+)})
W^{(n_{IJ})}_{\De, \ell, \De', \ell'; \De_1+1,\De_2, \De_3, \De_4, \De_5}\nonumber\\
& - (\scr{D}_{X_2}^{(- 0)}\cdot \scr{D}_{X_3}^{(0+)})
 W^{(n_{IJ})}_{\De, \ell, \De', \ell'; \De_1,\De_2+1, \De_3, \De_4, \De_5}
 \bigg)\bigg]\,,
\]
where the 3-point $6j$ symbols for the special crossing relation may be found in \Eq{special 6j symbols}. In our case, they are given by
\[
F^{(+ 0)}_{(+ 0)}=\,&\frac{(\De-1) \left(\De -\Delta _1+\Delta _2+\ell\right) 
\left(d-\De +\Delta _1-\Delta _2+\ell-2\right)}{2 \left(\Delta_1-2\right) \left(d-2 \Delta _1\right) \left(d-\Delta _1-1\right)}\,,\nonumber\\
F^{(+ 0)}_{(- 0)}=\,&-\frac{(\De-1) \left(\De +\Delta _1-\Delta _2+\ell\right) 
\left(-d+\De +\Delta _1-\Delta _2-\ell+2\right) }{2 \left(d-2 \Delta _1\right)}\nonumber\\
&\times
\left(-2d+\De +\Delta _1+\Delta _2-\ell+2\right) \left(-d+\De +\Delta _1+\Delta _2+\ell\right)\,,\nonumber\\
F^{(0+)}_{(+ 0)}=\,&\frac{\Delta -\Delta _1+\Delta _2+\ell}{2 \left(\Delta _1-2\right) (\ell+1) \left(d-2 \Delta _1\right) \left(d-\Delta _1-1\right)}\,,\nonumber\\
F^{(0+)}_{(- 0)}=\,&-\frac{\left(\Delta +\Delta _1-\Delta _2+\ell\right) \left(-\Delta +\Delta _1+\Delta _2+\ell\right) }{2 (\ell+1) \left(d-2 \Delta _1\right)}
\left(\Delta +\Delta
   _1+\Delta _2+\ell-d\right)\,,\nonumber\\
F^{(0-)}_{(+ 0)} =\,&\frac{\ell \left(d-\Delta +\Delta _1-\Delta _2+\ell-2\right)}{2 \left(\Delta _1-2\right) \left(d-2 \Delta _1\right)
   \left(d-\Delta _1-1\right)}\,,\nonumber\\
F^{(0-)}_{(- 0)}  =\,&-\frac{ \left(2 d-\Delta -\Delta _1-\Delta _2+\ell-2\right) \left(d+\Delta -\Delta _1-\Delta _2+\ell-2\right) }{2 \left(d-2 \Delta _1\right)}\nonumber\\
&\times
\ell\left(d-\Delta -\Delta _1+\Delta _2+\ell-2\right)\,.
\]
Specifically, we convert the relations \Eq{equation Q1} - \Eq{equation Q2}
into the corresponding relations for the conformal blocks by gluing them between the structures $\vev{\phi_{\De_1}(X_1)\phi_{\De_2}(X_2)\scr{O}_{\De,\ell}}$ and $  \vev{\scr{O}'_{\De',\ell'}
\phi_{\De_4}(X_4)\phi_{\De_5}(X_5)}$ via
\[
W_{\De, \ell; \De^\prime, \ell^\prime; \De_1, \De_2, \De_3, \De_4, \De_5}^{(V)(i, n_{IJ})}=
\vev{\phi_{\De_1}(X_1)\phi_{\De_2}(X_2)\scr{O}_{\De,\ell}}\bowtie   
Q_{(\ell,1,\ell')}^{(i, n_{IJ})} \bowtie
  \vev{\scr{O}'_{\De',\ell'}
\phi_{\De_4}(X_4)\phi_{\De_5}(X_5)}\,,
\]
for $i=1,2,3$, with the replacements \Eq{replacements 1}-\Eq{replacements 3}.

Combining everything together, we ultimately arrive at the results
\[
\eql{block vector 1}
W_{\De, \ell; \De^\prime, \ell^\prime; \De_1, \De_2, \De_3, \De_4, \De_5}^{(V)(1, n_{IJ})} =\,&
\frac{\left(\ell '-n_{IJ}\right)}{\Delta '-\Delta +\Delta _3-\ell '+\ell -1}W_{\De, \ell; \De^\prime, \ell^\prime; \De_1, \De_2, \De_3, \De_4, \De_5}^{(V)(2, n_{IJ})}\nonumber\\
&+
\sum_{m=n_{IJ}}^{n_{IJ}+1} 
\mathscr{A}_{ (+0)(0+)}^{(1)(m)}
(\scr{D}_{X_1}^{(+0) }\cdot \scr{D}_{X_3}^{(0+)})
 W^{(m)}_{\De, \ell, \De', \ell';\De_1-1,\De_2, \De_3, \De_4, \De_5} \nonumber\\
&+\mathscr{A}_{ (+0)(0+)}^{(2)(m)} (\scr{D}_{X_2}^{(+ 0)}\cdot \scr{D}_{X_3}^{(0+)})
W^{(m)}_{\De, \ell, \De', \ell'; \De_1,\De_2-1, \De_3, \De_4, \De_5}\nonumber\\
&+\mathscr{A}_{ (-0)(0+)}^{(1)(m)}(\scr{D}_{X_1}^{(- 0) }\cdot \scr{D}_{X_3}^{(0+)})
W^{(m)}_{\De, \ell, \De', \ell'; \De_1+1,\De_2, \De_3, \De_4, \De_5}
\nonumber\\
&+\mathscr{A}_{ (-0)(0+)}^{(2)(m)}(\scr{D}_{X_2}^{(- 0)}\cdot \scr{D}_{X_3}^{(0+)})
 W^{(m)}_{\De, \ell, \De', \ell'; \De_1,\De_2+1, \De_3, \De_4, \De_5}\,,
\]

\[
\eql{block vector 3}
W_{\De, \ell; \De^\prime, \ell^\prime; \De_1, \De_2, \De_3, \De_4, \De_5}^{(V)(3, n_{IJ})}
=\,&\sum_{m=n_{IJ}}^{n_{IJ}+1}
\mathscr{C}_{(+0) (0+)}^{(1)(m)}(\scr{D}_{X_1}^
{(+  0) }\cdot \scr{D}_{X_3}^{(0+)})
 W^{(m)}_{\De, \ell, \De', \ell';\De_1-1,\De_2, \De_3, \De_4, \De_5} \nonumber\\
&+\mathscr{C}_{(+ 0)(0+)}^{(2)(m)} 
(\scr{D}_{X_2}^{(+ 0)}\cdot \scr{D}_{X_3}^{(0+)})
W^{(m)}_{\De, \ell, \De', \ell'; \De_1,\De_2-1, \De_3, \De_4, \De_5}\nonumber\\
&+\mathscr{C}_{(- 0)(0+)}^{(1)(m)}
(\scr{D}_{X_1}^{(- 0) }\cdot \scr{D}_{X_3}^{(0+)})
W^{(m)}_{\De, \ell, \De', \ell'; \De_1+1,\De_2, \De_3, \De_4, \De_5}
\nonumber\\
&+\mathscr{C}_{(- 0)(0+)}^{(2)(m)}(\scr{D}_{X_2}^{(- 0)}\cdot \scr{D}_{X_3}^{(0+)})
 W^{(m)}_{\De, \ell, \De', \ell'; \De_1,\De_2+1, \De_3, \De_4, \De_5}\,,
\]

and
\[
\eql{block vector 2}
W_{\De, \ell; \De^\prime, \ell^\prime; \De_1, \De_2, \De_3, \De_4, \De_5}^{(V)(2, n_{IJ})}
=\,&\frac{(n_{IJ}-\ell ) \left(n_{IJ}-\ell '+1\right)}{(n_{IJ}+1) \left(\Delta '-\Delta
   +\Delta _3-2 n_{IJ}+\ell '+\ell -1\right)}W_{\De, \ell; \De^\prime, \ell^\prime; \De_1, \De_2, \De_3, \De_4, \De_5}^{(V)(2, n_{IJ}+1)}
   \nonumber\\
&+\sum_{m=n_{IJ}}^{n_{IJ}+2}
\mathscr{B}_{(+0) (0+)}^{(1)(m)}(\scr{D}_{X_1}^
{(+  0) }\cdot \scr{D}_{X_3}^{(0+)})
 W^{(m)}_{\De, \ell, \De', \ell';\De_1-1,\De_2, \De_3, \De_4, \De_5}\nonumber\\ 
&+\mathscr{B}_{(+ 0)(0+)}^{(2)(m)} 
(\scr{D}_{X_2}^{(+ 0)}\cdot \scr{D}_{X_3}^{(0+)})
W^{(m)}_{\De, \ell, \De', \ell'; \De_1,\De_2-1, \De_3, \De_4, \De_5}\nonumber\\
&+\mathscr{B}_{(- 0)(0+)}^{(1)(m)}
(\scr{D}_{X_1}^{(- 0) }\cdot \scr{D}_{X_3}^{(0+)})
W^{(m)}_{\De, \ell, \De', \ell'; \De_1+1,\De_2, \De_3, \De_4, \De_5}\nonumber\\
&+\mathscr{B}_{(- 0)(0+)}^{(2)(m)}(\scr{D}_{X_2}^{(- 0)}\cdot \scr{D}_{X_3}^{(0+)})
 W^{(m)}_{\De, \ell, \De', \ell'; \De_1,\De_2+1, \De_3, \De_4, \De_5}\,.
\]

The respective $\mathscr{A}$, $\mathscr{B}$, and $\mathscr{C}$ coefficients are all assembled in Appendix~\ref{app:F}.  

\subsection{Promoting $\Phi$ to a tensor operator}
\label{sec:phitensor}
 
Armed with the above results, we next promote $\Phi$ to a spin-2 operator $T^{AB}$ by taking
\[
\vev{\scr{O}_{\De, \ell}(X_1)\Phi_{\De_3}(X_2) \scr{O}^\prime_{\De', \ell'}(X_3)}\to \vev{\scr{O}_{\De, \ell}(X_1)T^{AB}(X_2) \scr{O}^\prime_{\De', \ell'}(X_3)}\,,
\]  
which again maps a (spin)-(scalar)-(spin) type 3-point function to a (spin)-(spin)-(spin) type one.

In this case, the set of allowed 3-point structures is 
\[ 
\eql{spin-2 structures 1}
Q_{(\ell,2,\ell')}(X_1, X_2, X_3; Z_1, Z_2, Z_3)=\sum_{i=1}^6 \kappa_{i, n_{IJ}}  Q^{(i, n_{IJ})}_{(\ell,2,\ell')}    
\]
with 
\[
Q^{(i, n_{IJ})}_{(\ell,2,\ell')}  =
\dfrac{ q^{(i, n_{IJ})}_{(\ell,2,\ell')}}{(X_{12})^{\sfrac{1}{2}( \De+\De_3-\De'+ \ell - \ell^\prime+ 2 )}
(X_{13})^{\sfrac{1}{2}( \De-\De_3+\De'+ \ell + \ell^\prime -2)}(X_{23})^{\sfrac{1}{2}(- \De+\De_3 +\De'- \ell+ \ell^\prime
+ 2 )}}\,,
\]
where the six structures $q^{(a, n_{IJ})}_{(\ell,2,\ell')}$ are given by
\[
\eql{spin-2 structures 2}
q^{(1, n_{IJ})}_{(\ell,2,\ell')} &= V_1^{\ell -n_{IJ} } V_2^2 V_3^{\ell^\prime - n_{IJ} } H_{13}^{n_{IJ}}\,,\nonumber\\
q^{(2, n_{IJ})}_{(\ell,2,\ell')}&=V_1^{\ell -n_{IJ} } V_2 V_3^{(\ell^\prime-1) - n_{IJ} }  
H_{13}^{n_{IJ} } H_{23}\,,\nonumber\\
q^{(3, n_{IJ})}_{(\ell,2,\ell')} &=V_1^{\ell - n_{IJ} }V_3^{(\ell^\prime-2) - n_{IJ}}  
H_{13}^{n_{IJ}} H_{23}^2\,,\nonumber\\
q^{(4, n_{IJ})}_{(\ell,2,\ell')} &=V_1^{(\ell -1)-n_{IJ}} V_2 V_3^{\ell^\prime - n_{IJ} } H_{12}
H_{13}^{n_{IJ} }\,,\nonumber\\
q^{(5, n_{IJ})}_{(\ell,2,\ell')} &=V_1^{(\ell -1)-n_{IJ} } V_3^{(\ell^\prime -1) - n_{IJ}} H_{12}
H_{13}^{n_{IJ} } H_{23}\,,\nonumber\\
q^{(6, n_{IJ})}_{(\ell,2,\ell')} &=V_1^{(\ell -2)-n_{IJ} }V_3^{\ell^\prime - n_{IJ} } H_{12}^2 
H_{13}^{n_{IJ}}\,.  
\]
The structures $q^{(1, n_{IJ})}_{(\ell,2,\ell')}$, 
$q^{(2, n_{IJ})}_{(\ell,2,\ell')}$,   
$q^{(3, n_{IJ})}_{(\ell,2,\ell')}$, $q^{(4, n_{IJ})}_{(\ell,2,\ell')}$,
$q^{(5, n_{IJ})}_{(\ell,2,\ell')}$, and 
$q^{(6, n_{IJ})}_{(\ell,2,\ell')}$        
exist for $n_{IJ}$ in the ranges $[0, \min(\ell, \ell')]$,  $[0, \min(\ell, \ell'-1)]$,  $[0, \min(\ell, \ell'-2)]$,
 $[0, \min(\ell-1, \ell')]$,  $[0, \min(\ell-1, \ell'-1)]$, and $[0, \min(\ell-2, \ell')]$,  respectively.\footnote{Note that in $d=3$ there are redundancies between the structures so that $H_{12} H_{13} H_{23}$ is not independent~\cite{Costa:2011mg}. This means that the structures $q^{(5,n_{IJ})}_{(\ell,2,\ell')}$ are not independent and this should be taken into account when applying the results in $d=3$.}

Our hope is to recycle much of the calculation for the spin-1 case here. In particular, we intend to act with the three distinct weight-shifting operator combinations once again, but this time, on the three structures corresponding to the (spin)-(vector)-(spin) 3-point structures, rather than on the (spin)-(scalar)-(spin) ones. That is, in this case, we take the blocks computed in the previous section with $\Phi\to v^A$ to be our seed blocks.

Proceeding in a manner analogous to the spin-$1$ case, we find three sets of equations similar to \Eq{one} - \Eq{three}, corresponding to taking the derivatives
\[
(\scr{D}_{X_3}^{(0+)}\cdot \scr{D}_{X_I}^{(-0)})Q_{(\ell,1,\ell')}^{(i, n_{IJ})}\,,\nonumber\\  
(\scr{D}_{X_3}^{(0+)}\cdot \scr{D}_{X_I}^{(0+)}){} Q_{(\ell, 1,\ell')}^{(i, n_{IJ})}\,, \nonumber\\
(\scr{D}_{X_3}^{(0+)}\cdot \scr{D}_{X_I}^{(0-)})Q_{(\ell,1,\ell')}^{(i, n_{IJ})}\,,
\] 
for $i= 1, 2, 3$. 

However, since we have six different structures here, we just require six independent equations. In particular, we find that it is sufficient to consider
\[ 
(\scr{D}_{X_3}^{(0+)}\cdot \scr{D}_{X_I}^{(-0)})Q_{(\ell,1,\ell')}^{(1, n_{IJ})}&=  \xi_{11} Q_{(\ell,2,\ell')}^{(1, n_{IJ})}+ \xi_{12}  Q_{(\ell,2,\ell')}^{(2, n_{IJ})} + \xi_{14}  Q_{(\ell,2,\ell')}^{(4, n_{IJ})}\,,\eql{one tensor1}
\\
(\scr{D}_{X_3}^{(0+)}\cdot \scr{D}_{X_I}^{(-0)})Q_{(\ell,1,\ell')}^{(2, n_{IJ})} &= \xi_{22}  Q_{(\ell,2,\ell')}^{(2, n_{IJ})}+\xi_{23} Q_{(\ell,2,\ell')}^{(3, n_{IJ})} +\xi_{25} Q_{(\ell,2,\ell')}^{(5, n_{IJ})}\,,\eql{one tensor2}
\\
(\scr{D}_{X_3}^{(0+)}\cdot \scr{D}_{X_I}^{(-0)}) Q_{(\ell,1,\ell')}^{(3, n_{IJ})}&=  \xi_{34} Q_{(\ell,2,\ell')}^{(4, n_{IJ})}+\xi_{35} Q_{(\ell,2,\ell')}^{(5, n_{IJ})} +\xi_{36}  Q_{(\ell,2,\ell')}^{(6, n_{IJ})}\,,\eql{one tensor3}
\]
\[
 (\scr{D}_{X_3}^{(0+)}\cdot \scr{D}_{X_I}^{(0+)})Q_{(\ell,1,\ell')}^{(2, n_{IJ})} =\,& \ka_{22}  Q_{(\ell,2,\ell')}^{(2, n_{IJ})}+\la_{22}  Q_{(\ell,2,\ell')}^{(2, n_{IJ}+1)}+\ka_{23} Q_{(\ell,2,\ell')}^{(3, n_{IJ})} \nonumber\\ 
 & +\la_{23} Q_{(\ell,2,\ell')}^{(3, n_{IJ}+1)} 
 +\ka_{25} Q_{(\ell,2,\ell')}^{(5, n_{IJ})} +\la_{25} Q_{(\ell,2,\ell')}^{(5, n_{IJ}+1)}\,,\eql{two tensor1}
\\
 (\scr{D}_{X_3}^{(0+)}\cdot \scr{D}_{X_I}^{(0+)})Q_{(\ell,1,\ell')}^{(3, n_{IJ})} =\,&  \ka_{34} Q_{(\ell,2,\ell')}^{(4, n_{IJ})}+\la_{34} Q_{(\ell,2,\ell')}^{(4, n_{IJ}+1)}+\ka_{35} Q_{(\ell,2,\ell')}^{(5, n_{IJ})}\nonumber\\
 &  +\la_{35} Q_{(\ell,2,\ell')}^{(5, n_{IJ}+1)} 
 +\ka_{36}  Q_{(\ell,2,\ell')}^{(6, n_{IJ})} +\la_{36}  Q_{(\ell,2,\ell')}^{(6, n_{IJ}+1)}\,,\eql{two tensor2}
\]
and
\[
\eql{three tensor}
(\scr{D}_{X_3}^{(0+)}\cdot \scr{D}_{X_I}^{(0-)})Q_{(\ell,1,\ell')}^{(2, n_{IJ})}  =\,& 
\rho_{22}  Q_{(\ell,2,\ell')}^{(2, n_{IJ}-1)}
+\si_{22}  Q_{(\ell,2,\ell')}^{(2, n_{IJ})}
+\tau_{22}  Q_{(\ell,2,\ell')}^{(2, n_{IJ}+1)}\nonumber\\
&+\rho_{23} Q_{(\ell,2,\ell')}^{(3, n_{IJ}-1)} 
+\si_{23} Q_{(\ell,2,\ell')}^{(3, n_{IJ})} 
+\tau_{23} Q_{(\ell,2,\ell')}^{(3, n_{IJ}+1)} \nonumber\\
&+\rho_{25} Q_{(\ell,2,\ell')}^{(5, n_{IJ}-1)}
+\si_{25} Q_{(\ell,2,\ell')}^{(5, n_{IJ})}
+\tau_{25} Q_{(\ell,2,\ell')}^{(5, n_{IJ}+1)}\,.
\]
The respective coefficients may be found in Appendix~\ref{app:D}. 

We next adopt a procedure analogous to the one employed for the vector case discussed above. In particular, we first apply the three equations in \Eq{one tensor1} - \Eq{one tensor3} to solve for $\{ Q_{(\ell,2,\ell')}^{(2, n_{IJ})}, Q_{(\ell,2,\ell')}^{(4, n_{IJ})}, Q_{(\ell,2,\ell')}^{(5, n_{IJ})}\}$ in terms of $\{Q_{(\ell,2,\ell')}^{(1, n_{IJ})}$, $Q_{(\ell,2,\ell')}^{(3, n_{IJ})}$, $Q_{(\ell,2,\ell')}^{(6, n_{IJ})}\}$. We may then generate the corresponding equivalent statements for the structures labeled by $n_{IJ}-1$ and $n_{IJ}+1$. We subsequently insert these expressions into \Eq{two tensor1} - \Eq{two tensor2} and solve for $Q_{(\ell,2,\ell')}^{(1, n_{IJ})}$ and $Q_{(\ell,2,\ell')}^{(6, n_{IJ})}$. At this point, the structures $Q_{(\ell,2,\ell')}^{(i, n_{IJ})}$ for $i=1,2, 4, 5, 6$ are expressed purely in terms of differential operators acting on the spin-$1$ structures $\{Q_{(\ell,1,\ell')}^{(1, n_{IJ})},Q_{(\ell,1,\ell')}^{(2, n_{IJ})},Q_{(\ell,1,\ell')}^{(3, n_{IJ})}\}$ as well as the spin-$2$ structure $Q_{(\ell,2,\ell')}^{(3, n_{IJ})}$. This remaining structure $Q_{(\ell,2,\ell')}^{(3, n_{IJ})}$ may be extracted directly by solving the last of the above equations, namely \Eq{three tensor}. This allows us to express it in the form of a recursion relation, which terminates at $n_{IJ} = \min(\ell, \ell'-2)$.

The explicit results for the structures $Q_{(\ell,2,\ell')}^{(i, n_{IJ})}$ are given below:
\[\eql{Q21}
Q_{(\ell,2,\ell')}^{(1, n_{IJ})}=\,&
\frac{  \left(-n_{IJ}+\ell '-1\right) \left(\ell '-n_{IJ}\right)}{\left(\Delta '-\Delta
   +\Delta _3-\ell '+\ell \right) \left(\Delta '-\Delta +\Delta _3-\ell '+\ell +2\right)}Q_{(\ell,2,\ell')}^{(3, n_{IJ})}\nonumber\\
&+a_1'(\scr{D}_{X_3}^{(0+)}\cdot \scr{D}_{X_I}^{(-0)})Q_{(\ell,1,\ell')}^{(1, n_{IJ})} +
a_2'(\scr{D}_{X_3}^{(0+)}\cdot \scr{D}_{X_I}^{(0+)})Q_{(\ell,1,\ell')}^{(3, n_{IJ})}\nonumber\\
   &+a_3'
   (\scr{D}_{X_3}^{(0+)}\cdot \scr{D}_{X_I}^{(-0)}) Q_{(\ell,1,\ell')}^{(3, n_{IJ}+1)} 
   +a_4'(\scr{D}_{X_3}^{(0+)}\cdot \scr{D}_{X_I}^{(0+)})Q_{(\ell,1,\ell')}^{(2, n_{IJ})} \nonumber\\
   &+a_5'   (\scr{D}_{X_3}^{(0+)}\cdot \scr{D}_{X_I}^{(-0)})Q_{(\ell,1,\ell')}^{(2, n_{IJ}+1)} 
   +a_6'(\scr{D}_{X_3}^{(0+)}\cdot \scr{D}_{X_I}^{(-0)}) Q_{(\ell,1,\ell')}^{(3, n_{IJ})}   
  +a_7'
   (\scr{D}_{X_3}^{(0+)}\cdot \scr{D}_{X_I}^{(-0)})Q_{(\ell,1,\ell')}^{(2, n_{IJ})}\,,
\]
\[
Q_{(\ell,2,\ell')}^{(2, n_{IJ})}=\,&
\frac{ \left(-n_{IJ}+\ell '-1\right)}{\Delta '-\Delta +\Delta _3-\ell '+\ell +2}Q_{(\ell,2,\ell')}^{(3, n_{IJ})}
\nonumber\\
&+
b_1'(\scr{D}_{X_3}^{(0+)}\cdot \scr{D}_{X_I}^{(0+)})Q_{(\ell,1,\ell')}^{(2, n_{IJ})} 
+ b_2'
(\scr{D}_{X_3}^{(0+)}\cdot \scr{D}_{X_I}^{(-0)})Q_{(\ell,1,\ell')}^{(2, n_{IJ}+1)}+b_3'  (\scr{D}_{X_3}^{(0+)}\cdot \scr{D}_{X_I}^{(-0)})Q_{(\ell,1,\ell')}^{(2, n_{IJ})}\,,
\]
\[
Q_{(\ell,2,\ell')}^{(3, n_{IJ})}=\,&
\frac{ (n_{IJ}-\ell ) \left(n_{IJ}-\ell '+2\right)}{(n_{IJ}+1) \left(\Delta '-\Delta
   +\Delta _3-2 n_{IJ}+\ell '+\ell \right)}
   Q_{(\ell,2,\ell')}^{(3, n_{IJ}+1)}\nonumber\\
   &+c_1'
   (\scr{D}_{X_3}^{(0+)}\cdot \scr{D}_{X_I}^{(-0)})Q_{(\ell,1,\ell')}^{(2,n_{IJ}+2)}
   +c_2'
   (\scr{D}_{X_3}^{(0+)}\cdot \scr{D}_{X_I}^{(0+)})Q_{(\ell,1,\ell')}^{(2, n_{IJ})} \nonumber\\
   &+c_3'
    (\scr{D}_{X_3}^{(0+)}\cdot \scr{D}_{X_I}^{(0+)})Q_{(\ell,1,\ell')}^{(2, n_{IJ}+1)} 
   +c_4'
  (\scr{D}_{X_3}^{(0+)}\cdot \scr{D}_{X_I}^{(0-)})Q_{(\ell,1,\ell')}^{(2, n_{IJ}+1)}  \nonumber\\
  &+c_5' (\scr{D}_{X_3}^{(0+)}\cdot \scr{D}_{X_I}^{(-0)})Q_{(\ell,1,\ell')}^{(2, n_{IJ})} 
  +c_6' (\scr{D}_{X_3}^{(0+)}\cdot \scr{D}_{X_I}^{(-0)})Q_{(\ell,1,\ell')}^{(2, n_{IJ}+1)}\,,
 \]
\[
Q_{(\ell,2,\ell')}^{(4, n_{IJ})}=\,&
d_1' (\scr{D}_{X_3}^{(0+)}\cdot \scr{D}_{X_I}^{(0+)})Q_{(\ell,1,\ell')}^{(3, n_{IJ})} +d_2'
 (\scr{D}_{X_3}^{(0+)}\cdot \scr{D}_{X_I}^{(-0)}) Q_{(\ell,1,\ell')}^{(3, n_{IJ}+1)}\nonumber\\
 &+d_3'
   (\scr{D}_{X_3}^{(0+)}\cdot \scr{D}_{X_I}^{(0+)})Q_{(\ell,1,\ell')}^{(2, n_{IJ})}  
  +d_4'(\scr{D}_{X_3}^{(0+)}\cdot \scr{D}_{X_I}^{(-0)})Q_{(\ell,1,\ell')}^{(2, n_{IJ}+1)}\nonumber\\
   &+d_5'(\scr{D}_{X_3}^{(0+)}\cdot \scr{D}_{X_I}^{(-0)})Q_{(\ell,1,\ell')}^{(2, n_{IJ})} 
 +d_6' (\scr{D}_{X_3}^{(0+)}\cdot \scr{D}_{X_I}^{(-0)}) Q_{(\ell,1,\ell')}^{(3, n_{IJ})}\,,
\]
 \[
Q_{(\ell,2,\ell')}^{(5, n_{IJ})}=\,&
e_1'(\scr{D}_{X_3}^{(0+)}\cdot \scr{D}_{X_I}^{(-0)})Q_{(\ell,1,\ell')}^{(2, n_{IJ})}  
   +e_2'(\scr{D}_{X_3}^{(0+)}\cdot \scr{D}_{X_I}^{(-0)})Q_{(\ell,1,\ell')}^{(2, n_{IJ}+1)} 
   +e_3'(\scr{D}_{X_3}^{(0+)}\cdot \scr{D}_{X_I}^{(0+)})Q_{(\ell,1,\ell')}^{(2, n_{IJ})}\,, 
\]
\[
Q_{(\ell,2,\ell')}^{(6, n_{IJ})}=\,&f_1'(\scr{D}_{X_3}^{(0+)}\cdot \scr{D}_{X_I}^{(-0)}) Q_{(\ell,1,\ell')}^{(3, n_{IJ})}
   +f_2' (\scr{D}_{X_3}^{(0+)}\cdot \scr{D}_{X_I}^{(-0)}) Q_{(\ell,1,\ell')}^{(3, n_{IJ}+1)}
+f_3'(\scr{D}_{X_3}^{(0+)}\cdot \scr{D}_{X_I}^{(0+)})Q_{(\ell,1,\ell')}^{(3, n_{IJ})}\,.\eql{Q26}
\]
The coefficients can be found in Appendix~\ref{app:E}.

We can then proceed to convert these relations into the corresponding statements for the conformal blocks exactly as before, this time with
\[
W_{\De, \ell; \De^\prime, \ell^\prime; \De_1, \De_2, \De_3, \De_4, \De_5}^{(T)(a, n_{IJ})}=
\vev{\phi_{\De_1}(X_1)\phi_{\De_2}(X_2)\scr{O}_{\De,\ell}}\bowtie   
Q_{(\ell,2,\ell')}^{(a, n_{IJ})} \bowtie
  \vev{\scr{O}'_{\De',\ell'}
\phi_{\De_4}(X_4)\phi_{\De_5}(X_5)}\,,
\]
for each of $a = 1, \ldots, 6$. Again, we implement a set of substitutions directly analogous to those in \Eq{replacements 1}-\Eq{replacements 3}, with the difference that here we map each of 
\[
(\scr{D}_{X_3}^{(0+)}\cdot \scr{D}_{X_I}^{(-0)})Q_{(\ell,1,\ell')}^{(i, n_{IJ})}\,, \quad (\scr{D}_{X_3}^{(0+)}\cdot \scr{D}_{X_I}^{(0+)})Q_{(\ell,1,\ell')}^{(i, n_{IJ})}\,,\quad
 (\scr{D}_{X_3}^{(0+)}\cdot \scr{D}_{X_I}^{(0-)})Q_{(\ell,1,\ell')}^{(i, n_{IJ})}\,,
 \]
 to the set 
 \[
\bigg\{ (\scr{D}_{X_1}^{(+ 0) }\cdot \scr{D}_{X_3}^{(0+)})
 W^{(V)(i, n_{IJ})}_{\De, \ell, \De', \ell'; \De_1-1,\De_2, \De_3, \De_4, \De_5}\,,
\quad
 (\scr{D}_{X_2}^{(+ 0)}\cdot \scr{D}_{X_3}^{(0+)})
W^{(V)(i, n_{IJ})}_{\De, \ell, \De', \ell'; \De_1,\De_2-1, \De_3, \De_4, \De_5}\,,
 \nonumber\\ 
  (\scr{D}_{X_1}^{(- 0)}\cdot\scr{D}_{X_3}^{(0+)})
 W^{(V)(i, n_{IJ})}_{\De, \ell, \De', \ell'; \De_1+1,\De_2, \De_3, \De_4, \De_5}\,, 
\quad
  (\scr{D}_{X_2}^{(-0)}\cdot\scr{D}_{X_3}^{(0+)})
W^{(V)(i, n_{IJ})}_{\De, \ell, \De', \ell'; \De_1,\De_2+1, \De_3, \De_4, \De_5} 
\bigg\}\,,
\]
with the selfsame $6j$ coefficients featured in \Eq{replacements 1}-\Eq{replacements 3}. That is, these relations are identical to the ones used in the vector case, up to the replacements
\[
\vev{\scr{O}_{\De,\ell}(X_I)\Phi_{\De_3}(X_3)\scr{O}_{\De^\prime,\ell^\prime}(X_J)}^{(n_{IJ})} \mapsto& Q_{(\ell,1,\ell')}^{(i, n_{IJ})}\,,\\\nonumber
W^{(n_{IJ})}_{\De, \ell, \De', \ell'; \De_1,\De_2, \De_3, \De_4, \De_5}\mapsto& W^{(V)(i, n_{IJ})}_{\De, \ell, \De', \ell'; \De_1,\De_2, \De_3, \De_4, \De_5}\,,
\]
for $i = 1,2,3$.

With this, it is effortless to write down the expressions for the six distinct conformal block structures for the case where $\Phi$ is a spin-2 tensor operator. The results take the form:
\[\eql{block tensor 1}
W_{\De, \ell; \De^\prime, \ell^\prime; \De_1, \De_2, \De_3, \De_4, \De_5}^{(T)(1, n_{IJ})}
=\,&
\frac{  \left(-n_{IJ}+\ell '-1\right) \left(\ell '-n_{IJ}\right)}{\left(\Delta '-\Delta
   +\Delta _3-\ell '+\ell \right) \left(\Delta '-\Delta +\Delta _3-\ell '+\ell +2\right)}W_{\De, \ell; \De^\prime, \ell^\prime; \De_1, \De_2, \De_3, \De_4, \De_5}^{(T)(3, n_{IJ})}\nonumber\\
&+\mathscr{A}' {}_{(+0)(0+)}^{(1)(1, n_{IJ})}
(\scr{D}_{X_1}^{(+0) }\cdot \scr{D}_{X_3}^{(0+)})
 W^{(V)(1,n_{IJ})}_{\De, \ell, \De', \ell';\De_1-1,\De_2, \De_3, \De_4, \De_5} \nonumber\\
&+\mathscr{A}'{}_{ (+0)(0+)}^{(2)(1, n_{IJ})} (\scr{D}_{X_2}^{(+ 0)}\cdot \scr{D}_{X_3}^{(0+)})
W^{(V)(1, n_{IJ})}_{\De, \ell, \De', \ell'; \De_1,\De_2-1, \De_3, \De_4, \De_5}\nonumber\\
&+\mathscr{A}' {}_{ (-0)(0+)}^{(1)(1, n_{IJ})}(\scr{D}_{X_1}^{(- 0) }\cdot \scr{D}_{X_3}^{(0+)})
W^{(V)(1, n_{IJ})}_{\De, \ell, \De', \ell'; \De_1+1,\De_2, \De_3, \De_4, \De_5}
\nonumber\\
&+\mathscr{A}' {}_{ (-0)(0+)}^{(2)(1, n_{IJ})}(\scr{D}_{X_2}^{(- 0)}\cdot \scr{D}_{X_3}^{(0+)})
 W^{(V)(1,n_{IJ})}_{\De, \ell, \De', \ell'; \De_1,\De_2+1, \De_3, \De_4, \De_5}\nonumber\\   
&+\sum_{i=2}^3
\sum_{m=n_{IJ}}^{n_{IJ}+1} 
\mathscr{A}' {}_{(+0)(0+)}^{(1)(i, m)}
(\scr{D}_{X_1}^{(+0) }\cdot \scr{D}_{X_3}^{(0+)})
 W^{(V)(i,m)}_{\De, \ell, \De', \ell';\De_1-1,\De_2, \De_3, \De_4, \De_5} \nonumber\\
&+\mathscr{A}'{}_{ (+0)(0+)}^{(2)(i, m)} (\scr{D}_{X_2}^{(+ 0)}\cdot \scr{D}_{X_3}^{(0+)})
W^{(V)(i, m)}_{\De, \ell, \De', \ell'; \De_1,\De_2-1, \De_3, \De_4, \De_5}\nonumber\\
&+\mathscr{A}' {}_{ (-0)(0+)}^{(1)(i, m)}(\scr{D}_{X_1}^{(- 0) }\cdot \scr{D}_{X_3}^{(0+)})
W^{(V)(i, m)}_{\De, \ell, \De', \ell'; \De_1+1,\De_2, \De_3, \De_4, \De_5}
\nonumber\\
&+\mathscr{A}' {}_{ (-0)(0+)}^{(2)(i, m)}(\scr{D}_{X_2}^{(- 0)}\cdot \scr{D}_{X_3}^{(0+)})
 W^{(V)(i,m)}_{\De, \ell, \De', \ell'; \De_1,\De_2+1, \De_3, \De_4, \De_5}\,,
\]
\[
W_{\De, \ell; \De^\prime, \ell^\prime; \De_1, \De_2, \De_3, \De_4, \De_5}^{(T)(2, n_{IJ})}=\,&
\frac{ \left(-n_{IJ}+\ell '-1\right)}{\Delta '-\Delta +\Delta _3-\ell '+\ell +2}W_{\De, \ell; \De^\prime, \ell^\prime; \De_1, \De_2, \De_3, \De_4, \De_5}^{(T)(3, n_{IJ})}
\nonumber\\
&+
\sum_{m=n_{IJ}}^{n_{IJ}+1} 
\mathscr{B}' {}_{(+0)(0+)}^{(1)(2, m)}
(\scr{D}_{X_1}^{(+0) }\cdot \scr{D}_{X_3}^{(0+)})
 W^{(V)(2,m)}_{\De, \ell, \De', \ell';\De_1-1,\De_2, \De_3, \De_4, \De_5} \nonumber\\
&+\mathscr{B}'{}_{ (+0)(0+)}^{(2)(2,m)} (\scr{D}_{X_2}^{(+ 0)}\cdot \scr{D}_{X_3}^{(0+)})
W^{(V)(2, m)}_{\De, \ell, \De', \ell'; \De_1,\De_2-1, \De_3, \De_4, \De_5}\nonumber\\
&+\mathscr{B}' {}_{ (-0)(0+)}^{(1)(2,m)}(\scr{D}_{X_1}^{(- 0) }\cdot \scr{D}_{X_3}^{(0+)})
W^{(V)(2, m)}_{\De, \ell, \De', \ell'; \De_1+1,\De_2, \De_3, \De_4, \De_5}
\nonumber\\
&+\mathscr{B}' {}_{ (-0)(0+)}^{(2)(2,m)}(\scr{D}_{X_2}^{(- 0)}\cdot \scr{D}_{X_3}^{(0+)})
 W^{(V)(2,m)}_{\De, \ell, \De', \ell'; \De_1,\De_2+1, \De_3, \De_4, \De_5}\,,
\]
\[
W_{\De, \ell; \De^\prime, \ell^\prime; \De_1, \De_2, \De_3, \De_4, \De_5}^{(T)(3, n_{IJ})}
=\,&
\frac{ (n_{IJ}-\ell ) \left(n_{IJ}-\ell '+2\right)}{(n_{IJ}+1) \left(\Delta '-\Delta
   +\Delta _3-2 n_{IJ}+\ell '+\ell \right)}W_{\De, \ell; \De^\prime, \ell^\prime; \De_1, \De_2, \De_3, \De_4, \De_5}^{(T)(3, n_{IJ}+1)}\nonumber\\
   &+\sum_{m=n_{IJ}}^{n_{IJ}+2} 
\mathscr{C}' {}_{(+0)(0+)}^{(1)(2, m)}
(\scr{D}_{X_1}^{(+0) }\cdot \scr{D}_{X_3}^{(0+)})
 W^{(V)(2,m)}_{\De, \ell, \De', \ell';\De_1-1,\De_2, \De_3, \De_4, \De_5} \nonumber\\
&+\mathscr{C}'{}_{ (+0)(0+)}^{(2)(2,m)} (\scr{D}_{X_2}^{(+ 0)}\cdot \scr{D}_{X_3}^{(0+)})
W^{(V)(2, m)}_{\De, \ell, \De', \ell'; \De_1,\De_2-1, \De_3, \De_4, \De_5}\nonumber\\
&+\mathscr{C}' {}_{ (-0)(0+)}^{(1)(2,m)}(\scr{D}_{X_1}^{(- 0) }\cdot \scr{D}_{X_3}^{(0+)})
W^{(V)(2, m)}_{\De, \ell, \De', \ell'; \De_1+1,\De_2, \De_3, \De_4, \De_5}
\nonumber\\
&+\mathscr{C}' {}_{ (-0)(0+)}^{(2)(2,m)}(\scr{D}_{X_2}^{(- 0)}\cdot \scr{D}_{X_3}^{(0+)})
 W^{(V)(2,m)}_{\De, \ell, \De', \ell'; \De_1,\De_2+1, \De_3, \De_4, \De_5}\,,
 \]
 \[
 W_{\De, \ell; \De^\prime, \ell^\prime; \De_1, \De_2, \De_3, \De_4, \De_5}^{(T)(4, n_{IJ})}
=\,&
\sum_{i=2}^3
\sum_{m=n_{IJ}}^{n_{IJ}+1} 
\mathscr{D}' {}_{(+0)(0+)}^{(1)(i, m)}
(\scr{D}_{X_1}^{(+0) }\cdot \scr{D}_{X_3}^{(0+)})
 W^{(V)(i,m)}_{\De, \ell, \De', \ell';\De_1-1,\De_2, \De_3, \De_4, \De_5} \nonumber\\
&+\mathscr{D}'{}_{ (+0)(0+)}^{(2)(i, m)} (\scr{D}_{X_2}^{(+ 0)}\cdot \scr{D}_{X_3}^{(0+)})
W^{(V)(i, m)}_{\De, \ell, \De', \ell'; \De_1,\De_2-1, \De_3, \De_4, \De_5}\nonumber\\
&+\mathscr{D}' {}_{ (-0)(0+)}^{(1)(i, m)}(\scr{D}_{X_1}^{(- 0) }\cdot \scr{D}_{X_3}^{(0+)})
W^{(V)(i, m)}_{\De, \ell, \De', \ell'; \De_1+1,\De_2, \De_3, \De_4, \De_5}
\nonumber\\
&+\mathscr{D}' {}_{ (-0)(0+)}^{(2)(i, m)}(\scr{D}_{X_2}^{(- 0)}\cdot \scr{D}_{X_3}^{(0+)})
 W^{(V)(i,m)}_{\De, \ell, \De', \ell'; \De_1,\De_2+1, \De_3, \De_4, \De_5}\,,
\]
\[
 W_{\De, \ell; \De^\prime, \ell^\prime; \De_1, \De_2, \De_3, \De_4, \De_5}^{(T)(5, n_{IJ})}=\,&
\sum_{m=n_{IJ}}^{n_{IJ}+1} 
\mathscr{E}' {}_{(+0)(0+)}^{(1)(2, m)}
(\scr{D}_{X_1}^{(+0) }\cdot \scr{D}_{X_3}^{(0+)})
 W^{(V)(2,m)}_{\De, \ell, \De', \ell';\De_1-1,\De_2, \De_3, \De_4, \De_5} \nonumber\\
&+\mathscr{E}'{}_{ (+0)(0+)}^{(2)(2,m)} (\scr{D}_{X_2}^{(+ 0)}\cdot \scr{D}_{X_3}^{(0+)})
W^{(V)(2, m)}_{\De, \ell, \De', \ell'; \De_1,\De_2-1, \De_3, \De_4, \De_5}\nonumber\\
&+\mathscr{E}' {}_{ (-0)(0+)}^{(1)(2,m)}(\scr{D}_{X_1}^{(- 0) }\cdot \scr{D}_{X_3}^{(0+)})
W^{(V)(2, m)}_{\De, \ell, \De', \ell'; \De_1+1,\De_2, \De_3, \De_4, \De_5}
\nonumber\\
&+\mathscr{E}' {}_{ (-0)(0+)}^{(2)(2,m)}(\scr{D}_{X_2}^{(- 0)}\cdot \scr{D}_{X_3}^{(0+)})
 W^{(V)(2,m)}_{\De, \ell, \De', \ell'; \De_1,\De_2+1, \De_3, \De_4, \De_5}\,,
\]
\[
W_{\De, \ell; \De^\prime, \ell^\prime; \De_1, \De_2, \De_3, \De_4, \De_5}^{(T)(6, n_{IJ})}=\,&\sum_{m=n_{IJ}}^{n_{IJ}+1} 
\mathscr{F}' {}_{(+0)(0+)}^{(1)(3, m)}
(\scr{D}_{X_1}^{(+0) }\cdot \scr{D}_{X_3}^{(0+)})
 W^{(V)(3,m)}_{\De, \ell, \De', \ell';\De_1-1,\De_2, \De_3, \De_4, \De_5} \nonumber\\
&+\mathscr{F}'{}_{ (+0)(0+)}^{(2)(3,m)} (\scr{D}_{X_2}^{(+ 0)}\cdot \scr{D}_{X_3}^{(0+)})
W^{(V)(3, m)}_{\De, \ell, \De', \ell'; \De_1,\De_2-1, \De_3, \De_4, \De_5}\nonumber\\
&+\mathscr{F}' {}_{ (-0)(0+)}^{(1)(3,m)}(\scr{D}_{X_1}^{(- 0) }\cdot \scr{D}_{X_3}^{(0+)})
W^{(V)(3, m)}_{\De, \ell, \De', \ell'; \De_1+1,\De_2, \De_3, \De_4, \De_5}
\nonumber\\
&+\mathscr{F}' {}_{ (-0)(0+)}^{(2)(3,m)}(\scr{D}_{X_2}^{(- 0)}\cdot \scr{D}_{X_3}^{(0+)})
 W^{(V)(3,m)}_{\De, \ell, \De', \ell'; \De_1,\De_2+1, \De_3, \De_4, \De_5}\,.\eql{block tensor 6}
\]
The coefficients are given explicitly in Appendix~\ref{app:F}. 

In the next section, we consider applying the above results for the 5-point conformal blocks for the purpose of extracting a set of novel sum rules for the CFT data.

\section{The averaged null energy condition: an application}\label{sec:anec}

All QFTs are now known to respect a special positivity condition known as the averaged null energy condition (ANEC). This condition states that the energy flux operator, defined as
\[
\eql{flux defn}
\scr{E} = \int_{-\infty}^{\infty} dx^- \, \,T_{--}(x^-, 0)\,,
\]
where the integral is over a complete null line, has a nonnegative expectation value in any state, $\vev{\Psi|\scr{E}|\Psi}\geq 0$.

This positivity condition was originally explored at length in \cite{Hofman:2008ar} for the purpose of deriving universal bounds on 3-point functions. Since then, the ANEC has been rigorously established by means of two distinct methods, namely through arguments based on causality in \cite{Hartman:2016lgu} and through monotonicity of relative entropy in \cite{Faulkner:2016mzt}.

In the context of CFTs, this statement enforces nontrivial bounds on OPE coefficients. For 4d CFTs, demanding that the ANEC hold for the energy flux operator $\scr{E}$ in wavepacket states of the form 
\[
\ket{\Psi} = \myint d^4 x \,\, e^{-i q t} e^{-(t^2+\vec{x}^2))/R^2} \epsilon_{\mu \nu} T^{\mu\nu}(x)\ket{0}\,, \qquad R q \gg 1\,,
\]
where $q>0$ is the state's energy, $\epsilon_{\mu\nu}$ is a polarization tensor, and $R\gg q^{-1}$ is a long-distance cutoff,
leads to the famous ``conformal collider" constraints originally proposed by Hofman and Maldacena \cite{Hofman:2008ar}. 

The energy flux may be determined by considering the 3-point function of the energy-momentum tensor $\vev{TTT}$. Since $\vev{TTT}$ contains three linearly independent tensor structures by conformal invariance, it may be parameterized in terms of the 3-point structures which appear in a free field theory:
\[
\eql{free field}
\vev{TTT} = n_b \vev{TTT}_b+n_f \vev{TTT}_f+n_v \vev{TTT}_v\,,
\]
where $\vev{TTT}_b$, $\vev{TTT}_f$, and $\vev{TTT}_v$ correspond to 3-point structures for the free scalar, free Dirac fermion, and free vector theories. Here the OPE coefficients $n_b$, $n_f$, and $n_v$ are the numbers of bosons, fermions, and vectors in a free CFT but are arbitrary parameters in an interacting theory. The collider bounds may be simply and suggestively stated as
\[
n_b, n_f, n_v \geq 0\,.
\] 
These bounds are respected by any unitary parity-preserving 4d CFT. Effectively, they are in place to ensure that the energy flux measured by an idealized calorimeter cell at infinity is nonnegative. Such bounds were generalized to other dimensions in~\cite{Buchel:2009sk,Chowdhury:2012km,Chowdhury:2017vel}. An independent proof of the bounds was given in~\cite{Hofman:2016awc,Chowdhury:2018uyv}, where they were shown to follow from the conformal bootstrap. More recently, a higher-spin version of the ANEC and corresponding bounds were established in~\cite{Hartman:2016lgu, Meltzer:2018tnm} and a detailed discussion of more general light-ray operators was given in~\cite{Kravchuk:2018htv,Kologlu:2019mfz,Chang:2020qpj}.

One obtains these and similar constraints by studying 3-point conformal correlation functions, such as $\vev{\scr{O} J^{(s)} \scr{O}}$, where in general $J^{(s)}$ denotes the lightest even spin-$s$ operator appearing on the Regge trajectory of the stress tensor~\cite{Meltzer:2018tnm}.  Additional constraints on couplings $\vev{TT \scr{O}}$ were shown to arise from considering states corresponding to a superposition of the stress tensor $T$ and a scalar $\scr{O}$~\cite{Cordova:2017zej}.  
Here we wish to explore the question of whether one may derive novel constraints by studying ANEC positivity in higher-point correlation functions. The analysis of the 5-point function, which we sketch in this section, will be a first step in this direction. 

Both the ANEC and its higher spin generalizations may be conveniently couched in a Minkowski space setup \cite{Hartman:2016lgu} via the notion of Rindler positivity. In particular, one may state the ANEC by invoking the positivity properties of Rindler symmetric correlation functions in Minkowski space. 

Correlation functions in Minkowski space restricted to the left and right Rindler wedges satisfy a positivity property akin to reflection positivity in Euclidean space referred to as Rindler positivity. The Rindler reflection for scalar operators is defined by
\[
\bar{x} =\overline{(t, y, \vec{x})} = (-t^\ast, -y^\ast, \vec{x})\,, \quad \bar{\scr{O}}(x) =\scr{O}^\dagger(\bar{x})\,,
\]
where the transverse coordinate $\vec{x}$ is taken to be real but we allow complex $(t,y)$. For real $(t, y)$, its action is to map operators from one Rindler wedge to the other. When acting on a spinning operator $\scr{O}_{\mu\nu\dots}$, the Rindler reflection maps
\[
\scr{O}_{\mu\nu\dots}(t, y, \vec{x})\mapsto \overline{\scr{O}_{\mu\nu\dots}(t, y, \vec{x})} =(-1)^P \scr{O}^\dagger_{\mu\nu\dots}
(-t^\ast, -y^\ast, \vec{x})\,,
\]
where $P$ denotes the sum of $t$- and $y$-indices.

Rindler positivity applied to 4-point functions is then the statement
\[
\vev{\bar{\scr{O}}_i\bar{\scr{O}}_j
\scr{O}_i\scr{O}_j}\geq 0\,,
\]
where the unbarred operators $\scr{O}_i$ and $\scr{O}_j$ are inserted in the right Rindler wedge, and Rindler reflection leaves the order of the operators unchanged.

Here we consider the expectation value $\vev{\Psi|\scr{E}|\Psi}$ for states created by pairs of real (Hermitian) scalar operators. We are then interested in studying the implications of Rindler positivity of the 5-point function
\[
\vev{\phi_i(\bar{x}_1) \phi_j(\bar{x}_2) \scr{E} \phi_i(x_1) \phi_j(x_2)} \geq 0\,,
\]
where $x_1$ and $x_2$ are taken to be in the right Rindler wedge.

Now, while the energy flux $\scr{E}$ is traditionally defined as in \Eq{flux defn}, it is convenient to alternatively define it via a covariant form as in \cite{Zhiboedov:2013opa, Meltzer:2018tnm}:
\[
\eql{flux defn covariant}
\scr{E}(n)=\int_{-\infty}^{\infty} d(x\cdot n) \lim_{x\cdot \bar{n}\to \infty} (x\cdot \bar{n})^{d-2} T_{\mu \nu}(x\cdot n, x\cdot \bar{n}, \vec{0}) \bar{n}^\mu \bar{n}^\nu, \qquad n= (1,\vec{n}), \,\, \bar{n}=(1,-\vec{n})\,.
\]
The advantage of this form is that it is manifestly Lorentz invariant. In CFTs, the two definitions are equivalent, but the form \Eq{flux defn covariant} turns out to be simpler to handle in practice.

With this definition, we may re-express the positivity condition $\vev{\Psi|\scr{E}|\Psi}  \geq 0$ as:
\[
\eql{matrix element}
0 \leq \,& \int_{-\infty}^{\infty} d(x_3\cdot n) \lim_{x_3\cdot \bar{n}\to \infty} (x_3\cdot \bar{n})^{d-2}
\\
& \times\vev{\phi_i(\bar{x}_1)\phi_j(\bar{x}_2)T_{\mu \nu}(x_3\cdot n, x_3\cdot \bar{n}, \vec{0}) \bar{n}^\mu \bar{n}^\nu\phi_i(x_1)\phi_j(x_2)}\,, \qquad n= (1,\vec{n}), \,\, \bar{n}=(1,-\vec{n})\,.\nonumber
\]
Now we have expressed the condition directly in terms of the 5-point functions studied in the previous sections of this paper, which allow us to compute the integrand in terms of the coordinate differences $x_{ij}$.

To calculate the relevant integrals of the type \Eq{matrix element} in practice, we may follow the approach of \cite{Komargodski:2016gci}. Since we need to take the limit $x_3\cdot \bar{n}\to \infty$, we can first use that $x_3= \frac{x_3\cdot n}{2}\bar{n} +\frac{x_3\cdot \bar{n} }{2}n$, so we have
\[
\eql{results limit}
\lim_{x_3\cdot \bar{n}\to \infty} x_{13}^2 &=
- (x_3\cdot \bar{n})(x_{13}\cdot n)\,, \quad
\lim_{x_3\cdot \bar{n}\to \infty} x_{23}^2 =
- (x_3\cdot \bar{n})(x_{23}\cdot n)\,, \nonumber\\
\lim_{x_3\cdot \bar{n}\to \infty} x_{34}^2 &=
 (x_3\cdot \bar{n})(x_{34}\cdot n)\,, \quad
\lim_{x_3\cdot \bar{n}\to \infty} x_{35}^2 =
 (x_3\cdot \bar{n})(x_{35}\cdot n)\,.
\]
To perform the light-ray integrals after taking the limit, we can then make use of the formula:
\[
\int_{-\infty}^{\infty} d(x_3\cdot n) \dfrac{1}{(x_{23}\cdot n)^a (x_{34}\cdot n)^b}=\dfrac{2\pi i}{(x_{24}\cdot n)^{a+b-1}}\dfrac{\Gamma(a+b-1)}{\Gamma(a)\Gamma(b)}\,.
\]  

Having set the stage, let us take a closer look at the positivity condition for the special case of identical operators $\phi_i = \phi_j = \phi$. The $\phi \times \phi$ OPE is given by  
\[
\phi\times\phi =\,& \mathds{1}+  \sum_{\Delta, \ell\, \text{even}} \la_{\phi \phi \scr{O}_{\De, \ell}}\scr{O}_{\De, \ell}\,,
\]
where we have separated off the identity contribution and restricted the sum to only even-spin symmetric traceless operators.  In fact, identity exchange will give no contribution to the 5-point function $\vev{\phi\phi \scr{E} \phi\phi}$ because, as explained in~\cite{Kologlu:2019mfz}, light-ray operators such as $\scr{E}$ annihilate the vacuum. 

Thus, we generally expect that the OPE limit $x_{12}, x_{45} \rightarrow 0$ will be dominated by the stress tensor $T$ or low-dimension scalars. The contribution due to stress energy tensor exchange in both channels takes the form
\[
\vev{\phi \phi T \phi \phi} \supset\,&  \vev{\phi(X_1)\phi(X_2)T}\bowtie   
\vev{TTT} \bowtie
  \vev{T\phi(X_4)\phi(X_5)} \nonumber\\
  =\,& |\la_{\phi  \phi  T}|^2 \sum_{a=1}^3 \lambda_{TTT}^{(a)} W_{d, 2; d,  2; \De_\phi,\De_\phi, d, \De_\phi, \De_\phi}^{(T) a }\,,
\]
where we encounter the square of the OPE coefficient $|\la_{\phi  \phi  T}|^2$ and the OPE coefficients featured in $\vev{TTT}$. Upon invoking conservation of $T$ and symmetry under the interchange of the three $T$s, there are a total of three independent structures corresponding to the exchange of identical stress energy tensors $T$. In $d\geq 4$ spacetime dimensions, the 3-point function of stress energy tensors $\vev{TTT}$ may be parameterized in terms of three independent structures, namely the free boson, free fermion, and free $(d-2)/2$ form 3-point structures as in \Eq{free field}. We may therefore equivalently express the coefficients $\lambda_{TTT}^{(1)}$, $\lambda_{TTT}^{(2)}$, and $\lambda_{TTT}^{(3)}$ in terms of $n_b$, $n_f$, and $n_v$. 

Given this, we end up with an inequality
\[ \label{eq:ineqTT}
 0 \leq\,& |\la_{\phi  \phi  T}|^2  \int_{-\infty}^{\infty} d(x_3\cdot n) \lim_{x_3\cdot \bar{n}\to \infty} (x_3\cdot \bar{n})^{d-2}
 \sum_{a=1}^3 \lambda_{TTT}^{(a)} W_{d, 2; d,  2; \De_\phi,\De_\phi, d, \De_\phi, \De_\phi}^{(T) a } \bigg|_{X_3 = X_3(x_3 \cdot n, x_3 \cdot \bar{n}, \vec{0}), Z_3 = Z_3(\bar{n})}  \nonumber\\
&+ \sum_{\scr{O}_{\De, \ell}}
\sum_{ \scr{O}^\prime_{\De', \ell'}} \la_{\phi \phi \scr{O}_{\De, \ell}} 
\la_{\scr{O}^\prime_{\De', \ell'}\phi \phi }  \int_{-\infty}^{\infty} d(x_3\cdot n) \lim_{x_3\cdot \bar{n}\to \infty} (x_3\cdot \bar{n})^{d-2}
 \nonumber\\
&\times \bigg[
\sum_{n_{IJ}=0}^{\min[\ell, \ell^\prime ]}\la^{(1)(n_{IJ})}_{\scr{O}_{\De, \ell} T \scr{O}^\prime_{\De', \ell'}}W_{\De, \ell; \De^\prime, \ell^\prime; \De_\phi \De_\phi, d, \De_\phi, \De_\phi}^{(T)(1, n_{IJ})}%
+\sum_{n_{IJ}=0}^{\min[\ell, \ell^\prime -1]}\la^{(2) (n_{IJ})}_{\scr{O}_{\De, \ell} T \scr{O}^\prime_{\De', \ell'}}W_{\De, \ell; \De^\prime, \ell^\prime; \De_\phi \De_\phi, d, \De_\phi, \De_\phi}^{(T)(2, n_{IJ})}\nonumber\\
&+\sum_{n_{IJ}=0}^{\min[\ell, \ell^\prime -2]} \la^{(3) (n_{IJ})}_{\scr{O}_{\De, \ell} T \scr{O}^\prime_{\De', \ell'}}W_{\De, \ell; \De^\prime, \ell^\prime; \De_\phi \De_\phi, d, \De_\phi, \De_\phi}^{(T)(3, n_{IJ})}
+\sum_{n_{IJ}=0}^{\min[\ell -1, \ell^\prime]} \la^{(4) (n_{IJ})}_{\scr{O}_{\De, \ell} T \scr{O}^\prime_{\De', \ell'}}W_{\De, \ell; \De^\prime, \ell^\prime; \De_\phi \De_\phi, d, \De_\phi, \De_\phi}^{(T)(4, n_{IJ})}\nonumber\\
&+\sum_{n_{IJ}=0}^{\min[\ell -1, \ell^\prime -1]} \la^{(5)(n_{IJ})}_{\scr{O}_{\De, \ell} T \scr{O}^\prime_{\De', \ell'}}W_{\De, \ell; \De^\prime, \ell^\prime; \De_\phi \De_\phi, d, \De_\phi, \De_\phi}^{(T)(5, n_{IJ})}\nonumber\\
&+\sum_{n_{IJ}=0}^{\min[\ell -2, \ell^\prime]}  \la^{(6)(n_{IJ})}_{\scr{O}_{\De, \ell} T \scr{O}^\prime_{\De', \ell'}}W_{\De, \ell; \De^\prime, \ell^\prime; \De_\phi \De_\phi, d, \De_\phi, \De_\phi}^{(T)(6, n_{IJ})}
\bigg]\bigg|_{X_3 = X_3(x_3 \cdot n, x_3 \cdot \bar{n}, \vec{0}), Z_3 = Z_3(\bar{n})}\,. 
\]
Here the conservation of $T$ imposes constraints on the coefficients $\la^{(i)(n_{IJ})}_{\scr{O}_{\De, \ell} T \scr{O}^\prime_{\De', \ell'}}$ so that they are not all independent. This statement relates the contributions from $(T, T)$ exchange to the rest of the expansion. In a limit where the first term dominates, we expect this to reproduce the Hofman-Maldacena bounds, while including the other terms in the OPE gives a more complicated sum rule. 

It is likely that the sum rules obtained in this way are not as strong as those obtained using the logic of~\cite{Cordova:2017zej,Meltzer:2018tnm}, which considers states built from arbitrary linear combinations of a certain set of operators. Here we have considered states $\phi(x_1) \phi(x_2)\ket{0}$, which correspond to a particular family of linear combinations of operators parametrized by $(x_{1}, x_{2})$, motivated by the OPE. On the other hand, they yield a very natural way to incorporate the sum over higher-spin operators into the inequalities. It would be interesting to better understand how the bounds~(\ref{eq:ineqTT}) compare with those obtained using the logic of~\cite{Cordova:2017zej,Meltzer:2018tnm}, and which regions of position space lead to the strongest constraints. We defer exploration of these questions to future work.

Beyond an examination of states of the type $\phi(x_1) \phi(x_2) \ket{0}$, one can also consider smeared states:
\[
\eql{simplest state}
\ket{\Phi}\equiv\ket{\phi_i \phi_j}_f = \myint d^d x_1 \myint d^d x_2 \,\, f(x_1, x_2) \phi_i(x_1)\phi_j(x_2)\ket{0}\,.
\]
Here we would want to choose $f$ to have support in some localized region of the right Rindler wedge such that convergence of the $\phi_i \times \phi_j$ OPE is preserved, along with other convenient properties. E.g., one might wish to choose $f(x_1,x_2) \propto e^{-i q (t_1+ t_2)}$ to correspond to approximate energy eigenstates.
In the discussion below, we will keep $f$ general, but one could always reproduce the position-localized states discussed above by choosing $f$ to correspond to a product of $\delta$ functions.

One can also analyze more general mixed states created by linear combinations of operators acting on the vacuum. Here we will give an initial discussion of the kinds of bounds that could arise from considering mixed states which combine the bilocal states of \Eq{simplest state} with one or more local operators.

For example, we may consider mixing with a state created by the stress tensor 
\[
\ket{T(q, \epsilon)}= \mathscr{N}\myint d^d x\,\, e^{-i q t}\epsilon_{\mu\nu} T^{\mu\nu}(x)\ket{0}
\]
and superpose it with a state created by a pair of Hermitian scalar operators $(\phi_i, \phi_j)$. We remark that the state $\ket{T(q, \epsilon)}$ is a momentum eigenstate labeled by a particular choice of polarization tensor $\epsilon_{\mu\nu}$ and the energy $q$. Different choices of $\epsilon_{\mu\nu}$ produce different forms of the resulting bound. 

We parameterize such a mixed state in terms of normalized coefficients $\al_i$ as 
\[
\al_1\ket{\phi_i\phi_j}_{f}+ \al_2\ket{T(q, \epsilon)}\,.
\]
Upon evaluating the energy one-point function in this state, we find a $2\times 2$ matrix. Imposing the ANEC for all $\alpha_i$ then translates to the requirement that this matrix must be positive definite:
\[
\eql{matrix 1}
\left(\begin{matrix}
{}_f\bra{\phi_i\phi_j}\scr{E}\ket{\phi_i\phi_j}_f
 &  {}_f\bra{\phi_i\phi_j}\scr{E}\ket{T(q, \epsilon)}\\
 \bra{T(q, \epsilon)}\scr{E}\ket{\phi_i\phi_j}_f
& \bra{T(q, \epsilon)}\scr{E}\ket{T(q, \epsilon)} 
\end{matrix}\right)\succcurlyeq 0\,.
\]
We expect such a requirement to be a stronger condition than merely demanding that the diagonal entries be nonnegative; hence, it will introduce new restrictions on the OPE coefficients. We note that the above matrix elements all involve known objects. In particular, here we encounter the 3-point function $\vev{TTT}$, the 4-point functions $\vev{\phi\phi TT}$ and $\vev{TT\phi\phi}$, and the 5-point function $\vev{\phi\phi T \phi\phi}$. 

By making different choices for the polarization tensor $\epsilon$, one can additionally isolate each of the coefficients $n_b, n_f, n_v$ appearing in $\vev{TTT}$. If we label the polarizations using their spin under the $SO(d-2)$ symmetry group which preserves the direction of future null infinity $\vec{n}$, they are picked out using the spin zero, one, and two polarizations ($\epsilon^{(0)}$, $\epsilon^{(1)}$, $\epsilon^{(2)}$), respectively. E.g., in~\cite{Cordova:2017zej} the authors considered mixed states involving scalars $\scr{O}$ and the stress tensor $T$, and used the off-diagonal terms to derive a sum rule constraining the OPE coefficients $\lambda_{TT\scr{O}}$. In this case, there were nontrivial interference terms for the polarization $\epsilon^{(0)}$, which picks out the coefficient $n_b$, so the bound is expressed in terms of this quantity.

Now let us turn to the bounds generated by the $2\times 2$ matrix in \Eq{matrix 1}. 
In this case, we may enforce the requirement that the matrix be positive definite by demanding
\[\label{eq:simplest bound}
{}_f\bra{\phi_i\phi_j}\scr{E}\ket{\phi_i\phi_j}_f \geq 0
\] 
along with the determinant condition
\[
{}_f\bra{\phi_i\phi_j}\scr{E}\ket{\phi_i\phi_j}_f \bra{T(q, \epsilon)}\scr{E}\ket{T(q, \epsilon)} -| {}_f\bra{\phi_i\phi_j}\scr{E}\ket{T(q, \epsilon)}|^2 \geq 0\,.
\]
We therefore see that enforcing the ANEC in this special state not only implies \Eq{simplest bound} but also the nonnegativity of the determinant. This additional constraint allows us to improve on the bound coming from \Eq{simplest bound}. 

We may understand this more explicitly as follows. If we consider taking the polarization tensor to be the spin zero tensor $\epsilon_{\mu\nu}^{(0)}$, then we find that the ``determinant constraint" is 
\[
\frac{n_b}{c_T} \rho_0(d){}_f\bra{\phi_i\phi_j}\scr{E}\ket{\phi_i\phi_j}_f \geq  |{}_f\bra{\phi_i\phi_j}\scr{E}\ket{T(q, \epsilon^{(0)})}|^2\,,
\]
where we have used that $\bra{T(q, \epsilon^{(0)})}\scr{E}\ket{T(q, \epsilon^{(0)})} = \rho_0(d) \frac{n_b}{c_T}$, where $\rho(d)$ is a positive function of $d$, while $n_b\geq 0$ by the collider constraints and $c_T \geq 0$ by unitarity. For the case of nonzero $n_b>0$, it is evident that this type of constraint improves upon the bound in \Eq{simplest bound}. Then, not only is the matrix element ${}_f\bra{\phi_i\phi_j}\scr{E}\ket{\phi_i\phi_j}_f$ nonnegative, but it is also bounded from below by a nonzero quantity related to the 4-point function $\vev{\phi_i \phi_j T T}$. 
Unlike the situation in~\cite{Cordova:2017zej}, here we may also choose the polarization tensors $\epsilon^{(1)}$ or $\epsilon^{(2)}$, such that the constraints involve either $n_f$ or $n_v$, and we still in general expect a nontrivial constraint.

We could also consider studying states of the form 
\[
\be_1\ket{\phi_i\phi_j}_f+\be_2\ket{\chi(q)}\,,
\]
where we mix with a state created by a scalar $\chi$, with
\[
\ket{\chi(q)}= \mathscr{N}\myint d^d x\,\, e^{-i q t} \chi(x)\ket{0}\,.
\]

Once again, upon computing the expectation value of $\scr{E}$, we expect to find a positive definite $2\times 2$ matrix of the form 
\[
\left(\begin{matrix}
{}_f\bra{\phi_i \phi_j}\scr{E}\ket{\phi_i \phi_j}_f
 &   {}_f\bra{\phi_i \phi_j}\scr{E}\ket{\chi(q)}
 \\
\bra{\chi(q)}\scr{E}\ket{\phi_i \phi_j}_f 
&\bra{\chi(q)}\scr{E}\ket{\chi(q)} 
\end{matrix}\right)\succcurlyeq 0\,,
\]
which again involves known 5-point, 4-point, and 3-point functions. In addition, it is straightforward to write down the form of the entry $\bra{\chi(q)}\scr{E}\ket{\chi(q)}$. This is the expectation value in the scalar state $\ket{\chi(q)}$, which gives rise to a uniform energy distribution:
\[
\bra{\chi(q)}\scr{E}\ket{\chi(q)}=\dfrac{q}{\Omega_{d-2}}\,.
\]
Thus, the determinant condition on this matrix will also strengthen the bound~(\ref{eq:simplest bound}) in a nontrivial way, as determined by the matrix elements ${}_f\bra{\phi_i \phi_j}\scr{E}\ket{\chi(q)}$ and $\bra{\chi(q)}\scr{E}\ket{\phi_i \phi_j}_f$.

One may continue on in this vein by considering even more general linear combinations of scalar and stress tensor states. For example, we could try combining the above two cases:
\[
\ga_1\ket{\phi_i \phi_j}_f+\ga_2\ket{T(q)}+\ga_3\ket{\chi(q)}\,.
\]
Such a state would accordingly lead to an even larger set of constraints, namely a $3\times 3$ matrix
\[
\left(\begin{matrix}
{}_f\bra{\phi_i\phi_j}\scr{E}\ket{\phi_i \phi_j}{}_f
 &  {}_f \bra{\phi_i\phi_j}\scr{E}\ket{T(q, \epsilon)} & {}_f\bra{\phi_i\phi_j}\scr{E}\ket{\chi(q)}\\
  \bra{T(q, \epsilon)}\scr{E}\ket{\phi_i\phi_j}_f& \bra{T(q, \epsilon)}\scr{E}\ket{T(q, \epsilon)} & \bra{T(q, \epsilon)}\scr{E}\ket{\chi(q)} 
\\
 \bra{\chi(q)}\scr{E}\ket{\phi_i\phi_j}{}_f&  \bra{\chi(q)}\scr{E}\ket{T(q, \epsilon)} &\bra{\chi(q)}\scr{E}\ket{\chi(q)} \end{matrix}\right)\succcurlyeq 0\,.
\]
Demanding the positivity of the one-point function of $\scr{E}$ in this state can then generate even stronger constraints on the 5-point function.

In this fashion, we may use the properties of the chosen state(s) to extract various bounds relating the 5-point function to lower-point functions, and in turn obtain constraints on the OPE coefficients and scaling dimensions appearing in their conformal block expansions. It may be interesting to perform a more in-depth analysis of the expectation values of $\scr{E}$ in the above states and numerically study the corresponding constraints. E.g., it could be enlightening to check how the corresponding bounds involving the 5-point $\vev{\sigma\sigma T \sigma \sigma}$ function look in the 3d Ising model. We leave such explorations to future work.

\section{Conclusion}
\label{sec:conclusion}

In this paper we have presented a concrete and practical approach to computing conformal blocks appearing in 5-point functions of arbitrary scalar operators in general CFTs. In particular, by making use of weight-shifting operators, we have derived a simple set of recursion relations which can be used to directly relate arbitrary scalar blocks with spinning operators exchanged to those with scalars exchanged.

We additionally considered promoting one of the external operators to have spin 1 or 2, deriving additional recursion relations which relate their conformal blocks to those for external scalars. One possible application of these results is that they allow us to use the OPE to compute the expectation value of the ANEC operator in bilocal states, which must be positive. In this work we gave an initial discussion of the resulting constraints, deferring further exploration to future work.

In this paper we focused on computing the blocks that would appear in parity-even 5-point functions in parity-preserving CFTs. Additional dimension-specific computations would be needed to compute the blocks appearing in parity-odd 5-point functions or parity-violating theories, e.g.~$d=3$ Chern-Simons theories, which could be pursued in future work.

An alternate approach to computing conformal blocks is to take advantage of the pole structure in the exchanged operator dimensions, which leads to Zamolodchikov-like recursion relations. In the future it might be interesting to develop this approach for higher-point functions such as the 5-point functions considered in this work. 

The 5-point functions satisfy various crossing-symmetry constraints which we have not explored in this paper. For example, $\vev{\sigma(x_1) \sigma(x_2) \epsilon(x_3) \sigma(x_4) \sigma(x_5)}$ could be expanded in the $(12)(45)$ OPE, the $(14)(25)$ OPE, the $(13)(45)$ OPE, and so on. It could be interesting to study how these crossing constraints are satisfied in concrete CFTs such as the 3d Ising model, and whether they can be used to learn about couplings like $\lambda_{\scr{O}_{\De,\ell} \epsilon \scr{O}'_{\De', \ell'}}$ that can't be easily accessed using 4-point functions of scalar operators. It could also be interesting to study the 5-point function and associated crossing constraints in various limits such as the lightcone or Regge limits. The blocks computed in this paper should allow such analyses to be pursued.

\acknowledgments
We thank Soner Albayrak, Jean-Fran\c{c}ois Fortin, Vasco Gon\c{c}alves, Petr Kravchuk, David Meltzer, Eric Perlmutter, Vladimir Rosenhaus, David Simmons-Duffin, and Witold Skiba for discussions. We are especially grateful to Petar Tadi\'c for pointing out a number of mistakes in the earlier version of the paper and giving many useful comments. DP was supported by Simons Foundation grant 488651 (Simons Collaboration on the Nonperturbative Bootstrap) and DOE grants DE-SC0020318 and DE-SC0017660. VP was supported in part by NSERC.

\appendix
\section{Standard box tensor basis}
\label{app:A}

Here we will review the basis of 3-point tensor structures used in the body of the paper. As originally described in \cite{Costa:2011mg}, the most general form of the embedding space 3-point function of symmetric traceless tensor operators with spins $\ell_i$ and dimensions $\De_i$ can be expressed as follows:
\[
\eql{general STT 3-point}
\vev{\Phi_1(X_1; Z_1)\Phi_2(X_2; Z_2)\Phi_3(X_3; Z_3)}
=\sum_{n_{12}, n_{13}, n_{23} \geq 0} \la_{n_{12}, n_{13}, n_{23}} 
\left[
\begin{matrix} 
\De_1 & \De_2 &  \De_3\\
\ell_1 & \ell_2 & \ell_3\\
n_{23} & n_{13} & n_{12}
\end{matrix}\right] + O(Z_i^2, Z_i\cdot X_i)\,,
\]
where the sum runs over all possible elementary 3-point function tensor structures. In this expression, we can choose to work in the so-called standard box tensor basis, where the individual 3-point structures are given by 
\[
\eql{3-point structures}
\left[
\begin{matrix} 
\De_1 & \De_2 &  \De_3\\
\ell_1 & \ell_2 & \ell_3\\
n_{23} & n_{13} & n_{12}
\end{matrix}\right] \equiv \dfrac{V_1^{m_1} V_2^{m_2}V_3^{m_3} H_{12}^{n_{12}}
H_{13}^{n_{13}} H_{23}^{n_{23}}}{X_{12}^{\sfrac{1}{2}(\tau_1+\tau_2-\tau_3)}
X_{13}^{\sfrac{1}{2}(\tau_1+\tau_3-\tau_2)}X_{23}^{\sfrac{1}{2}(\tau_2+\tau_3-\tau_1)}
}\,,
\]
where $\tau_i \equiv \De_i+\ell_i$. 

Here the basic constituent building blocks $H_{ij}$ and $V_{i,jk}$ are defined as
\[
H_{ij} &\equiv   -2[(Z_i\cdot Z_j)(X_i\cdot X_j) 
-(Z_i\cdot X_j)(Z_j\cdot X_i)]\,, \nonumber\\
V_{i,jk}&\equiv 
\dfrac{(Z_i \cdot X_j)(X_i\cdot X_k)- (Z_i\cdot X_k)(X_i\cdot X_j)}
{(X_j\cdot X_k)}\,,
\] 
which are transverse objects. Due to the relations $V_{i, jk} = - V_{i, kj}$ and $H_{ij} = H_{ji}$, 
not all of the $V_{i, jk}$ and $H_{ij}$ are linearly independent. We therefore expect to have three distinct constituent objects, which we can choose to be
\[
V_1 &\equiv V_{1,23}\,, \quad V_2 \equiv V_{2,31}\,, \quad
V_3 \equiv V_{3, 12}\,.
\]   

The exponents $m_i$ and $n_{ij}$ are nonnegative integers which respect the following three constraints:
\[
m_i + \sum_{j\neq i} n_{ij} = \ell_i\,.
\]
The content of \Eq{3-point structures} is that for general spins $\ell_i$, there are generically several inequivalent 3-point structures consistent with conformal invariance. The total number of such elementary 3-point structures precisely corresponds to the number of nonnegative integer points $(n_{12}, n_{13}, n_{23})$ which comprise a three-dimensional polyhedron defined by the constraints
\[
n_{12} + n_{13} \leq \ell_1\,, \quad n_{12} + n_{23} \leq \ell_2\,, \quad n_{13} + n_{23} \leq \ell_3\,.
\]

This number may be determined in closed form to be \cite{Costa:2011mg,Kravchuk:2016qvl}
\[
N(\ell_1,\ell_2,\ell_3)=\dfrac{(\ell_1+1)(\ell_1+2)(3\ell_2-\ell_1+3)}{6}-\dfrac{p(p+2)(2p+5)}{24}
-\dfrac{1-(-1)^p}{16},
\]
where the spins have been ordered such that $\ell_1\leq \ell_2\leq \ell_3$ and $p\equiv \text{max}(0,\ell_1+\ell_2-\ell_3)$. 

Thus, these independent elementary 3-point structures are completely and uniquely specified by a particular choice of three nonnegative integers $n_{ij}$ subject to
\[
m_1 \equiv \ell_1 -n_{12}-n_{13} \geq 0\,, \quad 
m_2 \equiv \ell_2 - n_{12} -n_{23} \geq 0\,, \quad
m_3 \equiv \ell_3 - n_{13} -n_{23} \geq 0\,.
\]   

In the present case, we have a 3-point function of the type (scalar)-(scalar)-(spin-$\ell$), which just contains the unique tensor structure
\[
\left[
\begin{matrix} 
\De_1 & \De_2 &  \De_3\\
0 & 0 & \ell\\
0 & 0 & 0
\end{matrix}\right] = \dfrac{  V_3^\ell
}{X_{12}^{\sfrac{1}{2}(\De_1+\De_2-\De_3-\ell)}
X_{13}^{\sfrac{1}{2}(\De_1+\De_3-\De_2+\ell)}X_{23}^{\sfrac{1}{2}(\De_2+\De_3-\De_1+\ell)}
}\,.
\]
Next, for 3-point functions of the type (scalar)-(spin-$\ell_2$)-(spin-$\ell_3$), we have several conformally invariant 3-point structures. Since $\ell_1= 0$ and $-n_{12}-n_{13} < 0$, the assumption of nonnegative $n_{ij}$ leads us to the restricted set of conditions
\[
n_{12} = n_{13} = 0\,, \quad
m_2 = \ell_2  -n_{23} \geq 0\,, \quad
m_3 = \ell_3  -n_{23} \geq 0\,.
\]
With this, we find that we can write the 3-point structures as follows:
\[
\left[
\begin{matrix} 
\De_1 & \De_2 &  \De_3\\
0 & \ell_2 & \ell_3\\
n_{23} & 0 & 0
\end{matrix}\right] = \dfrac{ V_2^{m_2}V_3^{m_3} H_{23}^{n_{23}}}{(X_{12})^{\sfrac{1}{2}(\De_1+\tau_2-\tau_3)}
(X_{13})^{\sfrac{1}{2}(\De_1+\tau_3-\tau_2)}(X_{23})^{\sfrac{1}{2}(\tau_2+\tau_3-\De_1)}
}\,.
\]
In this case, the number of inequivalent structures is given by
\[
N(0, \ell_2,  \ell_3)=(\ell_2+1)-\dfrac{p(p+2)(2p+5)}{24}
-\dfrac{1-(-1)^p}{16}\,,
\]
where the spins are ordered as in $0\leq  \ell_2 \leq \ell_3$ with
$p\equiv \text{max}(0, \ell_2-\ell_3)$. Thus, we find the simple result
\[
\ell_3 \geq \ell_2: \ggap N(0, \ell_2,  \ell_3) = (\ell_2+1)\,.  
\]

\section{Recursion relation coefficients} 
\label{app:B}
The coefficients appearing in \Eq{13 rel} are explicitly given by
\[
s_{n_{IJ}} &= -\frac{\left(\De -\De _{12}+\ell-2\right) \left(\De+\De ' -\De _3+\ell-\ell ' +2 n_{IJ}-2\right)}{2 (\De +\ell-2) (\De +\ell-1 )}\,,\\
s_{n_{IJ}+1}&=\frac{\left(\ell'-n_{IJ}\right)\left(\De-\De_{12}+\ell-2\right) }{2(\De +\ell-2) (\De +\ell-1 )}\,,\\
t_{n_{IJ}-1} &=\frac{n_{IJ}(d+2(n_{IJ}-2)) \left(\De _{12}+d-\De+\ell-2\right) \left(2(d- n_{IJ})-\De'-\De+\De _3+\ell '+3(\ell-2)\right)}{2(d+2(\ell-3))(d+2 (\ell-2)) (d-\De +\ell-3) (d-\De +\ell-2)}\,,\\
t_{n_{IJ}} &= -\frac{(\ell-1 -n_{IJ}) \left(d-\De +\De _{12}+\ell-2\right) }{2 (d+2(\ell-3)) (d+2( \ell-2)) (d-\De +\ell-3) (d-\De +\ell-2)} \nonumber\\
   &\big[d(d -2)+n_{IJ} \left(d-4 n_{IJ}+3(\ell+\ell'-1) -1\right) \nonumber\\
&-\left(\De+\De ' -\De_3\right) (d+n_{IJ}+\ell-4)
+(\ell-2) \left(\ell-\ell' -4+2 d \right)\big]\,,\\
t_{n_{IJ}+1}&=\frac{(\ell-2-n_{IJ} ) (\ell-1-n_{IJ}) \left(d-\De +\De _{12}+\ell-2\right) \left(\ell '-n_{IJ}\right)}{2 (d+2 (\ell-3)) (d+2( \ell-2)) (d-\De +\ell-3) (d-\De +\ell-2)}\,,\\
u_{n_{IJ}-1} &=
\frac{n_{IJ} (d+2(n_{IJ}-2)) \left(\De -\De _{12}+\ell-2\right) \left(d-\De +\De _{12}+\ell-2\right)}{4 (d-2 \De ) (d-2 (\De +1)) (\De +\ell-2) (\De +\ell-1 ) (d-\De +\ell-3) (d-\De +\ell-2)}\nonumber\\
& \times \left(\De-\De ' +\De _3-2 n_{IJ}+\ell  +\ell'  \right) \left(2
   d-\De '-3 \De +\De _3-2 n_{IJ}+\ell +\ell' -2\right) \,,
\]
\[
u_{n_{IJ}} &=-\frac{\left(\De -\De _{12}+\ell-2\right) \left(d-\De +\De _{12}+\ell-2\right)}{4 (d-2 \De ) (d-2 (\De +1)) (\De +
\ell-2) (\De +\ell-1 ) (d-\De +\ell-3) (d-\De+\ell-2)}\nonumber\\
&\times \bigg(\De  \left[d^2 (\De -2)-d \left(2 \De^2-\De -4\right)+\De 
    \left(\De ^2+\De -3\right)-1\right] \nonumber\\
& +(\ell-1)  \ell' \big[2 (d^2-4 n_{IJ}^2)-d \left(\De '-\De _3+5 (\De+1) \right) \nonumber\\
 &+2 n_{IJ} \left(d-2 \left(\De '+\De -\De _3\right)\right)+2  \left(\De '+\De  (2 \De +3)-\De _3+1\right)\big]\nonumber\\
& +\left(\ell +\ell' -1\right) \big[16 n_{IJ}^3 -2 n_{IJ}^2 \left(d-3(\De + \De' - \De _3)+4\right) +(d-2) (\De -1) (d-\De -2)\nonumber\\
&-n_{IJ} \left(3 d^2-2 d \left(\De '+3 \De -\De _3+5\right)+2 \left(3 \De'+\De  (2 \De +5)-3 \De _3+2\right)\right)\big]
\nonumber\\
&+\left(\ell +\ell'-1 \right)^2 (d (\De -n_{IJ}-1)-(\De -1) (\De +2)+n_{IJ}(3-5 n_{IJ}) )+(\ell-1)  \ell ' \left(\ell +\ell' -1\right)
(d+2(2 n_{IJ}-1)) \nonumber\\
 &-(\De -1) (d-\De -2)   \left(\De _3-\De'\right) \left(d-\De '-2 \De +\De_3\right)+(d-2) (d-1)\nonumber\\
&+n_{IJ}^2 \big[2 d (2 d-7)+1-\left(\De _3-\De'\right) \left(-2(d-6)-\De '-2 \De +\De _3\right) - \De(10(d -2)-7 \De)
 \big] \nonumber\\ 
 &-n_{IJ} \big[\left(\De _3-\De'\right) \left(d \left(d-\De ' +\De _3-2( \De+1)\right)
 +3( \De '+2 \De - \De _3)-4\right)\nonumber\\
 &+\De(d (d-\De -8)+5 \De +12)
+d(d-3) -1 \big] +4 n_{IJ}^3 \left((d+2)-2(\De+\De '-\De _3 ) \right)-12 n_{IJ}^4\bigg)\,,
\]
\[
 u_{n_{IJ}+1}&=\frac{(\ell-1 -n_{IJ}) \left(\ell '-n_{IJ}\right) \left(\De -\De _{12}+\ell-2\right) \left(d-\De +\De _{12}+
 \ell-2\right) }{4 (d-2 \De)(d-2(\De +1))(\De +\ell-2) (\De +\ell-1 ) (d-\De +\ell-3) (d-\De +\ell-2)}\nonumber\\
&\times   \big[((\ell-1)(\ell'-2)+ \ell'(\ell-3))-d \left(d-\De '-3 \De +\De _3+n_{IJ}-2\right)\nonumber\\
  &-2 \left(\De '+\De  (\De +2)-\De _3-1\right)
  -2n_{IJ} \left(2 (\ell+ \ell'-1)-3 n_{IJ}- \left(\De +\De '-\De _3+2\right)\right)\big]\,,\\
u_{n_{IJ}+2}&= \frac{(\ell-1 -n_{IJ}) \left(\ell '-n_{IJ}\right) (\ell-n_{IJ} -2)\left(\ell '-n_{IJ}-1\right) \left(\De -\De _{12}
+\ell-2\right) \left(d-\De +\De _{12}+\ell-2\right) }{4 (d-2\De ) (d-2 (\De +1)) (\De +\ell-2) (\De +\ell-1 ) (d-\De +\ell-3) (d-\De +\ell-2)}\,.
\]
\subsection{Recursion relation coefficients for the block with $n_{IJ}=\ell'=\ell$}
The coefficients appearing in \Eq{special case} are given by
\label{app:B.1}
\[
\bar{r}_{0\,(\ell-1)}^{(-0)}&=s_{\ell-1} (s_{2\,\ell-1}^{(0+)})^{-1},\nonumber\\
\bar{r}_{1\,(\ell-1)}^{(-0)}&=r_{1\,(\ell-1)}^{(-0)} s_{\ell-1} (s_{2\,\ell-1}^{(0+)})^{-1} - 1,\nonumber\\
\bar{r}_{2\,(\ell-2)}^{(-0)}&= s_{\ell-1}(s_{2\,\ell-1}^{(0+)})^{-1}
r_{2\,(\ell-2)}^{(-0)},\nonumber\\
\bar{r}_{3\,(\ell-2)}^{(-0)}&= s_{\ell-1}(s_{2\,\ell-1}^{(0+)})^{-1}r_{3\,(\ell-2)}^{(-0)},\nonumber\\
\bar{r}_{3\,(\ell-1)}^{(-0)}&= s_{\ell-1}(s_{2\,\ell-1}^{(0+)})^{-1}r_{3\,(\ell-1)}^{(-0)},\nonumber\\
 \bar{s}_{1\,(\ell-1)}^{(0+)} &= s_{\ell-1}(s_{2\,\ell-1}^{(0+)})^{-1}s_{1\,(\ell-1)}^{(0+)},\nonumber\\
\bar{s}_{2\,\ell}^{(0+)}&=s_{\ell}     +  s_{\ell-1}s_{2\,\ell}^{(0+)}(-s_{2\,\ell-1}^{(0+)})^{-1}, \nonumber\\
\bar{s}_{3\,(\ell-2)}^{(0+)}&=s_{\ell-1}(s_{2\,\ell-1}^{(0+)})^{-1}s_{3\,(\ell-2)}^{(0+)},\nonumber\\
\bar{s}_{4\,(\ell-2)}^{(0+)}&= s_{\ell-1}(s_{2\,\ell-1}^{(0+)})^{-1}s_{4\,(\ell-2)}^{(0+)},\nonumber\\
\bar{s}_{4\,(\ell-1)}^{(0+)}&=s_{\ell-1}(s_{2\,\ell-1}^{(0+)})^{-1}s_{4\,(\ell-1)}^{(0+)},\nonumber\\
\bar{t}_{1\,(\ell-2)}^{(0-)}&= s_{\ell-1}(s_{2\,\ell-1}^{(0+)})^{-1}t_{1\,(\ell-2)}^{(0-)},\nonumber\\
\bar{t}_{2\,(\ell-2)}^{(0-)}&=   t_{2\,(\ell-2)}^{(0-)}  s_{\ell-1} (s_{2\,\ell-1}^{(0+)})^{-1}- t_{\ell-2},\nonumber\\
 \bar{t}_{3\,(\ell-3)}^{(0-)} &=s_{\ell-1}(s_{2\,\ell-1}^{(0+)})^{-1}t_{3\,(\ell-3)}^{(0-)},\nonumber\\
 \bar{t}_{3\,(\ell-2)}^{(0-)} &=s_{\ell-1}(s_{2\,\ell-1}^{(0+)})^{-1}t_{3\,(\ell-2)}^{(0-)},\nonumber\\ 
\bar{t}_{4\,(\ell-3)}^{(0-)}&=s_{\ell-1}(s_{2\,\ell-1}^{(0+)})^{-1}t_{4\,(\ell-3)}^{(0-)},\nonumber\\ 
\bar{t}_{4\,(\ell-2)}^{(0-)}&=s_{\ell-1}(s_{2\,\ell-1}^{(0+)})^{-1}t_{4\,(\ell-2)}^{(0-)},\nonumber\\ 
\bar{u}_{1\,(\ell-2)}^{(+0)}&=s_{\ell-1}(s_{2\,\ell-1}^{(0+)})^{-1}u_{1\,(\ell-2)}^{(+0)},\nonumber\\ 
\bar{u}_{1\,(\ell-1)}^{(+0)}&=s_{\ell-1}(s_{2\,\ell-1}^{(0+)})^{-1}u_{1\,(\ell-1)}^{(+0)},\nonumber\\ 
\bar{u}_{2\,(\ell-2)}^{(+0)}&=   u_{2\,(\ell-2)}^{(+0)} s_{\ell-1} (s_{2\,\ell-1}^{(0+)})^{-1}-u_{\ell-2},\nonumber\\ 
\bar{u}_{2\,(\ell-1)}^{(+0)}&=u_{2\,(\ell-1)}^{(+0)}  s_{\ell-1} (s_{2\,\ell-1}^{(0+)})^{-1}-u_{\ell-1},\nonumber\\
\bar{u}_{3\,(\ell-3)}^{(+0)}&=s_{\ell-1}(s_{2\,\ell-1}^{(0+)})^{-1}u_{3\,(\ell-3)}^{(+0)},\nonumber\\ 
\bar{u}_{3\,(\ell-2)}^{(+0)}&=s_{\ell-1}(s_{2\,\ell-1}^{(0+)})^{-1}u_{3\,(\ell-2)}^{(+0)},\nonumber\\ 
\bar{u}_{4\,(\ell-3)}^{(+0)}&=s_{\ell-1}(s_{2\,\ell-1}^{(0+)})^{-1}u_{4\,(\ell-3)}^{(+0)},\nonumber\\ 
\bar{u}_{4\,(\ell-2)}^{(+0)}&=s_{\ell-1}(s_{2\,\ell-1}^{(0+)})^{-1}u_{4\,(\ell-2)}^{(+0)},
\nonumber\\
\bar{u}_{4\,(\ell-1)}^{(+0)}&=s_{\ell-1}(s_{2\,\ell-1}^{(0+)})^{-1}u_{4\,(\ell-1)}^{(+0)}\,.
\]
\section{Symmetry-transformed recursion relations}
\label{app:C}

Here we write explicitly the recursion relations obtained by applying the symmetry transformations \Eq{symmetry 12} and \Eq{symmetry 45} to the fixed-$\Delta_3$ recursion \Eq{recursion 15}. We choose the conventions of \Eq{convention 1 prefactor} and \Eq{convention 1 cross-ratios}. We first recast \Eq{recursion 15} in the form
\[
\eql{15 orig}
&(u_1 u_2)^{-1/2} G_{(\ell-1, \ell'-1; 0, 0)}^{(n_{IJ})}(\De_{12}+1, \De_{45}-1) = 
G_{(\ell-1, \ell'-1; -1, - 1)}^{(n_{IJ})}  \nonumber\\
&+ \sum_{m_{IJ}=n_{IJ}}^{n_{IJ}+1} 
r_{1\,m_{IJ}}^{(-0)}G_{(\ell-1, \ell'; -1, 0)}^{(m_{IJ})}
+\sum_{m_{IJ}=n_{IJ}-1}^{n_{IJ}+1} 
r_{2\,m_{IJ}}^{(-0)} G_{(\ell-1, \ell'-2; -1, 0)}^{(m_{IJ})}
+ \sum_{m_{IJ}=n_{IJ}-1}^{n_{IJ}+2}  
r_{3\,m_{IJ}}^{(-0)}G_{(\ell-1, \ell'-1; -1, 1)}^{(m_{IJ})}  \nonumber\\   
&+   \sum_{m_{IJ}=n_{IJ}}^{n_{IJ}+1} 
s_{1\,m_{IJ}}^{(0+)}G_{(\ell, \ell'-1; 0,  - 1)}^{(m_{IJ})}
+\sum_{m_{IJ}=n_{IJ}}^{n_{IJ}+2} 
s_{2\,m_{IJ}}^{(0+)}G_{(\ell, \ell'; 0, 0)}^{(m_{IJ})}   
+\sum_{m_{IJ}=n_{IJ}-1}^{n_{IJ}+2} 
s_{3\,m_{IJ}}^{(0+)}G_{(\ell, \ell'-2; 0, 0)}^{(m_{IJ})}   \nonumber\\
&+ \sum_{m_{IJ}=n_{IJ}-1}^{n_{IJ}+3}  
s_{4\,m_{IJ}}^{(0+)}G_{(\ell, \ell'-1; 0, 1)}^{(m_{IJ})} 
+ \sum_{m_{IJ}=n_{IJ}-1}^{n_{IJ}+1} 
t_{1\,m_{IJ}}^{(0-)}G_{(\ell-2, \ell'-1; 0, - 1)}^{(m_{IJ})} \
+ \sum_{m_{IJ}=n_{IJ}-1}^{n_{IJ}+2} 
t_{2\,m_{IJ}}^{(0-)}G_{(\ell-2, \ell'; 0, 0)}^{(m_{IJ})}    \nonumber\\
&+\sum_{m_{IJ}=n_{IJ}-2}^{n_{IJ}+2} 
t_{3\,m_{IJ}}^{(0-)} G_{(\ell-2, \ell'-2; 0,0)}^{(m_{IJ})}
+ \sum_{m_{IJ}=n_{IJ}-2}^{n_{IJ}+2}  
t_{4\,m_{IJ}}^{(0-)}G_{(\ell-2, \ell'-1; 0, 1)}^{(m_{IJ})} 
+  \sum_{m_{IJ}=n_{IJ}-1}^{n_{IJ}+2} 
u_{1\,m_{IJ}}^{(+0)}G_{(\ell-1, \ell'-1; 1, - 1)}^{(m_{IJ})}     \nonumber\\
&+\sum_{m_{IJ}=n_{IJ}-1}^{n_{IJ}+3}
u_{2\,m_{IJ}}^{(+0)}G_{(\ell-1, \ell'; 1, 0)}^{(m_{IJ})}   
+\sum_{m_{IJ}=n_{IJ}-2}^{n_{IJ}+3} 
u_{3\,m_{IJ}}^{(+0)}G_{(\ell-1, \ell'-2; 1, 0)}^{(m_{IJ})}  
+ \sum_{m_{IJ}=n_{IJ}-2}^{n_{IJ}+4}  
u_{4\,m_{IJ}}^{(+0)}G_{(\ell-1, \ell'-1; 1,1)}^{(m_{IJ})}.
\]
Using the symmetry under the interchange $1\leftrightarrow 2$, namely \Eq{symmetry 12}, we obtain
\[
\eql{15 sym 12}
&- ( u_1 u_2 )^{-1/2} G_{(\ell-1, \ell'-1; 0, 0)}^{(n_{IJ})}(\De_{12}-1, \De_{45}-1) =
- G_{(\ell-1, \ell'-1; -1, - 1)}^{(n_{IJ})}\nonumber\\  
&- \sum_{m_{IJ}=n_{IJ}}^{n_{IJ}+1} r_{1\,m_{IJ}}^{(-0)}(-\De_{12}, \De_{45})G_{(\ell-1, \ell'; -1, 0)}^{(m_{IJ})}
-\sum_{m_{IJ}=n_{IJ}-1}^{n_{IJ}+1} 
r_{2\,m_{IJ}}^{(-0)}(-\De_{12}, \De_{45}) G_{(\ell-1, \ell'-2; -1, 0)}^{(m_{IJ})}   \nonumber\\  
&- \sum_{m_{IJ}=n_{IJ}-1}^{n_{IJ}+2}  
r_{3\,m_{IJ}}^{(-0)}(-\De_{12}, \De_{45})G_{(\ell-1, \ell'-1; -1, 1)}^{(m_{IJ})}
+   \sum_{m_{IJ}=n_{IJ}}^{n_{IJ}+1} 
s_{1\,m_{IJ}}^{(0+)}(-\De_{12}, \De_{45})G_{(\ell, \ell'-1; 0,  - 1)}^{(m_{IJ})}  \nonumber\\
&+\sum_{m_{IJ}=n_{IJ}}^{n_{IJ}+2} 
s_{2\,m_{IJ}}^{(0+)}(-\De_{12}, \De_{45})G_{(\ell, \ell'; 0, 0)}^{(m_{IJ})}   
+\sum_{m_{IJ}=n_{IJ}-1}^{n_{IJ}+2} 
s_{3\,m_{IJ}}^{(0+)}(-\De_{12}, \De_{45})G_{(\ell, \ell'-2; 0, 0)}^{(m_{IJ})} \nonumber\\
&+ \sum_{m_{IJ}=n_{IJ}-1}^{n_{IJ}+3}  
s_{4\,m_{IJ}}^{(0+)}(-\De_{12}, \De_{45})G_{(\ell, \ell'-1; 0, 1)}^{(m_{IJ})}
+ \sum_{m_{IJ}=n_{IJ}-1}^{n_{IJ}+1} 
t_{1\,m_{IJ}}^{(0-)}(-\De_{12}, \De_{45})G_{(\ell-2, \ell'-1; 0, - 1)}^{(m_{IJ})} \nonumber\\
&+ \sum_{m_{IJ}=n_{IJ}-1}^{n_{IJ}+2} 
t_{2\,m_{IJ}}^{(0-)}(-\De_{12}, \De_{45})G_{(\ell-2, \ell'; 0, 0)}^{(m_{IJ})} 
+\sum_{m_{IJ}=n_{IJ}-2}^{n_{IJ}+2} 
t_{3\,m_{IJ}}^{(0-)}(-\De_{12}, \De_{45}) G_{(\ell-2, \ell'-2; 0,0)}^{(m_{IJ})} \nonumber\\
&+ \sum_{m_{IJ}=n_{IJ}-2}^{n_{IJ}+2}  
t_{4\,m_{IJ}}^{(0-)}(-\De_{12}, \De_{45})G_{(\ell-2, \ell'-1; 0, 1)}^{(m_{IJ})} -  \sum_{m_{IJ}=n_{IJ}-1}^{n_{IJ}+2} 
u_{1\,m_{IJ}}^{(+0)}(-\De_{12}, \De_{45})G_{(\ell-1, \ell'-1; 1, - 1)}^{(m_{IJ})}   \nonumber\\
&-\sum_{m_{IJ}=n_{IJ}-1}^{n_{IJ}+3}
u_{2\,m_{IJ}}^{(+0)}(-\De_{12}, \De_{45})G_{(\ell-1, \ell'; 1, 0)}^{(m_{IJ})}   -\sum_{m_{IJ}=n_{IJ}-2}^{n_{IJ}+3} 
u_{3\,m_{IJ}}^{(+0)}(-\De_{12}, \De_{45})G_{(\ell-1, \ell'-2; 1, 0)}^{(m_{IJ})} \nonumber\\
&- \sum_{m_{IJ}=n_{IJ}-2}^{n_{IJ}+4}  
u_{4\,m_{IJ}}^{(+0)}(-\De_{12}, \De_{45})G_{(\ell-1, \ell'-1; 1,1)}^{(m_{IJ})}\,.
\]
Similarly, using the symmetry under the interchange $4 \leftrightarrow 5$, namely \Eq{symmetry 45}, we obtain
\[
\eql{15 sym 45}
&- ( u_1 u_2 )^{-1/2} G_{(\ell-1, \ell'-1; 0, 0)}^{(n_{IJ})}(\De_{12}+1, \De_{45}+1)  =
-G_{(\ell-1, \ell'-1; -1, - 1)}^{(n_{IJ})} \nonumber\\  
&+ \sum_{m_{IJ}=n_{IJ}}^{n_{IJ}+1} 
r_{1\,m_{IJ}}^{(-0)}(\De_{12}, -\De_{45})G_{(\ell-1, \ell'; -1, 0)}^{(m_{IJ})}  
+\sum_{m_{IJ}=n_{IJ}-1}^{n_{IJ}+1} 
r_{2\,m_{IJ}}^{(-0)}(\De_{12}, -\De_{45}) G_{(\ell-1, \ell'-2; -1, 0)}^{(m_{IJ})} \nonumber\\  
&- \sum_{m_{IJ}=n_{IJ}-1}^{n_{IJ}+2}  
r_{3\,m_{IJ}}^{(-0)}(\De_{12}, -\De_{45})G_{(\ell-1, \ell'-1; -1, 1)}^{(m_{IJ})} 
-   \sum_{m_{IJ}=n_{IJ}}^{n_{IJ}+1} 
s_{1\,m_{IJ}}^{(0+)}(\De_{12}, -\De_{45})G_{(\ell, \ell'-1; 0,  - 1)}^{(m_{IJ})}  \nonumber\\
&+\sum_{m_{IJ}=n_{IJ}}^{n_{IJ}+2} 
s_{2\,m_{IJ}}^{(0+)}(\De_{12}, -\De_{45})G_{(\ell, \ell'; 0, 0)}^{(m_{IJ})}    
+\sum_{m_{IJ}=n_{IJ}-1}^{n_{IJ}+2} 
s_{3\,m_{IJ}}^{(0+)}(\De_{12}, -\De_{45})G_{(\ell, \ell'-2; 0, 0)}^{(m_{IJ})} \nonumber\\
&- \sum_{m_{IJ}=n_{IJ}-1}^{n_{IJ}+3}  
s_{4\,m_{IJ}}^{(0+)}(\De_{12}, -\De_{45})G_{(\ell, \ell'-1; 0, 1)}^{(m_{IJ})} 
-\sum_{m_{IJ}=n_{IJ}-1}^{n_{IJ}+1} 
t_{1\,m_{IJ}}^{(0-)}(\De_{12}, -\De_{45})G_{(\ell-2, \ell'-1; 0, - 1)}^{(m_{IJ})} \nonumber\\  
&+ \sum_{m_{IJ}=n_{IJ}-1}^{n_{IJ}+2} 
t_{2\,m_{IJ}}^{(0-)}(\De_{12}, -\De_{45})G_{(\ell-2, \ell'; 0, 0)}^{(m_{IJ})}  
+\sum_{m_{IJ}=n_{IJ}-2}^{n_{IJ}+2} 
t_{3\,m_{IJ}}^{(0-)}(\De_{12},-\De_{45}) G_{(\ell-2, \ell'-2; 0,0)}^{(m_{IJ})} \nonumber\\  
&- \sum_{m_{IJ}=n_{IJ}-2}^{n_{IJ}+2}  
t_{4\,m_{IJ}}^{(0-)}(\De_{12}, -\De_{45})G_{(\ell-2, \ell'-1; 0, 1)}^{(m_{IJ})} 
-  \sum_{m_{IJ}=n_{IJ}-1}^{n_{IJ}+2} 
u_{1\,m_{IJ}}^{(+0)}(\De_{12}, -\De_{45})G_{(\ell-1, \ell'-1; 1, - 1)}^{(m_{IJ})}     \nonumber\\ 
&+\sum_{m_{IJ}=n_{IJ}-1}^{n_{IJ}+3}
u_{2\,m_{IJ}}^{(+0)}(\De_{12}, -\De_{45})G_{(\ell-1, \ell'; 1, 0)}^{(m_{IJ})}   
+\sum_{m_{IJ}=n_{IJ}-2}^{n_{IJ}+3} 
u_{3\,m_{IJ}}^{(+0)}(\De_{12}, -\De_{45})G_{(\ell-1, \ell'-2; 1, 0)}^{(m_{IJ})} \nonumber\\  
&- \sum_{m_{IJ}=n_{IJ}-2}^{n_{IJ}+4}  
u_{4\,m_{IJ}}^{(+0)}(\De_{12}, -\De_{45})G_{(\ell-1, \ell'-1; 1,1)}^{(m_{IJ})}\,.
\]
Finally, using both \Eq{symmetry 12} and \Eq{symmetry 45}, we have
\[
\eql{15 sym 1245}
&( u_1u_2 )^{-1/2} w_{ 2;4 }  G_{(\ell-1, \ell'-1; 0, 0)}^{(n_{IJ})}(\De_{12}-1, \De_{45}+1)=
G_{(\ell-1, \ell'-1; -1, - 1)}^{(n_{IJ})} \nonumber\\  
&- \sum_{m_{IJ}=n_{IJ}}^{n_{IJ}+1} 
r_{1\,m_{IJ}}^{(-0)}(-\De_{12}, -\De_{45})G_{(\ell-1, \ell'; -1, 0)}^{(m_{IJ})}
-\sum_{m_{IJ}=n_{IJ}-1}^{n_{IJ}+1} 
r_{2\,m_{IJ}}^{(-0)}(-\De_{12}, -\De_{45}) G_{(\ell-1, \ell'-2; -1, 0)}^{(m_{IJ})}   \nonumber\\  
&+ \sum_{m_{IJ}=n_{IJ}-1}^{n_{IJ}+2}  
r_{3\,m_{IJ}}^{(-0)}(-\De_{12}, -\De_{45})G_{(\ell-1, \ell'-1; -1, 1)}^{(m_{IJ})}
-   \sum_{m_{IJ}=n_{IJ}}^{n_{IJ}+1} 
s_{1\,m_{IJ}}^{(0+)}(-\De_{12}, -\De_{45})G_{(\ell, \ell'-1; 0,  - 1)}^{(m_{IJ})}  \nonumber\\
&+\sum_{m_{IJ}=n_{IJ}}^{n_{IJ}+2} 
s_{2\,m_{IJ}}^{(0+)}(-\De_{12}, -\De_{45})G_{(\ell, \ell'; 0, 0)}^{(m_{IJ})}   
+\sum_{m_{IJ}=n_{IJ}-1}^{n_{IJ}+2} 
s_{3\,m_{IJ}}^{(0+)}(-\De_{12}, -\De_{45})G_{(\ell, \ell'-2; 0, 0)}^{(m_{IJ})}   \nonumber\\
&- \sum_{m_{IJ}=n_{IJ}-1}^{n_{IJ}+3}  
s_{4\,m_{IJ}}^{(0+)}(-\De_{12}, -\De_{45})G_{(\ell, \ell'-1; 0, 1)}^{(m_{IJ})} 
- \sum_{m_{IJ}=n_{IJ}-1}^{n_{IJ}+1} 
t_{1\,m_{IJ}}^{(0-)}(-\De_{12}, -\De_{45})G_{(\ell-2, \ell'-1; 0, - 1)}^{(m_{IJ})}  \nonumber\\
&+ \sum_{m_{IJ}=n_{IJ}-1}^{n_{IJ}+2} 
t_{2\,m_{IJ}}^{(0-)}(-\De_{12}, -\De_{45})G_{(\ell-2, \ell'; 0, 0)}^{(m_{IJ})} 
+\sum_{m_{IJ}=n_{IJ}-2}^{n_{IJ}+2} 
t_{3\,m_{IJ}}^{(0-)}(-\De_{12}, -\De_{45}) G_{(\ell-2, \ell'-2; 0,0)}^{(m_{IJ})}  \nonumber\\  
&- \sum_{m_{IJ}=n_{IJ}-2}^{n_{IJ}+2}  
t_{4\,m_{IJ}}^{(0-)}(-\De_{12}, -\De_{45})G_{(\ell-2, \ell'-1; 0, 1)}^{(m_{IJ})} 
+  \sum_{m_{IJ}=n_{IJ}-1}^{n_{IJ}+2} 
u_{1\,m_{IJ}}^{(+0)}(-\De_{12}, -\De_{45})G_{(\ell-1, \ell'-1; 1, - 1)}^{(m_{IJ})}     \nonumber\\  
&-\sum_{m_{IJ}=n_{IJ}-1}^{n_{IJ}+3}
u_{2\,m_{IJ}}^{(+0)}(-\De_{12}, -\De_{45})G_{(\ell-1, \ell'; 1, 0)}^{(m_{IJ})}   
-\sum_{m_{IJ}=n_{IJ}-2}^{n_{IJ}+3} 
u_{3\,m_{IJ}}^{(+0)}(-\De_{12}, -\De_{45})G_{(\ell-1, \ell'-2; 1, 0)}^{(m_{IJ})}  \nonumber\\  
&+ \sum_{m_{IJ}=n_{IJ}-2}^{n_{IJ}+4}  
u_{4\,m_{IJ}}^{(+0)}(-\De_{12}, -\De_{45})G_{(\ell-1, \ell'-1; 1,1)}^{(m_{IJ})}\,.
\]

We may combine the original fixed-$\Delta_3$ recursion relation with the above symmetry-transformed recursion relations in the following manner. We take 
(\Eq{15 orig} + \Eq{15 sym 12})
$+\, c$ (\Eq{15 sym 45} + 
\Eq{15 sym 1245}), where $c$ is chosen such that the blocks in the set \[
\big\{
G_{(\ell-1, \ell'-1; 1,1)}^{(n_{IJ}-2)}\,,
G_{(\ell-1, \ell'-1; 1,1)}^{(n_{IJ}-1)}\,, 
G_{(\ell-1, \ell'-1; 1,1)}^{(n_{IJ})}\,, 
G_{(\ell-1, \ell'-1; 1,1)}^{(n_{IJ}+1)}\,,
G_{(\ell-1, \ell'-1; 1,1)}^{(n_{IJ}+2)}\,, 
G_{(\ell-1, \ell'-1; 1,1)}^{(n_{IJ}+3)}\,, 
G_{(\ell-1, \ell'-1; 1,1)}^{(n_{IJ}+4)}
\big\}
\]
are eliminated. This occurs when $c$ takes the value 
\[
c=\frac{\left(\Delta '+\Delta _{45}+\ell '-2\right) \left(d-\Delta '-\Delta _{45}+\ell
   '-2\right)}{\left(\Delta '-\Delta _{45}+\ell '-2\right) \left(d-\Delta '+\Delta
   _{45}+\ell '-2\right)}\,.
\]

With this, we obtain the attractively simpler relation
\[\eql{eq:15 simpler}
G_{(\ell, \ell'; 0, 0)}^{(n_{IJ}+1)}  
=\frac{1}{\bar{s}_{2\,n_{IJ}+1}^{(0+)}}\bigg(&(u_1 u_2)^{-1/2} \left[
G_{(\ell-1, \ell'-1; 0, 0)}^{(n_{IJ})}(\De_{12}+1, \De_{45}-1) \right.\nonumber\\
&
-  G_{(\ell-1, \ell'-1; 0, 0)}^{(n_{IJ})}(\De_{12}-1, \De_{45}-1)\nonumber\\
&- c  G_{(\ell-1, \ell'-1; 0, 0)}^{(n_{IJ})}(\De_{12}+1, \De_{45}+1) \nonumber\\
& \left.
+c w_{ 2;4 }  G_{(\ell-1, \ell'-1; 0, 0)}^{(n_{IJ})}(\De_{12}-1, \De_{45}+1)   \right]
\nonumber\\&- \left(   \sum_{m_{IJ}=n_{IJ}}^{n_{IJ}+1} 
\bar{s}_{1\,m_{IJ}}^{(0+)}G_{(\ell, \ell'-1; 0,  - 1)}^{(m_{IJ})}+
\bar{s}_{2\,n_{IJ}}^{(0+)}G_{(\ell, \ell'; 0, 0)}^{(n_{IJ})} +
\bar{s}_{2\,n_{IJ}+2}^{(0+)}G_{(\ell, \ell'; 0, 0)}^{(n_{IJ}+2)} \right.
 \nonumber\\
&
+\sum_{m_{IJ}=n_{IJ}-1}^{n_{IJ}+2} 
\bar{s}_{3\,m_{IJ}}^{(0+)}G_{(\ell, \ell'-2; 0, 0)}^{(m_{IJ})} + \sum_{m_{IJ}=n_{IJ}-1}^{n_{IJ}+1} 
\bar{t}_{1\,m_{IJ}}^{(0-)}G_{(\ell-2, \ell'-1; 0, - 1)}^{(m_{IJ})}\nonumber\\
&
 \
+ \sum_{m_{IJ}=n_{IJ}-1}^{n_{IJ}+2} 
\bar{t}_{2\,m_{IJ}}^{(0-)}G_{(\ell-2, \ell'; 0, 0)}^{(m_{IJ})}+\sum_{m_{IJ}=n_{IJ}-2}^{n_{IJ}+2} 
\bar{t}_{3\,m_{IJ}}^{(0-)} G_{(\ell-2, \ell'-2; 0,0)}^{(m_{IJ})}\nonumber\\
&+  \sum_{m_{IJ}=n_{IJ}-1}^{n_{IJ}+2} 
\bar{u}_{1\,m_{IJ}}^{(+0)}G_{(\ell-1, \ell'-1; 1, - 1)}^{(m_{IJ})}   +\sum_{m_{IJ}=n_{IJ}-1}^{n_{IJ}+3}
\bar{u}_{2\,m_{IJ}}^{(+0)}G_{(\ell-1, \ell'; 1, 0)}^{(m_{IJ})}    
 \nonumber\\
&
\left. +\sum_{m_{IJ}=n_{IJ}-2}^{n_{IJ}+3} 
\bar{u}_{3\,m_{IJ}}^{(+0)}G_{(\ell-1, \ell'-2; 1, 0)}^{(m_{IJ})} \right) \bigg)\,.
\]
The coefficients are given in the attached \texttt{Mathematica} notebook.
 
\section{3-point coefficients for the vector and tensor cases}
\label{app:D}
The coefficients in \Eq{one} - \Eq{three} for promoting the scalar $\Phi$ to a vector $v^A$ using weight-shifting operators applied to 3-point functions are given by:
\[
\al_1=&\frac{1}{2} \left(-\Delta '+\Delta -\Delta _3+\ell '-\ell +1\right)\,,\\
\be_1=& \frac{1}{2} \left(\ell '-n_{IJ}\right)\,,\\
\ga_1=& \frac{n_{IJ}-\ell }{2}\,,\\
\al_2=& \frac{1}{2} \left(\Delta '-\Delta +\Delta _3-\ell '+\ell -1\right) \left(\Delta '+\Delta
   -\Delta _3+2 n_{IJ}-\ell '+\ell -1\right)\,,\\
\be_2 =& \frac{1}{2} \left(\ell '-n_{IJ}\right) \left(-\Delta '-\Delta +\Delta _3-2 n_{IJ}+\ell '-\ell
   +1\right)\,,\\
    \ga_2 =& \frac{1}{2} \left(2 (\Delta -1) \left(\Delta _3-1\right)-2 n_{IJ}^2 \right. \nonumber\\
    & \left.+n_{IJ} \left(-\Delta '-\Delta
   +\Delta _3+\ell '+1\right)+\ell  \left(\Delta '+\Delta +\Delta _3+n_{IJ}-\ell
   '-3\right)+\ell ^2\right)\,,\\
   \al_3=& \frac{1}{2} \left(\ell '-n_{IJ}\right) \left(-\Delta '+\Delta -\Delta _3+\ell '-\ell +1\right)\,,\\
   \be_3=& \frac{1}{2} \left(-n_{IJ}+\ell '-1\right) \left(\ell '-n_{IJ}\right)\,, \\
   \ga_3 =& -\frac{1}{2} \left(\ell '-n_{IJ}\right)(-n_{IJ}+\ell -1) \,,\\
\al_4=&\frac{1}{2} n_{IJ} \big[2 d^2 \left(\Delta '-\Delta -\Delta _3-\ell '+\ell +1\right) \nonumber\\
&+d \big(-\Delta ' \left(\Delta '+8\right)+\Delta _3^2+\Delta  (\Delta +6)
   +2  \Delta _3(\Delta+1)+3 \ell ^2\nonumber\\
   &-2 \ell  \left(-\Delta '+2 \Delta +4 \Delta _3+\ell
   '\right)+2 \left(\Delta '+4\right) \ell '-\ell'{}^2-3\big)\nonumber\\
   &+4\big(-\Delta ^2-\Delta _3^2-\left(2 \Delta _3+1\right) \ell ^2+2 \ell  \left(
   -\Delta'+\Delta(\Delta _3  +1)+\ell '\right)+\left(\ell '-\Delta'\right)^2+1\big)\big]\nonumber\\
   &+n_{IJ}^2 \left(\Delta '-\Delta +\Delta _3-\ell '+\ell -1\right)
   \left(d-\Delta '-\Delta +\Delta _3+\ell '+3 \ell +3\right)\nonumber\\
   &-2 n_{IJ}^3 \left(\Delta '-\Delta
   +\Delta _3-\ell '+\ell -1\right)\,,\\
   \be_4=&\frac{1}{2} n_{IJ}^2 \big(2 d^2+d \left(-\Delta '-\Delta +\Delta _3-\ell '+3 \ell
   -11\right)+2 \Delta ' \left(\ell '+3\right)\nonumber\\
   &-2 \left(3( \Delta _3- \Delta) +\ell '
   \left(-\Delta +\Delta _3+\ell '+6\right)+3 \ell  \left(\ell
   '+3\right)+1\right)\big)+\frac{1}{2} n_{IJ} \big[-2 d^2 \left(\Delta _3+\ell '\right)
   \nonumber\\
   &+d
   \big(\Delta '-\Delta +2 \Delta  \Delta _3+5 \Delta _3+\left(\Delta '+\Delta -\Delta
   _3+8\right) \ell '-3 \ell  \left(2 \Delta _3+\ell '-1\right)-\left(\ell
   '\right)^2+3\big)\nonumber\\
   &-4 \left(\Delta '+\left(\Delta _3-1\right) \ell ^2+\Delta ' \ell
   '-\ell  \left(-\Delta +(\Delta +2) \Delta _3+3 \ell '+1\right)-\left(\ell '-\Delta
   \right) \left(\Delta _3+\ell '\right)\right)\big]
   \nonumber\\
   &+n_{IJ}^3 \left(d-\Delta '-\Delta
   +\Delta _3+3 \ell '+3 \ell +5\right)-2 n_{IJ}^4\,,
   \]
   \[
\ga_4=&
\frac{1}{2} n_{IJ}^2 \big[-2 d^2+d \left(\Delta '+\Delta +3 \Delta _3-\ell '-\ell +7\right)\nonumber\\
&+2 \left(3( \ell ^2-\Delta ')-\Delta -2 \Delta  \Delta _3+\Delta _3+\ell  \left(-\Delta
   '-\Delta +3 \Delta _3+\ell '+10\right)+3 \ell '+3\right)\big]\nonumber\\
  & +\frac{1}{2} n_{IJ} (\ell
   +1) \big[2 d^2+d \left(-\Delta '-\Delta -3 \Delta _3+\ell '+3 \ell -5\right)
   \nonumber\\
   &-4
   \left(-\Delta '-\Delta  \Delta _3+\ell '+\left(\Delta _3+2\right) \ell
   \right)\big]+n_{IJ}^3 \left(-d+\Delta '+\Delta -\Delta _3-\ell '-5 \ell -5\right)+2 n_{IJ}^4\,,\\
\al_5=&\frac{1}{2} n_{IJ} \big(d^2 \left(\Delta '-\Delta -\Delta _3-\ell '+\ell +1\right)+d
   \big(\left(-\Delta '+\Delta _3-1\right) \left(\Delta '+\Delta _3+2\right)
   \nonumber\\
   &+\Delta (\Delta +3)+\ell^2+\left(\Delta '+\Delta -\Delta _3+4\right) \ell '
   -\ell  \left(2\Delta +4 \Delta _3+\ell '-1\right)\big)\nonumber\\
   &-3 \Delta ^2-2 \Delta'+\Delta _3(2\Delta -3 \Delta _3)+\ell  \big(-2 \ell^2
   +6 \Delta -2 \Delta _3+4
   \left(- \Delta '+\Delta  \Delta _3+\left(-\Delta '+\Delta -\Delta _3+2\right) \ell '+
\ell'{}^2\right)\big)\nonumber\\
   &+3 \left(\ell '-\Delta '\right)^2+2 \ell '-\ell ^2 \left(
   2(\Delta '- \Delta) +6 \Delta _3+2 \ell '+1\right)+3\big)
-\frac{1}{2} (\ell +1)
   \big(d^2 \left(\Delta '-\Delta -\Delta _3-\ell '+\ell +1\right)\nonumber\\
   &+d \big(\Delta  (\Delta +2)-\Delta'\left(\Delta '+2\right)+\Delta _3 \left(\Delta _3+2\right)
   +\ell\left(2 \ell +\Delta'-3( \Delta + \Delta _3)-2 \ell '+1\right)\nonumber\\
   &+\left(\Delta '+\Delta-\Delta _3+3\right) \ell '-3\big)
+\ell ^3+\ell  \left(
   -\Delta'{}^2+\left(\Delta +\Delta _3\right){}^2+2 \left( \Delta + \Delta _3+\Delta
   -\Delta _3+2\right) \ell'+\ell '{}^2-3\right)\nonumber\\
   &-2 \left(\Delta^2-\Delta '{}^2+\Delta _3^2+\left(\Delta'+\Delta -\Delta _3+1\right) \ell
   '-1\right)
   -2 \ell ^2 \left(\Delta +\Delta _3+\ell '\right)\big)\nonumber\\
   &+\frac{1}{2} n_{IJ}^2\left(\Delta'
   -\Delta +\Delta _3-\ell '+\ell -1\right) \left(d-\Delta '-\Delta +\Delta
   _3+3 \ell '+7 \ell +5\right)-2 n_{IJ}^3 \left(\Delta '-\Delta +\Delta _3-\ell '+\ell
   -1\right)\,,\\
    \ga_5= & 2 n_{IJ}^4+\frac{1}{2} n_{IJ}^3 \left(-d+\Delta '+\Delta -\Delta _3-3 \ell '-11 \ell -5\right)\nonumber\\
    &+\frac{1}{2} n_{IJ}^2 \big(-d^2+d \left(\Delta '+\Delta +\Delta _3+2\right)
-3 (\Delta '- \ell ')  \nonumber\\
&+\Delta _3( -2 \Delta  +1)+\ell  \left(7 \ell '+9 \ell +10-\Delta '-\Delta +3 \Delta _3\right)-\Delta  +1\big)\nonumber\\
   &-\frac{1}{2} \ell  (\ell +1) \big(d^2+d \left(-\Delta '-\Delta -\Delta _3+2 \ell
   -1\right) +2 \left(\Delta '+\Delta  \Delta _3\right)+\ell ^2-\ell  \left(\Delta
   '+\Delta +\Delta _3+\ell '+1\right)\big)\nonumber\\
   &+\frac{1}{2} n_{IJ} \big(d^2 (2 \ell +1)+d \left(-\Delta '-\Delta -\Delta _3+3 \ell ^2-\ell 
   \left(2 \Delta '+2 \Delta +2 \Delta _3+1\right)-1\right) +2 \left(\Delta '+\Delta 
   \Delta _3\right)\nonumber\\
   &-\ell ^3-2 \ell  \left(-2 \Delta '+\Delta _3(1-2 \Delta  )+2
   \ell '+1\right)-\ell ^2 \left(\Delta '+\Delta +3 \Delta _3+5 \ell '+5\right)\big)\,,\\
   \be_5=&   -2 n_{IJ}^4+\frac{1}{2} n_{IJ}^3 \left(d-\Delta '-\Delta +\Delta _3
   +7( \ell '+ \ell) +5\right)
   +\frac{1}{2} n_{IJ}^2 \big(d^2+d \left(-\Delta '-\Delta +\Delta _3-\ell '+\ell -4\right)
   \nonumber\\
   &+3(\Delta '+ \Delta - \Delta _3)-2 \ell ^2+\ell ' \left(\Delta '+\Delta -\Delta _3-3
   \ell '-8\right)-\ell  \left(11 \ell '+7\right)+1\big)
\nonumber\\
&+\frac{1}{2} \ell ' (\ell +1) \left(d^2+d \left(-\Delta '-\Delta +\Delta _3+2 \ell
   -3\right)+2 \left(\Delta '+\Delta -\Delta _3+1\right)+\ell ^2-\ell  \left(\Delta
   '+\Delta -\Delta _3+\ell '+3\right)\right)\nonumber\\
   &+\frac{1}{2} n_{IJ} \big(-d^2 \left(\ell '+\ell +1\right)+d \big(\Delta '+\Delta -\Delta _3-2
   \ell ^2+\ell  \left(\Delta '+\Delta -\Delta _3-\ell '+1\right)\nonumber\\
   &+\left(\Delta '+\Delta-\Delta _3+4\right) \ell '+3\big)-\ell ^3+\ell  \left(-\Delta '-\Delta +\Delta _3+4
   \ell ' \left(\ell '+2\right)+1\right)\nonumber\\
   &+\left(3 \ell '+2\right) \left(-\Delta '-\Delta
   +\Delta _3+\ell '-1\right)+\ell ^2 \left(\Delta '+\Delta -\Delta _3+3 \ell
   '+2\right)\big)\,,
   \]
   \[
     \al_6=& \frac{1}{2} (\ell -n_{IJ}) (-n_{IJ}+\ell +1) \left(\ell '-n_{IJ}\right) \left(\Delta '-\Delta +\Delta
   _3-\ell '+\ell -1\right)\,, \\
   \be_6=&-\frac{1}{2} (\ell -n_{IJ}) (-n_{IJ}+\ell +1) \left(n_{IJ}-\ell '\right) \left(n_{IJ}-\ell '+1\right)\,,\\
   \ga_6=&\frac{1}{2} (\ell -n_{IJ}) \left(\ell '-n_{IJ}\right)(-n_{IJ}+\ell -1)  (-n_{IJ}+\ell +1)\,.
   \]

The coefficients in \Eq{one tensor1} - \Eq{three tensor} for promoting the vector $v^A$ to a tensor $T^{AB}$ using weight-shifting operators applied to 3-point functions are:
\[
\xi_{11}=&\frac{1}{2} \left(-\Delta '+\Delta-\Delta _3+\ell '-\ell \right)\,,\\
\xi_{12} =&\frac{1}{2} \left(\ell '-n_{IJ}\right)\,,\\
\xi_{14} =&\frac{1}{2}(n_{IJ}-\ell)\,,\\
\xi_{22} =&\frac{1}{2} \left(-\Delta '+\Delta
   -\Delta _3+\ell '-\ell -2\right)\,,\\
\xi_{23} =&\frac{1}{2} \left(-n_{IJ}+\ell '-1\right)\,,\\
\xi_{25} =&\frac{1}{2}(n_{IJ}-\ell)\,, \\
\xi_{34} =&\frac{1}{2} \left(-\Delta '+\Delta
   -\Delta _3+\ell '-\ell +2\right)\,,\\
   \xi_{35} =&\frac{1}{2} \left(\ell '-n_{IJ}\right)\,,\\
   \xi_{36} =&\frac{1}{2} (n_{IJ}-\ell +1)\,,\\
   \ka_{22}=&\frac{1}{2} \left(\Delta '-\Delta
   +\Delta _3-\ell '+\ell +2\right)
   \left(\Delta '+\Delta -\Delta
   _3+2 n_{IJ}-\ell '+\ell \right)\,,\\
   \la_{22}=&-\frac{1}{2} \left(-n_{IJ}+\ell
   '-1\right) \left(\Delta '-\Delta
   +\Delta _3-\ell '+\ell +2\right)\,,\\
   \ka_{23}=&-\frac{1}{2} \left(-n_{IJ}+\ell
   '-1\right) \left(\Delta '+\Delta
   -\Delta _3+2 n_{IJ}-\ell '+\ell
   \right)\,,\\
   \la_{23}=&\frac{1}{2} \left(-n_{IJ}+\ell '-2\right)
   \left(-n_{IJ}+\ell '-1\right)\,,\\
   \ka_{25}=& -n_{IJ}^2+\frac{1}{2} n_{IJ} \left(-\Delta
   '-\Delta +\Delta _3+\ell '+\ell
   \right) \nonumber\\
   &+\frac{1}{2} \left(2
   (\Delta -1) \Delta _3+\ell
   ^2+\ell  \left(\Delta '+\Delta
   +\Delta _3-\ell '\right)\right)\,,\\
   \la_{25}=&\frac{1}{2} (-n_{IJ}+\ell -1)
   \left(n_{IJ}-\ell '+1\right)\,,\\
   \ka_{34}=&\frac{1}{2} \left(\Delta '-\Delta
   +\Delta _3-\ell '+\ell -2\right)
   \left(\Delta '+\Delta -\Delta
   _3+2 n_{IJ}-\ell '+\ell -2\right)\,,\\
   \la_{34}=&\frac{1}{2} \left(\ell '-n_{IJ}\right)
   \left(-\Delta '+\Delta -\Delta
   _3+\ell '-\ell +2\right)\,,
   \]
   \[
   \ka_{35}=&\frac{1}{2} \left(\ell '-n_{IJ}\right)
   \left(-\Delta '-\Delta +\Delta
   _3-2 n_{IJ}+\ell '-\ell +2\right)\,,\\
    \la_{35}=&\frac{1}{2} \left(-n_{IJ}+\ell '-1\right)
   \left(\ell '-n_{IJ}\right)\,,\\
   \ka_{36}=&-n_{IJ}^2+\frac{1}{2} n_{IJ} \left(-\Delta
   '-\Delta +\Delta _3+\ell '+\ell
   \right)\nonumber\\
   &+\frac{1}{2} \left(-\Delta
   '-\Delta 
   +\Delta_3(2 \Delta  -1)+\ell ^2+\ell 
   \left(\Delta '+\Delta +\Delta
   _3-\ell '-3\right)+\ell
   '+2\right)\,,\\
   \la_{36}=&\frac{1}{2} (-n_{IJ}+\ell -2)
   \left(n_{IJ}-\ell '\right)\,,\\
\rho_{22}=&
-2 n_{IJ}^3 \left(\Delta '-\Delta +\Delta _3-\ell '+\ell +2\right)+n_{IJ}^2 \left(\Delta '-\Delta +\Delta _3-\ell '+\ell +2\right) \left(d-\Delta '-\Delta
   +\Delta _3+\ell '+3 \ell +2\right)\nonumber\\
   &+n_{IJ}\bigg(d^2 \left(\Delta '-\Delta -\Delta _3-\ell '+\ell +2\right)
   -2 \big(\Delta ^2-\left(\Delta '+2\right)^2+\Delta _3 \left(\Delta _3+2\right)+\left(2
   \Delta _3+3\right) \ell ^2
\nonumber\\   
&-2 \ell  \left(-\Delta '+\Delta  \left(\Delta
   _3+2\right)+\ell '-2\right)+2 \left(\Delta '+2\right) \ell '
   - \ell'{}^2\big)
  \nonumber\\
  &+\frac{1}{2} d \big(-\left(\Delta '+2\right) \left(\Delta '+10\right)+\Delta _3^2+\Delta 
   (\Delta +8)+2 \Delta  \Delta _3+4 \Delta _3+3 \ell ^2\nonumber\\
   &-2 \ell  \left(-\Delta '+2 \Delta
   +4 \Delta _3+\ell '+2\right)+2 \left(\Delta '+6\right) \ell '- \ell'{}^2\big)
   \bigg)
\,,\\
\si_{22}=&
\frac{1}{2} n_{IJ}^2 \left(\Delta '-\Delta +\Delta _3-\ell '+\ell +2\right) \left(d-\Delta
   '-\Delta +\Delta _3+3 \ell '+7 \ell +2\right)-2 n_{IJ}^3 \left(\Delta '-\Delta +\Delta
   _3-\ell '+\ell +2\right)\nonumber\\
   &+\frac{1}{2} n_{IJ} \bigg(d^2 \left(\Delta '-\Delta -\Delta _3-\ell '+\ell +2\right)    -3 \Delta ^2-3 \Delta _3^2+2 \Delta
    \Delta _3-2 \ell ^3\nonumber\\
    &+d
   \left(-\left(\Delta '+2\right) \left(\Delta '+4\right)+\Delta  (\Delta +2)+\Delta _3
   \left(\Delta _3+4\right) \right.\nonumber\\
   &\qquad\left.+\left(\Delta '+\Delta -\Delta _3+4\right) \ell '-\ell
    \left(2 \Delta +4 \Delta _3+\ell ' - \ell+2\right)\right)\nonumber\\
&+2 \ell  \left(3 \Delta +\Delta _3+2 \left(\Delta  \Delta
   _3+\left(-\Delta '+\Delta -\Delta _3-2\right) \ell '+\ell'{}^2\right)-1\right)\nonumber\\
   &+3 \left(\ell '-\Delta '\right)^2+2 \left(5 \Delta '+\Delta
   -3 \Delta _3-5 \ell '+4\right)-\ell ^2 \left(2 \Delta '-2 \Delta +6 \Delta _3+2 \ell
   '+7\right)\bigg)\nonumber\\
   &-\frac{1}{2} (\ell +1) \bigg(d^2 \left(\Delta '-\Delta -\Delta _3-\ell '+\ell +2\right)   -2   \ell ^2 \left(\Delta +\Delta _3+\ell '\right)\nonumber\\
   &+d
   \big(-\left(\Delta '+2\right) \left(\Delta '+3\right)+\Delta  (\Delta +1)+\Delta _3
   \left(\Delta _3+5\right)+2 \ell ^2\nonumber\\
   &+\ell  \left(\Delta '-3 \Delta -3 \Delta _3-2 \ell
   '+1\right)+\left(\Delta '+\Delta -\Delta _3+3\right) \ell '\big)\nonumber\\
   &+\ell ^3+\ell 
   \left(-\Delta ' \left(\Delta '+2\right)+\Delta _3^2+6 \Delta _3+2 \Delta +2 \Delta _3
   \left(\Delta -\ell '\right)+\left(\Delta +\ell '\right)^2-2\right)\nonumber\\
   &-2
   \left(-\left(\Delta '+1\right) \left(\Delta '+2\right)+(\Delta -1) \Delta +\Delta _3
   \left(\Delta _3+3\right)+\ell '\left(\Delta '+\Delta -\Delta _3+1\right) 
   \right)
   \bigg)\,,
   \]
   \[
   \tau_{22}=&-\frac{1}{2} (\ell -n_{IJ}) (-n_{IJ}+\ell +1) \left(n_{IJ}-\ell '+1\right) \left(\Delta '-\Delta +\Delta
   _3-\ell '+\ell +2\right)\,,\\
     \rho_{23}=&n_{IJ} \bigg(n_{IJ}^2 \left(d-\Delta '-\Delta +\Delta _3+3 \ell
   +2\right)+ \ell '\left(\frac{d}{2}+n_{IJ}-2\right) \left(-2 d+\Delta '+\Delta -\Delta _3+3
   n_{IJ}-3 \ell +2\right)\nonumber\\
   &-\frac{1}{2}  \ell '{}^2(d+2 n_{IJ}-4) 
 +\frac{1}{2} (d-4) n_{IJ}
   \left(2 d-\Delta '-\Delta +\Delta _3+3 \ell -2\right) \nonumber\\
   &-\Delta _3 (d+2 \ell -2)
   (d-\Delta +\ell -1)-2 n_{IJ}^3\bigg)\,,\\
   \si_{23}=&\frac{1}{2} (-n_{IJ}+\ell +1) \left(-n_{IJ}+\ell '-1\right) \bigg(d^2+d \left(-\Delta '-\Delta
   +\Delta _3+n_{IJ}+2 \ell -3\right)+2 \left(\Delta '+\Delta -\Delta _3+1\right)\nonumber\\
   &-n_{IJ}
   \left(\Delta '+\Delta -\Delta _3+4 n_{IJ}-3 \ell '+2\right)-\ell  \left(\Delta '+\Delta
   -\Delta _3-3 n_{IJ}+\ell '+2\right)+\ell ^2\bigg)\,,\\
   \tau_{23}=&-\frac{1}{2} (\ell -n_{IJ}) (-n_{IJ}+\ell +1) \left(n_{IJ}-\ell '+1\right) \left(n_{IJ}-\ell '+2\right)\,,\\
\rho_{25}=&
n_{IJ} (-n_{IJ}+\ell +1) \bigg(\frac{1}{2} \big(2 d^2+d \left(-\Delta '-\Delta -3 \Delta _3+\ell
   '+3 \ell -10\right)\nonumber\\
   &+4 \left(\Delta '+\Delta _3 \Delta +\Delta -\ell '-\left(\Delta
   _3+3\right) \ell +2\right)\big)
   -2 n_{IJ}^2+n_{IJ} \left(d-\Delta '-\Delta +\Delta _3+\ell '+3 \ell
   +2\right)\bigg)\,,\\
\sigma_{25}=&-\frac{1}{2} (\ell -n_{IJ}) (-n_{IJ}+\ell +1) \bigg(d^2+n_{IJ} \left(d-\Delta '-\Delta +\Delta _3+3 \ell
   '+3 \ell -2\right)\nonumber\\
   &+d \left(-\Delta '-\Delta -\Delta _3+2 \ell -3\right)+2 \left(\Delta
   '+\Delta _3 \Delta +\Delta +1\right)-4 n_{IJ}^2+\ell ^2-\ell  \left(\Delta '+\Delta +\Delta
   _3+\ell '+2\right)\bigg)\,,\\
\tau_{25}=&-\frac{1}{2} (-n_{IJ}+\ell -1) (\ell -n_{IJ}) (-n_{IJ}+\ell +1) \left(n_{IJ}-\ell '+1\right)\,.
\]

\section{Recursion coefficients for 3-point structures}
\label{app:E}
The coefficients appearing in the recursion relations \Eq{equation Q1} - \Eq{equation Q2} for the 3-point structures $Q_{(\ell,1,\ell')}^{(i, n_{IJ})}$ are given by:
\[
 a_1=\,&\frac{(n_{IJ}-\ell )}{\left(\Delta _3-1\right) (\Delta +\ell -1) \left(\Delta
   '-\Delta +\Delta _3-\ell '+\ell -1\right)}\,,\\
   a_2=\,&\frac{(\ell -n_{IJ}) \left(\ell '-n_{IJ}\right)}{\left(\Delta _3-1\right) (\Delta
   +\ell -1) \left(\Delta '-\Delta +\Delta _3-\ell '+\ell -1\right)}\,,\\
   a_3=\,&-\frac{ \left(\frac{(\ell -n_{IJ}) \left(\Delta '+\Delta -\Delta _3+2 n_{IJ}-\ell
   '+\ell -1\right)}{\left(\Delta _3-1\right) (\Delta +\ell -1)}+2\right)}{\Delta
   '-\Delta +\Delta _3-\ell '+\ell -1}\,,
   \]
   \[
b_1 =\,&\frac{ (-n_{IJ}+\ell -1) (\ell -n_{IJ}) \left(-n_{IJ}+\ell '-1\right) \left(d-\Delta
   '-\Delta _3+\ell '\right)}{\left(\Delta _3-1\right) (n_{IJ}+1) (\Delta +\ell -1) (d-\Delta
   +\ell -1) \left(\Delta '-\Delta +\Delta _3-2 n_{IJ}+\ell '+\ell -1\right)}\,,\\
b_2 =\,&\frac{ \left(-\Delta '+\Delta -\Delta _3+\ell '-\ell
   +1\right)}{\left(\Delta _3-1\right) (n_{IJ}+1) (d+2 \ell -2) (d-\Delta +\ell -1)
   \left(\Delta '-\Delta +\Delta _3-2 n_{IJ}+\ell '+\ell -1\right)}\,,\\
b_3 =\,&\frac{2  (\ell -n_{IJ}) \left(d-\Delta '+\Delta -\Delta _3+\ell '+\ell
   -1\right)}{\left(\Delta _3-1\right) (d+2 \ell -2) (\Delta +\ell -1) \left(\Delta
   '-\Delta +\Delta _3-2 n_{IJ}+\ell '+\ell -1\right)}\,,\\
b_4=\,&     -\frac{ (-n_{IJ}+\ell -1) (\ell -n_{IJ}) \left(d-\Delta '+\Delta -\Delta _3+\ell
   '+\ell -1\right)}{\left(\Delta _3-1\right) (n_{IJ}+1) (d+2 \ell -2) (\Delta +\ell -1)
   \left(\Delta '-\Delta +\Delta _3-2 n_{IJ}+\ell '+\ell -1\right)}\,,\\
b_5=\,&  \frac{   \left(-\Delta '+\Delta +\Delta _3-2 n_{IJ}+\ell '+\ell -1\right)}{\left(\Delta _3-1\right) (\Delta +\ell -1)
   (d-\Delta +\ell -1) \left(\Delta '-\Delta +\Delta _3-2 n_{IJ}+\ell '+\ell -1\right)}\nonumber\\
&\times \bigg(2 d
   (\Delta +n_{IJ}-1)+\Delta '-\Delta  \left(\Delta '+\Delta +\Delta _3\right)+\Delta _3\nonumber\\
   &-2 n_{IJ}
   \left(\Delta '+\Delta _3\right)+(\Delta +2 n_{IJ}) \ell '+\ell  \left(\Delta '+\Delta
   _3-\ell '+\ell -2\right)-\ell '+1\bigg)\,,\\
b_6=\,&    -\frac{(\ell -n_{IJ})  }{\left(\Delta _3-1\right) (n_{IJ}+1) (\Delta +\ell -1)
   (d-\Delta +\ell -1) \left(\Delta '-\Delta +\Delta _3-2 n_{IJ}+\ell '+\ell -1\right)}\nonumber\\
    &\times 
    \bigg((d+2) \Delta ^2-3 d \Delta +4 n_{IJ}^2 \left(-d+\Delta '+\Delta _3-\ell '\right)\nonumber\\
    &+n_{IJ} \big(-d
   \Delta -\Delta ' \left(d+4 \ell '+2 \ell -6\right)+\Delta _3 \left(d-2 \ell '-2 \ell
   +6\right)+\ell ' \left(3 d+3 \ell '+2 \ell -6\right)\nonumber\\
   &+d (\ell -5)+\Delta
   ^2+ \Delta '{}^2-\Delta _3^2-(\ell -1)^2\big)+\Delta _3 \left(d \Delta
   +(\Delta +\ell -3) \ell '-2 \ell +2\right)\nonumber\\
   &+\ell ' \left(d (\Delta +2)-\Delta ^2+\ell
   ^2+\Delta ' (-\Delta +\ell -3)-1\right) +\Delta ' \left(-d \Delta +\Delta  \Delta '-2
   \ell +2\right)\nonumber\\
   &+\Delta  \ell  (d+\ell -2)-\Delta ^3-\Delta _3^2 \Delta +\Delta -(\ell
   -3) \ell '{}^2-2 (\ell -2) \ell -2\bigg)\,,\\
c_1=\,&\frac{\left(\Delta '+\Delta -\Delta _3+2 n_{IJ}-\ell '+\ell
   -1\right)}{\left(\Delta _3-1\right)
   (\Delta +\ell -1)}\,,\\
c_2=\,&\frac{\left(n_{IJ}-\ell '\right)}{\left(\Delta _3-1\right)
   (\Delta +\ell -1)}\,,\\
c_3=\,&\frac{1}{\left(\Delta _3-1\right)
   (\Delta +\ell -1)}\,,
\] 
The coefficients appearing in the recursion relations \Eq{Q21} - \Eq{Q26} for the 3-point structures $Q_{(\ell,2,\ell')}^{(a, n_{IJ})}$ are given by:
\[
a_1'=\,&-\frac{2 }{\Delta '-\Delta +\Delta _3-\ell '+\ell }\,,\\
a_2'=\,&\frac{ (-n_{IJ}+\ell -1) (\ell -n_{IJ})}{\Delta _3 (\Delta +\ell -1) \left(\Delta
   '-\Delta +\Delta _3-\ell '+\ell -2\right) \left(\Delta '-\Delta +\Delta _3-\ell '+\ell
   \right)}\,,\\
   a_3'=\,&\frac{(-n_{IJ}+\ell -1) (\ell -n_{IJ}) \left(n_{IJ}-\ell '\right)}{\Delta _3 (\Delta
   +\ell -1) \left(\Delta '-\Delta +\Delta _3-\ell '+\ell -2\right) \left(\Delta '-\Delta
   +\Delta _3-\ell '+\ell \right)}\,,\\ 
   a_4'=\,&\frac{2  (\ell -n_{IJ}) \left(n_{IJ}-\ell '\right)}{\Delta _3 (\Delta +\ell -1)
   \left(\Delta '-\Delta +\Delta _3-\ell '+\ell -2\right) \left(\Delta '-\Delta +\Delta
   _3-\ell '+\ell +2\right)}\,,\\
   a_5'=\,&\frac{2  (\ell -n_{IJ}) \left(n_{IJ}-\ell '\right) \left(n_{IJ}-\ell '+1\right)}{\Delta
   _3 (\Delta +\ell -1) \left(\Delta '-\Delta +\Delta _3-\ell '+\ell -2\right)
   \left(\Delta '-\Delta +\Delta _3-\ell '+\ell +2\right)}\,,\\
   a_6'=\,& \frac{ (\ell -n_{IJ}) \left(\Delta '-\Delta +\Delta _3-\ell '+\ell \right)}{\Delta _3 (\Delta +\ell -1) \left(\Delta '-\Delta +\Delta
   _3-\ell '+\ell -2\right) }\nonumber\\
   &\times \left(-2 n_{IJ}^2+n_{IJ} \left(-\Delta '-\Delta +\Delta _3+\ell
   '+\ell \right)+(\ell -1) \left(\Delta '+\Delta -\ell '+\ell -2\right)+\Delta _3 (2
   \Delta +\ell -1)\right)\,,\\
 a_7'=\,&   -\frac{1}{\Delta _3 (\Delta +\ell -1)
   \left(\Delta '-\Delta +\Delta _3-\ell '+\ell -2\right) \left(\Delta '-\Delta +\Delta
   _3-\ell '+\ell \right) \left(\Delta '-\Delta +\Delta _3-\ell '+\ell +2\right)} 
 \nonumber\\
 &\times \bigg(2  \left(\ell '-n_{IJ}\right) \big((\Delta -1) \Delta _3^2-2 n_{IJ}^2
   \left(\Delta '-\Delta +\Delta _3-\ell '+\ell \right)\nonumber\\
   &+n_{IJ} \left(-\Delta '-\Delta +\Delta
   _3+\ell '+\ell \right) \left(\Delta '-\Delta +\Delta _3-\ell '+\ell \right) +\ell 
   \left(\Delta '-\Delta -\ell '+\ell \right) \left(\Delta '+\Delta -\ell '+\ell
   \right)\nonumber\\
   &+\Delta _3 \left(-\Delta ^2-\Delta -(\Delta +\ell -1) \ell '+\Delta ' (\Delta
   +\ell -1)+2 \Delta  \ell +(\ell -3) \ell +2\right)\big)
   \bigg)\,,\\
b_1'=\,&\frac{ (n_{IJ}-\ell )}{\Delta _3 (\Delta +\ell -1) \left(\Delta '-\Delta +\Delta
   _3-\ell '+\ell +2\right)}\,,\\
   b_2'=\,&\frac{ (\ell -n_{IJ}) \left(-n_{IJ}+\ell '-1\right)}{\Delta _3 (\Delta +\ell -1)
   \left(\Delta '-\Delta +\Delta _3-\ell '+\ell +2\right)}\,,\\
   b_3'=\,&-\frac{ \left(2 (\Delta -1) \Delta _3-2 n_{IJ}^2+n_{IJ} \left(-\Delta '-\Delta +\Delta
   _3+\ell '+\ell \right)+\ell ^2+\ell  \left(\Delta '+\Delta +\Delta _3-\ell
   '\right)\right)}{\Delta _3 (\Delta +\ell -1) \left(\Delta '-\Delta +\Delta _3-\ell
   '+\ell +2\right)}\,,
\]
 \[
 c_1'=\,&\frac{(-n_{IJ}+\ell -1) (\ell -n_{IJ}) \left(-n_{IJ}+\ell '-2\right) \left(d-\Delta
   '-\Delta _3+\ell '-3\right)}{\Delta _3 (n_{IJ}+1) (\Delta +\ell -1) (d-\Delta +\ell -1)
   \left(\Delta '-\Delta +\Delta _3-2 n_{IJ}+\ell '+\ell \right)}\,,\\
   c_2'=\,&\frac{2 (\ell -n_{IJ}) \left(d-\Delta '+\Delta -\Delta _3+\ell '+\ell
   -4\right)}{\Delta _3 (d+2 \ell -2) (\Delta +\ell -1) \left(\Delta '-\Delta +\Delta
   _3-2 n_{IJ}+\ell '+\ell \right)}\,,\\ 
   c_3'=\,&-\frac{(-n_{IJ}+\ell -1) (\ell -n_{IJ}) \left(d-\Delta '+\Delta -\Delta _3+\ell
   '+\ell -4\right)}{\Delta _3 (n_{IJ}+1) (d+2 \ell -2) (\Delta +\ell -1) \left(\Delta
   '-\Delta +\Delta _3-2 n_{IJ}+\ell '+\ell \right)}\,,\\ 
   c_4'=\,& -\frac{ \left(\Delta '-\Delta +\Delta _3-\ell '+\ell +2\right)}{\Delta _3
   (n_{IJ}+1) (d+2 \ell -2) (d-\Delta +\ell -1) \left(\Delta '-\Delta +\Delta _3-2 n_{IJ}+\ell
   '+\ell \right)}\,,\\
   c_5'=\,&\frac{1}{\Delta _3 (\Delta
   +\ell -1) (d-\Delta +\ell -1) \left(\Delta '-\Delta +\Delta _3-2 n_{IJ}+\ell '+\ell
   \right)}\nonumber\\
   &\times \bigg(
   \left(-\Delta '+\Delta +\Delta _3-2 n_{IJ}+\ell '+\ell -2\right) \big(2 d
   (\Delta +n_{IJ}-1)+\Delta '-\Delta  \left(\Delta '+\Delta +\Delta _3\right)\nonumber\\
   &-3 \Delta
   +\Delta _3-2 n_{IJ} \left(\Delta '+\Delta _3+3\right)+(\Delta +2 n_{IJ}) \ell '+\ell 
   \left(\Delta '+\Delta _3-\ell '+\ell +1\right)-\ell '+4\big)
   \bigg)\,,\\ 
   c_6'=\,& -\frac{(\ell -n_{IJ})}{\Delta _3 (n_{IJ}+1)
   (\Delta +\ell -1) (d-\Delta +\ell -1) \left(\Delta '-\Delta +\Delta _3-2 n_{IJ}+\ell '+\ell
   \right)}\nonumber\\
&\times  \bigg((d+3) \Delta ^2-4 n_{IJ}^2 \left(d-\Delta '-\Delta _3+\ell
   '-3\right) +n_{IJ} \big(-\Delta ' \left(d+4 \ell '+2 \ell \right)+\Delta _3 \left(d-2 \ell
   '-2 \ell +6\right)\nonumber\\
   &+\ell ' \left(3 d+3 \ell '+2 \ell -18\right)+d (-\Delta +\ell
   -8)+\Delta ^2+ \Delta '(\Delta '+12)-\Delta _3^2-\ell  (\ell
   +4)+26\big)\nonumber\\
   &+\ell ' \left(d (\Delta +2)-\Delta ^2+\Delta ' (-\Delta +\ell -3)+\ell 
   (\ell +4)-13\right)\nonumber\\
   &+\Delta _3 \left((d-3) \Delta +(\Delta +\ell -3) \ell '-3 \ell
   +5\right) +\Delta ' \left( \Delta (\Delta '-d+3 ) -3 \ell +5\right)\nonumber\\
   &+\Delta
    (d (\ell -4)+(\ell -2) \ell )-2 d
    +\Delta(-\Delta ^2  -\Delta _3^2  +1) -\ell '{}^2(\ell -3)
 -3 \ell  (\ell +1)+12\bigg) \,,\\
d_1'=\,&\frac{(n_{IJ}-\ell +1)}{\Delta _3 (\Delta +\ell -1) \left(\Delta '-\Delta
   +\Delta _3-\ell '+\ell -2\right)}\,,\\
d_2'=\,&   \frac{ (-n_{IJ}+\ell -1) \left(\ell '-n_{IJ}\right)}{\Delta _3 (\Delta +\ell -1)
   \left(\Delta '-\Delta +\Delta _3-\ell '+\ell -2\right)}\,,\\
d_3'=\,&\frac{ \left(\ell '-n_{IJ}\right)}{\Delta _3 (\Delta +\ell -1) \left(\Delta
   '-\Delta +\Delta _3-\ell '+\ell -2\right)}\,,\\
d_4'=\,&    -\frac{\left(-n_{IJ}+\ell '-1\right) \left(\ell '-n_{IJ}\right)}{\Delta _3 (\Delta
   +\ell -1) \left(\Delta '-\Delta +\Delta _3-\ell '+\ell -2\right)}\,,\\
d_5'=\,&\frac{ \left(\ell '-n_{IJ}\right) \left(\Delta '+\Delta -\Delta _3+2 n_{IJ}-\ell
   '+\ell \right)}{\Delta _3 (\Delta +\ell -1) \left(\Delta '-\Delta +\Delta _3-\ell
   '+\ell -2\right)}\,,\\
d_6'=\,&\frac{ \left(-\frac{2 (\Delta +n_{IJ}) \left(\Delta _3-n_{IJ}+\ell -1\right)}{\Delta
   '-\Delta +\Delta _3-\ell '+\ell -2}+n_{IJ}-\ell +1\right)}{\Delta _3 (\Delta +\ell -1)} \,,   
\]  
\[
e_1'=\,&\frac{ \left(\Delta '+\Delta -\Delta _3+2 n_{IJ}-\ell '+\ell \right)}{\Delta _3
   (\Delta +\ell -1)}\,,\\
   e_2'=\,&\frac{ \left(n_{IJ}-\ell '+1\right)}{\Delta _3 (\Delta
   +\ell -1)}\,,\\
      e_3'=\,&\frac{1}{\Delta _3 (\Delta +\ell -1)}\,,\\
f_1'=\,&\frac{ \left(\Delta '+\Delta -\Delta _3+2 n_{IJ}-\ell '+\ell -2\right)}{\Delta _3
   (\Delta +\ell -1)}\,,\\ 
f_2'=\,&   \frac{ \left(n_{IJ}-\ell '\right)}{\Delta _3 (\Delta
   +\ell -1)}\,,\\ 
f_3'=\,&\frac{1}{\Delta _3 (\Delta +\ell -1)}\,.
\]

\section{Conformal block coefficients}
\label{app:F}
The coefficients appearing in the vector conformal block recursion relations \Eq{block vector 1} - \Eq{block vector 2} are given by:
\[
\mathscr{A}_{ (+0)(0+)}^{(1)(n_{IJ})}=\,&a_1 C^{(0+)(0-)} F^{(0+)}_{(+0)}+a_3 C^{(-0)(+0)} F^{(+0)}_{(+0)}\,,
\nonumber\\
\mathscr{A}_{ (+0)(0+)}^{(1)(n_{IJ}+1)}=\,&a_2 C^{(-0)(+0)}F^{(+0)}_{(+0)}\,, \nonumber\\
\mathscr{A}_{ (+0)(0+)}^{(2)(n_{IJ})} =\,& a_3 C^{(-0)(+0)} F^{(+0)}_{(+0)}\,, \nonumber\\
\mathscr{A}_{ (+0)(0+)}^{(2)(n_{IJ}+1)} =\,&a_2 C^{(-0)(+0)} F^{(+0)}_{(+0)}\,, \nonumber\\
\mathscr{A}_{ (-0)(0+)}^{(1)(n_{IJ})} =\,&a_1 C^{(0+)(0-)} F^{(0+)}_{(-0)} 
+a_3 C^{(-0)(+0)} F^{(+0)}_{(-0)}\,, \nonumber\\
\mathscr{A}_{ (-0)(0+)}^{(1)(n_{IJ}+1)} =\,&a_2 C^{(-0)(+0)} F^{(+0)}_{(-0)}\,, \nonumber\\
\mathscr{A}_{ (-0)(0+)}^{(2)(n_{IJ})}=\,&a_3 C^{(-0)(+0)} F^{(+0)}_{(-0)}
-a_1 C^{(0+)(0-)} F^{(0+)}_{(-0)}\,,\nonumber\\
\mathscr{A}_{ (-0)(0+)}^{(2)(n_{IJ}+1)}=\,&a_2 C^{(-0)(+0)} F^{(+0)}_{(-0)}\,,
\]
\[
\mathscr{B}_{(+0)(0+)}^{(1)(n_{IJ})}=\,&b_3 C^{(0+)(0-)} F^{(0+)}_{(+0)}+b_5 C^{(-0)(+0)} F^{(+0)}_{(+0)}\,,
\nonumber\\
 \mathscr{B}_{(+0)(0+)}^{(1)(n_{IJ}+1)}=\,&
b_2 C^{(0-)(0+)} F^{(0-)}_{(+0)}+b_4 C^{(0+)(0-)} 
F^{(0+)}_{(+0)}+b_6 C^{(-0)(+0)}F^{(+0)}_{(+0)}\,, 
\nonumber\\
\mathscr{B}_{ (+0)(0+)}^{(1)(n_{IJ}+2)}=\,&b_1 C^{(-0)(+0)}F^{(+0)}_{(+0)}\,, \nonumber\\
\mathscr{B}_{ (+0)(0+)}^{(2)(n_{IJ})} =\,& b_5 C^{(-0)(+0)} F^{(+0)}_{(+0)}\,, \nonumber\\
\mathscr{B}_{ (+0)(0+)}^{(2)(n_{IJ}+1)} =\,&b_6 C^{(-0)(+0)}F^{(+0)}_{(+0)}\,, \nonumber\\
\mathscr{B}_{ (+0)(0+)}^{(2)(n_{IJ}+2)} =\,&b_1 C^{(-0)(+0)}F^{(+0)}_{(+0)}\,, 
\nonumber\\
\mathscr{B}_{ (-0)(0+)}^{(1)(n_{IJ})} =\,&b_3 C^{(0+)(0-)} F^{(0+)}_{(-0)} +b_5 C^{(-0)(+0)} F^{(+0)}_{(-0)}\,, \nonumber\\
\mathscr{B}_{ (-0)(0+)}^{(1)(n_{IJ}+1)} =\,&b_2 C^{(0-)(0+)} F^{(0-)}_{(-0)}+b_4 C^{(0+)(0-)} 
F^{(0+)}_{(-0)}+b_6 C^{(-0)(+0)} F^{(+0)}_{(-0)}\,,\nonumber\\
\mathscr{B}_{ (-0)(0+)}^{(1)(n_{IJ}+2)} =\,&b_1 C^{(-0)(+0)}F^{(+0)}_{(-0)}\,,\nonumber\\
\mathscr{B}_{ (-0)(0+)}^{(2)(n_{IJ})} =\,&-b_3 C^{(0+)(0-)} F^{(0+)}_{(-0)}+b_5 C^{(-0)(+0)} F^{(+0)}_{(-0)}\,, \nonumber\\
\mathscr{B}_{ (-0)(0+)}^{(2)(n_{IJ}+1)}=\,&-b_2 C^{(0-)(0+)} F^{(0-)}_{(-0)}-b_4 C^{(0+)(0-)} 
F^{(0+)}_{(-0)}+b_6 C^{(-0)(+0)} F^{(+0)}_{(-0)}\,,\nonumber\\
\mathscr{B}_{ (-0)(0+)}^{(2)(n_{IJ}+2)}=\,&b_1 C^{(-0)(+0)} F^{(+0)}_{(-0)}\,,\\\nonumber\\
\mathscr{C}_{ (+0)(0+)}^{(1)(n_{IJ})}=\,&c_3 C^{(0+)(0-)} F^{(0+)}_{(+0)}+c_1 C^{(-0)(+0)} F^{(+0)}_{(+0)}\,,
\nonumber\\
\mathscr{C}_{ (+0)(0+)}^{(1)(n_{IJ}+1)}=\,&c_2 C^{(-0)(+0)}F^{(+0)}_{(+0)}\,, \nonumber\\
\mathscr{C}_{ (+0)(0+)}^{(2)(n_{IJ})} =\,& c_1 C^{(-0)(+0)} F^{(+0)}_{(+0)}\,, \nonumber\\
\mathscr{C}_{ (+0)(0+)}^{(2)(n_{IJ}+1)} =\,&c_2 C^{(-0)(+0)} F^{(+0)}_{(+0)}\,, \nonumber\\
\mathscr{C}_{ (-0)(0+)}^{(1)(n_{IJ})} =\,&c_3 C^{(0+)(0-)} F^{(0+)}_{(-0)} 
+c_1 C^{(-0)(+0)} F^{(+0)(-0)}\,, \nonumber\\
\mathscr{C}_{ (-0)(0+)}^{(1)(n_{IJ}+1)} =\,&c_2 C^{(-0)(+0)} F^{(+0)}_{(-0)}\,, \nonumber\\
\mathscr{C}_{ (-0)(0+)}^{(2)(n_{IJ})}=\,&c_1 C^{(-0)(+0)} F^{(+0)}_{(-0)}
-c_3 C^{(0+)(0-)} F^{(0+)}_{(-0)}\,,\nonumber\\
\mathscr{C}_{ (-0)(0+)}^{(2)(n_{IJ}+1)}=\,&c_2 C^{(-0)(+0)} F^{(+0)}_{(-0)}\,.
\]
The coefficients appearing in the tensor conformal block recursion relations \Eq{block tensor 1} - \Eq{block tensor 6} are given by:
\[
\mathscr{A}' {}_{(+0)(0+)}^{(1)(1, n_{IJ})}=\,&
a_1' C^{(-0)(+0)} F^{(+0)}_{(+0)}\,,\nonumber\\
\mathscr{A}'{}_{ (+0)(0+)}^{(2)(1, n_{IJ})} =\,&a_1' C^{(-0)(+0)} F^{(+0)}_{(+0)}\,,\nonumber\\
\mathscr{A}' {}_{ (-0)(0+)}^{(1)(1, n_{IJ})}=\,&a_1' C^{(-0)(+0)} F^{(+0)}_{(-0)}\,,\nonumber\\
\mathscr{A}' {}_{ (-0)(0+)}^{(2)(1, n_{IJ})}=\,&a_1' C^{(-0)(+0)} F^{(+0)}_{(-0)}\,,\nonumber\\
\mathscr{A}' {}_{(+0)(0+)}^{(1)(2, n_{IJ})}=\,&
a_4' C^{(0+)(0-)}F^{(0+)}_{(+0)}+a_7'
  C^{(-0)(+0)} F^{(+0)}_{(+0)}\,,
   \nonumber\\
\mathscr{A}' {}_{(+0)(0+)}^{(1)(2, n_{IJ}+1)}=\,&a_5' C^{(-0)(+0)}F^{(+0)}_{(+0)}\,,\nonumber\\
\mathscr{A}' {}_{(+0)(0+)}^{(1)(3, n_{IJ})}=\,&a_2' C^{(0+)(0-)}F^{(0+)}_{(+0)}
+a_6'   C^{(-0)(+0)} F^{(+0)}_{(+0)}\,,\nonumber\\
\mathscr{A}' {}_{(+0)(0+)}^{(1)(3, n_{IJ}+1)}=\,&a_3' C^{(-0)(+0)} F^{(+0)}_{(+0)}\,,\nonumber\\
\mathscr{A}'{}_{ (+0)(0+)}^{(2)(2, n_{IJ})}=\,& a_7' C^{(-0)(+0)} F^{(+0)}_{(+0)}\,,\nonumber\\
\mathscr{A}'{}_{ (+0)(0+)}^{(2)(2, n_{IJ}+1)}=\,&  a_5' C^{(-0)(+0)} F^{(+0)}_{(+0)}\,,\nonumber\\
\mathscr{A}'{}_{ (+0)(0+)}^{(2)(3, n_{IJ})}=\,& a_6' C^{(-0)(+0)} F^{(+0)}_{(+0)}\,,\nonumber\\
\mathscr{A}'{}_{ (+0)(0+)}^{(2)(3, n_{IJ}+1)}=\,&  a_3' C^{(-0)(+0)} F^{(+0)}_{(+0)}\,,\nonumber\\
\mathscr{A}' {}_{ (-0)(0+)}^{(1)(2, n_{IJ})}=\,&a_4' C^{(0+)(0-)}
F^{(0+)}_{(-0)}+a_7'
C^{(-0)(+0)} F^{(+0)}_{(-0)}\,,\nonumber\\
\mathscr{A}' {}_{ (-0)(0+)}^{(1)(2, n_{IJ}+1)}=\,&a_5' C^{(-0)(+0)}F^{(+0)}_{(-0)}\,,\nonumber\\
\mathscr{A}' {}_{ (-0)(0+)}^{(1)(3, n_{IJ})}=\,&a_2' C^{(0+)(0-)}F^{(0+)}_{(-0)}+a_6'
C^{(-0)(+0)} F^{(+0)}_{(-0)}\,,\nonumber\\
\mathscr{A}' {}_{ (-0)(0+)}^{(1)(3, n_{IJ}+1)}=\,&a_3' C^{(-0)(+0)} F^{(+0)}_{(-0)}\,,\nonumber\\
\mathscr{A}' {}_{ (-0)(0+)}^{(2)(2, n_{IJ})}=\,&a_7' C^{(-0)(+0)}F^{(+0)}_{(-0)}-a_4'
C^{(0+)(0-)} F^{(0+)}_{(-0)}\,,\nonumber\\
\mathscr{A}' {}_{ (-0)(0+)}^{(2)(2, n_{IJ}+1)} 
=\,&a_5' C^{(-0)(+0)} F^{(+0)}_{(-0)}\,,\nonumber\\
\mathscr{A}' {}_{ (-0)(0+)}^{(2)(3, n_{IJ})} =\,&a_6' C^{(-0)(+0)}F^{(+0)}_{(-0)}-a_2'
 C^{(0+)(0-)} F^{(0+)}_{(-0)}\,,\nonumber\\
\mathscr{A}' {}_{ (-0)(0+)}^{(2)(3, n_{IJ}+1)}=\,&a_3' C^{(-0)(+0)} F^{(+0)}_{(-0)} \,,
\]
\[
\mathscr{B}' {}_{ (+0)(0+)}^{(1)(2, n_{IJ})}=\,&b_1' C^{(0+)(0-)} F^{(0+)}_{(+0)}+b_3' C^{(-0)(+0)} F^{(+0)}_{(+0)}\,,
\nonumber\\
\mathscr{B}' {}_{ (+0)(0+)}^{(1)(2, n_{IJ}+1)}=\,&b_2' C^{(-0)(+0)} F^{(+0)}_{(+0)}\,,
\nonumber\\
\mathscr{B}' {}_{ (+0)(0+)}^{(2)(2, n_{IJ})}=\,&b_3' C^{(-0)(+0)} F^{(+0)}_{(+0)}\,,
\nonumber\\
\mathscr{B}' {}_{ (+0)(0+)}^{(2)(2, n_{IJ}+1)}=\,&b_2' C^{(-0)(+0)} F^{(+0)}_{(+0)}\,,\nonumber\\
\mathscr{B}' {}_{ (-0)(0+)}^{(1)(2, n_{IJ})}=\,&b_1' C^{(0+)(0-)} F^{(0+)}_{(-0)}+b_3' C^{(-0)(+0)} F^{(+0)}_{(-0)}\,,\nonumber\\
\mathscr{B}' {}_{ (-0)(0+)}^{(1)(2, n_{IJ}+1)}=\,&b_2' C^{(-0)(+0)} F^{(+0)(-0)}\,,\nonumber\\
\mathscr{B}' {}_{ (-0)(0+)}^{(2)(2, n_{IJ})}=\,&b_3' C^{(-0)(+0)} F^{(+0)}_{(-0)}-b_1' C^{(0+)(0-)} F^{(0+)}_{(-0)}\,,\nonumber\\
\mathscr{B}' {}_{ (-0)(0+)}^{(2)(2, n_{IJ}+1)}=\,&b_2' C^{(-0)(+0)} F^{(+0)}_{(-0)}\,,\\\nonumber\\
\mathscr{C}' {}_{ (+0)(0+)}^{(1)(2, n_{IJ})}=\,&c_2'C^{(0+)(0-)}  F^{(0+)}_{(+0)}+c_5' C^{(-0)(+0)}F^{(+0)}_{(+0)}\,,\nonumber\\
\mathscr{C}' {}_{(+0)(0+)}^{(1)(2, n_{IJ}+1)}=\,&c_4' C^{(0-)(0+)} F^{(0-)}_{(+0)}+c_3' C^{(0+)(0-)} 
 F^{(0+)}_{(+0)}+c_6' C^{(-0)(+0)} F^{(+0)}_{(+0)}\,,
\nonumber\\
\mathscr{C}' {}_{ (+0)(0+)}^{(1)(2, n_{IJ}+2)}=\,&c_1' C^{(-0)(+0)} F^{(+0)}_{(+0)}\,,
\nonumber\\
\mathscr{C}' {}_{ (+0)(0+)}^{(2)(2, n_{IJ})}=\,& c_5' C^{(-0)(+0)} F^{(+0)}_{(+0)}\,,
\nonumber\\
\mathscr{C}' {}_{ (+0)(0+)}^{(2)(2, n_{IJ}+1)}=\,& c_6' C^{(-0)(+0)} F^{(+0)}_{(+0)}\,,\nonumber\\
\mathscr{C}' {}_{ (+0)(0+)}^{(2)(2, n_{IJ}+2)}=\,& c_1' C^{(-0)(+0)} F^{(+0)}_{(+0)}\,,\nonumber\\
\mathscr{C}' {}_{ (-0)(0+)}^{(1)(2, n_{IJ})}=\,&c_2' C^{(0+)(0-)} F^{(0+)}_{(-0)}+c_5' C^{(-0)(+0)} F^{(+0)}_{(-0)}\,,\nonumber\\
\mathscr{C}' {}_{ (-0)(0+)}^{(1)(2, n_{IJ}+1)}=\,&c_4' C^{(0-)(0+)}  F^{(0-)}_{(-0)}+c_3'C^{(0+)(0-)} 
  F^{(0+)}_{(-0)}+c_6' C^{(-0)(+0)} F^{(+0)}_{(-0)}\,,\nonumber\\
\mathscr{C}' {}_{ (-0)(0+)}^{(1)(2, n_{IJ}+2)}=\,&c_1' C^{(-0)(+0)} F^{(+0)}_{(-0)}\,,\nonumber\\
\mathscr{C}' {}_{ (-0)(0+)}^{(2)(2, n_{IJ})}=\,&c_5' C^{(-0)(+0)} F^{(+0)}_{(-0)}-c_2'C^{(0+)(0-)}  F^{(0+)}_{(-0)}\,,\nonumber\\
\mathscr{C}' {}_{ (-0)(0+)}^{(2)(2, n_{IJ}+1)}=\,&-c_4' 
C^{(0-)(0+)} F^{(0-)}_{(-0)}-c_3' C^{(0+)(0-)} F^{(0+)}_{(-0)}+c_6' C^{(-0)(+0)} F^{(+0)}_{(-0)}\,,\nonumber\\
\mathscr{C}' {}_{ (-0)(0+)}^{(2)(2, n_{IJ}+2)}=\,&c_1' C^{(-0)(+0)}F^{(+0)}_{(-0)} 
\]
\[
\mathscr{D}' {}_{ (+0)(0+)}^{(1)(2, n_{IJ})}=\,&d_3'  C^{(0+)(0-)} F^{(0+)(+0)}+d_5'C^{(-0)(+0)}  F^{(+0)(+0)}\,,
\nonumber\\
\mathscr{D}' {}_{ (+0)(0+)}^{(1)(2, n_{IJ}+1)}=\,& d_4' C^{(0-)(0+)} F^{(0-)}_{(+0)}\,,
\nonumber\\
\mathscr{D}' {}_{ (+0)(0+)}^{(2)(2, n_{IJ})}=\,& d_5' C^{(-0)(+0)} F^{(+0)}_{(+0)}\,,
\nonumber\\
\mathscr{D}' {}_{ (+0)(0+)}^{(2)(2, n_{IJ}+1)}=\,&0,\nonumber\\
\mathscr{D}' {}_{ (-0)(0+)}^{(1)(2, n_{IJ})}=\,& d_3' C^{(0+)(0-)} F^{(0+)}_{(-0)}+d_5'
C^{(-0)(+0)}  F^{(+0)}_{(-0)}\,,\nonumber\\
\mathscr{D}' {}_{ (-0)(0+)}^{(1)(2, n_{IJ}+1)}=\,&d_4' C^{(0-)(0+)} F^{(0-)}_{(-0)}\,,\nonumber\\
\mathscr{D}' {}_{ (-0)(0+)}^{(2)(2, n_{IJ})}=\,& d_5' C^{(-0)(+0)}F^{(+0)}_{(-0)}-d_3' C^{(0+)(0-)} F^{(0+)}_{(-0)}\,,\nonumber\\
\mathscr{D}' {}_{ (-0)(0+)}^{(2)(2, n_{IJ}+1)}=\,&-d_4' C^{(0-)(0+)} F^{(0-)}_{(-0)}\,,\nonumber\\
\mathscr{D}' {}_{ (+0)(0+)}^{(1)(3, n_{IJ})}=\,&d_1' C^{(0+)(0-)}  F^{(0+)}_{(+0)}+d_6' C^{(-0)(+0)} F^{(+0)}_{(+0)}\,,
\nonumber\\
\mathscr{D}' {}_{ (+0)(0+)}^{(1)(3, n_{IJ}+1)}=\,&d_2' C^{(-0)(+0)} F^{(+0)}_{(+0)}\,,
\nonumber\\
\mathscr{D}' {}_{ (+0)(0+)}^{(2)(3, n_{IJ})}=\,& d_6' C^{(-0)(+0)}F^{(+0)}_{(+0)} ,
\nonumber\\
\mathscr{D}' {}_{ (+0)(0+)}^{(2)(3, n_{IJ}+1)}=\,&d_2' C^{(-0)(+0)}  F^{(+0)}_{(+0)} ,\nonumber\\
\mathscr{D}' {}_{ (-0)(0+)}^{(1)(3, n_{IJ})}=\,&d_1'  C^{(0+)(0-)} F^{(0+)}_{(-0)}+d_6' C^{(-0)(+0)} F^{(+0)}_{(-0)}\,,\nonumber\\
\mathscr{D}' {}_{ (-0)(0+)}^{(1)(3, n_{IJ}+1)}=\,&d_2' C^{(-0)(+0)} F^{(+0)}_{(-0)}\,,\nonumber\\
\mathscr{D}' {}_{ (-0)(0+)}^{(2)(3, n_{IJ})}=\,& d_6' C^{(-0)(+0)}F^{(+0)}_{(-0)}-d_1' C^{(0+)(0-)}  F^{(0+)}_{(-0)}\,,\nonumber\\
\mathscr{D}' {}_{ (-0)(0+)}^{(2)(3, n_{IJ}+1)}=\,&d_2' C^{(-0)(+0)} F^{(+0)(-0)}\,,\\\nonumber\\
\mathscr{E}' {}_{ (+0)(0+)}^{(1)(2, n_{IJ})}=\,&e_3'
C^{(0+)(0-)}F^{(0+)}_{(+0)}+ e_1' 
C^{(-0)(+0)}F^{(+0)}_{(+0)}\,,
\nonumber\\
\mathscr{E}' {}_{ (+0)(0+)}^{(1)(2, n_{IJ}+1)}=\,&e_2' 
C^{(-0)(+0)} F^{(+0)}_{(+0)}\,,
\nonumber\\
\mathscr{E}' {}_{ (+0)(0+)}^{(2)(2, n_{IJ})}=\,&  e_1' C^{(-0)(+0)}F^{(+0)}_{(+0)}\,,
\nonumber\\
\mathscr{E}' {}_{ (+0)(0+)}^{(2)(2, n_{IJ}+1)}=\,& e_2'
C^{(-0)(+0)} F^{(+0)}_{(+0)}\,,\nonumber\\
\mathscr{E}' {}_{ (-0)(0+)}^{(1)(2, n_{IJ})}=\,&
e_3' C^{(0+)(0-)} F^{(0+)}_{(-0)}+ e_1' C^{(-0)(+0)}F^{(+0)}_{(-0)}\,,\nonumber\\
\mathscr{E}' {}_{ (-0)(0+)}^{(1)(2, n_{IJ}+1)}=\,&e_2' C^{(-0)(+0)} F^{(+0)}_{(-0)}\,,\nonumber\\
\mathscr{E}' {}_{ (-0)(0+)}^{(2)(2, n_{IJ})}=\,&e_1' C^{(-0)(+0)} F^{(+0)}_{(-0)}-e_3' C^{(0+)(0-)} F^{(0+)}_{(-0)}\,,\nonumber\\
\mathscr{E}' {}_{ (-0)(0+)}^{(2)(2, n_{IJ}+1)}=\,&e_2' C^{(-0)(+0)} F^{(+0)}_{(-0)}\,,
\]
\[
\mathscr{F}' {}_{ (+0)(0+)}^{(1)(3, n_{IJ})}=\,& f_3' C^{(0+)(0-)} F^{(0+)}_{(+0)}+f_1'  C^{(-0)(+0)} F^{(+0)}_{(+0)}\,,
\nonumber\\
\mathscr{F}' {}_{ (+0)(0+)}^{(1)(3, n_{IJ}+1)}=\,& f_2' C^{(-0)(+0)}F^{(+0)}_{(+0)}\,,
\nonumber\\
\mathscr{F}' {}_{ (+0)(0+)}^{(2)(3, n_{IJ})}=\,&  f_1' C^{(-0)(+0)} F^{(+0)}_{(+0)}\,,
\nonumber\\
\mathscr{F}' {}_{ (+0)(0+)}^{(2)(3, n_{IJ}+1)}=\,& f_2'
C^{(-0)(+0)}F^{(+0)}_{(+0)}\,,\nonumber\\
\mathscr{F}' {}_{ (-0)(0+)}^{(1)(3, n_{IJ})}=\,& f_3' C^{(0+)(0-)} F^{(0+)}_{(-0)}+f_1' C^{(-0)(+0)} F^{(+0)}_{(-0)}\,,\nonumber\\
\mathscr{F}' {}_{ (-0)(0+)}^{(1)(3, n_{IJ}+1)}=\,&f_2' C^{(-0)(+0)} F^{(+0)}_{(-0)}\,,\nonumber\\
\mathscr{F}' {}_{ (-0)(0+)}^{(2)(3, n_{IJ})}=\,& f_1' C^{(-0)(+0)}F^{(+0)}_{(-0)}-f_3' C^{(0+)(0-)} F^{(0+)}_{(-0)}\,,\nonumber\\
\mathscr{F}' {}_{ (-0)(0+)}^{(2)(3, n_{IJ}+1)}=\,& f_2' C^{(-0)(+0)} F^{(+0)}_{(-0)}\,.
\]

\bibliographystyle{jhep}
\bibliography{Biblio}

\end{document}